\documentclass[preprint,letterpaper,sort&compress,12pt]{elsarticle}

\usepackage[latin9]{inputenc}

\usepackage[latin9]{inputenc}
\usepackage{amsmath}
\usepackage{amssymb}
\usepackage{graphicx}
\usepackage{verbatim}
\usepackage{bm}
\usepackage{mathbbol}
\usepackage{color}
\setcitestyle{compress}

\usepackage{mathtools}

\usepackage{graphicx,epsfig,amsfonts,amssymb}
\usepackage{bm}
\usepackage{times}
\usepackage{lipsum}
\usepackage{verbatim}

\usepackage{multirow}

\usepackage{tikz}

\usepackage[caption=false]{subfig}
\usepackage{wrapfig}

\usepackage[linktocpage=true]{hyperref}
\hypersetup{
    colorlinks,
    citecolor=blue,
    filecolor=blue,
    linkcolor=blue,
    urlcolor=blue
}

\usepackage{bm} 
\usepackage{color}
\usepackage{stackrel}
\usepackage{accents}

\usepackage{latexsym}

\usepackage{mathtools}

\newcommand{\beg}{\begin{equation}}
\newcommand{\en}{\end{equation}}
\newcommand{\begs}{\begin{subequations}}
\newcommand{\ens}{\end{subequations}}
\newcommand \bea {\begin{eqnarray}}
\newcommand \eea {\end{eqnarray}}
\newcommand{\bem}{\begin{bmatrix}}
\newcommand{\enm}{\end{bmatrix}}
\newcommand{\bpm}{\begin{pmatrix}}
\newcommand{\epm}{\end{pmatrix}}
\newcommand{\bvm}{\begin{vmatrix}}
\newcommand{\evm}{\end{vmatrix}}
\newcommand{\ba}{\begin{array}}
\newcommand{\ea}{\end{array}}

\def\t{\tau}
\newcommand{\by}{\times}

\def\ra{\rightarrow}

\def\mean#1{\langle #1 \rangle}

\newcommand{\re}[1]{(\ref{#1})}

\newcommand{\eref}[1]{Eq.~(\ref{#1})}
\newcommand{\fref}[1]{Fig.~\ref{#1}}
\newcommand{\fsref}[1]{Figs.~\ref{#1}}

\newcommand{\Rref}[1]{Ref.~\citealp{#1}}

\newcommand{\sref}[1]{Sect.~\ref{#1}}

\newcommand{\esref}[1]{Eqs.~(\ref{#1})}

\newcommand{\tref}[1]{Table~\ref{#1}}

\newcommand{\Tr}{\mathrm{Tr}\,}
\def\Z2{\mathbb{Z}_{2}}

\def\R{\mathbb{R}}

\newcommand{\blue}[1]{{\color{blue}{#1}}}

\makeatletter
\renewcommand*\env@matrix[1][\arraystretch]{%
  \edef\arraystretch{#1}%
  \hskip -\arraycolsep
  \let\@ifnextchar\new@ifnextchar
  \array{*\c@MaxMatrixCols c}}
\makeatother

\usepackage[section]{placeins}

\begin{document}

\title{Driven-Dissipative Dynamics of Atomic Ensembles in a Resonant Cavity: Quasiperiodic Route to Chaos and Chaotic Synchronization}

\author{Aniket Patra$^*$
\footnote{Current address: Max-Planck-Institut f\"{u}r Physik komplexer Systeme, D-01187 Dresden, Germany}}
\author{Boris L. Altshuler$^\dagger$}
\author{Emil A. Yuzbashyan$^*$}

\address{$^*$Department of Physics and Astronomy, Rutgers University, Piscataway, NJ 08854, USA \\ 
$^\dagger$Department of Physics, Columbia University, New York, NY 10027, USA}

\begin{abstract}

We analyze the origin and properties of the chaotic dynamics of two atomic ensembles in a driven-dissipative experimental setup, where they are collectively damped by a bad cavity mode and incoherently pumped by a Raman laser. Starting from the mean-field equations, we explain the emergence of chaos by way of quasiperiodicity -- presence of two or more incommensurate frequencies. This   is known as the Ruelle-Takens-Newhouse route to chaos. The equations of motion have a $\mathbb{Z}_{2}$-symmetry with respect to the interchange of the two ensembles. However, some of the attractors of these equations spontaneously break this symmetry. To understand the emergence and subsequent properties of various attractors, we concurrently study the mean-field trajectories, Poincar\'{e} sections, maximum and conditional Lyapunov exponents, and power spectra. Using Floquet analysis, we show that quasiperiodicity is born out of non-$\mathbb{Z}_{2}$-symmetric oscillations via a supercritical Neimark-Sacker bifurcation. Changing the detuning between the level spacings in the two ensembles and the repump rate results in the synchronization of the two chaotic ensembles. In this regime, the chaotic intensity fluctuations of the light radiated by the two ensembles are identical. Identifying the synchronization manifold, we understand the origin of synchronized chaos  as a tangent bifurcation intermittency  of the $\mathbb{Z}_{2}$-symmetric oscillations. At its birth, synchronized chaos is unstable. The interaction of this attractor with other attractors causes on-off intermittency until the synchronization manifold becomes sufficiently attractive. We also show coexistence of different phases in small pockets near the boundaries. 
      
\end{abstract}
 
\maketitle

\tableofcontents

\section{Introduction}

\begin{figure*}[tbp!]
\centering
\subfloat[\qquad \textbf{(a)}]{\label{Full_Phase_Diagram}\includegraphics[scale=0.37]{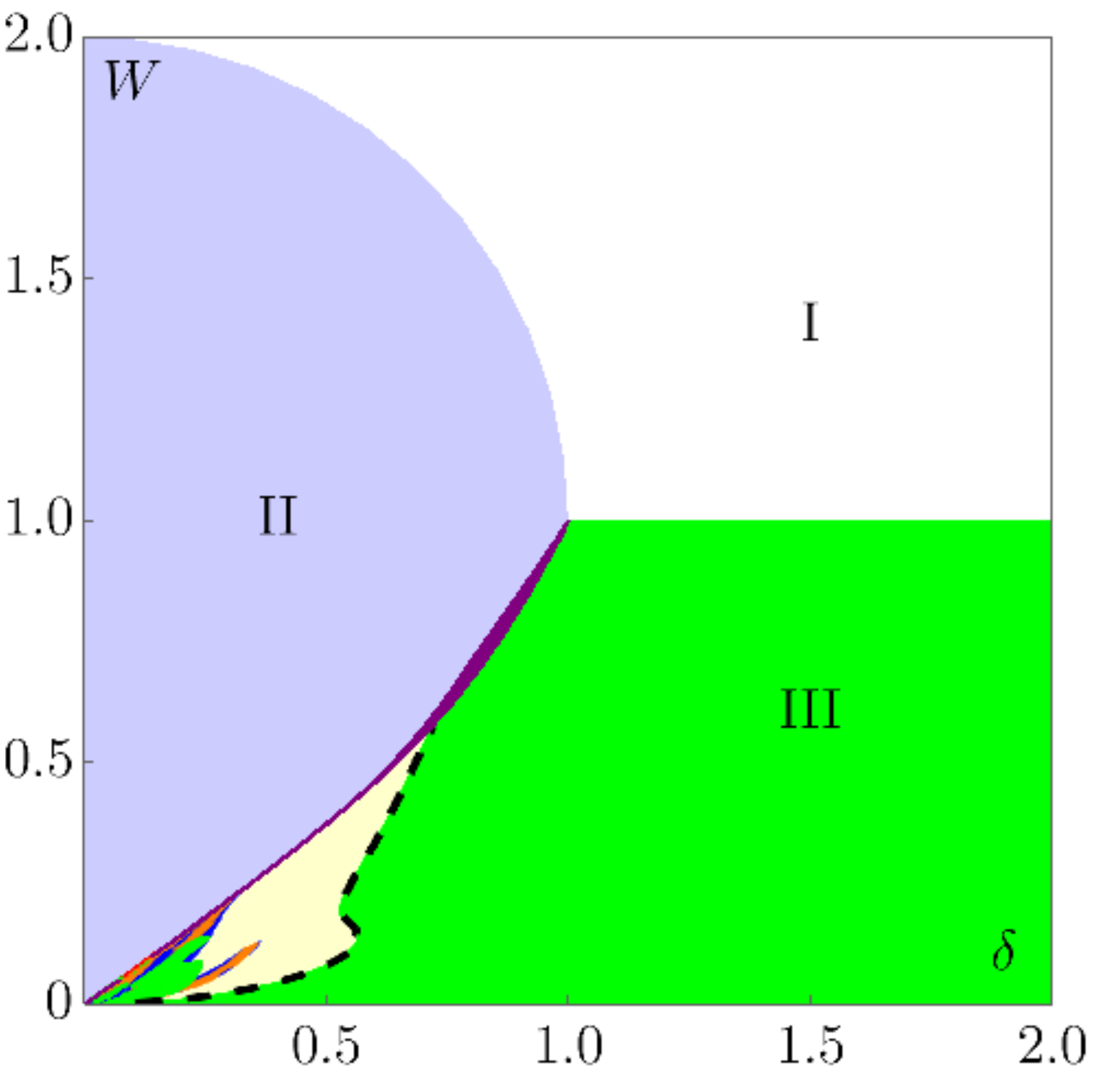}}\quad
\subfloat[\qquad \textbf{(b)}]{\label{Near_Origin}\includegraphics[scale=0.37]{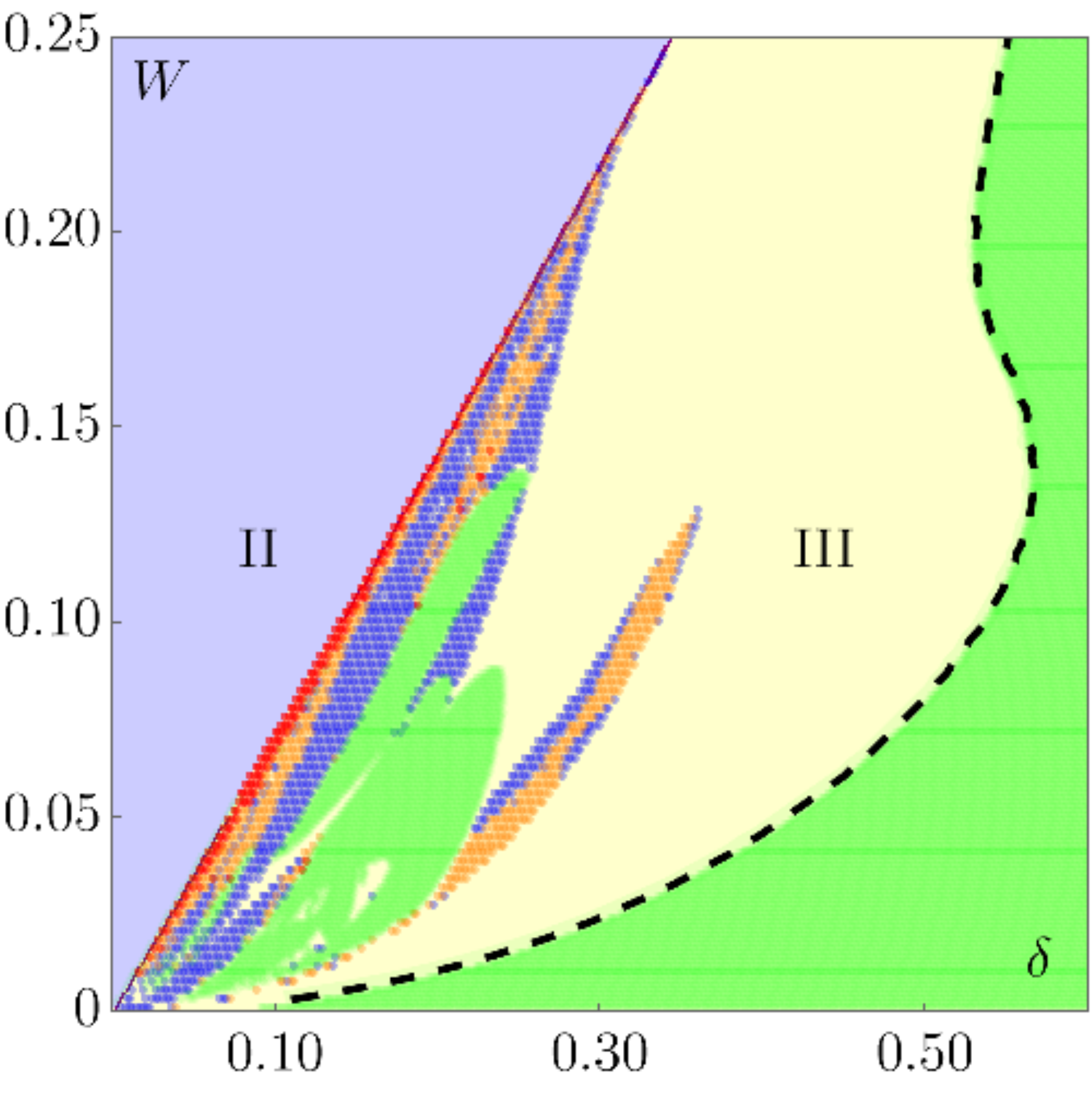}}
\caption{The phase diagram for two atomic ensembles Rabi coupled to a heavily damped cavity mode, and incoherently pumped by a transverse laser. Note, \textbf{(b)} is the magnified version of \textbf{(a)} near the origin. The two dimensionless parameters, detuning $\delta$ (difference between the level-spacings of the atoms belonging to the two ensembles) and repump rate $W$, are in the units of the collective decay rate $N\Gamma_{c}$ (see \sref{Review} for the definitions). The figures show the different nonequilibrium phases, e.g., I: normal non-superradiant phase, II: monochromatic superradiance, and III: different types of amplitude-modulated superradiance, including periodically modulated $\Z2$-symmetric (green) and symmetry-broken (yellow), quasiperiodic (blue), chaotic (orange), and synchronized chaotic (red) superradiance. The purple region in \textbf{(a)} denotes coexistence of monochromatic and amplitude-modulated superradiances. Within Phase III, moving from right to left, the $\Z2$-symmetry (involves the exchange between the two ensemble in the mean-field description) is first broken across the dashed line. Note, to the left of this line, this symmetry reappears in the green ($\Z2$-symmetric limit cycle) and red (synchronized chaos) subregions. Moreover, we observe coexistence of different types of amplitude-modulated superradiance in small pockets to the left of the dashed line, especially near the boundaries of two different behaviors (see \fsref{Coexist_(R)SBLC_QP} and \ref{Coexist_SLC_QP}). (For interpretation of the references to color in this figure legend, the reader is referred to the web version of this article.)}
\label{Phase_Diagram}
\end{figure*}

\begin{figure}[tbp!]
\begin{center}
\includegraphics[scale=0.45]{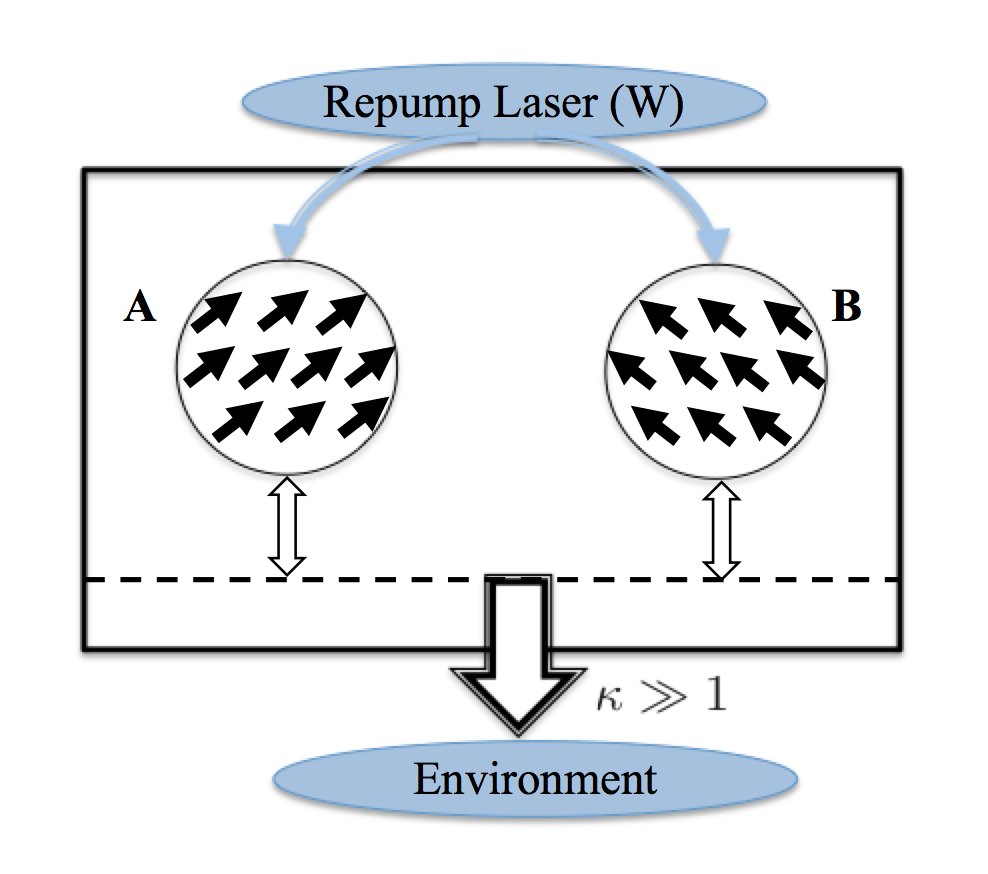}
\caption{Cartoon depicting the driven-dissipative experimental setup with two atomic ensembles inside a bad cavity. In ensembles `A' and `B', the solid arrows denote individual atoms. The double-headed block arrows correspond to the Rabi coupling between the ensembles and the cavity mode (dashed line). The rate of loss of photons from the cavity is $\kappa$.}\label{Setup}
\end{center}
\end{figure} 

Chaos appears in physics of all length scales -- on one hand at the cellular level in cancer growth \cite{Cancer_1, Cancer_2}, and on the other in various peculiarities of celestial and galactic dynamics \cite{Astro_1, Astro_2}. According to the Poincar\'{e}-Benedixon theorem, any phenomenon described by a system of three or more coupled nonlinear first-order differential equations can lead to chaos~\cite{Hilborn, Kuznetsov}. To predict long-term behavior of such systems, it is important to understand: \textbf{(1)} the mechanism via which chaos materializes, and \textbf{(2)} if the chaotic trajectories have any special properties.

 In this paper, we analyze the signatures of chaos in the light radiated by two atomic ensembles resonantly coupled to a ``bad" (fast leaking) cavity and incoherently pumped by a transverse laser~\cite{Holland_Two_Ensemble_Theory, Holland_Two_Ensemble_Expt, Patra_1, Patra_3}.  Each ensemble contains a large number $(N \approx 10^{6})$ of $\prescript{87}{}{\textrm{Rb}}^{}_{}$ or $\prescript{87}{}{\textrm{Sr}}^{}_{}$ atoms. This allows us to neglect the $\mathcal{O}(1/\sqrt{N})$ quantum fluctuations, and restrict ourselves to the mean-field picture \cite{Patra_1}. Representing the two ensembles with two classical spin $\bm{s}^{A}$ and $\bm{s}^{B}$, we describe this driven-dissipative setup in terms of six coupled nonlinear first-order differential equations (see \sref{Review}). Hence, occurrence of chaos is not surprising. 

 \fref{Phase_Diagram}  shows the full nonequilibrium phase diagram of the system, which we have already introduced    in \Rref{Patra_1}, and \fref{Setup} shows the cartoon of the experimental setup. Different nonequilibrium phases correspond to different asymptotic solutions (attractors) of the mean-field equations of motion.  Each phase has a unique signature in the spectrum of the light emitted by the cavity. Besides the character of the time-dependence, the most important criterion that we use to classify the  phases is the presence or absence of a $\Z2$-symmetry. This symmetry guarantees that in an appropriate rotating frame the mean-field equations remain unaltered when we interchange the ensembles while simultaneously flipping the sign of $s_{y}$, see \sref{Review} for details. It transpires that although the mean-field equations are $\Z2$-symmetric, some of the the nonequilibrium phases (asymptotic solutions of the mean-field equations) spontaneously break this symmetry.

In \Rref{Patra_1}, along with the normal non-superradiant Phase~I, we discussed in detail the monochromatic superradiance (Phase II) and periodic modulations of superradiance intensity in the green ($\Z2$-symmetric) and yellow (symmetry-broken) regions of Phase III. In the semiclassical description these periodic modulations are the periodic solutions, i.e., limit cycles of the mean-field equations. We summarize the main features of these solutions in \sref{Review}. In \Rref{Patra_3}  we reported the chaotic synchronization regime (red regions in \fref{Phase_Diagram}), where the chaotic dynamics of one ensemble  follows that of the other. 

Here we complete the description of the nonequilibrium phase diagram by analyzing the subregions in Phase III near the origin of \fref{Phase_Diagram}. Besides the mechanism that brings about chaos in the orange subregion of Phase III in \fref{Phase_Diagram}, we also study how it evolves into other phases. In particular, we explain how the nonlinear interaction between the two spins brings about chaos via quasiperiodicity (see \sref{Quasiperiodicity_Chaos}), and eventually gives rise to chaotic synchronization (see \sref{Chaotic_Synch}).   

\begin{table}[ht]
\begin{center}
\begin{tabular}{| c | c | }
\hline
Type of bifurcation & Subcategories \\
\hline \hline
\multirow{3}{*}{Local} & Period doubling \\
\cline{2-2} & Quasiperiodicity \\
\cline{2-2} &  \begin{tabular}{| c | c |} \multirow{4}{*}{Intermittency} & Tangent bifurcation int. (type I) \\
\cline{2-2} & Hopf bifurcation int. (type II) \\
\cline{2-2} &  Period doubling int. (type III) \\
\cline{2-2} & On-off int. \\
\end{tabular} \\ 
\hline
\multirow{2}{*}{Global}& Chaotic transients \\
\cline{2-2} & Crises \\
\hline
\end{tabular}
\end{center}
\caption{Different mechanisms for the manifestation of chaos \cite{Hilborn}.}
\label{Tab_Diff_Types_Chaos}
\end{table}

In \tref{Tab_Diff_Types_Chaos}, we list different scenarios of emergence of chaos \cite{Hilborn}. In our system, we only have examples of chaos via local bifurcations, where a limit cycle gives way to chaos as one changes a control parameter. In chaos via global bifurcation, on the other hand, it is necessary to follow trajectories for a significant range in the phase space, and see how various coexisting attractors affect these trajectories. The most commonly studied mechanism explaining the origin of chaos in deterministic systems is period doubling \cite{Hilborn, Mechanics}. In our system, however, we did not observe period doubling. In Phase III of our nonequilibrium phase diagram \fref{Phase_Diagram}, chaos arises via the quasiperiodic route as we move towards the boundary with Phase II. We observe on-off intermittency right before chaotic synchronization. Also, restricting the dynamics in the $\Z2$-symmetric submanifold, we observed tangent bifurcation intermittency before chaotic synchronization, see \fref{Reduced_Phase_Diagram}.

Universal features are observed in a few of the above scenarios, especially where the effective dynamics (e.g., the Poincar\'{e} map function) can be reduced to 1D. In the period doubling scenario, where the essential features of chaos are captured by a 1D iterated map, one obtains different Feigenbaum universality classes. Similar universal behavior is observed in quasiperiodic route to chaos where the essential dynamics is described by the 1D sine-circle map. Such dimensional reduction is not possible for the chaos in our system and we do not observe any universal features.

Inside Phase III, two incommensurate frequencies emerge from the symmetry-broken limit cycles via Neimark-Sacker or torus bifurcation, see \sref{Q-P}. One can use this quasiperiodic superradiance to generate unusual frequencies that are incommensurate to the carrier frequency. This way, it might have potential applications in building ultratunable lasers. This behavior quickly loses stability giving way to chaos, see \sref{QP->Chaos}.  

As we keep changing the experimental parameters, eventually in the red subregion of Phase III in \fref{Near_Origin} chaos becomes synchronized, i.e., the corresponding asymptotic solutions acquire the $\Z2$-symmetry. Irrespective of the chaotic nature of the dynamics, this allows one to obtain the full asymptotic behavior of some variables. For example, here the difference of the z-components of the classical spins $(s_{z}^{A} - s_{z}^{B})$ becomes identically zero after some time (cf. \fref{SC_Pictures}). Because of such predictive power in the face of complicated dynamics, and potential technological applications, e.g., in secure communication \cite{Uchida, Chaotic_Synch_1, Chaotic_Synch_2, Chaotic_Synch_3, Chaotic_Synch_6}, chaotic synchronization has been analyzed in different contexts that include electrical circuits \cite{Chaotic_Synch_2, Chaotic_Synch_3}, coupled lasers~\cite{Chaotic_Synch_4, Chaotic_Synch_5, Chaotic_Synch_6},  oscillators in laboratory plasma~\cite{Chaotic_Synch_7}, population dynamics \cite{Chaotic_Synch_8} and earthquake models \cite{Chaotic_Synch_9}. 

In this paper we open up new possibilities by merging the two fields -- chaotic synchronization and cavity QED. Moreover, in most of the above examples the coupling between the two chaotic components is \textit{directional}. In our case, however, the two atomic ensembles are \textit{mutually coupled}. We expect that upon spatially separating the two ensembles by a long distance while maintaining the coupling to a common cavity mode, one might be able to exploit the synchronized chaos for secure communication as well.    
  
To distinguish different phases (e.g., quasiperiodicity, chaos and synchronized chaos), we plot time evolution of different components of the classical spins in \fsref{QP_Pictures}, \ref{C_Pictures}, and \ref{SC_Pictures}. The behaviors of the 2D projections of the attractors on the $s_{\perp}^{A} - s_{\perp}^{B}$ plane (where $s_{\perp}^{\tau} = \sqrt{\big(s_{x}^{\tau}\big)^{2} + \big(s_{y}^{\tau}\big)^{2}}$, $\tau = A,B$), and the $s_{z}^{A} - s_{z}^{B}$ plane (see \fsref{QP_X_Sect}, \ref{C_X_Sect}, and \ref{SC_X_Sect}) are also instructive. Unlike the periodic ones, the lines in these projections develop nonzero thicknesses for the quasiperiodic trajectories. The chaotic projections fill up an extended region. For synchronized chaos, the projections are similar to the ones for the $\Z2$-symmetric limit cycles, cf. \cite{Patra_1}. The spontaneous restoration of the $\Z2$ symmetry restricts the synchronized chaotic attractors to the ``synchronization manifold", see \sref{Synch_Manifold}. We concurrently examine the Poincar\'{e} sections, maximum Lyapunov exponents and the power spectra for each of the behaviors. 

To define a Poincar\'{e} section \cite{Poincare_Paper} for a particular trajectory, we first introduce a Poincar\'{e} plane to be a transverse plane dividing the trajectory into two halves. For example, in \fsref{Traj_Sect_LC_Toy} and \ref{Traj_Sect_QP_Toy}, we show explicit examples of 3D periodic and quasiperiodic trajectories with the $X-Y$ planes as the corresponding Poincar\'{e} planes (shown in translucent red). Poincar\'{e} sections are then defined as the collection of successive directional intersections of the trajectories with the Poincar\'{e} plane, where we only select the intersections in a particular direction with respect to the plane. From \fsref{Poincare_LC_Toy} and \ref{Poincare_QP_Toy}, the qualitative difference between the respective  Poincar\'{e} sections for the periodic and quasiperiodic attractors is quite apparent. The first one is a collection of discrete points, whereas the latter is a continuous loop.   

For the mean-field dynamics, the attractors are 6D. A Poincar\'{e} section reduces the dimensionality by one and leaves us with a 5D object. This is hard to visualize. To get around this issue, we analyze the projected 2D Poincar\'{e} sections, henceforth simply referred to as the Poincar\'{e} sections, for $\bm{s}^{A}$ in different phases (see \fsref{Poincare_LC}, \ref{Poincare_QP}, \ref{Poincare_C}, and \ref{Poincare_SC}). In these pictures, the Poincar\'{e} planes are taken to be parallel to the $s_{x}^{A} - s_{y}^{A}$ plane. The equation for the planes is: $s_{z}^{A} = \textrm{const.} = \frac{1}{t_{1}}\int_{t_{0}}^{t_{0}+t_{1}}s^{A}_{z}dt$, with $t_{0}$ and $t_{1}$ being sufficiently large.  One notices similar qualitative differences in the Poincar\'{e} sections for periodic and quasiperiodic attractors here as was demonstrated earlier in \fsref{Poincare_LC_Toy} and \ref{Poincare_QP_Toy}. Moreover, we observe that the ones for the chaotic and synchronized chaotic attractors are also distinguishable from the former and among each other. Poincar\'{e} sections corresponding to $\bm{s}^{B}$ with Poincar\'{e} planes placed parallel to the $s_{x}^{B} - s_{y}^{B}$ plane produce similar figures. Note, since we are projecting 5D Poinacr\'{e} sections onto 2D planes, although the original Poincar\'{e} sections have no self-intersection, one can not rule out instances of self-intersections in the projected 2D ones that are shown here (see \fref{Poincare_QP}\blue{d}).

\begin{table}[tbt!]
\centering 
\begin{tabular}{| c | c | c | c |}
\hline
Attractor & Spectrum & Reflection & Peak at $f = 0$ \\ [0.5ex]
\hline\hline
Quasiperiodicity & Discrete & No & Yes \\
\hline
Chaos & Continuum & No & Yes \\
\hline
Synchronized Chaos & Continuum & Yes & No \\
\hline
\end{tabular}
\caption{Properties of power spectra of radiated light for three distinct amplitude-modulated superradiances. In the second, third and fourth columns we list whether the spectra are discrete or continuous, whether they possess a reflection symmetry about the $f = 0$ axis, and if they have the most prominent peak at $f = 0$, respectively. We show the spectra in \fsref{Spectrum_QP}, \ref{Spectrum_C_Full}, and \ref{Spectrum_SC_Full}. We show the presence and absence of a peak at the origin for the chaotic and synchronized chaotic attractors in \fsref{Spectrum_C_Mag} and \ref{Spectrum_SC_Mag}, respectively.}
\label{Tab_Power_Spectra}
\end{table}

Lyapunov exponents quantify how two infinitesimally close trajectories diverge in different directions. In particular, trajectories with positive maximum Lyapunov exponents entail chaos. Indeed, in \fref{Lyapunov}, the chaotic (including the synchronized chaotic one) attractor has positive maximum Lyapunov exponent. To differentiate synchronized chaos from non-synchronized, we also calculate the maximum Lyapunov exponent corresponding to the directions transverse to the synchronization manifold, alternatively known as the conditional Lyapunov exponent, in \sref{Synch_Manifold}. A negative conditional Lyapunov exponent and a positive maximum Lyapunov exponent at the same time indicates that the chaotic trajectory is attracted towards the synchronization manifold, i.e., the synchronization manifold is stable \cite{Uchida, Chaotic_Synch_1}. This is corroborated in \fref{Cond_Lyapunov}, where the conditional Lyapunov exponent is negative only for the synchronized chaotic trajectory. We provide algorithms for calculating the maximum Lyapunov exponent and conditional Lyapunov exponent in \ref{Appnd_Lya_Exp}.

The power spectrum of the radiated light $|\bm{E}(f)|^2$, where $\bm{E}(f)$ is the Fourier transform of the (complex) electric field, is experimentally measured with a Michelson interferometer. In our mean-field theory 
\beg
|\bm{E}(f)|^2\propto |l_{-}(f)|^{2},
\label{Expt_Power_Spectrum} 
\en 
where  $l_{-} = (s_{-}^{A}+s_{-}^{B})$. In \tref{Tab_Power_Spectra}, we compare the different properties of the quasiperiodic, chaotic and synchronized chaotic spectra.

\section{Mean-field Equations of Motion}\label{Review}

We describe the system (experimental setup shown in \fref{Setup}) in terms of the following master equation for the density matrix $\rho$:
\begs
\bea
\dot{\rho} &=& -\imath \big[\hat{H}, \rho\big] + \kappa\mathcal{L}[a]\rho + W\!\!\!\!\sum_{\tau = A,B}\sum_{j = 1}^{N}\mathcal{L}[\hat{\sigma}^{\tau}_{j+}]\rho, \label{Master}  \\ 
\hat{H} &=& \omega_{0}\hat{a}^{\dagger}\hat{a} + \!\!\! \sum_{\tau = A,B}\!\!\big[\omega_{\tau}\hat{S}_{\tau}^{z} + \frac{\Omega}{2}\big(\hat{a}^{\dagger}\hat{S}_{\tau}^{-}+\hat{a}\hat{S}_{\tau}^{+}\big)\big]. \label{H}  
\eea 
\label{Full_Master}%
\ens
The above density matrix $\rho$ contains both the atomic and the cavity degrees of freedom. The Hamiltonian  $\hat{H}$ conveys the Rabi coupling (Rabi frequency~$\Omega$) of two atomic ensembles ($\tau = A, B$) to the cavity mode $\omega_{0}$, where $\hat{a}^{\dagger} (\hat{a})$ creates (annihilates) a cavity photon. Since the coupling between the cavity mode and the nearest two atomic energy levels is the strongest, we consider individual atoms to be two-level systems and represent them by the Pauli matrices $\hat{\bm{\sigma}}_{j}$. It is possible to experimentally control the level spacings $\omega_{\tau}$ of the atoms in two different ensembles with two distinct Raman dressing lasers~\cite{Holland_Two_Ensemble_Expt}. Collectively the ensembles are then represented by the following operators:
\beg
\hat{S}^{A,B}_{z} = \frac{1}{2}\sum_{j = 1}^{N}\hat{\sigma}^{A,B}_{jz}, \quad \hat{S}^{A,B}_{\pm} = \sum_{j = 1}^{N}\hat{\sigma}^{A,B}_{j\pm}.
\label{Defn_Coll_Spins}
\en
Besides the Hamiltonian dynamics, we include two energy nonconserving processes in the master equation~\re{Full_Master} that are most relevant -- fast decay of the cavity intensity with a rate $\kappa (\gg 1)$, and incoherent pumping of the individual atoms at a rate $W$ with external lasers. We do so using the Lindblad superoperators, 
\beg 
\mathcal{L}[\hat{O}]\rho \equiv \frac{1}{2}\big(2\hat{O}\rho \hat{O}^{\dagger} - \hat{O}^{\dagger}\hat{O}\rho -\rho\hat{O}^{\dagger}\hat{O}\big).
\label{Lindblad_def}
\en%

We point out that these Lindblad operators are a result of starting with a Hamiltonian of the general form $H = H_{S} + H_{R} + H_{SR}$, and using the Born-Markov approximation \cite{Carmichael_1, Carmichael_2, Carmichael_3}. Here $H_{S}$, $H_{R}$, and $H_{SR}$ are the Hamiltonians for the system, the reservoir, and the interaction between them, respectively. Assuming the coupling between the system and the reservoir to be very weak, we employ the Born approximation and neglect higher order terms in $H_{SR}$. The Markovian approximation is appropriate when the reservoir is a large system maintained in the thermal equilibrium. In particular, one requires that the time required for a significant change in the system be very large compared to the reservoir correlation time. 

In the adiabatic limit $\kappa \rightarrow \infty$, it is possible to eliminate the cavity mode and write an effective master equation containing only the atomic operators \cite{Bad_cavity}. As shown in \Rref{Patra_1}, this is accomplished by  replacing 
\beg 
\hat{a}\to \frac{\imath \Omega}{\kappa}\sum_{\tau} \hat{S}_{\tau}^{-},
\label{Ad_Elim}
\en%
in \eref{Full_Master}. This produces the following effective master equation:
\begs
\bea
\dot{\rho}_\mathrm{at}  &=& -\imath \big[ \hat h, \rho_\mathrm{at} \big] + \Gamma_{c}\mathcal{L}[\hat{J}_{-}]\rho_\mathrm{at}  +W \sum_{\tau, j}^{N}\mathcal{L}[\hat{\sigma}^{\tau}_{j+}]\rho_\mathrm{at},\\ 
\hat h &=& \omega_{A}\hat{S}_{z}^{A} + \omega_{B}\hat{S}_{z}^{B},  
\eea
\label{Effective_Master}%
\ens 
where $\rho_\mathrm{at} =\Tr_{\! F} (\rho)$ (traced over the cavity mode) is the atomic density matrix, and $\Gamma_{c} = \frac{\Omega^{2}}{\kappa}$ is the collective decay rate.

The above partial trace of the original master equation \re{Full_Master} obtains a power series in $\Omega^{2}N/\kappa^{2}$ \cite{Bad_cavity}. The Markovian approximation is only valid in the bad cavity limit, when $\kappa \gg \Omega\sqrt{N}$ and we retain only the zeroth-order term. In a not so bad cavity, when it is not possible to neglect the higher order terms, one has to contend with memory effects. 

We then define the classical spins (as mentioned in the introduction) as the renormalized expectation value of the collective spin operator \re{Defn_Coll_Spins}:
\beg
\bm{s}^{\t} \equiv \frac{2}{N}\mean{\hat{\bm{S}}^{\tau}},
\label{Classical_Spin_Defn}
\en%
where $\tau = A, B$.  They obey the following mean-field equations:
\begs
\bea
\dot{s}^{\t}_{\pm} &=& \biggl(\pm\imath \omega_{\t}- \frac{W}{2}\biggr)s^{\tau}_{\pm} + \frac{1}{2}s^{\t}_{z}l_{\pm}, \\ \label{REOMPM}
\dot{s}^{\t}_{z} &=&  W\big(1 - s^{\t}_{z}\big) - \frac{1}{4}s^{\t}_{+}l_{-} - \frac{1}{4}s^{\tau}_{-}l_{+}. \label{REOMz} 
\eea 
\label{Mean-Field_1}%
\ens%
where $s^{\t}_{\pm} = s_{x}^{\tau} \pm \imath s_{y}^{\tau}$, and $\bm{l} = \sum_\tau \bm s^\tau$ is the total spin. The above equation stipulates that in the mean-field description the intensity of the emitted light ($\mean{\hat{a}^{\dagger}\hat{a}}$) is proportional to $|l_{-}|^{2}$, cf. \eref{Ad_Elim}. In a rotating reference frame, where all the energies are shifted by the mean level spacing $\frac{1}{2}\sum_{\tau}\omega_{\tau} \equiv 2\pi f_{\mathrm{mc}}$, one has
\beg
\omega_{A} =  -\omega_{B} = \frac{\delta}{2}.
\label{Detuning}
\en%
For $\prescript{87}{}{\textrm{Sr}}^{}_{}$, $f_{\mathrm{mc}} \approx 4.3 \times 10^{5}$ GHz \cite{3P0-1S0_87Sr}. As a result in such a reference frame the only two relevant energy scales are -- \textbf{(1)} detuning $\delta = \omega_{A} - \omega_{B}$, and \textbf{(2)} repump rate $W$. Also, note that in \eref{Mean-Field_1} and from now on we set
\beg 
N\Gamma_{c} = 1.
\label{Coll_Decay_Rate} 
\en%
This fixes the unit of energy and time. Starting from \eref{Mean-Field_1}, all the energies (including $\delta$ and $W$) are written in the units of $N\Gamma_{c}$, and time is written in the units of $(N\Gamma_{c})^{-1}$. For a typical $\prescript{87}{}{\textrm{Sr}}^{}_{}$ based experimental setup, $N\Gamma_{c} \approx 1.4$kHz \cite{Holland_One_Ensemble_Theory_1,Patra_1}.   

The above mean-field equations \re{Mean-Field_1} have the following two symmetries. First, they have an axial symmetry about the z-axis. Defining the rotation operator about z-axis 
\beg 
\R(\phi): (s^{\tau}_{\pm}, s^{\tau}_{z}) \longrightarrow (s^{\tau}_{\pm}e^{\pm \imath \phi}, s^{\tau}_{z}),
\label{Rot_Def}  
\en    
where $\phi$ is arbitrary, we observe that indeed \esref{Mean-Field_1} remain unchanged upon replacing $\bm{s}^{\tau}\ra \R(\phi)\cdot\bm{s}^{\tau}$. Secondly, \esref{Mean-Field_1} have the following $\Z2$ symmetry, where again they remain unchanged upon interchanging the A and B ensembles with altering the sign of $s_{y}^{\tau}$ via the operator
\beg  
\mathbb{\Sigma}:\big( s_{\pm}^{A}, s_{z}^{A}, s_{\pm}^{B}, s_{z}^{B} \big) \longrightarrow \big( s_{\mp}^{B}, s_{z}^{B}, s_{\mp}^{A}, s_{z}^{A} \big),
\label{Z2}
\en%
after rotating the spins about the the z-axis by an appropriate angle $\phi_{0}$. Thus a $\Z2$ symmetric attractor remains confined to a 4D synchronization submanifold in the full phase space that is defined by the relation
\beg 
\bm{s}^{\tau} = \mathbb{\Sigma}\circ \R(\phi_{0}) \cdot \bm{s}^{\tau}.
\label{Gen_Z2_Symm_MF}%
\en%
Here $\phi_{0}$ can take an arbitrary value between zero and $2\pi$. We write the two constraint relations for this submanifold explicitly as
\beg 
\begin{split} 
\big(s_{x}^{A}\big)^{2} + \big(s_{y}^{A}\big)^{2} = \big(s_{x}^{B}\big)^{2} + \big(s_{y}^{B}\big)^{2}\!\!,\quad s_{z}^{A} = s_{z}^{B}.
\end{split} 
\label{Gen_Z2_Symm_MF}%
\en%
In particular, after rotating the spins by an appropriate $\phi_{0}$, spin components obey $\bm{s}^{\tau} = \mathbb{\Sigma}\cdot \bm{s}^{\tau}$, i.e.,
\beg 
s^{A}_{x} = s^{B}_{x},\quad s^{A}_{y} = -s^{B}_{y},\quad s^{A}_{z} = s^{B}_{z}.
\label{Z2_Expl}
\en%
This defines a 3D $\Z2$-symmetric invariant submanifold, i.e., initial conditions on \re{Z2_Expl} confine the dynamics on the same. In the above 3D submanifold we write the decoupled evolution equations for $\bm{s}^{A}$, using $l_{x} = 2s^{A}_{x} = 2s^{B}_{x}$ and $l_{y} = 0$, as
\begs
\bea
\dot{s}_{x} &=& - \frac{\delta}{2}s_{y}- \frac{W}{2}s_{x} + s_{z}s_{x},\label{ReducedX} \\ 
\dot{s}_{y} &=& \frac{\delta}{2}s_{x} - \frac{W}{2}s_{y}, \label{ReducedY}\\ 
\dot{s}_{z} &=&  W\big(1 - s_{z}\big) - \big(s_{x}\big)^{2}, \label{ReducedZ} 
\eea 
\label{Symm_One_Spin_Eqn}%
\ens%
where we have dropped the superscript, and $\bm{s}^{B}$ is related to $\bm{s}^{A}$ by the \eref{Z2_Expl}. From now on, we use the same convention while describing an attractor that lies on this submanifold~\re{Z2_Expl}. 

Using the axial symmetry, it is further possible to separate \eref{Mean-Field_1} into two groups --  a closed set of five equations, and an equation for the overall phase. The latter, however, depends on the five former equations. To this end, we introduce the following new variables:
\beg 
 s^{\tau}_{\pm} = s^{\tau}_\perp e^{\pm\imath\phi_{\tau}}, \quad \phi_{A} = \Phi + \varphi, \quad \phi_{B} = \Phi - \varphi,
\label{New_Var_Rot}
\en%
where $\varphi$ is defined modulo $\pi$. The evolution equations for the five variables (other than $\Phi$) do not contain $\Phi$ in the right hand side, and are as follows:
\begin{gather}
\dot{s}^{A}_\perp = - \frac{W}{2}s^{A}_\perp + \frac{s^{A}_{z}}{2}\big(s^{A}_\perp + s^{B}_\perp\cos{2\varphi}\big), \nonumber \\ 
\dot{s}^{B}_\perp = - \frac{W}{2}s^{B}_\perp + \frac{s^{B}_{z}}{2}\big(s^{A}_\perp\cos{2\varphi} + s^{B}_\perp\big), \nonumber \\ 
\dot{s}^{A}_{z} =  W\big(1 - s^{A}_{z}\big) - \frac{s^{A}_{\perp}}{2}\big(s^{A}_\perp + s^{B}_\perp\cos{2\varphi}\big), \!\!\!\!\!\!\!\! \label{Mean-Field_Group5} \\ 
\dot{s}^{B}_{z} =  W\big(1 - s^{B}_{z}\big) - \frac{s^{B}_{\perp}}{2}\big(s^{A}_\perp\cos{2\varphi} + s^{B}_\perp\big), \nonumber \\ 
\dot{\varphi}= \frac{1}{2}\big(\omega_{A} - \omega_{B}\big) - \frac{\sin{2\varphi}}{4}\bigg(\frac{s^{A}_{z}s^{B}_\perp}{s^{A}_{\perp}} + \frac{s^{B}_{z}s^{A}_\perp}{s^{B}_{\perp}}\bigg). \nonumber
\end{gather}
This guarantees that the initial choice of $\Phi$ does not affect the values of $s^{\tau}_\perp, s_{z}^{\tau}$ and $\varphi$ at subsequent times. The overall phase $\Phi$, on the other hand, obeys the following equation:
\beg 
\dot{\Phi} = \frac{1}{2}\big(\omega_{A} + \omega_{B}\big) + \frac{\sin{2\varphi}}{4}\bigg(\frac{s^{A}_{z}s^{B}_\perp }{s^{A}_\perp} - \frac{s^{B}_{z}s^{A}_\perp }{s^{B}_\perp }\bigg).
\label{Mean-Field_Group1}
\en%

We obtain the trivial steady state, or TSS in region I of \fref{Phase_Diagram} corresponding to the normal phase (no radiation, atoms are maximally pumped)
\beg 
s^{\tau}_{x,y} = 0, \quad s^{\tau}_{z} = 1,
\label{TSS}
\en%
by setting the time derivatives in \eref{Mean-Field_1} to zero. This is the only asymptotic solution that retains both the symmetries. All the other solutions break at least the axial symmetry. As a result, for a particular $(\delta, W)$ one has infinitely many such attractors related to each other by rotations about z-axis. The TSS loses its stability on the boundary of regions I and II via supercritical pitchfork bifurcation \cite{Sup_PF}, bringing about the nontrivial steady state, or NTSS
\beg 
\begin{aligned}
s^{A,B}_{-} &= e^{-\imath(\Phi \pm \varphi)}\frac{l_{\perp}}{2}\sqrt{1 + \frac{\delta^{2}}{W^{2}}}, \\
s^{A}_{z} &= s^{B}_{z} = \frac{\delta^{2} + W^{2}}{2W},
\end{aligned}
\label{NTSS}
\en%
where 
\beg 
l_{\perp} = \sqrt{2\big(1- (W-1)^{2}- \delta^{2}\big)}, \quad
\varphi = \arctan{\frac{\delta}{W}},
\label{NTSS_1}
\en%
and $\Phi$ is an overall arbitrary phase. The nonzero intensity of the radiated light $\left(|l_{-}| \neq 0\right)$ in this phase indicates synchronization among the individual atoms in an ensemble, and ultimately between the two ensembles themselves \cite{Holland_Two_Ensemble_Theory}. For the NTSS in region II, the loss and pumping are balanced in a particular way leading to monochromatic superradiance. It loses its stability via subcritical Hopf bifurcation on the boundary between regions II and III, introducing no new solutions \cite{Patra_1}. However, this leads to the coexistence of the NTSS with different kinds of amplitude-modulated superradiance right before criticality in the purple subregion inside II. 

Across the $W = 1$, $\delta \geqslant 1$ half-line, the TSS loses its stability via supercritical Hopf bifurcation, giving rise to a $\Z2$-symmetric limit cycle in the green parts of region III \cite{Sup_H}. Right below this half-line, we obtain the following perturbative solution \cite{Patra_1} from \eref{Symm_One_Spin_Eqn} for the limit cycle when $\delta-W\gg 1-W$:
\begs
\bea 
s_{x} &\approx & \sqrt{2(1-W)}\cos{(\omega t - \alpha)}, \\
s_{y} &\approx & \sqrt{2(1-W)}\sin{\omega t}, \\ 
s_{z} &\approx &  1.
\eea
\label{W=1_Limit_LC}
\ens
Here $\omega = \frac{1}{2}\sqrt{\delta^{2} - 1},$ and $\tan{\alpha} = 1/\sqrt{\delta^{2} - 1}$. We mention in passing that similar perturbative solutions in terms of harmonic functions exist for the $\Z2$-symmetric limit cycle, when $W \ra 0, \delta \gtrapprox 1$ and when $\delta\gg W$. Close to the $\delta = W = 1$ point ($\delta-W \lessapprox 1-W$) the above perturbation theory in terms of harmonic functions [\eref{W=1_Limit_LC}] breaks down. Instead, we obtain a new perturbative solution in terms of the Jacobi elliptic function cn. The elliptic and the harmonic solutions coincide in the limit $\big[(1-W)/(\delta - 1)\big] \ra 0^+$ \cite{Patra_1}.   

As we move left in Phase III, two symmetry-broken limit cycles (related to each other by the $\Z2$ symmetry) are born after the $\Z2$-symmetric limit cycle loses stability across the dashed line. To obtain this line we employ the Floquet analysis. In our previous work \cite{Patra_1} we used a complementary numerical method to obtain the same line. First, we regroup the spin components in the following way -- \textbf{(1)} the symmetric ones on the $\Z2$-symmetric manifold~\re{Z2_Expl}
\beg 
\begin{aligned}
s_{x,z} = \frac{1}{2}\big(s^{A}_{x,z} + s^{B}_{x,z} \big), \quad s_{y} = \frac{1}{2}\big(s^{A}_{y} - s^{B}_{y} \big),
\end{aligned}
\label{Floquet_Symm_Var}
\en%
\textbf{(2)} the asymmetric ones covering the submanifold complementary to \re{Z2_Expl}
\beg 
\begin{aligned}
m_{x,z} = \big(s^{A}_{x,z} - s^{B}_{x,z} \big), \quad m_{y} = \big(s^{A}_{y} + s^{B}_{y} \big).
\end{aligned}
\label{Floquet_Asymm_Var}
\en%
For the $\Z2$-symmetric limit cycle $\bm{m} = 0$. In fact, to the first order in $\bm{m}$, the symmetric variables in \re{Floquet_Symm_Var} obey \eref{Symm_One_Spin_Eqn}. Close to the symmetry breaking, we obtain linearized equations for \re{Floquet_Asymm_Var}  from \eref{Mean-Field_1} as follows:
\begs
\bea
\dot{m}_{x} &=& - \frac{\delta}{2}m_{y}- \frac{W}{2}m_{x} + s_{x}m_{z}, \\ [7pt]
\dot{m}_{y} &=& \frac{\delta}{2}m_{x} + \big(s_{z} - \frac{W}{2}\big)m_{y}, \\[7pt] 
\dot{m}_{z} &=& - W m_{z} - s_{x}m_{x} - s_{y}m_{y}, 
\eea 
\label{Floquet_Eqn}
\ens%
Some of the coefficients of the components of $\bm{m}$ in the above equations \re{Floquet_Eqn} are linear functions of the symmetric spin components \re{Floquet_Symm_Var}. Therefore, they vary with time.  

We numerically obtain the monodromy matrix 
\beg 
\mathbb{B}_{m} = \big[\mathbb{M}(0)\big]^{-1}\cdot \mathbb{M}(T)
\en%
of the above system of differential equations \re{Floquet_Eqn}, where three independent solutions $\bm{m}$ of \eref{Floquet_Eqn} constiute the columns of $\mathbb{M}$. $T$ is the period of $\bm{s}$ in \re{Floquet_Symm_Var}. The eigenvalues of $\mathbb{B}_{m}$: $\rho_{i}\equiv e^{\nu_{i}T}$ ($i = 1,2,3$) are the Floquet multipliers, and $\nu_{i}$ are the Floquet exponents. Using the Floquet theorem we write the general solution for $\bm{m}$ as follows:
\beg 
\bm{m}(t) = \sum_{i = 1}^{3} C_{m, i}e^{\nu_{i}t}\bm{\xi}_{i}, \qquad \rho_{i}\equiv e^{\nu_{i}T}.
\label{Asymm_Floquet_Soln}
\en%
Here the period of $\bm{\xi}_{i}$ is also $T$, and $C_{m, i}$ ($i = 1, 2, 3$) are independent constants. 

Note,
\beg 
\bm{m}^{0} = \bpm s_{y} \\ -s_{x} \\ 0 \epm
\label{Special_m_TF}
\en 
is a particular solution for \eref{Floquet_Eqn}. This comes about due to the axial symmetry of \eref{Mean-Field_1}. The difference of $\R(\Delta \theta)\cdot~\bm{s}^{\tau}$ and $\bm{s}^{\tau}$ (here $\Delta \theta \rightarrow 0$), both of which are solutions of \eref{Mean-Field_1}, is $\Delta s_{x}^{\tau} = \Delta \theta s_{y}^{\tau}$ and $\Delta s_{x}^{\tau} = -\Delta \theta s_{x}^{\tau}$. As a result, $\bm{m} = 2\Delta\theta\bm{m}^{0}$ is indeed a solution of \eref{Floquet_Eqn}. Since this is a periodic function with the same period as $\bm{s}$, it corresponds to Floquet multiplier $\rho_{1} = 1$. Therefore we define $\bm{\xi}_{1} \equiv \bm{m}^{0}$. Using this, the general solution of \eref{Floquet_Eqn} takes the following form:
\beg 
\bm{m}(t) = C_{m, 1}\bm{\xi}_{1} + C_{m, 2}e^{\nu_{2}t}\bm{\xi}_{2} + C_{m, 3}e^{\nu_{3}t}\bm{\xi}_{3}.
\label{Asymm_Floquet_Soln_Expl}
\en%
The other two Floquet multipliers are real near the dashed line in \fref{Phase_Diagram}. The magnitude of one of them (say $|\rho_{2}|$) becomes more than one while $|\rho_{3}|$ remains less than one as we cross the line from the left. Thus, the fixed point of Poincar\'{e} map corresponding to the $\Z2$-symmetric limit cycle loses its stability by a supercritical pitchfork bifurcation \cite{Patra_1}.

\section{Quasiperiodic Route to Chaos}\label{Quasiperiodicity_Chaos}

\subsection{Quasiperiodicity} \label{Q-P}

\begin{figure*}[tbp!]
\centering
\subfloat[\textbf{(a)}]{\label{Traj_Sect_LC_Toy}\includegraphics[scale=0.225]{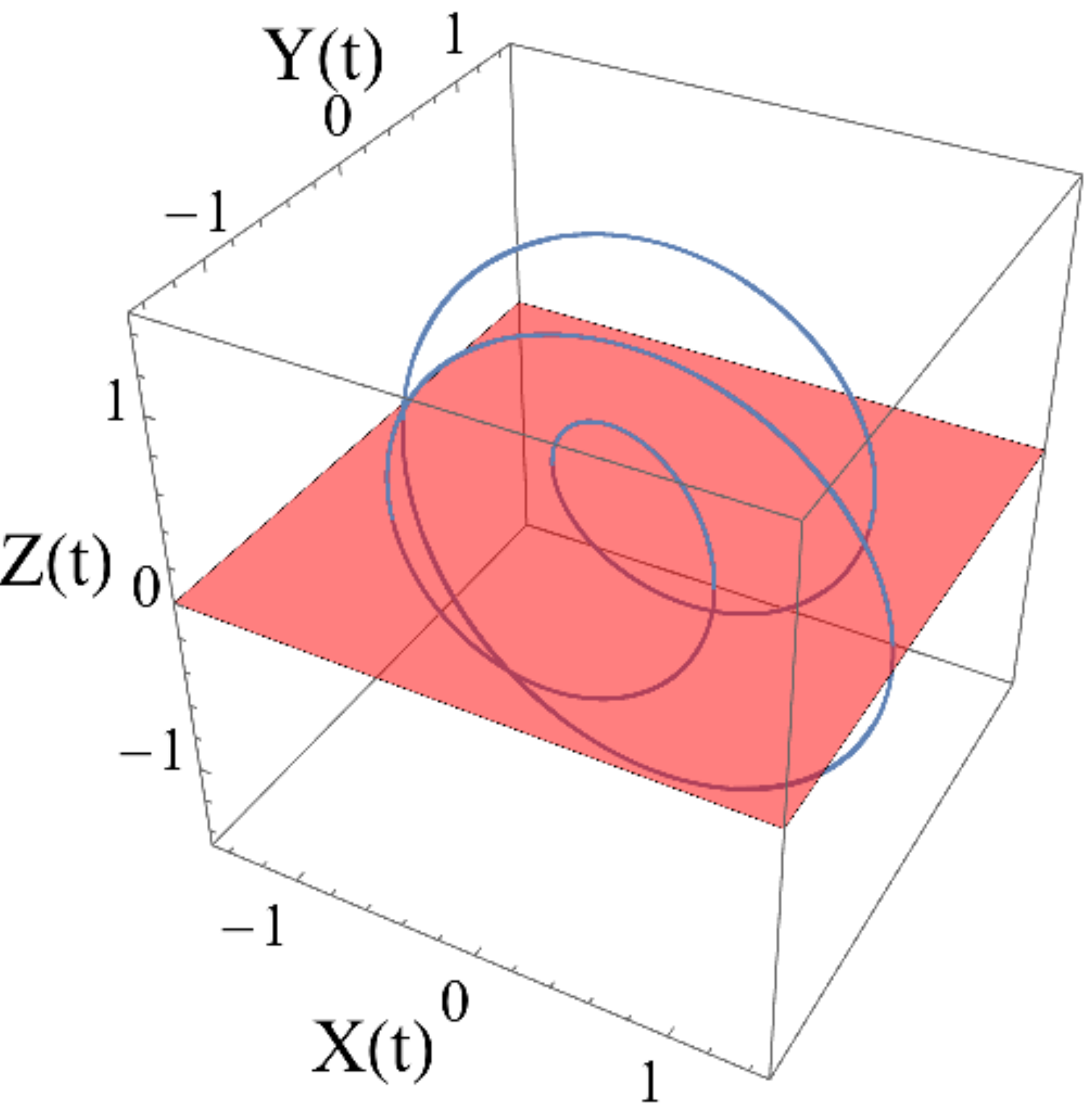}}\qquad
\subfloat[\qquad \textbf{(b)}]{\label{Traj_Sect_LC_Toy_X-Sect}\includegraphics[scale=0.225]{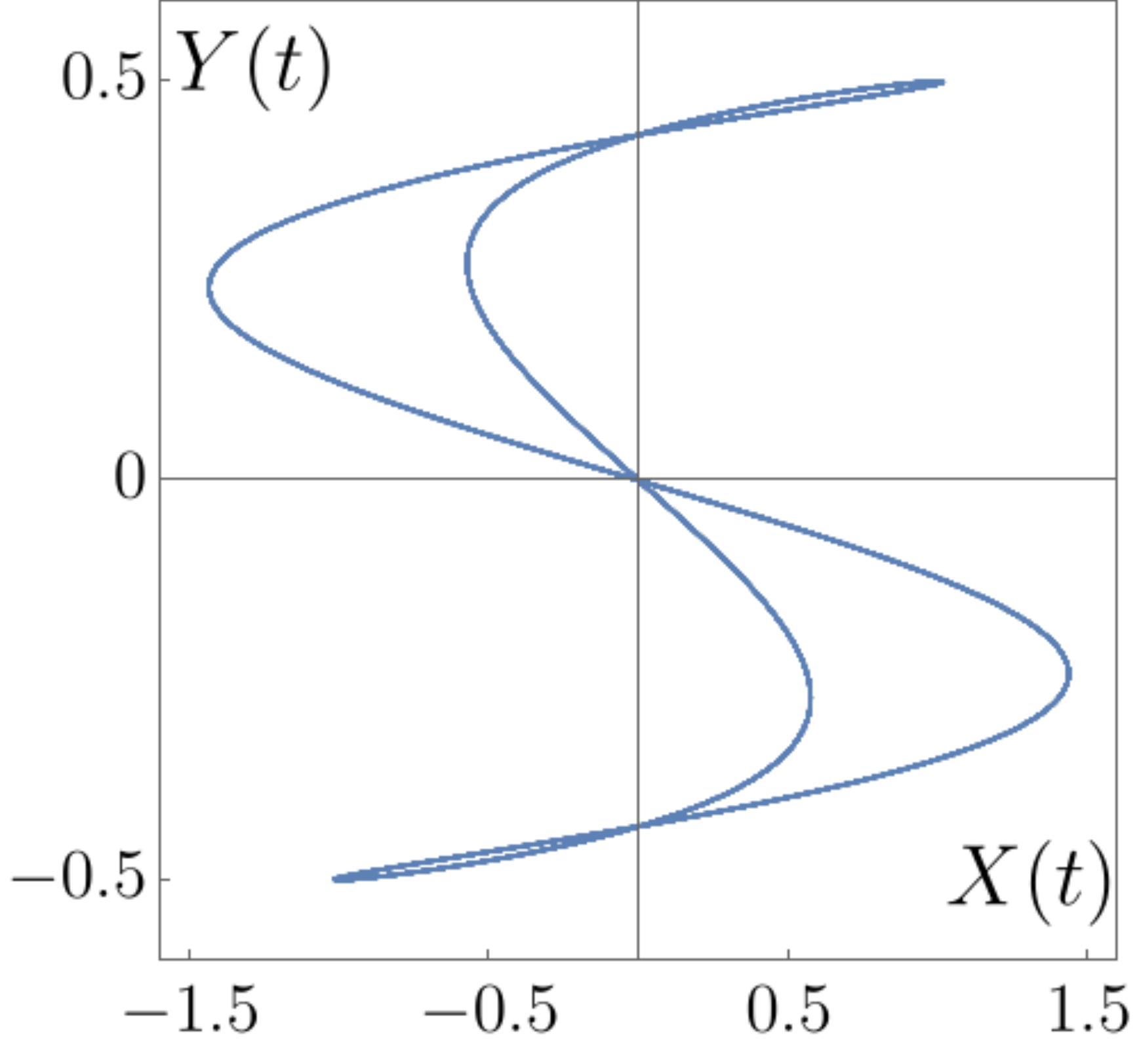}}\qquad
\subfloat[\qquad \textbf{(c)}]{\label{Poincare_LC_Toy}\includegraphics[scale=0.225]{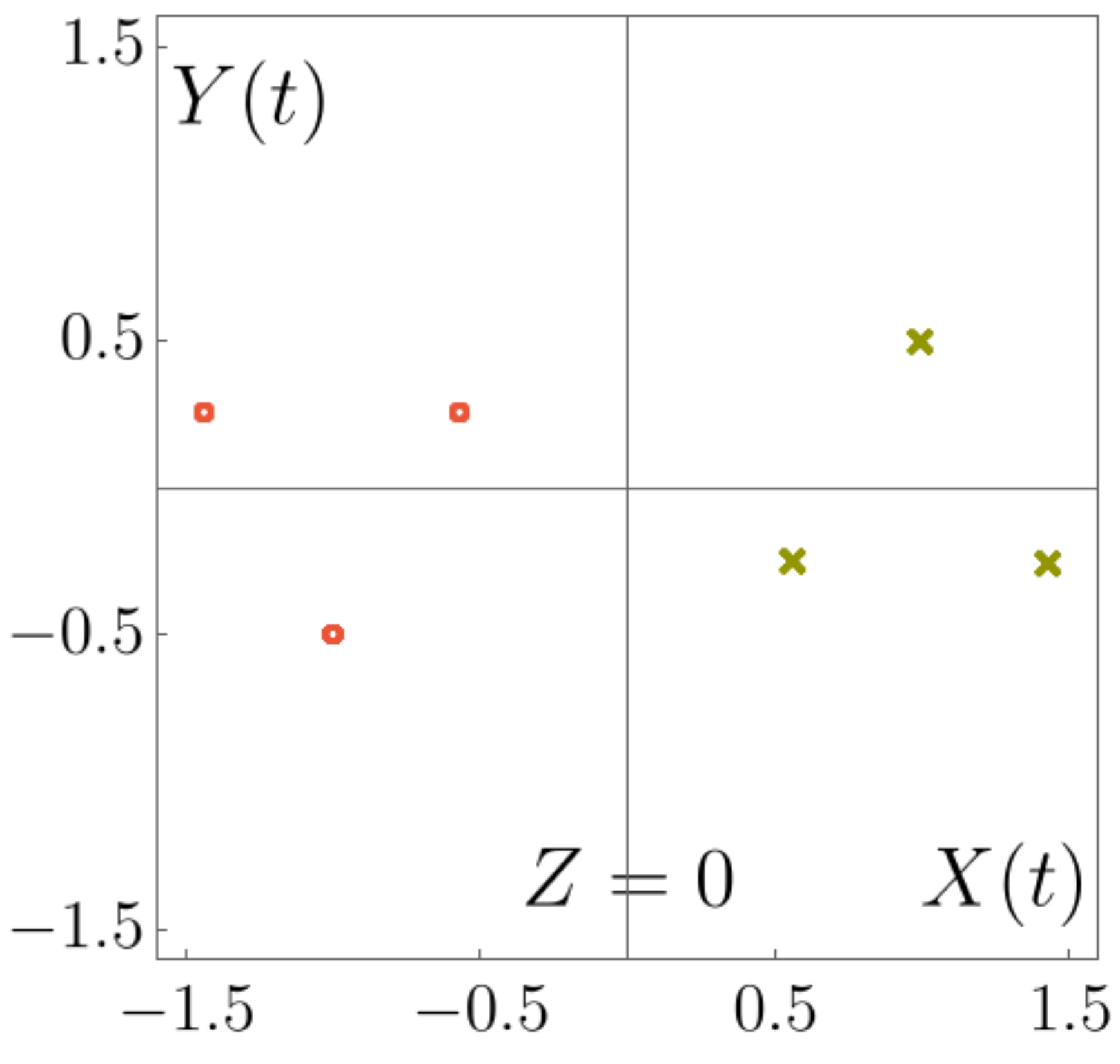}}\\
\caption{Simulation of \eref{T2} with $f_{1} = 1, f_{2} = 1/3, D_{1} = 1.0,$ and $D_{2} = 0.5$, depicting a periodic attractor. \textbf{(a)} The 3D trajectory with the transverse Poincar\'{e} plane (translucent red). This plane cuts the trajectory in two halves, and gives rise to the Poincar\'{e} sections. \textbf{(b)} The projection on the $X-Y$ plane is a single lined (self-intersecting) loop. The self-intersection is a result of the reduction of the number of dimensions. \textbf{(c)} We show two Poincar\'{e} sections -- a collection of a finite number of disconnected points -- generated by the orbit traversing the transverse plane from above and below in green crosses and red circles, respectively. Note the similarity with \fsref{Poincare_SLC} and \ref{Poincare_SBLC}. (For interpretation of the references to color in this figure legend, the reader is referred to the web version of this article.)}
\label{LC_Toy}

\end{figure*}

\begin{figure*}[tbp!]
\centering
\subfloat[\textbf{(a)}]{\label{Traj_Sect_QP_Toy}\includegraphics[scale=0.225]{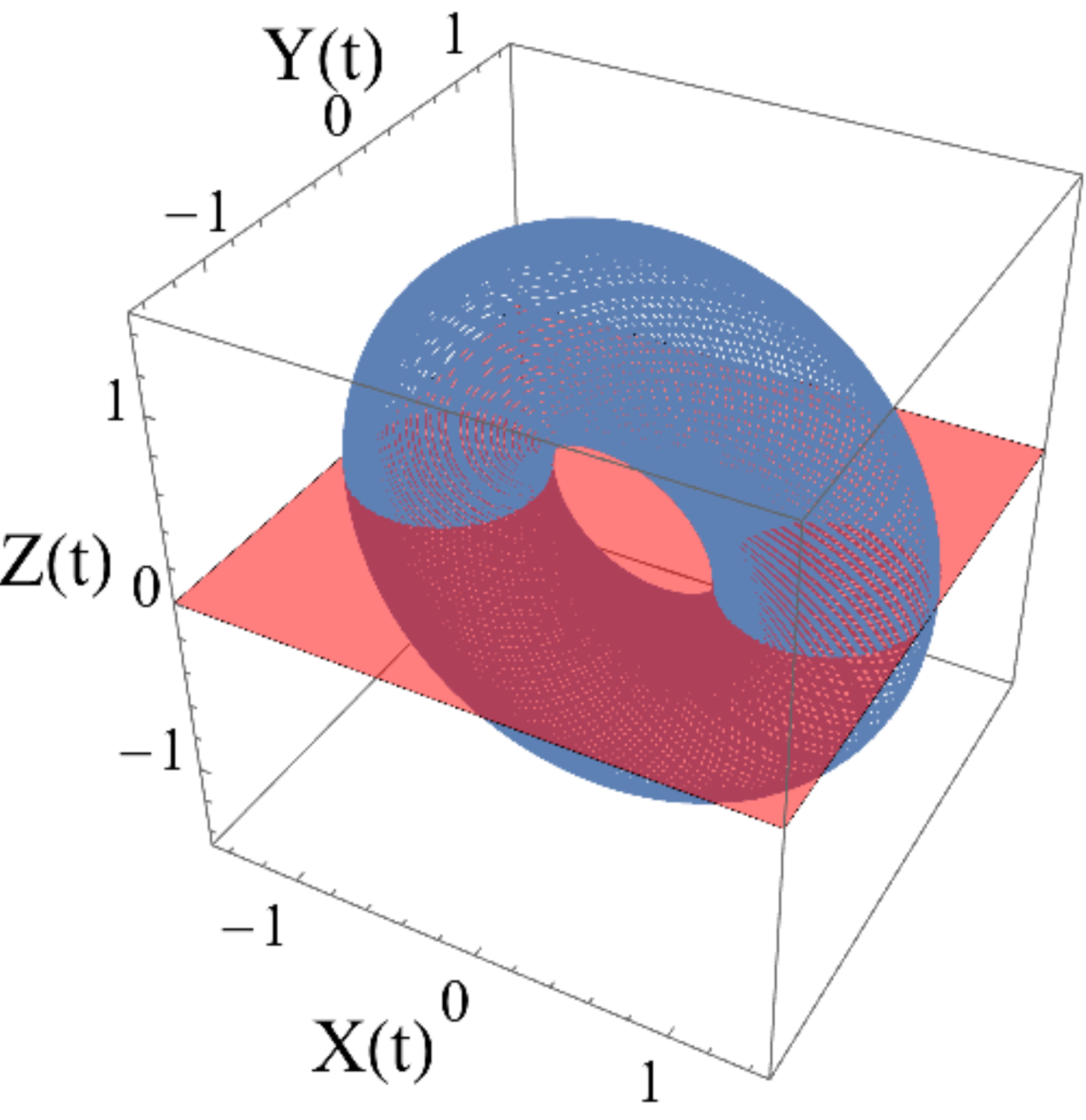}}\qquad
\subfloat[\qquad \textbf{(b)}]{\label{Traj_Sect_QP_Toy_X-Sect}\includegraphics[scale=0.225]{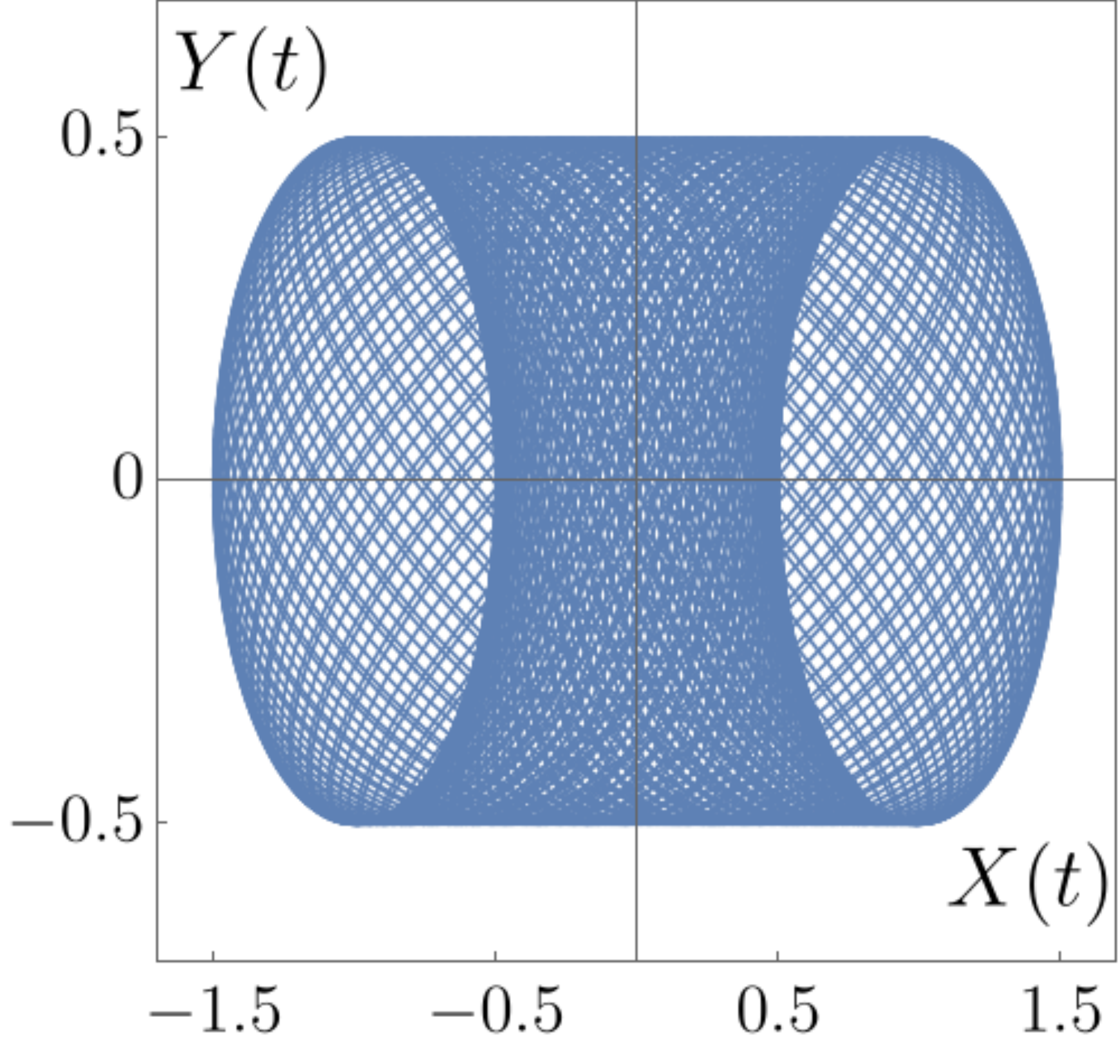}}\qquad
\subfloat[\qquad \textbf{(c)}]{\label{Poincare_QP_Toy}\includegraphics[scale=0.225]{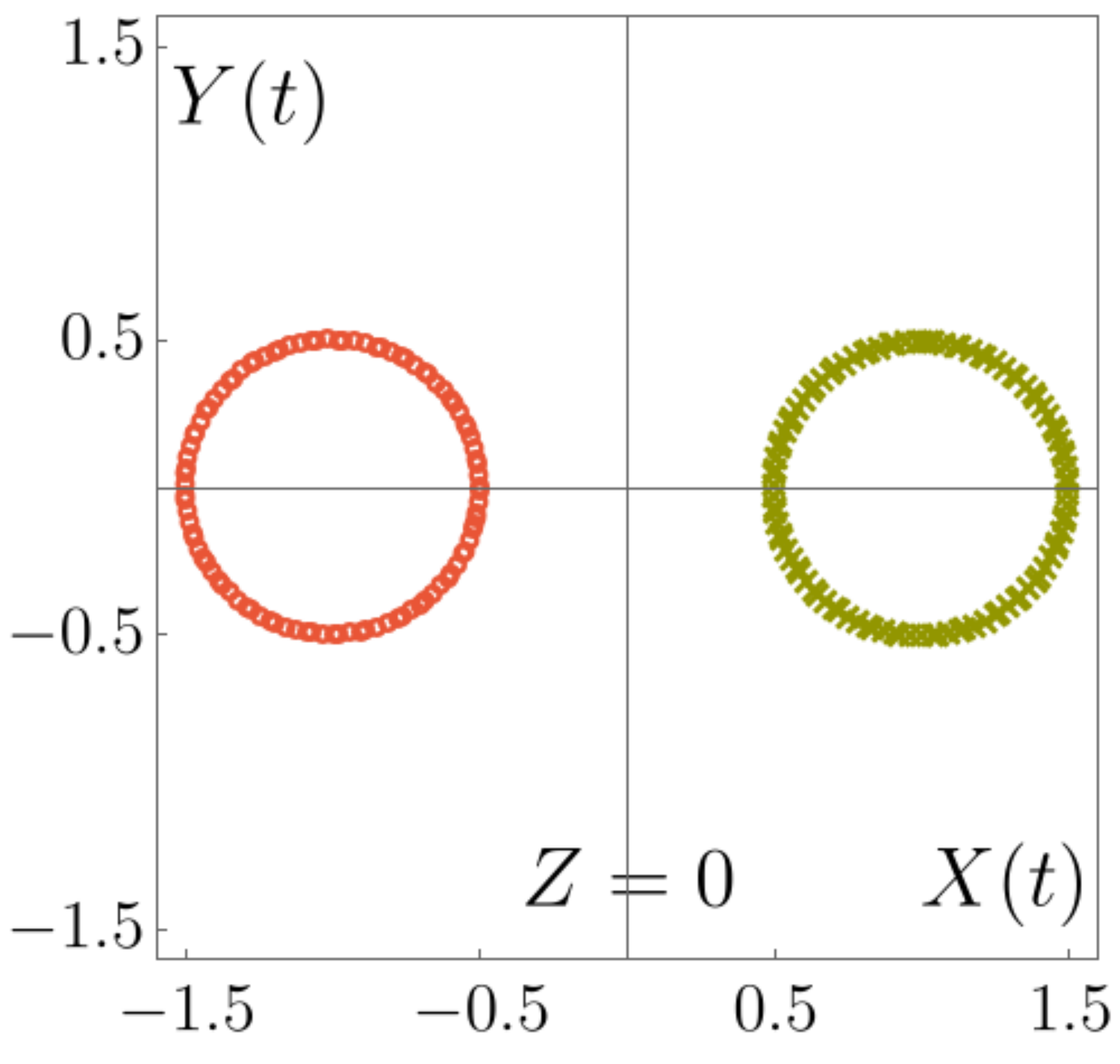}}
\caption{Quasiperiodicity with \eref{T2} for $f_{1} = 1, f_{2} = \sqrt{2}, D_{1} = 1.0,$ and $D_{2} = 0.5$. All the conventions are the same as in \fref{LC_Toy}. \textbf{(a)} The 3D trajectory with the transverse Poincar\'{e} plane. \textbf{(b)} The projection on the $X-Y$ plane. Note the qualitative similarity with \fref{QP_X_Sect}. \textbf{(c)} Two distinct Poincar\'{e} sections. Unlike the ones in \fref{Poincare_LC_Toy}, the Poincar\'{e} sections are two continuous circles. This is qualitatively similar to \fsref{Poincare_QP1} and \ref{Poincare_QP2}. (For interpretation of the references to color in this figure legend, the reader is referred to the web version of this article.)}
\label{QP_Toy}

\end{figure*}

\begin{figure*}[tbp!]
\centering
\subfloat[\qquad \textbf{(a)}]{\label{Sup_NS_SZA_vs_SZB_1}\includegraphics[scale=0.21]{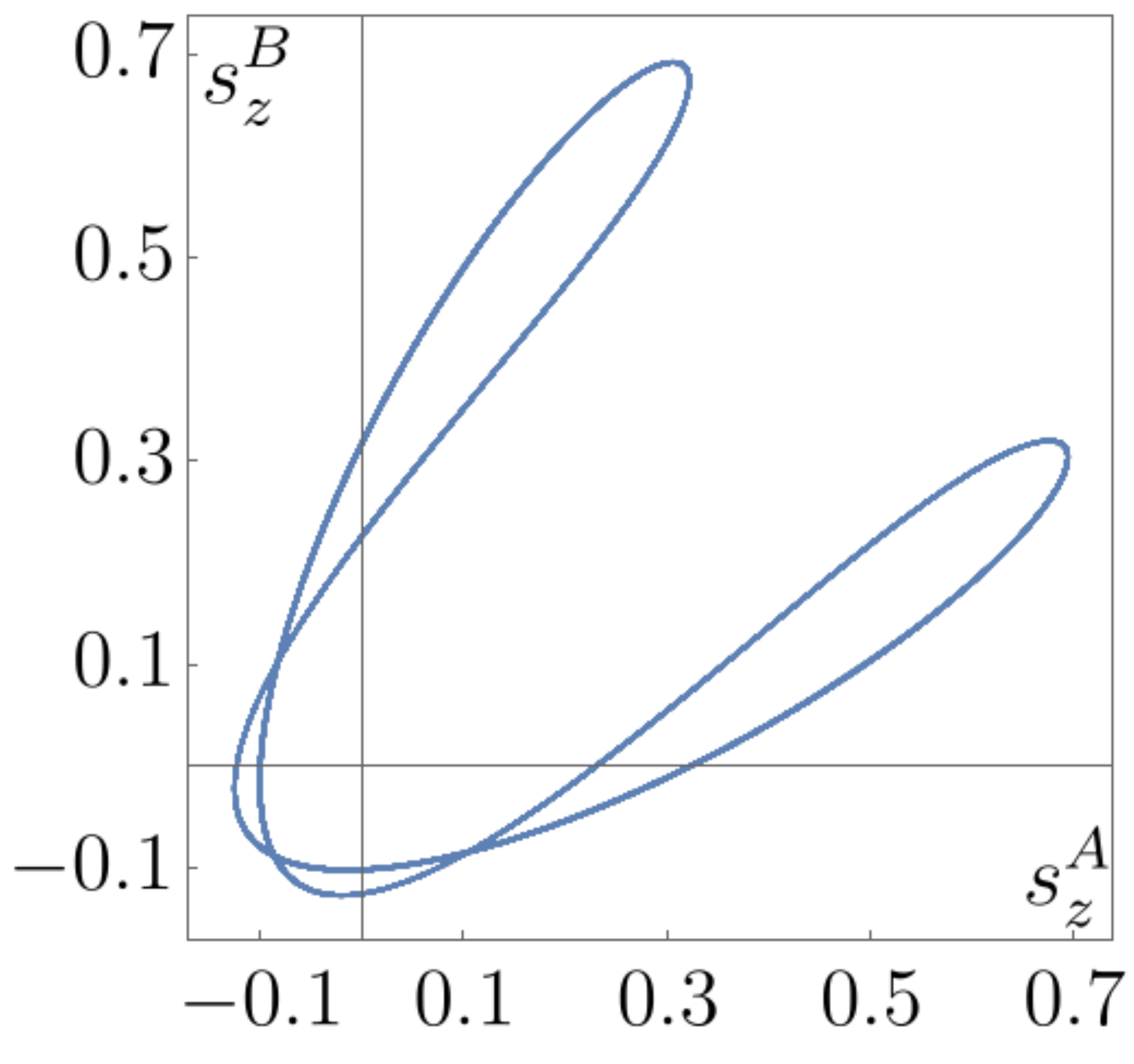}}\quad
\subfloat[\qquad \textbf{(b)}]{\label{Sup_NS_SZA_vs_SZB_2}\includegraphics[scale=0.21]{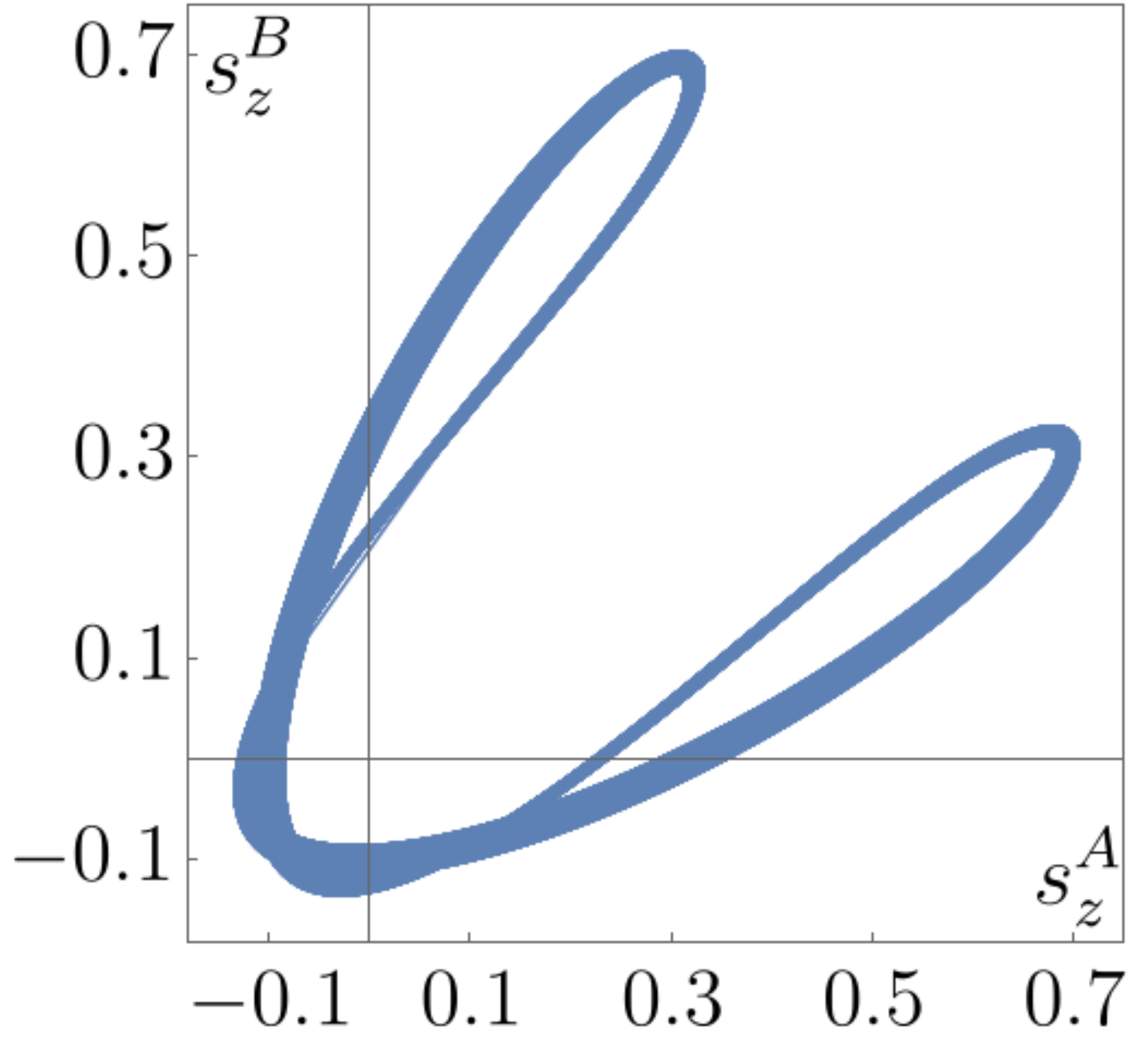}}\quad
\subfloat[\qquad \textbf{(c)}]{\label{Sup_NS_SZA_vs_SZB_3}\includegraphics[scale=0.21]{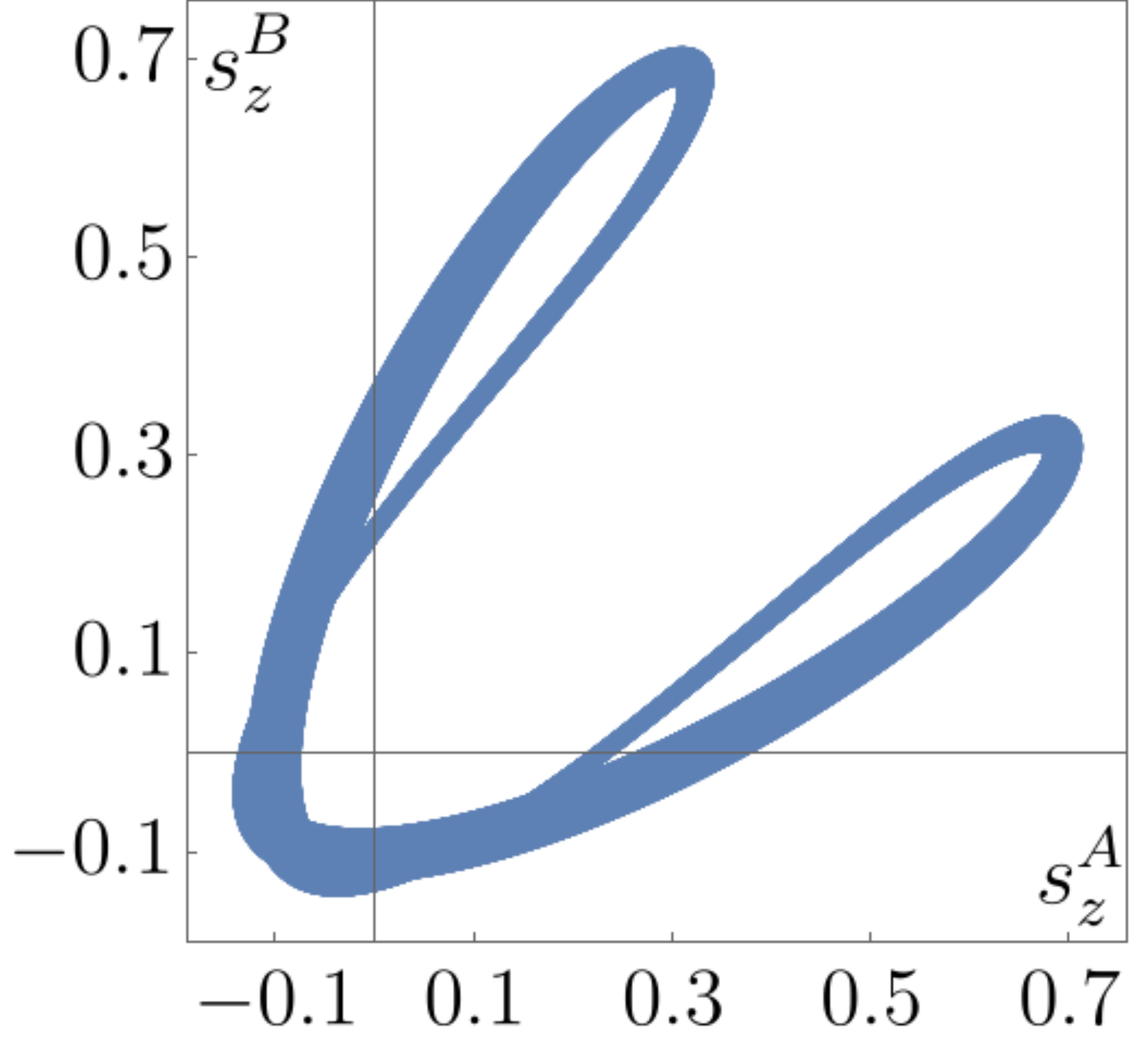}}
\caption{Gradual thickening of the $s_{z}^{A} - s_{z}^{B}$ projection showing supercritical Neimark-Sacker bifurcation at $(\delta \approx 0.2634, W = 0.1484)$. The projection in \textbf{(a)} belongs to a non-$\Z2$-symmetric limit cycle at $(\delta = 0.2635, W = 0.1484)$, whereas the other two correspond to quasiperiodic attractors. The parameters in \textbf{(b)} and \textbf{(c)} are $(\delta = 0.2633, W = 0.1484)$ and $(\delta = 0.2631, W = 0.1484)$, respectively.}
\label{Sup_NS}

\end{figure*}

\begin{figure*}[tbp!]
\centering
\subfloat[\qquad\textbf{(a)}]{\includegraphics[scale=0.3]{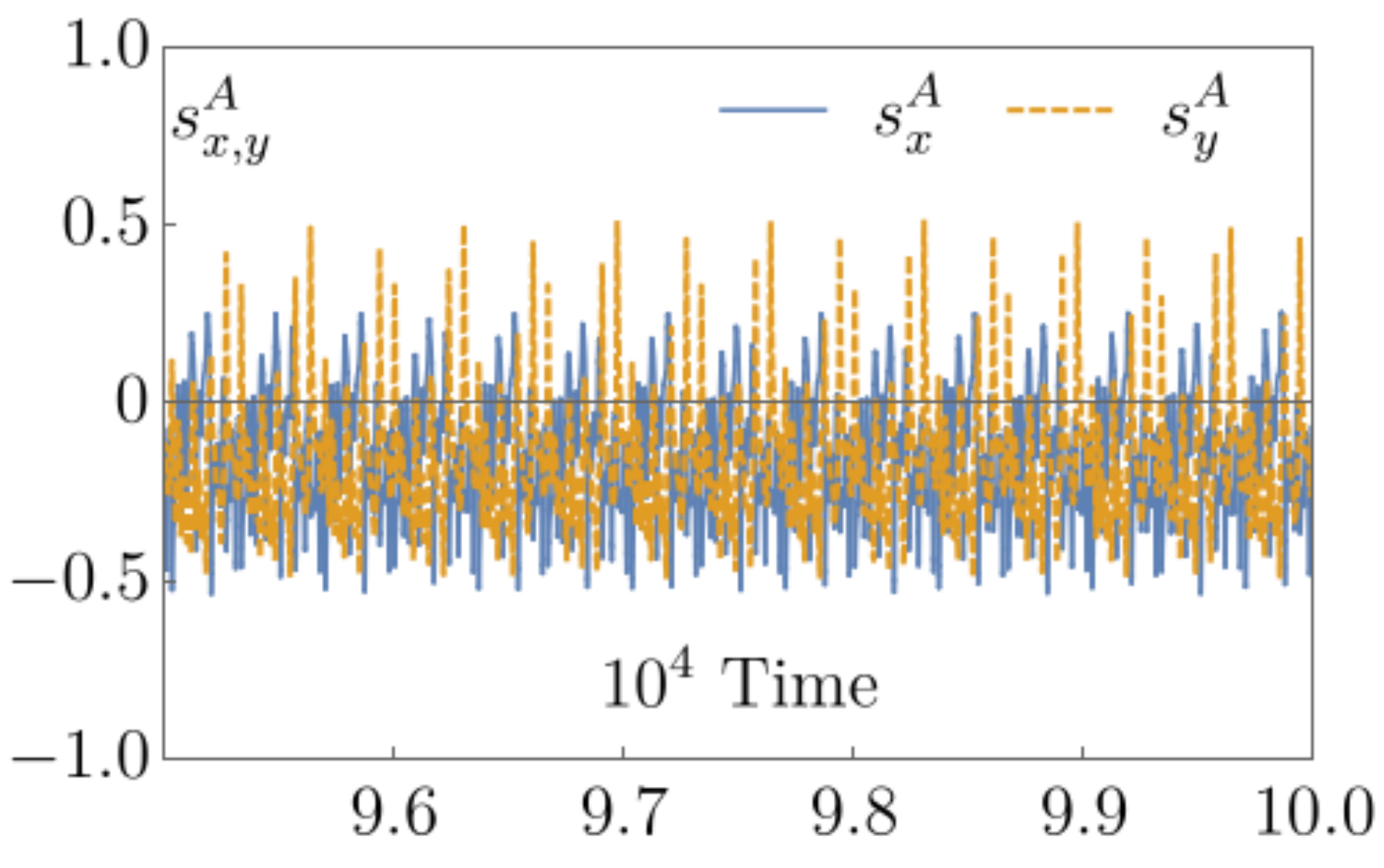}}\qquad\qquad
\subfloat[\qquad\textbf{(b)}]{\includegraphics[scale=0.3]{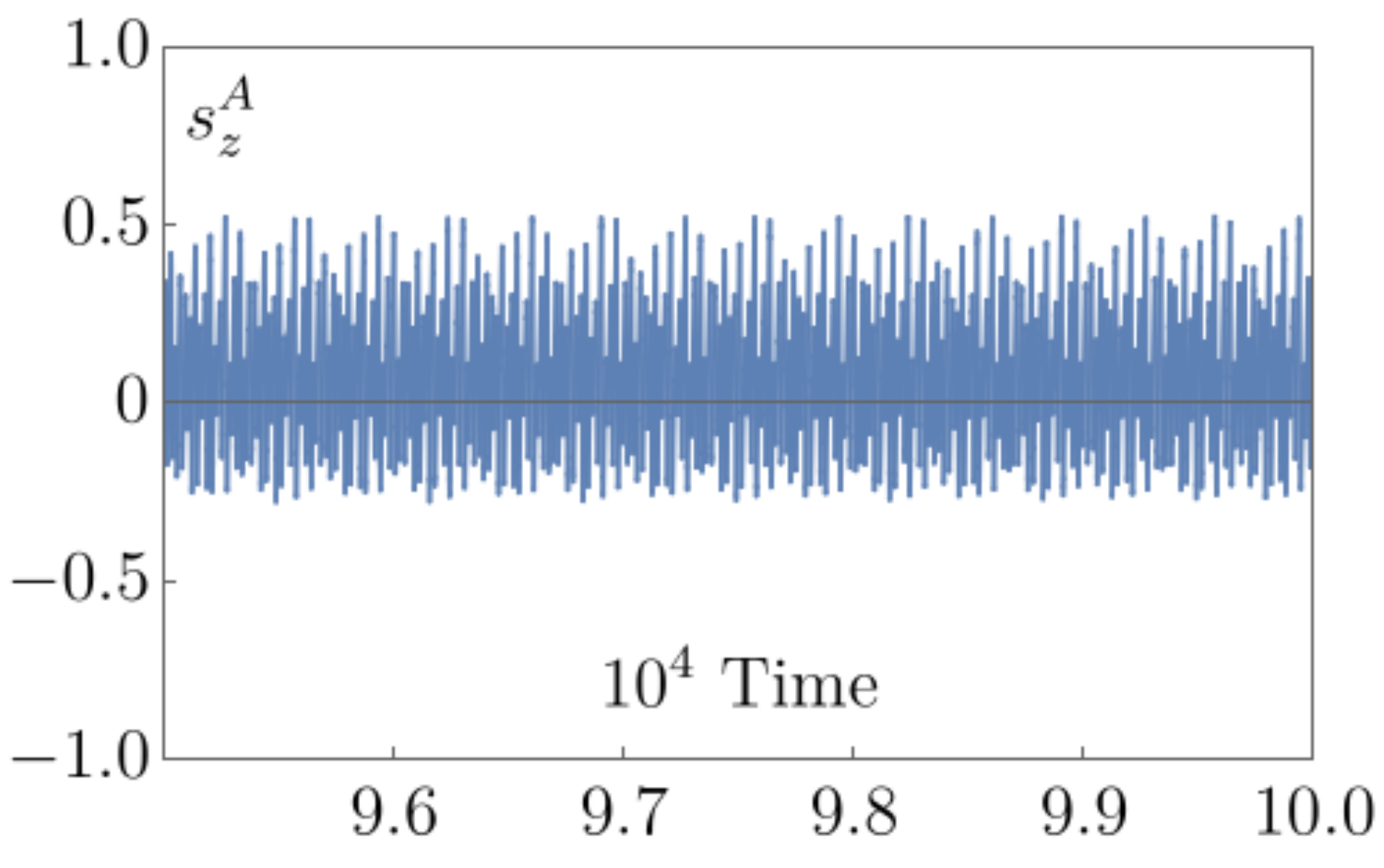}}\\
\subfloat[\qquad\textbf{(c)}]{\includegraphics[scale=0.3]{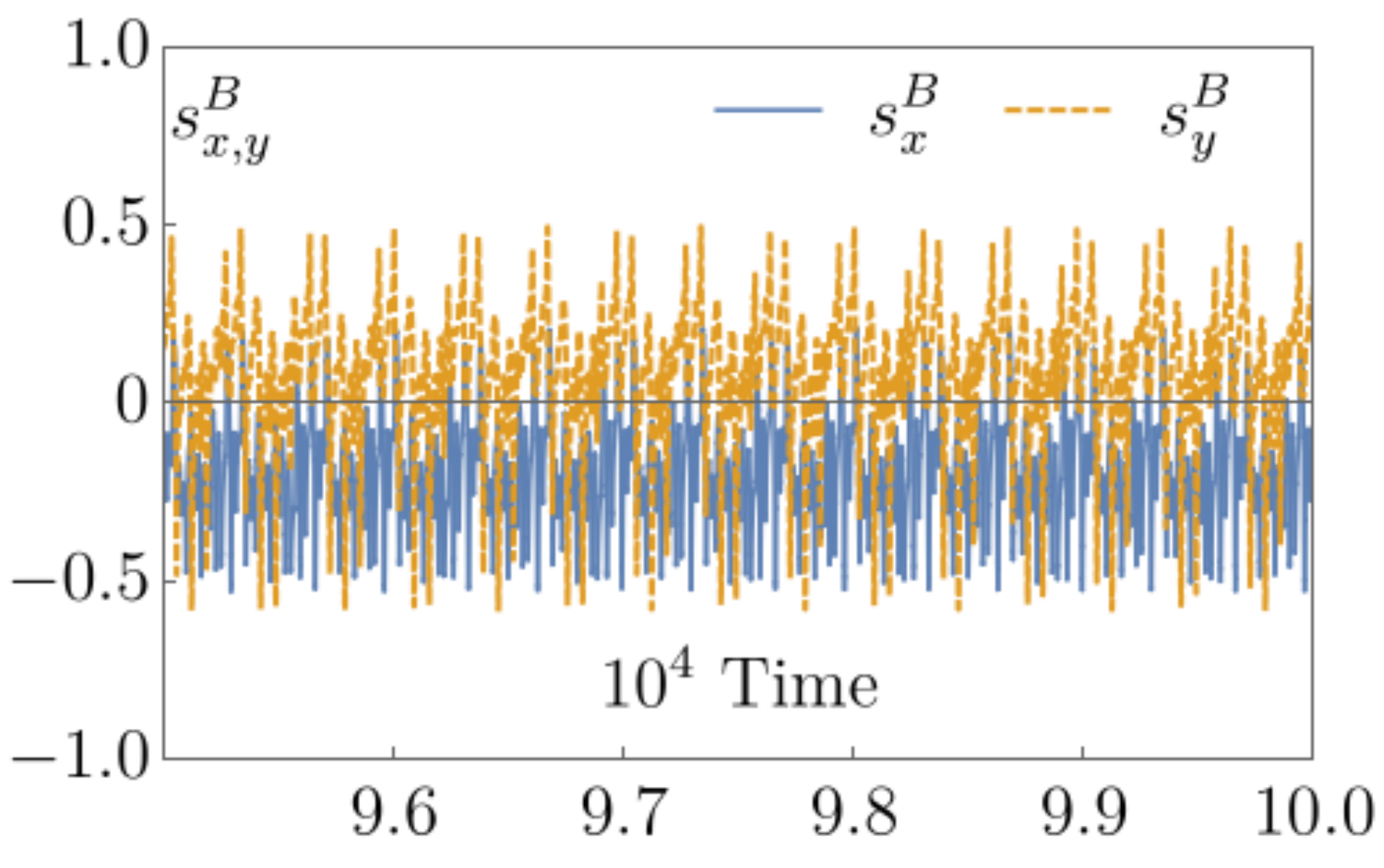}}\qquad\qquad
\subfloat[\qquad\textbf{(d)}]{\includegraphics[scale=0.3]{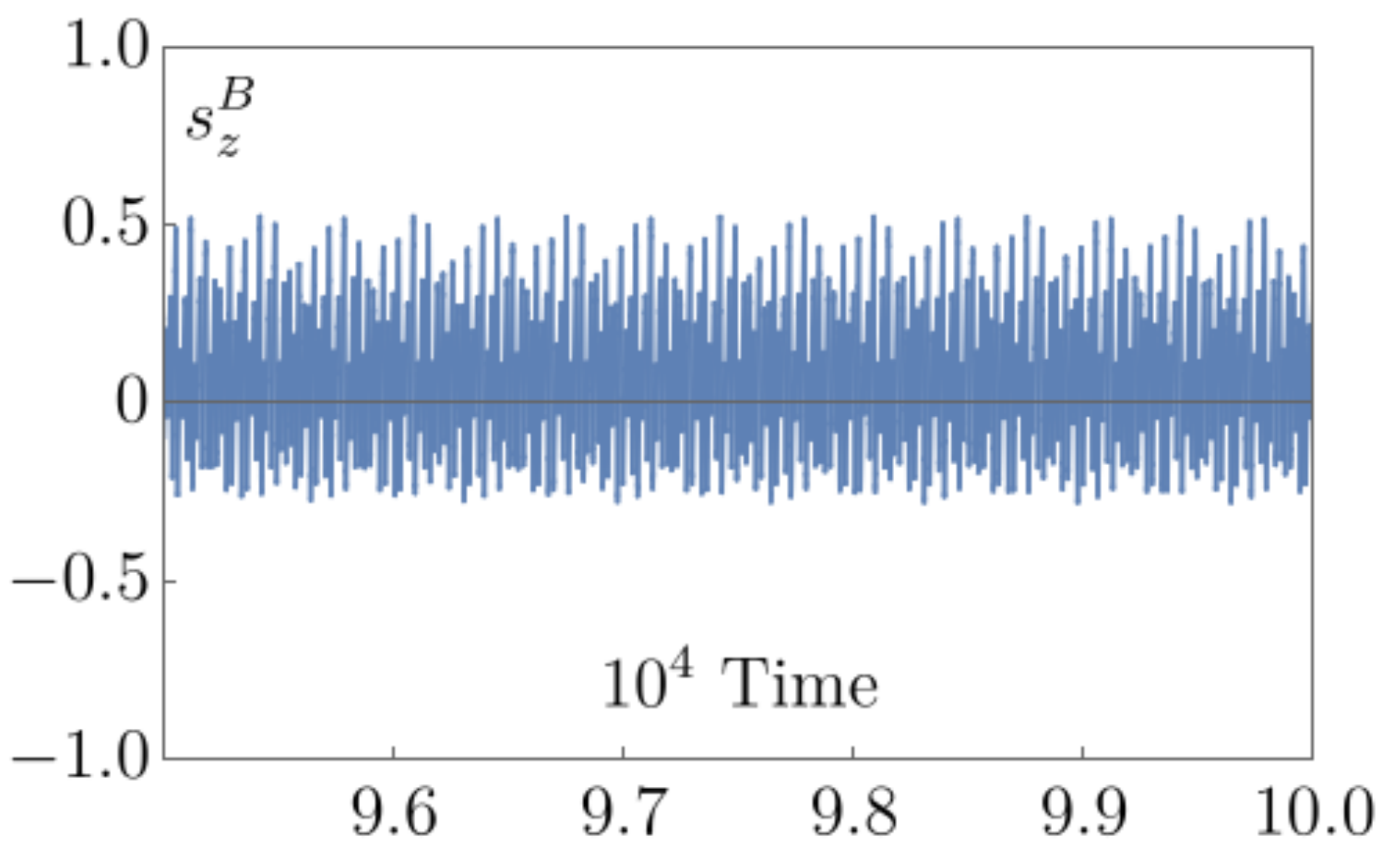}}\\
\subfloat[\qquad\textbf{(e)}]{\includegraphics[scale=0.3]{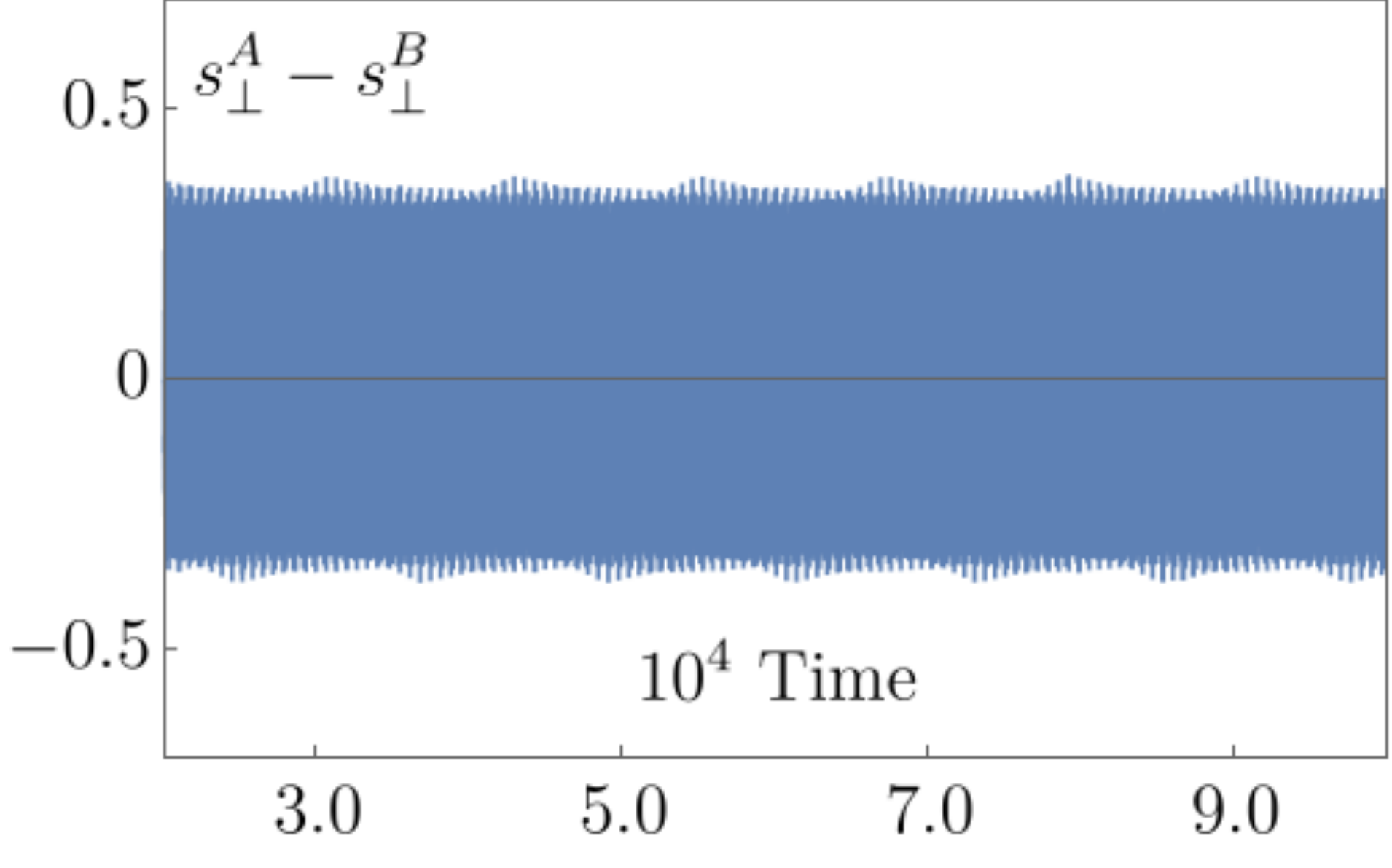}}\llap{\raisebox{2.4cm}{\includegraphics[height=2.0cm]{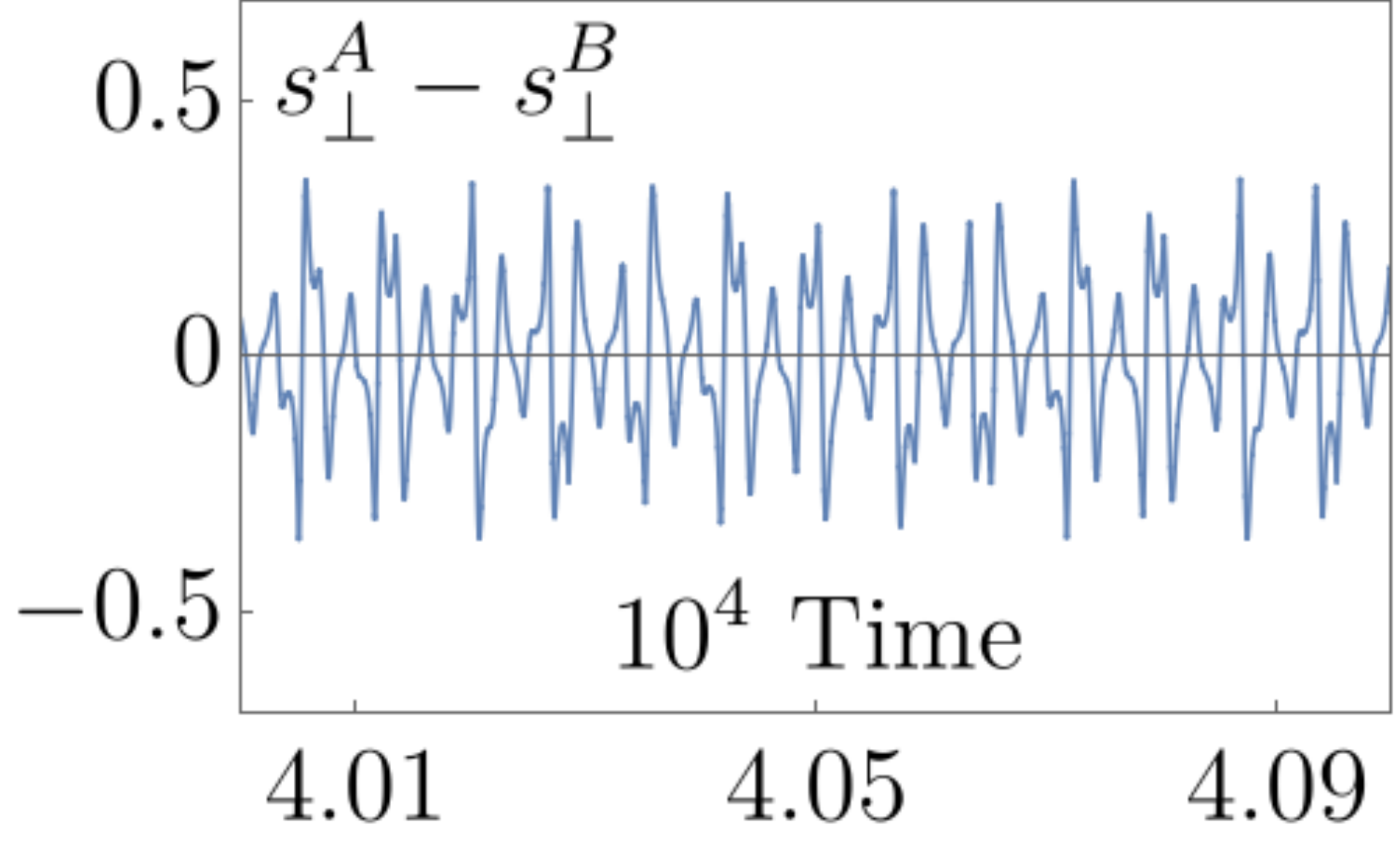}}}\qquad\qquad
\subfloat[\qquad\textbf{(f)}]{\includegraphics[scale=0.3]{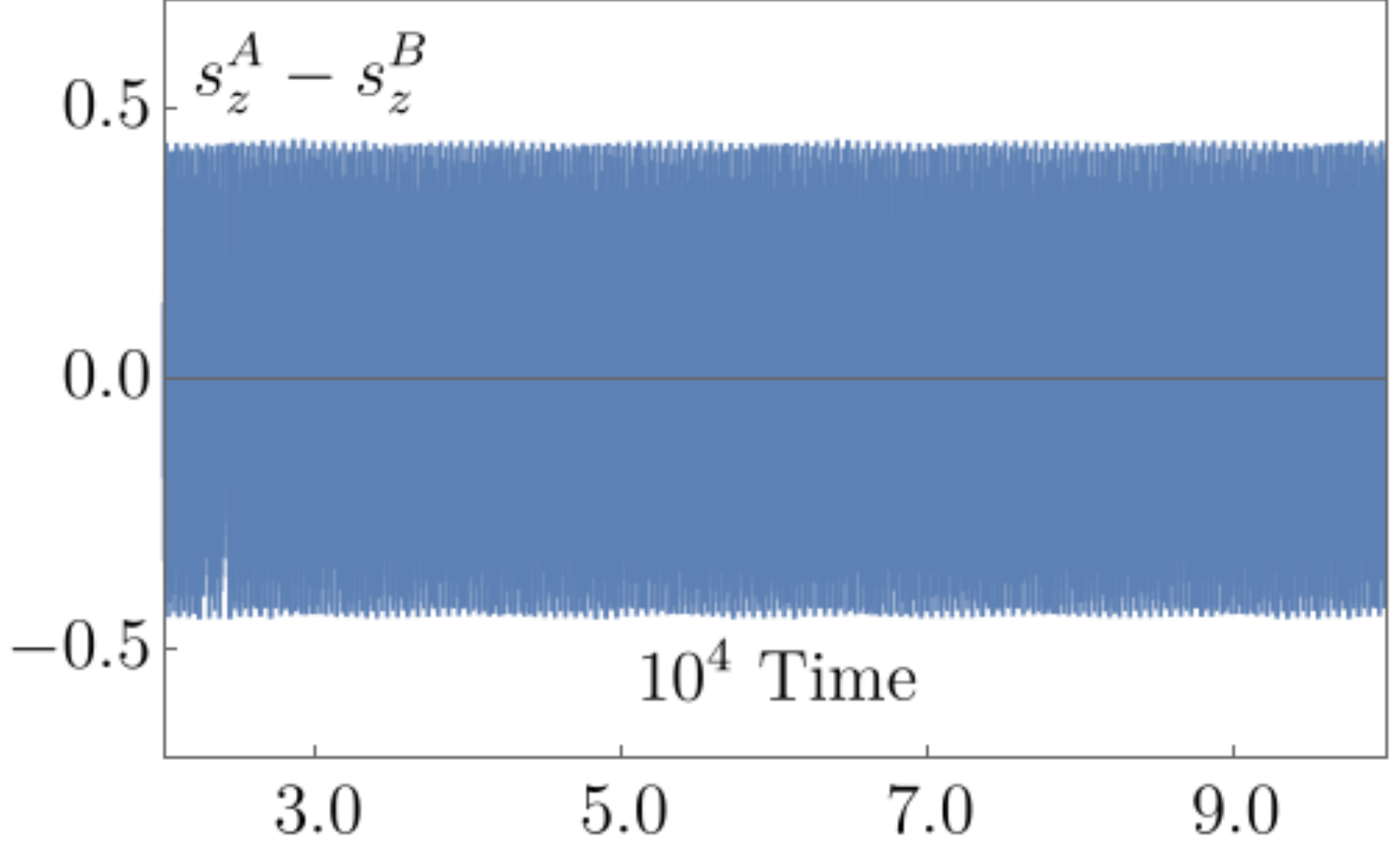}}\llap{\raisebox{2.4cm}{\includegraphics[height=2.0cm]{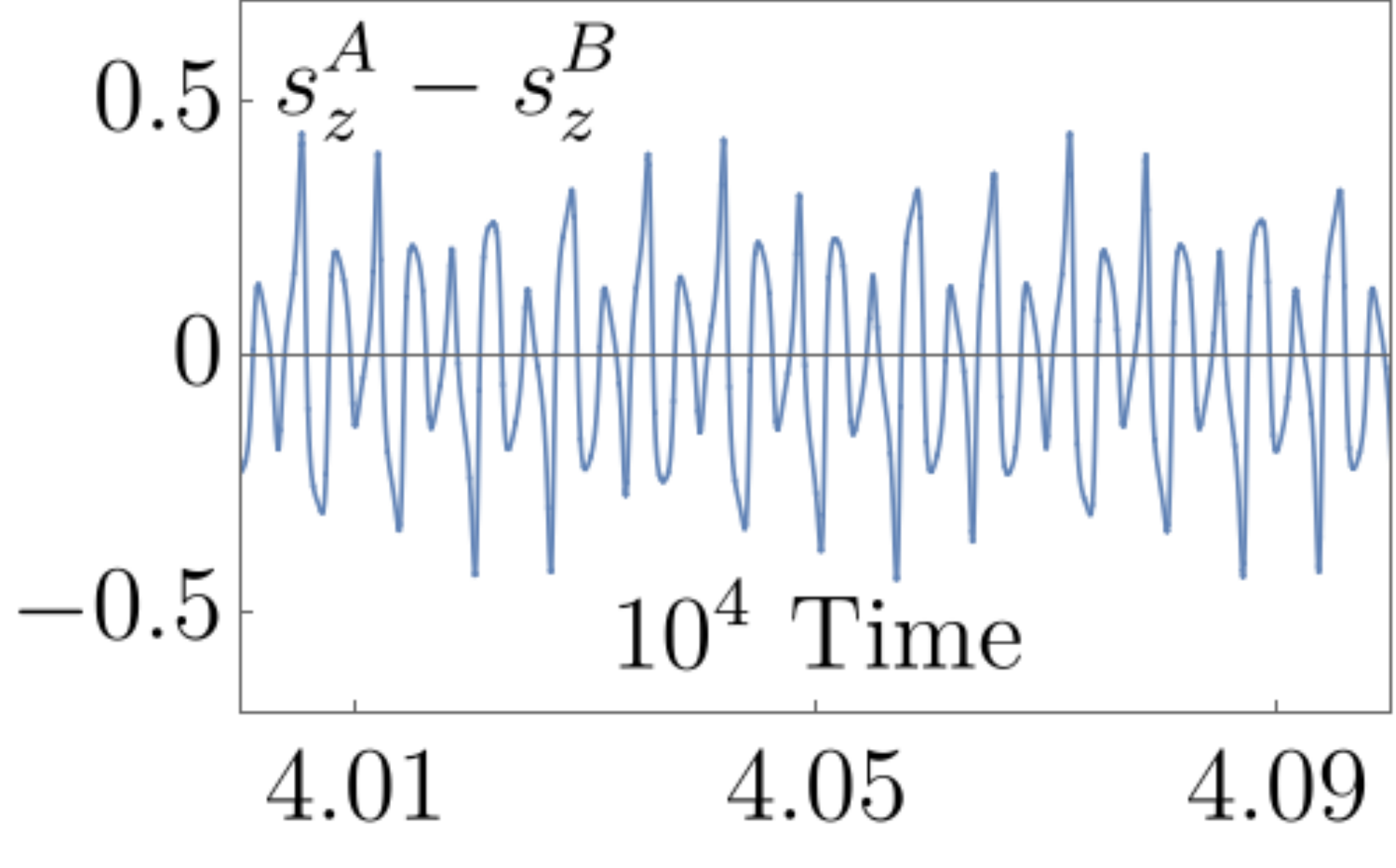}}}
\caption{Evolution of the collective spin operators $\bm{s}^{\tau}$ ($\tau = A, B$), for a quasiperiodic trajectory at $\delta = 0.115, W = 0.055$. In the corresponding spectrum for the radiated light, the bunching of discrete subpeaks due to the second incommensurate frequency around the dominant main harmonics is clearly seen in \fref{Spectrum_QP}. In the first and second row we show the time-variation of the three components of $\bm{s}^{A}$ and $\bm{s}^{B},$ respectively. In \textbf{(e)}, we show $(s_{\perp}^{A} - s_{\perp}^{B})$ vs. time, where $s_{\perp}^{\tau}$ is defined in the introduction. In \textbf{(f)}, we show $(s_{z}^{A} - s_{z}^{B})$ vs. time. The insets of \textbf{(e)} and \textbf{(f)} underscores the aperiodic nature of the attractor.}
\label{QP_Pictures}
\end{figure*}

\begin{figure*}[tbp!]
\centering
\subfloat[\qquad\textbf{(a)}]{\label{QP_SA_vs_SB}\includegraphics[scale=0.3]{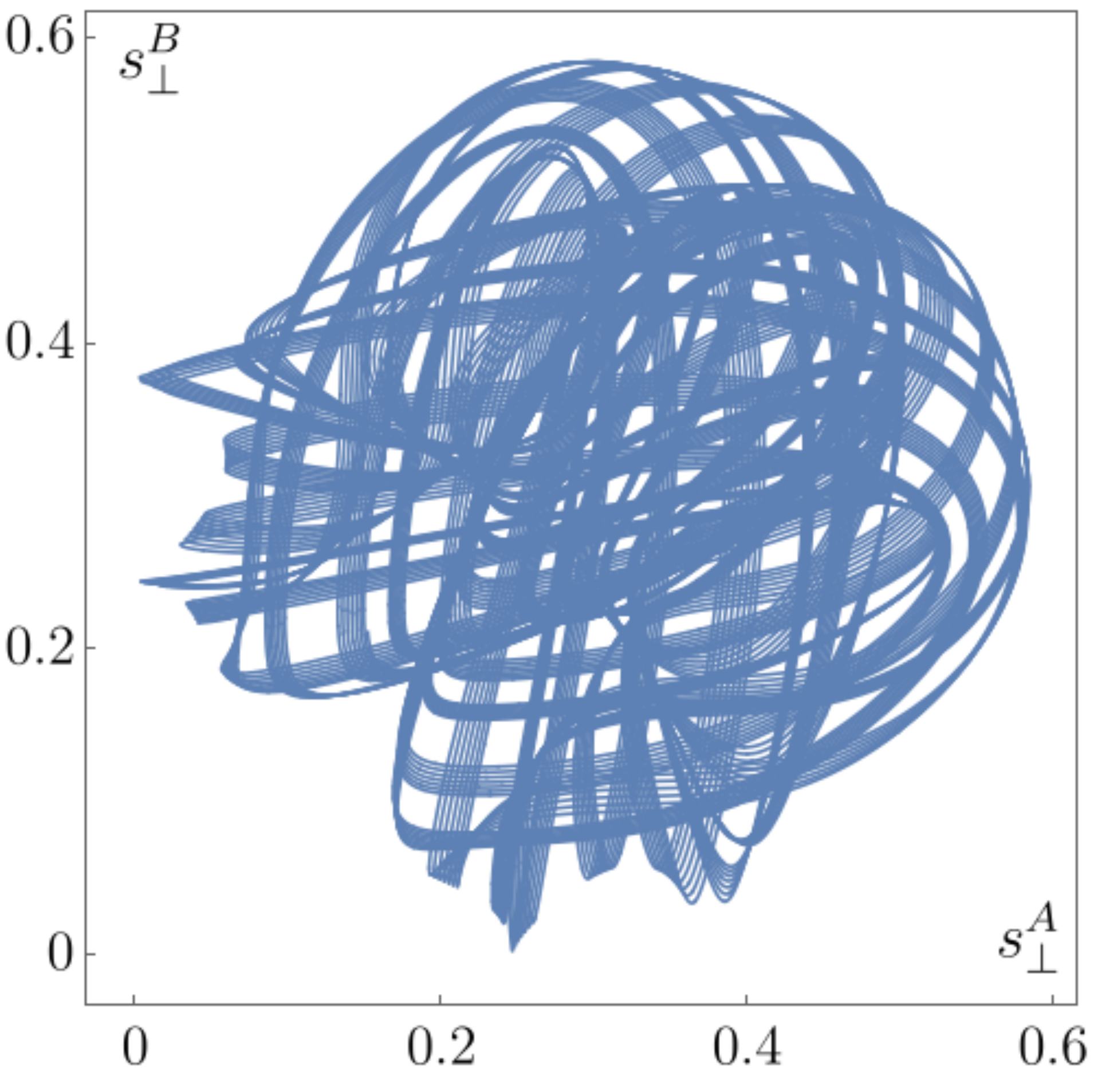}}\qquad\qquad
\subfloat[\qquad\textbf{(b)}]{\label{QP_SZA_vs_SZB}\includegraphics[scale=0.3]{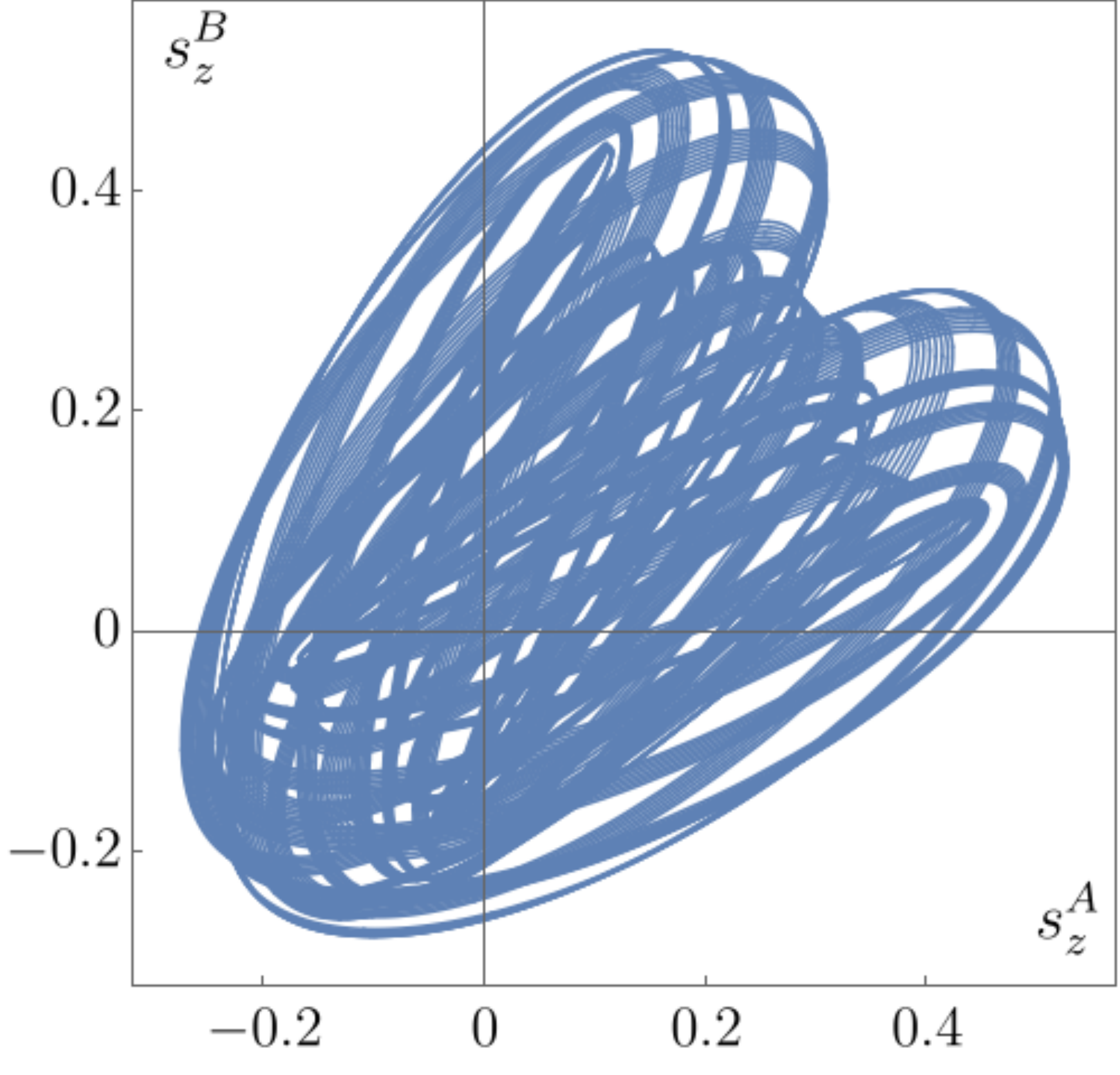}}
\caption{Different 2D projections of the 6D quasiperiodic trajectory. $\delta$ and $W$ are the same as in \fref{QP_Pictures}. In \textbf{(a)} and \textbf{(b)} we plot $s_{\perp}^{A}$ vs. $s_{\perp}^{B}$, and $s_{z}^{A}$ vs. $s_{z}^{B}$, respectively. Unlike the limit cycles, the lines here have acquired finite thickness. In the corresponding Poincar\'{e} section this is clearly demonstrated, see \fsref{Traj_Sect_QP2} and \ref{Poincare_QP2}.}
\label{QP_X_Sect}
\end{figure*}

\begin{figure*}[tbp!]
\centering
\subfloat[\textbf{(a)}]{\label{Traj_Sect_SLC}\includegraphics[scale=0.35]{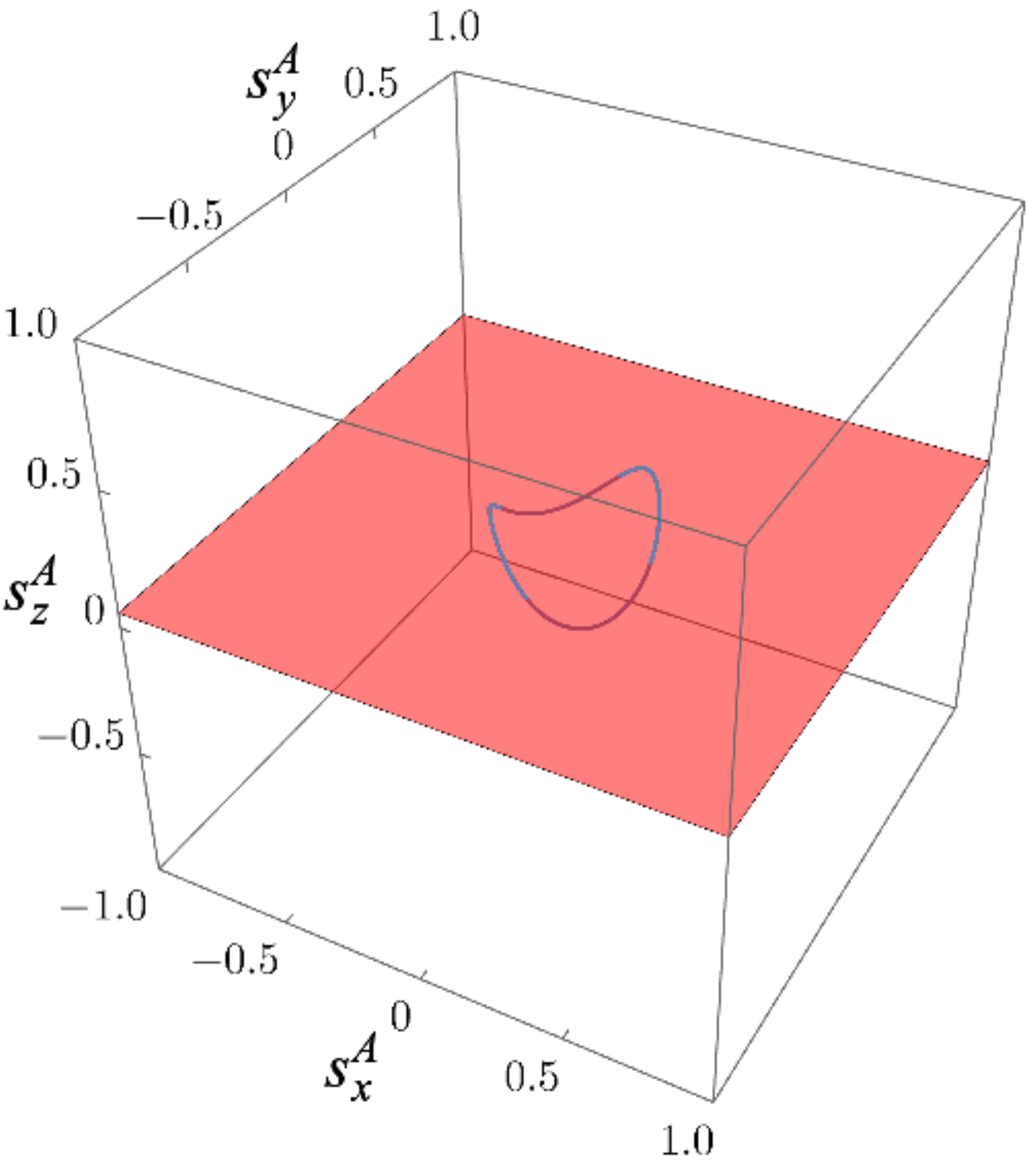}}\qquad\qquad
\subfloat[\qquad\textbf{(b)}]{\label{Poincare_SLC}\includegraphics[scale=0.33]{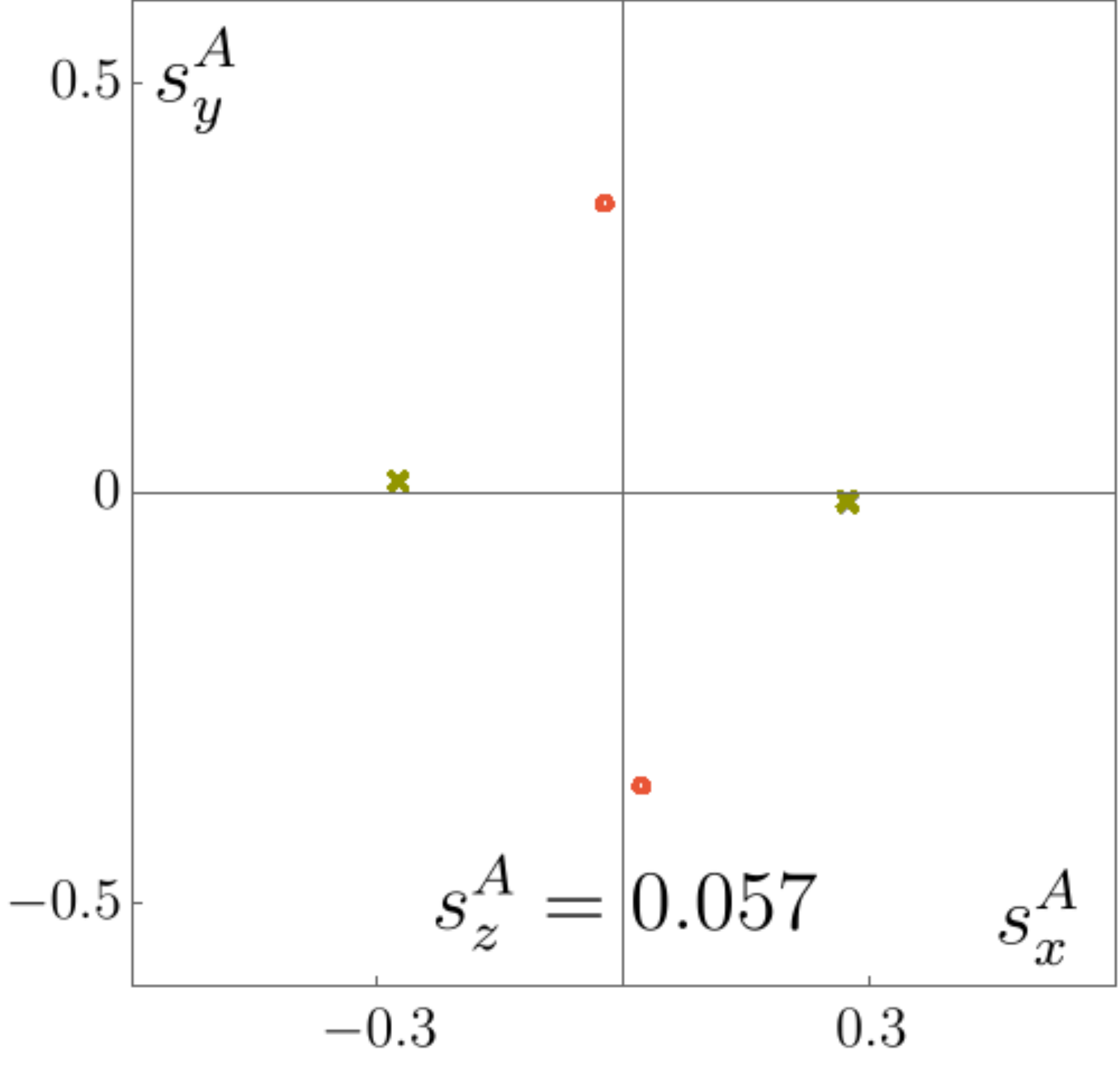}}\\
\subfloat[\textbf{(c)}]{\label{Traj_Sect_SBLC}\includegraphics[scale=0.35]{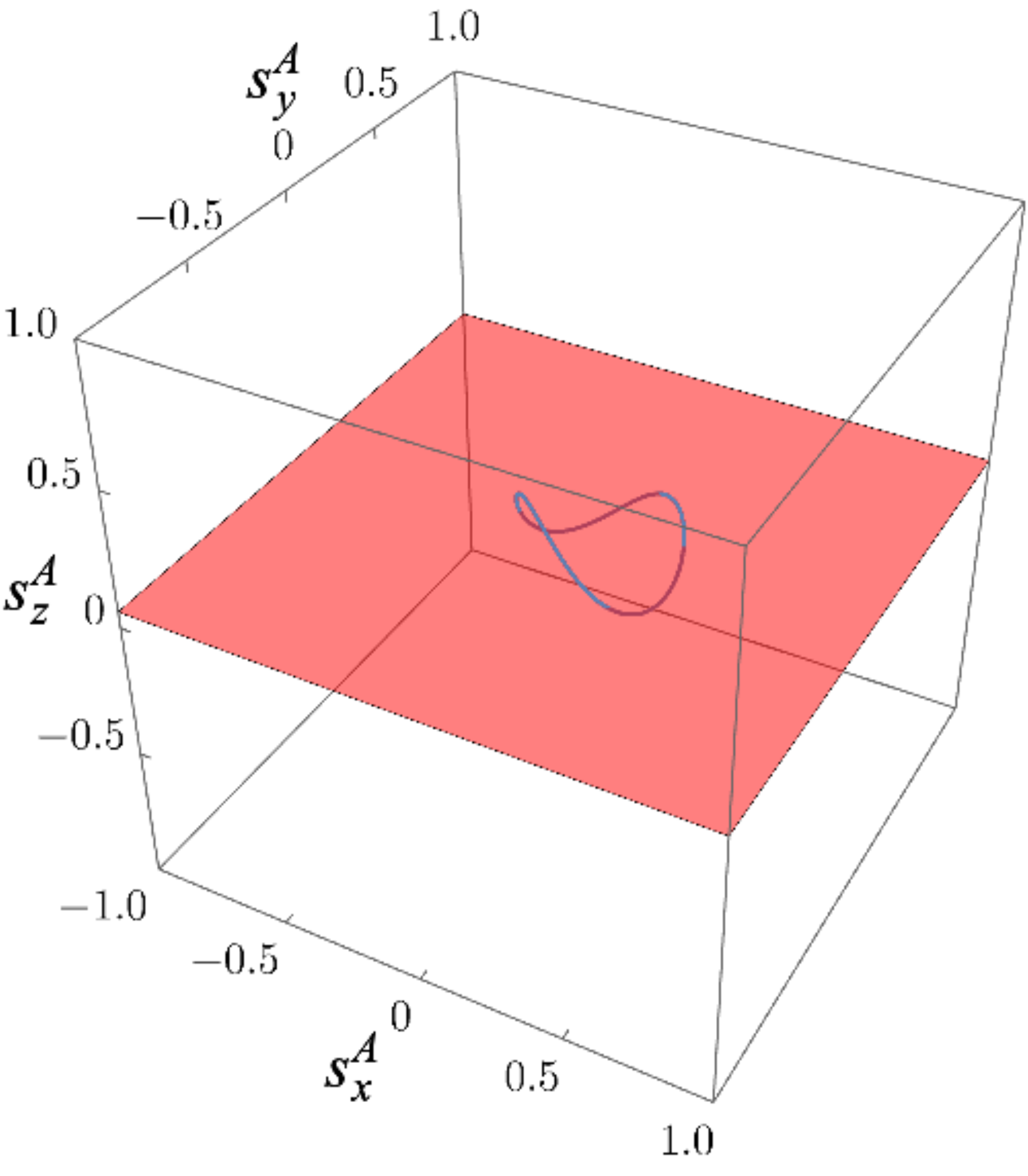}}\qquad\qquad
\subfloat[\qquad\textbf{(d)}]{\label{Poincare_SBLC}\includegraphics[scale=0.33]{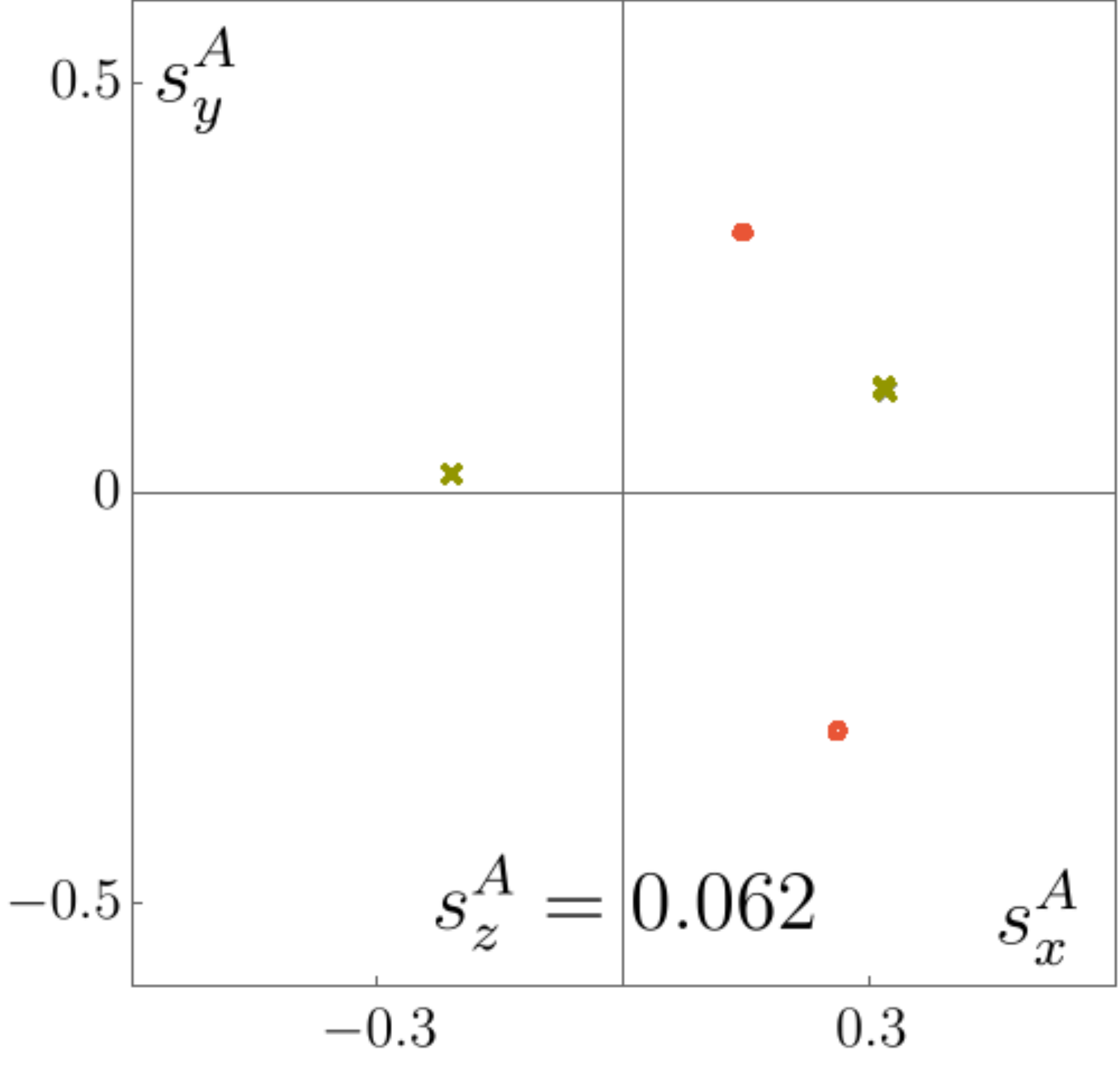}}
\caption{Poincar\'{e} sections for the periodic attractors with $\bm{s}^{A}$. Conventions are the same as in \fsref{Poincare_LC_Toy} and \ref{Poincare_QP_Toy}. In the first column, along with the trajectories traced by the tip of the vector $\bm{s}^{A}$, we show the transverse planes $(s_{z}^{A} = \textrm{const.})$ corresponding to the Poincar\'{e} sections. In the second column, we show the two distinct (pertaining to the orbits' crossing the plane from either above or below) Poincar\'{e} sections. Note the qualitative similarity with \fref{Poincare_LC_Toy}. The figures in \textbf{(a)} and \textbf{(b)} correspond to the $\Z2$-symmetric limit cycle at $\delta = 0.44, W = 0.055$. On the other hand, \textbf{(c)} and \textbf{(d)} correspond to the $\Z2$ symmetry-broken limit cycle at $\delta = 0.41, W = 0.055$. (For interpretation of the references to color in this figure legend, the reader is referred to the web version of this article.)}
\label{Poincare_LC}
\end{figure*}

\begin{figure*}[tbp!]
\centering
\subfloat[\textbf{(a)}]{\label{Traj_Sect_QP1}\includegraphics[scale=0.35]{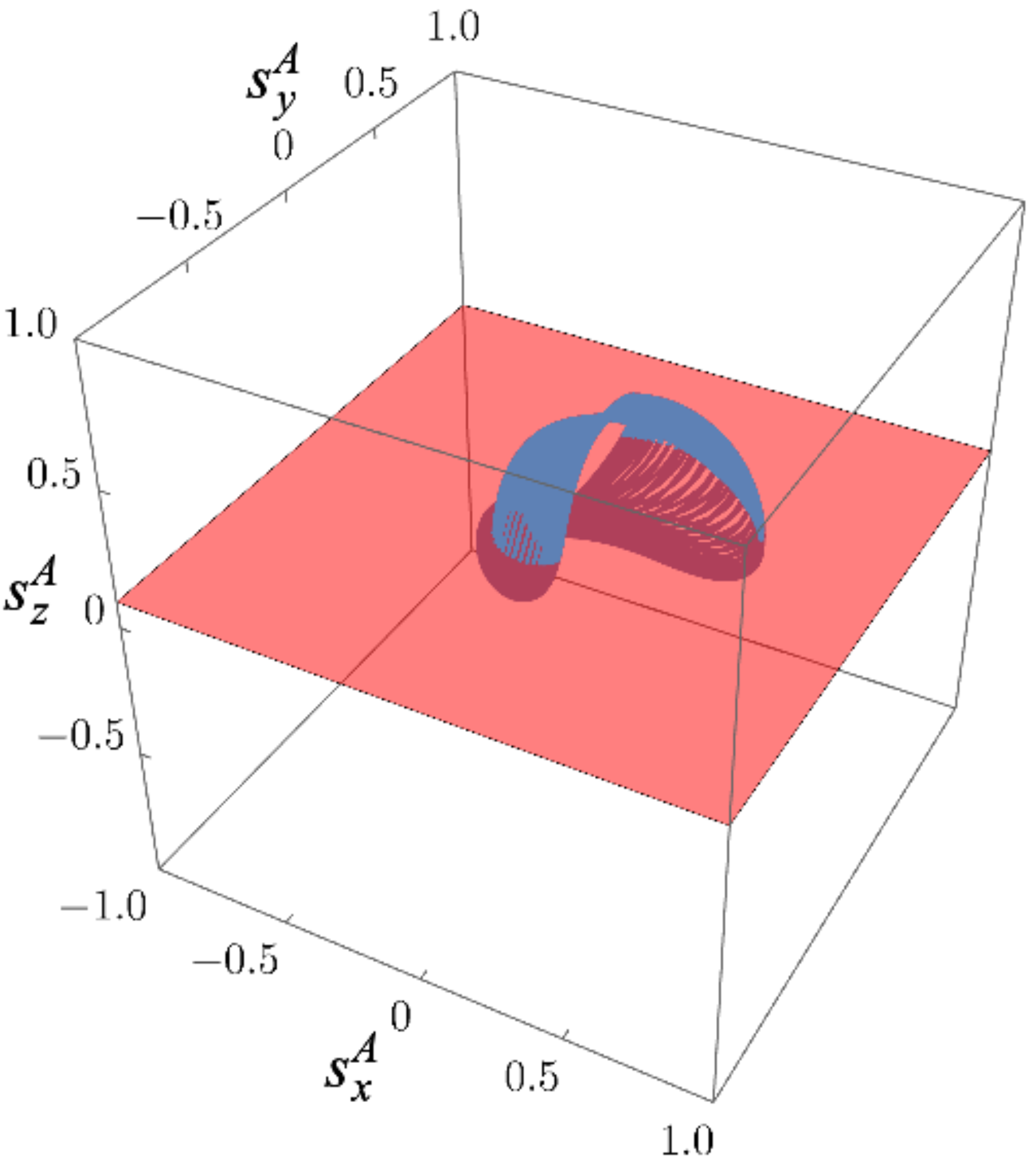}}\qquad\qquad
\subfloat[\qquad\textbf{(b)}]{\label{Poincare_QP1}\includegraphics[scale=0.33]{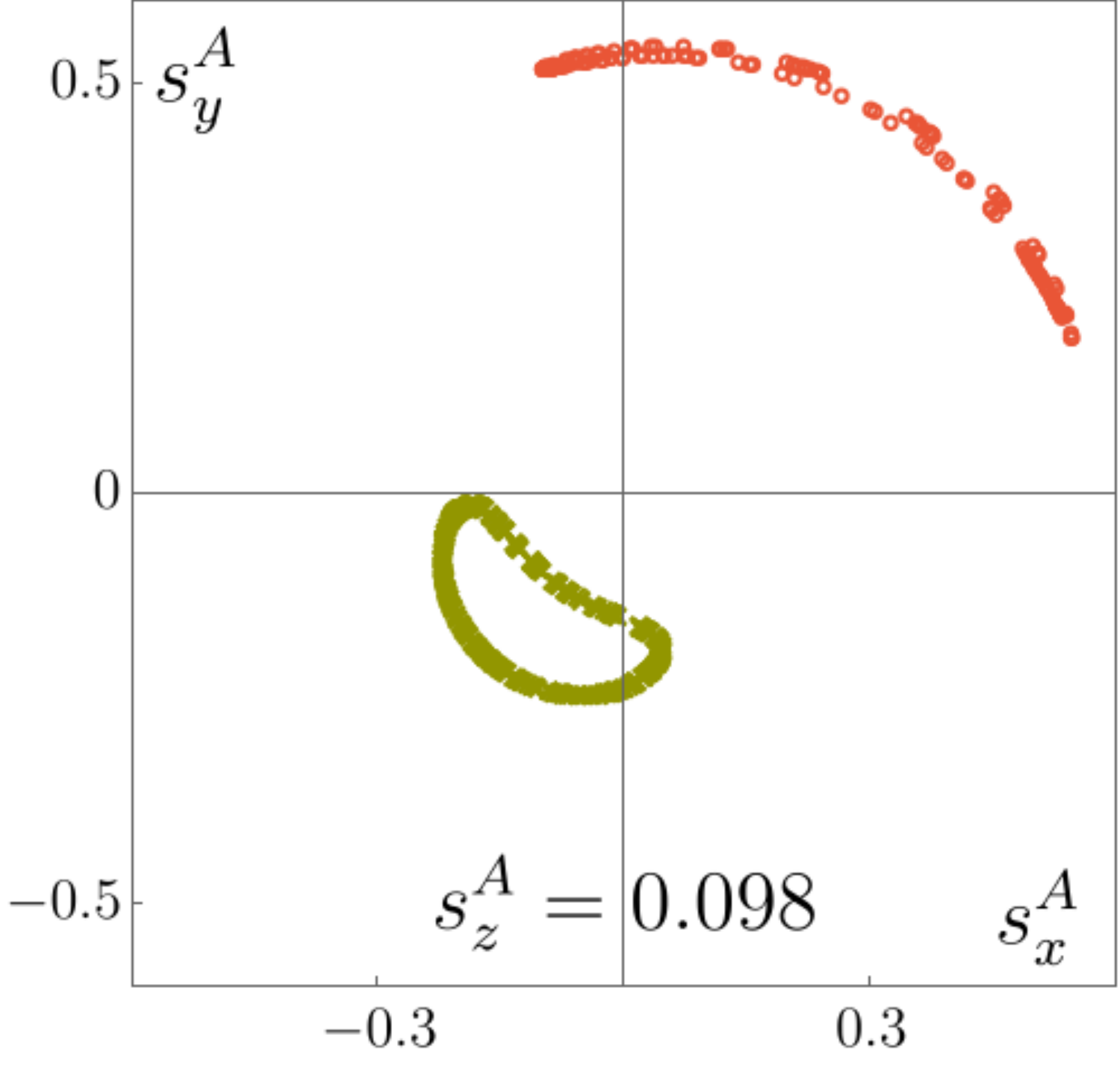}}\\
\subfloat[\textbf{(c)}]{\label{Traj_Sect_QP2}\includegraphics[scale=0.35]{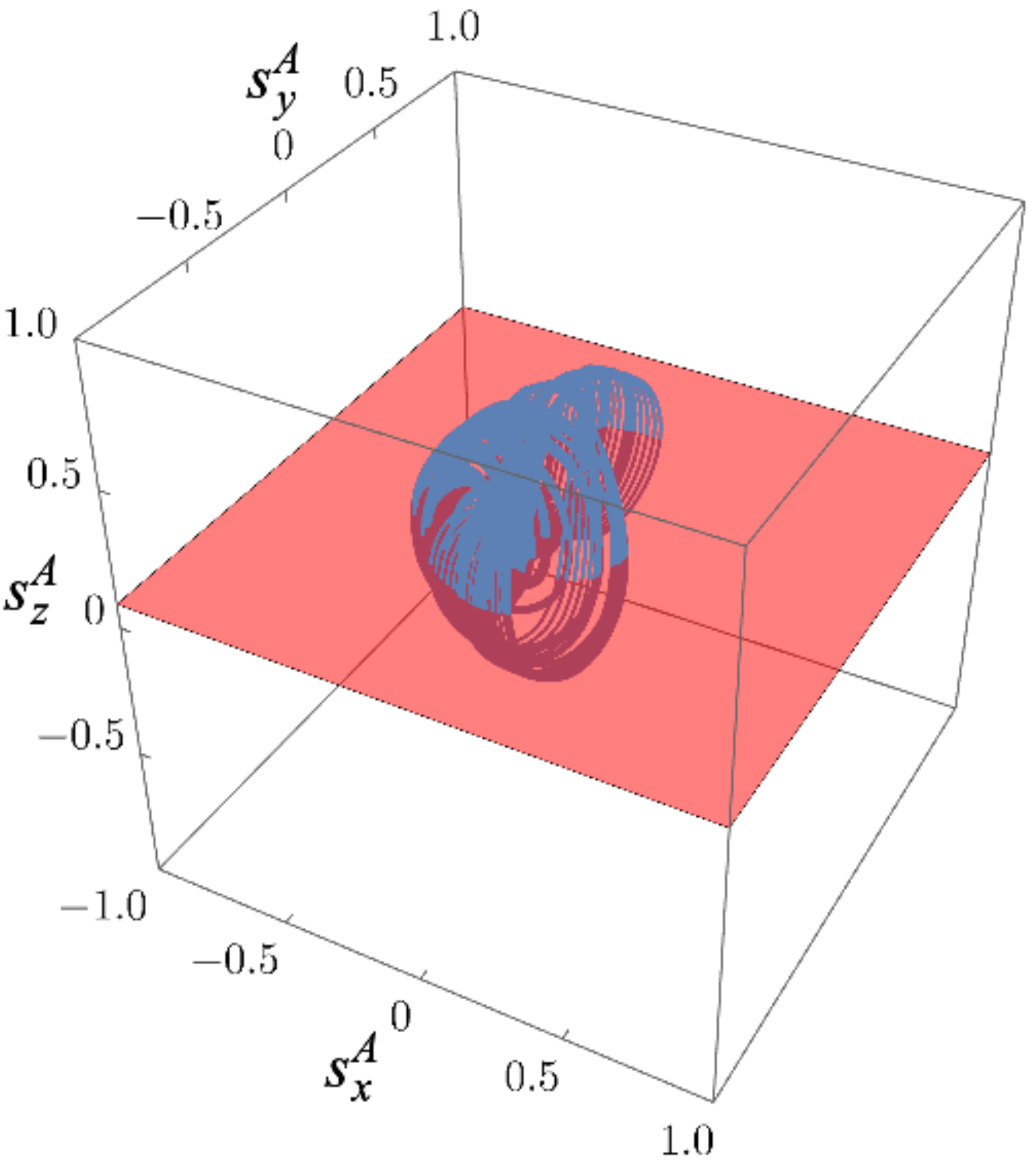}}\qquad\qquad
\subfloat[\qquad\textbf{(d)}]{\label{Poincare_QP2}\includegraphics[scale=0.33]{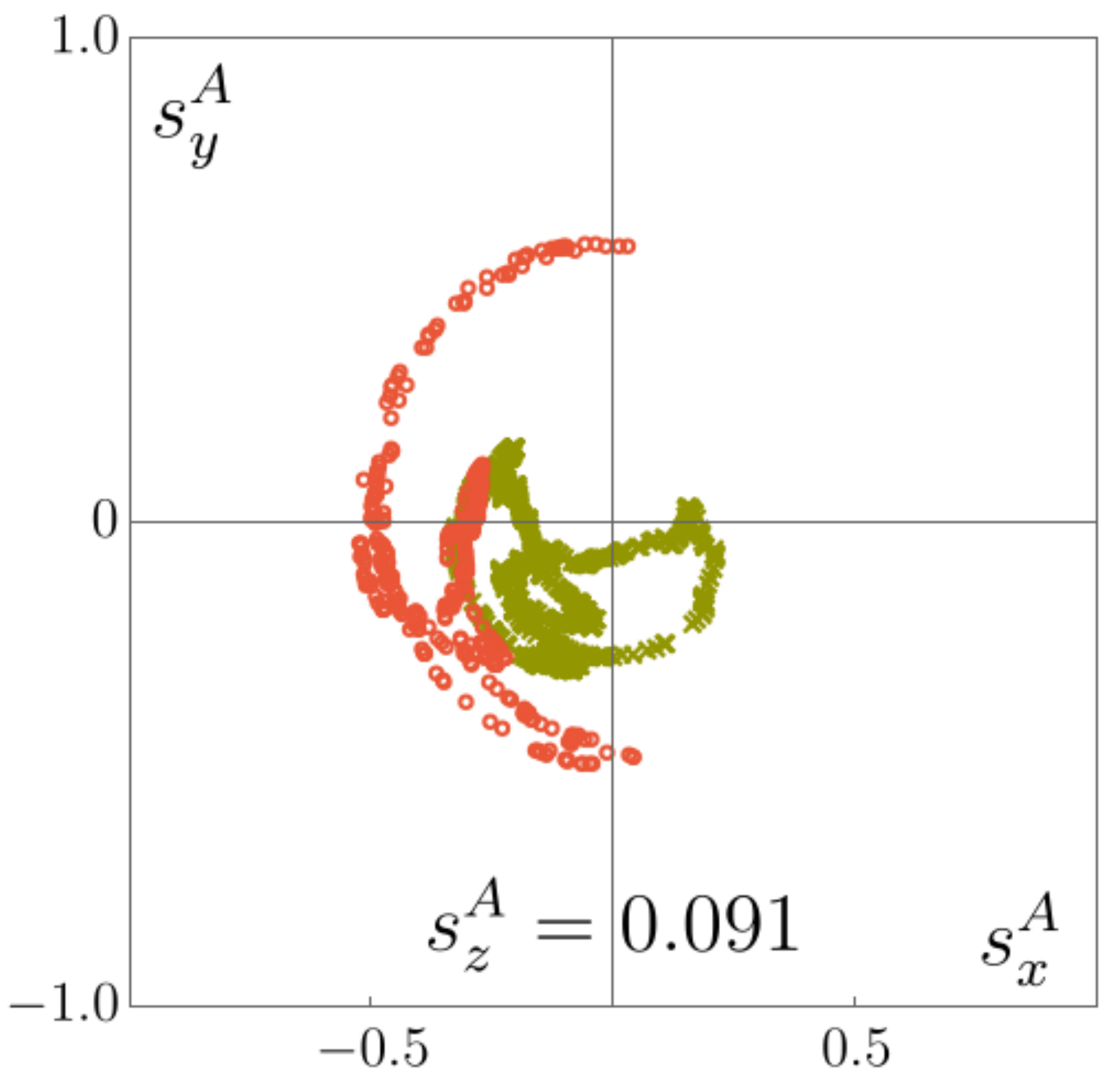}}
\caption{Poincar\'{e} sections for the quasiperiodic attractors with $\bm{s}^{A}$. We use the same conventions as before.  The parameters in \textbf{(a)} and \textbf{(b)} are: $\delta = 0.24, W = 0.055$, whereas in \textbf{(c)} and \textbf{(d)} we have $\delta = 0.115, W = 0.055$. Note the self-intersecting Poincar\'{e} section in \textbf{(d)}. This is unlike the one in \textbf{(b)}, which is topologically similar to \fref{Poincare_QP_Toy}. (For interpretation of the references to color in this figure legend, the reader is referred to the web version of this article.)}
\label{Poincare_QP}
\end{figure*}

\begin{figure}[tbp!]

\begin{center}
\includegraphics[scale=0.45]{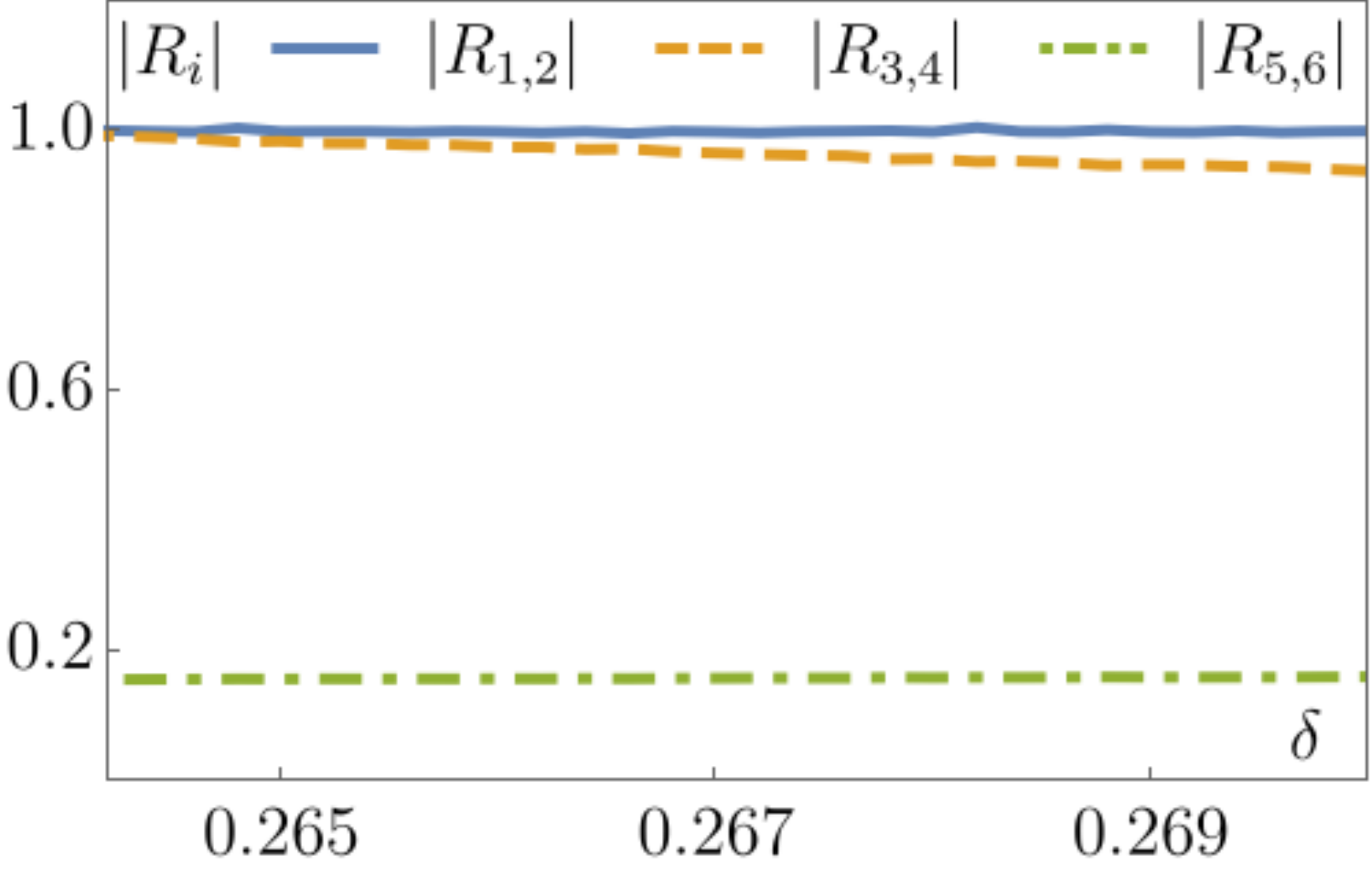}
\caption{The absolute value of the Floquet multipliers using \eref{Full_Floquet_Eqn} for $W = 0.1484$. The multipliers $R_{1}$ and $R_{2}$ (both $= 1$) correspond to the particular solutions $\Delta\textbf{K}^{01} = \dot{\textbf{K}}$ and $\Delta\textbf{K}^{02} = \widetilde{\textbf{K}}_{\perp}$ in \re{Special_m_FF}, where the latter stems from the axial symmetry of \eref{Mean-Field_1}. The complex conjugate multipliers are denoted as $R_{3,4}$, whose absolute values being greater than one ushers in quasiperiodicity at $\delta = 0.264$ via Neimark-Sacker bifurcation. Here $R_{5}$ and $R_{6}$ turns out to be complex conjugate of each other. Their norms remain less than one across the bifurcation. Since it is impossible to numerically obtain the limit cycle for $\delta<0.264$, we cannot continue the Floquet analysis past this point.}\label{Floquet_Full}
\end{center}

\end{figure}

\begin{figure}[tbp!]

\begin{center}
\includegraphics[scale=0.45, trim=0 0cm -0.8cm -0.8cm, clip]{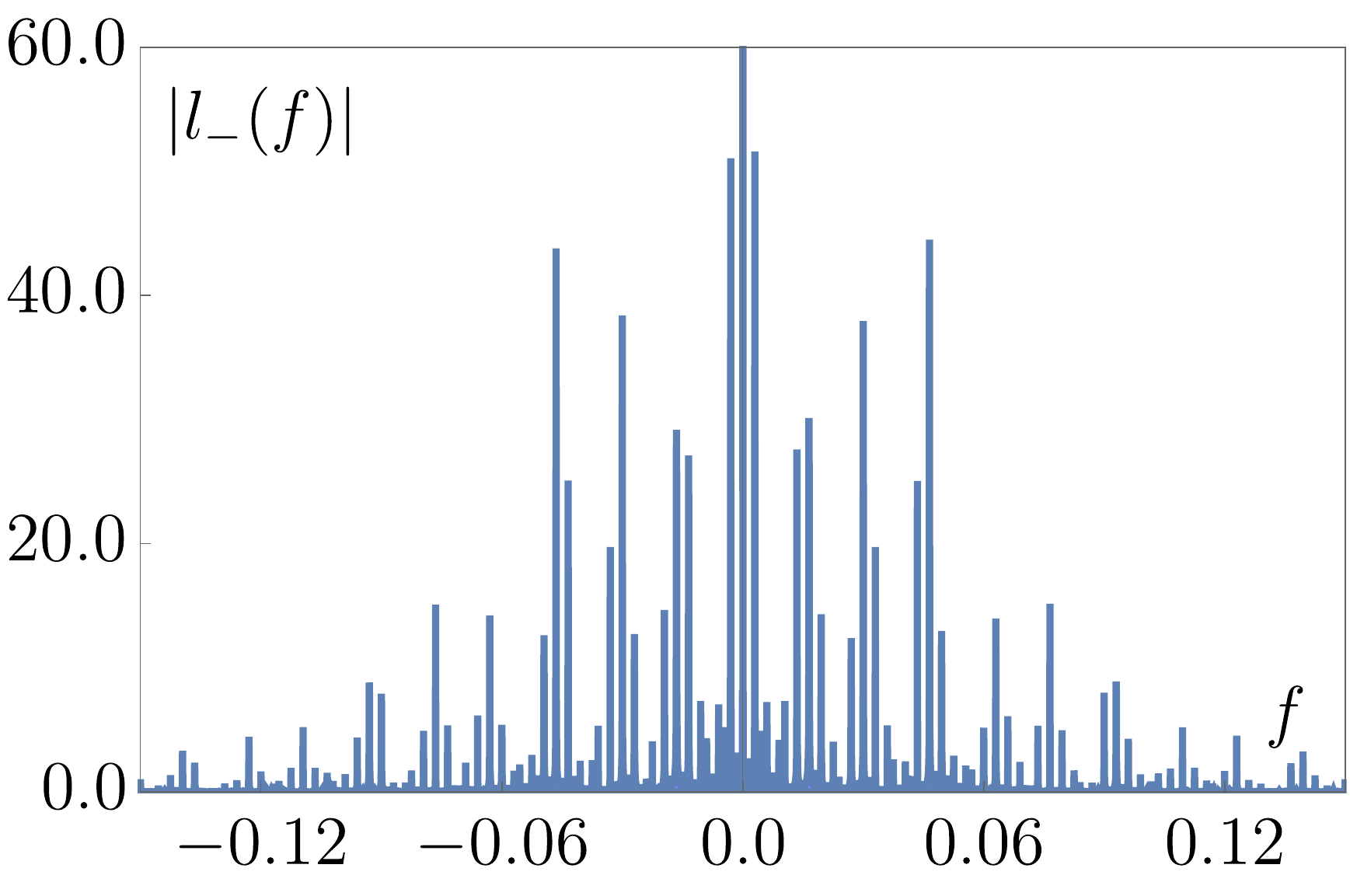}
\caption{Power spectrum of radiated light in the quasiperiodic superradiant phase, at $\delta = 0.115, W = 0.055.$ The spectrum is comprised of discrete peaks. The fundamental frequency $f_{1} \approx 1.6\times 10^{-2}$ and its higher harmonics denote the main peaks, whereas the auxiliary peaks are bunched around the main ones with spacings $f_{2} \approx 3.0\times 10^{-2}$. Note, the spectrum has no reflection symmetry about $f = 0$ axis. Moreover, the most prominent peak is at the origin.}\label{Spectrum_QP}
\end{center}

\end{figure}

We refer to an amplitude-modulated superradiance with more than one discrete incommensurate frequency as quasiperiodicity. The presence of $m$ such frequencies in the spectrum (in \fref{Spectrum_QP} we discern two distinct incommensurate frequencies) leads to a motion on an $m$-torus. For example, consider the dynamics of a particle described by 
\begs
\bea
X(t) &=& (D_{1} + D_{2}\sin{2\pi f_{2}t})\cos{2\pi f_{1}t}, \\ [7pt]
Y(t) &=& D_{2}\cos{2\pi f_{2}t}, \\[7pt] 
Z(t) &=& (D_{1} + D_{2}\sin{2\pi f_{2}t})\sin{2\pi f_{1}t}, 
\eea 
\label{T2}
\ens%
If $f_{1}:f_{2}$ is rational, \eref{T2} depicts a 3D periodic (also called mode-locked) attractor with period $1/f$, where $f$ is the lowest common multiple of $f_{1}$ and $f_{2}$, see \fref{LC_Toy}. On the other hand, the same \esref{T2} depict a quasiperiodic trajectory on a $2$-torus, when $f_{1}:f_{2}$ is irrational, see \fref{QP_Toy}.         

\subsubsection{Emergence of Quasiperiodicity: Neimark-Sacker Bifurcation of the Symmetry-broken Limit Cycle}
\label{Sect_NS}

While evolving into a quasiperiodic (\fref{Poincare_QP}) from a periodic (\fref{Poincare_LC}) attractor, the corresponding Poincar\'{e} map undergoes a supercritical Neimark-Sacker bifurcation -- where a fixed point of the map gives rise to an infinitesimal limit cycle in the center manifold. 

Note that in the yellow subregion of \fref{Phase_Diagram}, \eref{Mean-Field_1} sometimes (as in \fref{Coexist_(R)SBLC_QP}\blue{a}) leads to a 2-torus depicting an axially symmetric trivially quasiperiodic attractor. These attractors are periodic in $s^{\tau}_\perp, s_{z}^{\tau}$ and $\varphi$ (mod $\pi$). In the above 5D space they correspond to closed curves. The second incommensurate frequency $\omega_{q}$ originates from the Fourier series of the second term on the right hand side of \eref{Mean-Field_Group1}. We write a solution to this equation as
\beg
\Phi(t)=\Theta t+H(t),\quad \Theta=\frac{1}{2}(\omega_A+\omega_B)+ \omega_q.
\label{Theta}
\en
Here $H(t)$ and the limit cycle in the 5D space have the same period. In an appropriate rotating frame, where $\Theta = 0$, the axial symmetry is broken and a single 2-torus gives rise to a one parameter (characterized by $\Phi$) family of non-$\Z2$-symmetric limit cycles \cite{Patra_1}. In this paper, however, we do not regard such attractors is quasiperiodic. In particular, here the trajectories denoted as quasiperiodic are quasiperiodic even in the aforementioned 5D space.

One proves the occurrence of Neimark-Sacker bifurcation with the Floquet analysis, where we linearize the mean-field equations \re{Mean-Field_1} about the (non-$\Z2$-symmetric) periodic attractor as follows: 
\begs
\bea
\frac{\mathrm{d}\Delta s_{x}^{\tau}}{\mathrm{d}t} &=& -W\Delta s_{x}^{\tau} - \omega_{\tau}\Delta s_{y}^{\tau} 
+ \frac{1}{2}\big(l_{x}\Delta s_{z}^{\tau} + s_{z}^{\tau}\Delta l_{x}\big), \\
\frac{\mathrm{d}\Delta s_{y}^{\tau}}{\mathrm{d}t} &=&  \omega_{\tau}\Delta s_{x}^{\tau} -W\Delta s_{y}^{\tau}
+ \frac{1}{2}\big(l_{y}\Delta s_{z}^{\tau} + s_{z}^{\tau}\Delta l_{y}\big), \\
\frac{\mathrm{d}\Delta s_{z}^{\tau}}{\mathrm{d}t} &=& -\frac{1}{2}\big(l_{x}\Delta s_{x}^{\tau} + l_{y}\Delta s_{y}^{\tau}\big)  - \frac{1}{2}\big(s_{x}^{\tau}\Delta l_{x} + s_{y}^{\tau}\Delta l_{y}\big) - W\Delta s_{z}^{\tau}, 
\eea
\label{Full_Floquet_Eqn}
\ens%
where $\Delta l_{x,y} = \Delta s_{x,y}^{A} + \Delta s_{x,y}^{B}.$ Also, we introduce 
\beg 
\textbf{K}\equiv \big(\bm{s}^{A}, \bm{s}^{B}\big), \qquad \Delta\textbf{K}\equiv \big(\Delta\bm{s}^{A}, \Delta\bm{s}^{B}\big),
\en%
to respectively denote the 6D attractor and the small deviations about the same.
The Floquet multipliers (other than the trivial one, $R_{1} = 1$, that corresponds to the longitudinal perturbation along the limit cycle) obtained below are one-to-one correlated with the characteristic values of the Poincar\'{e} map near the bifurcation. We explain this in more detail in the context of a 3D periodic attractor in \sref{Tan_Bifurc_Int}.

Similar to the Floquet analysis discussed in \sref{Review}, we calculate the $6\by 6$ monodromy matrix for \eref{Full_Floquet_Eqn} as 
\beg 
\mathbb{B}_{K} = \big[\mathcal{K}(0)\big]^{-1}\cdot \mathcal{K}(T),
\en%
where the columns of $\mathcal{K}(t)$ consist of
independent solutions $\Delta\textbf{K}$ of \eref{Full_Floquet_Eqn}, and $T$ is the period of $\textbf{K}$. The six Floquet multipliers $R_{i} \equiv e^{\mu_{i}T}$, for $i = 1, \cdots, 6,$ are the eigenvalues of $\mathbb{B}_{K}$. The Floquet exponents are denoted as $\mu_{i}$. We then write the general solution for \eref{Full_Floquet_Eqn} as follows: 
\beg 
\Delta\textbf{K}(t) = \sum_{i = 1}^{6} C_{K,i}e^{\mu_{i}t}\textbf{P}_{i}(t), \qquad R_{i}\equiv e^{\mu_{i}T},
\label{Full_Floquet_Soln}
\en%
where $\textbf{P}_{i}(t+T) = \textbf{P}_{i}(t)$, and $C_{K,i}$ are arbitrary constants for $i = 1, \cdots, 6$. 

For \eref{Full_Floquet_Eqn}, we identify two particular solutions
\begs
\bea 
\Delta\textbf{K}^{01} &=& \dot{\textbf{K}} \equiv \big(\dot{\bm{s}}^{A}, \dot{\bm{s}}^{B}\big), \label{m01_FF} \\
\Delta\textbf{K}^{02} &=& \widetilde{\textbf{K}}_{\perp} \equiv \big(s_{y}^{A}, -s_{x}^{A}, 0, s_{y}^{B}, -s_{x}^{B}, 0\big), \label{m02_FF}
\eea
\label{Special_m_FF}
\ens%
that are periodic with the same period as $\textbf{K}$. Substituting \eref{m01_FF} for $\Delta \textbf{K}$, trivially satisfies \eref{Full_Floquet_Eqn}. On the other hand, the solution in \re{m02_FF} comes about due to the axial symmetry of \eref{Mean-Field_1}. In particular, considering the difference between two slightly rotated (about the z-axis) spins, we obtain 
\beg 
\R(\Delta \theta)\cdot \bm{s}^{\tau} - \bm{s}^{\tau} = \Delta \theta\big(s_{y}^{\tau}, -s_{x}^{\tau}, 0\big),
\label{Full_Floquet_Eigen = 1_Rot}
\en%
which explains the form of $\widetilde{\textbf{K}}_{\perp}$.

Defining $\textbf{P}_{1,2} \equiv \Delta\textbf{K}^{0(1,2)}$, we identify $R_{1} = R_{2} = 1$. This simplifies \eref{Full_Floquet_Soln} as follows: 
\beg 
\Delta\textbf{K} = C_{K,1}\textbf{P}_{1} + C_{K,2}\textbf{P}_{2} + \sum_{i = 3}^{6} C_{K,i}e^{\mu_{i}t}\textbf{P}_{i}.
\label{Full_Floquet_Soln_Simple}
\en%
When we decrease $\delta$ by keeping $W$ fixed (e.g., see \fref{Floquet_Full}), close to the bifurcation -- for a fixed $W$, the bifurcation occurs at $\delta = \delta_{c}$ -- we have at least two complex conjugate multipliers 
\beg 
R_{3,4} = \big(1 + a(\delta)\big)e^{\pm2\pi \imath \beta(1 + b(\delta))},
\label{CC_NS}
\en%
where
\beg
0<\beta<\frac{1}{2},\quad
a(\delta_{c}) = b(\delta_{c}) = 0, \quad
\frac{\mathrm{d}a(\delta)}{\mathrm{d}\delta}\bigg|_{\delta = \delta_{c}} > 0.
\en%
For $W = 0.1484$, we observe $\delta_{c} = 0.264$, and $|R_{3,4}(\delta_{c})| $ is indeed equal to one in \fref{Floquet_Full}. $|R_{i}|\leqslant 1$ for $i = 1,2,5$ and $6$.

\subsubsection{Coexistence}
\label{Sect_QP_Coexistence}

In our system, quasiperiodicity generally appears via supercritical Neimark-Sacker bifurcation, where the periodic attractor gradually morphs into a quasiperiodic one. Note the incremental increase in thickness in the $s_{z}^{A}-s_{z}^{B}$ projections after the bifurcation in \fref{Sup_NS}.

\begin{figure*}[tbp!]
\centering
\subfloat[\qquad\textbf{(a)}]{\label{Coexist_Full_(R)SBLC}\includegraphics[scale=0.3]{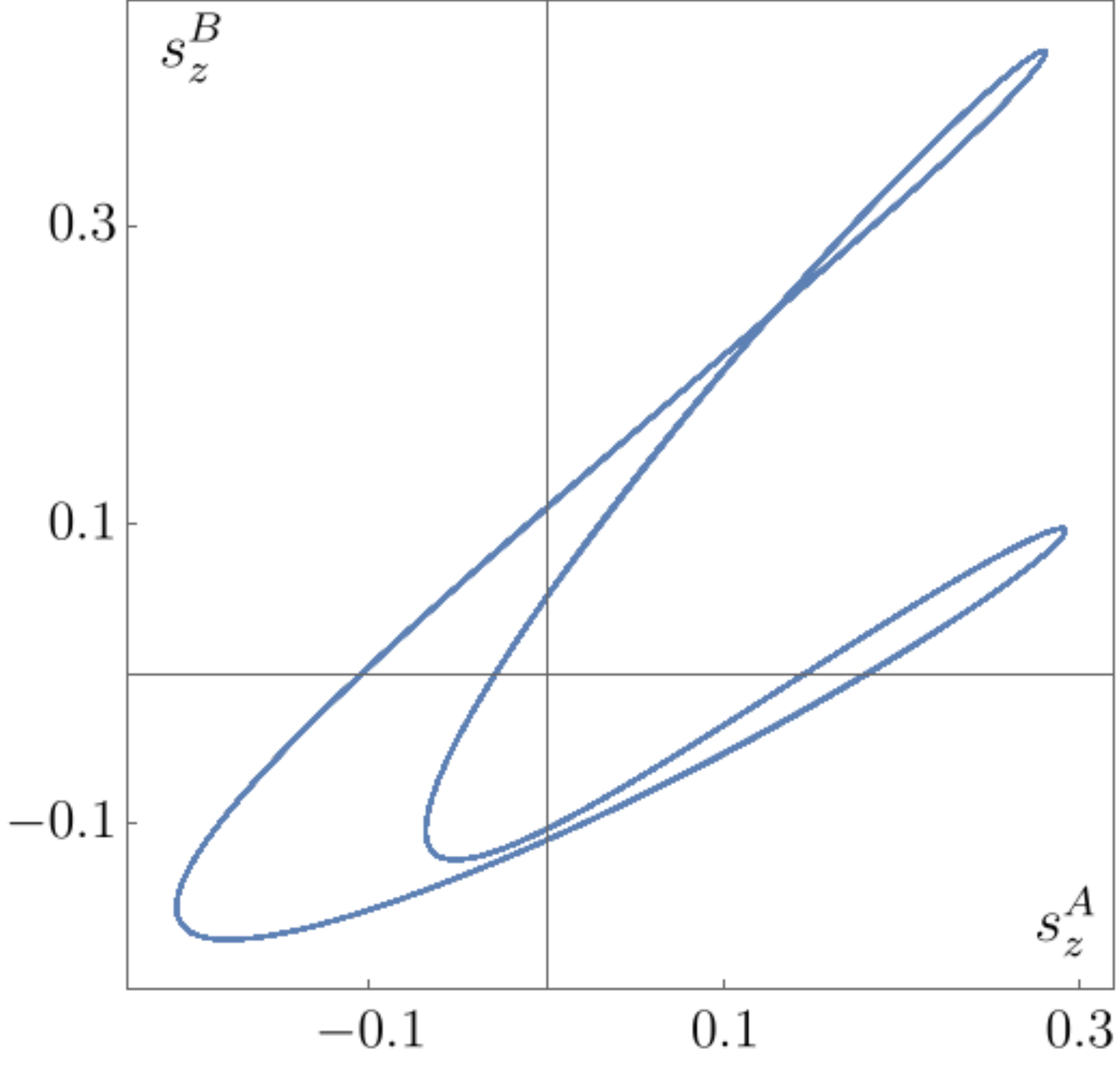}}\begin{picture}(0,0)
\put(-140,95){\includegraphics[height=2.5cm]{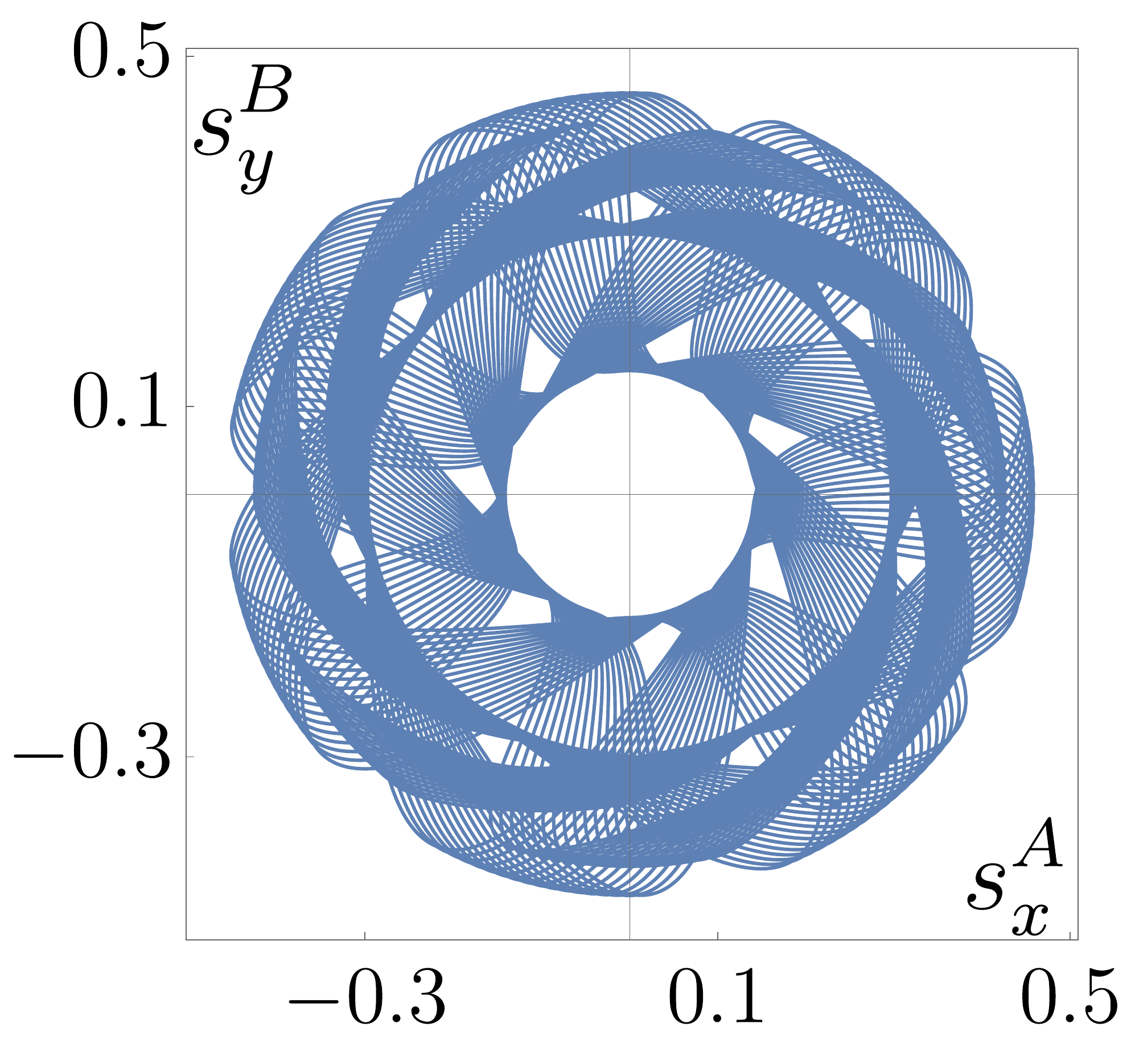}}
\end{picture}\qquad
\subfloat[\qquad\textbf{(b)}]{\label{Coexist_Full_QP}\includegraphics[scale=0.3]{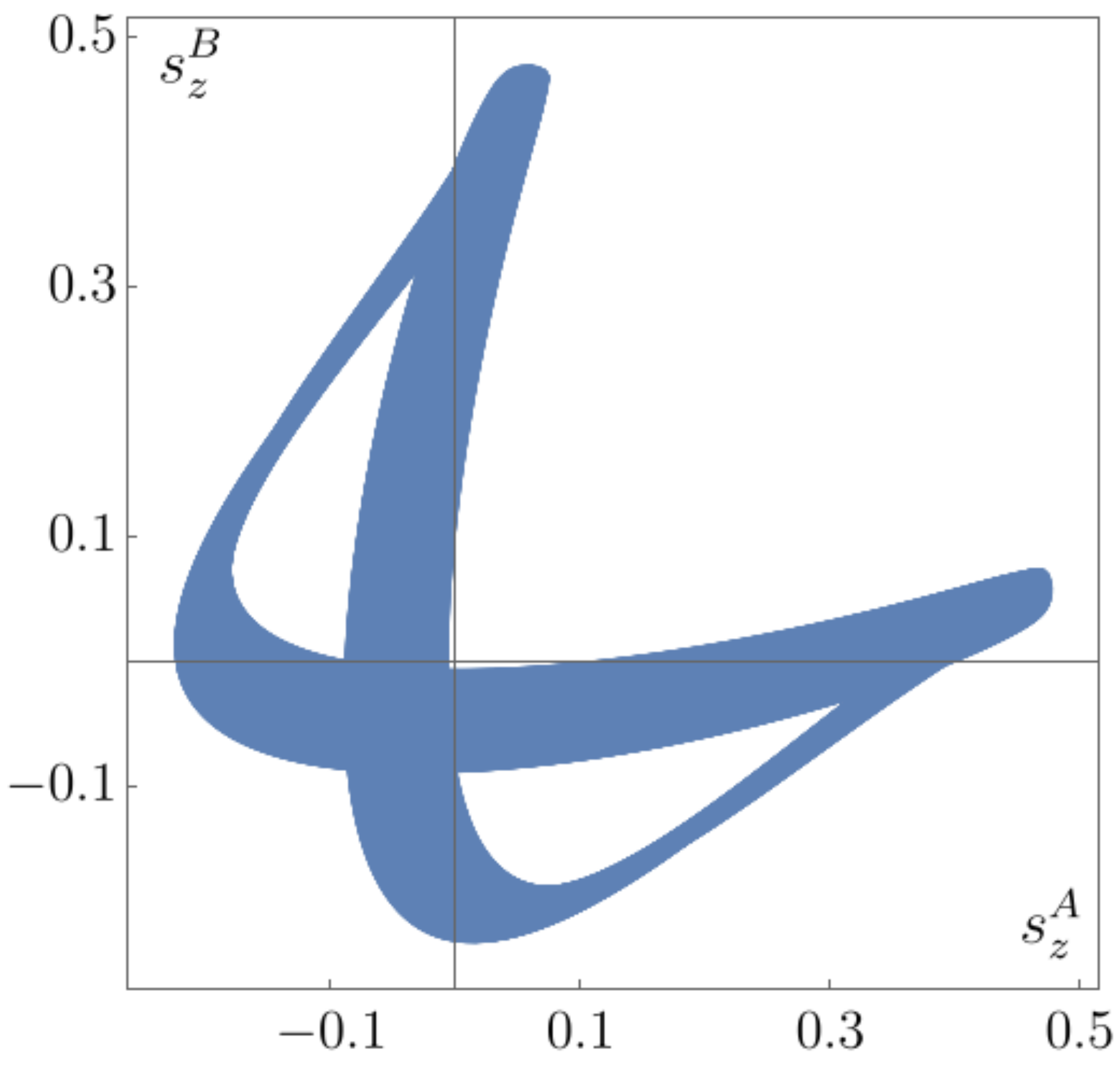}}
\caption{Coexistence to the left of the $\Z2$ symmetry breaking line (dashed line in \fref{Phase_Diagram}) at $\delta = 0.225,$ $W = 0.05$. For initial condition $\textbf{K}_{0} = (0.378594, -0.0256723, 0.439373, -0.224928, 0.85278, -0.530865)$, the asymptotic solution in \textbf{(a)} is a symmetry-broken limit cycle (in the 5D space of $s^{\tau}_\perp, s_{z}^{\tau}$ and $\varphi$)  with a visibly reflection symmetry broken power spectrum \cite{Patra_1}. This is, in fact, the trivial quasiperiodic attractor discussed around \eref{Theta}, see the inset of \textbf{(a)}. On the other hand we end up with a quasiperiodic (even in the aforementioned 5D space) attractor in \textbf{(b)} for $\textbf{K}_{0} = (-0.729897, 0.538791, -0.298485, -0.117912, -0.0486406, -0.668281)$.}
\label{Coexist_(R)SBLC_QP}
\end{figure*}

\begin{figure*}[tbp!]
\centering
\subfloat[\qquad\textbf{(a)}]{\label{Coexist_Full_SLC}\includegraphics[scale=0.3]{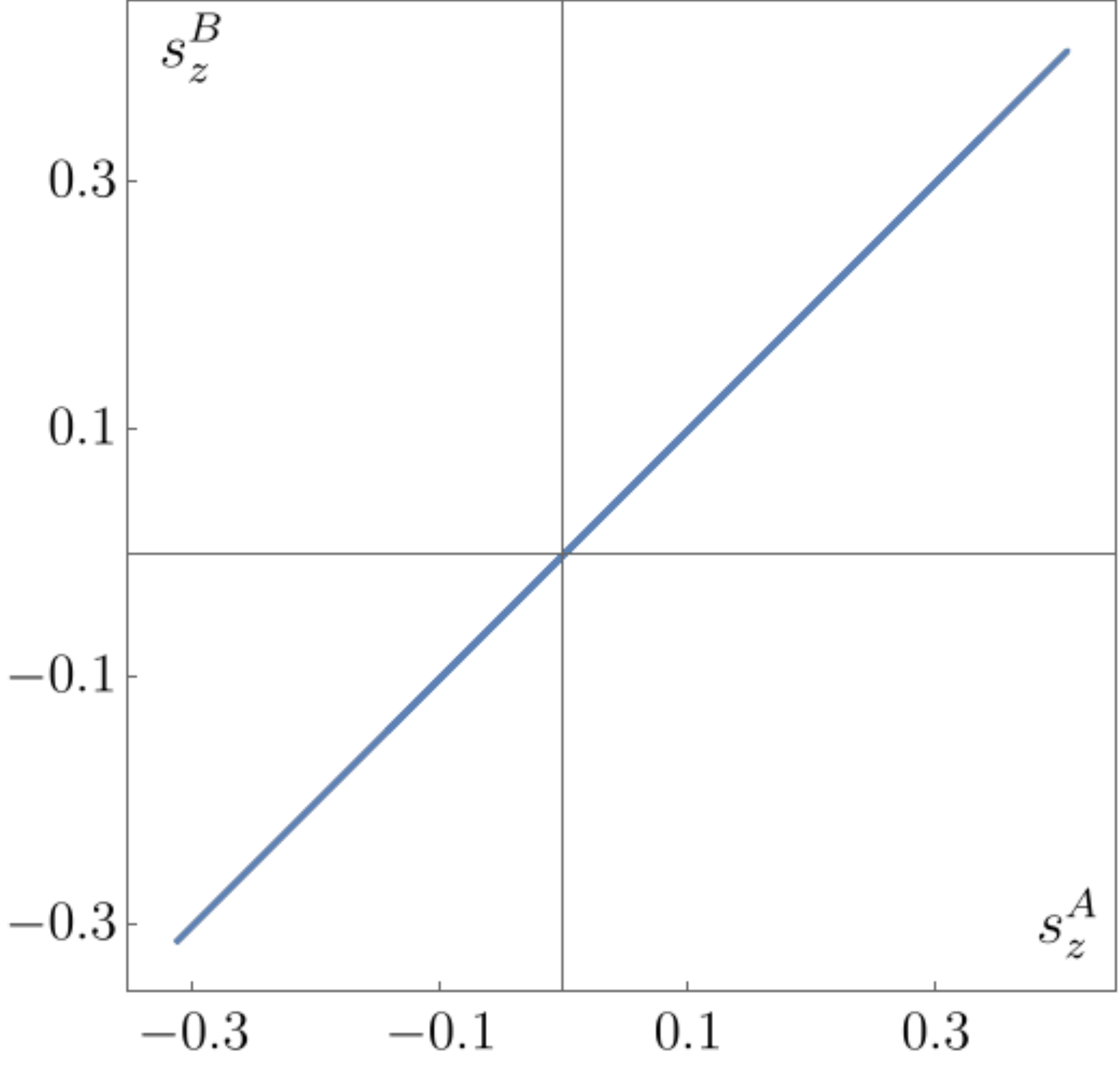}}
\llap{\raisebox{1.5cm}{\includegraphics[height=2.4cm]{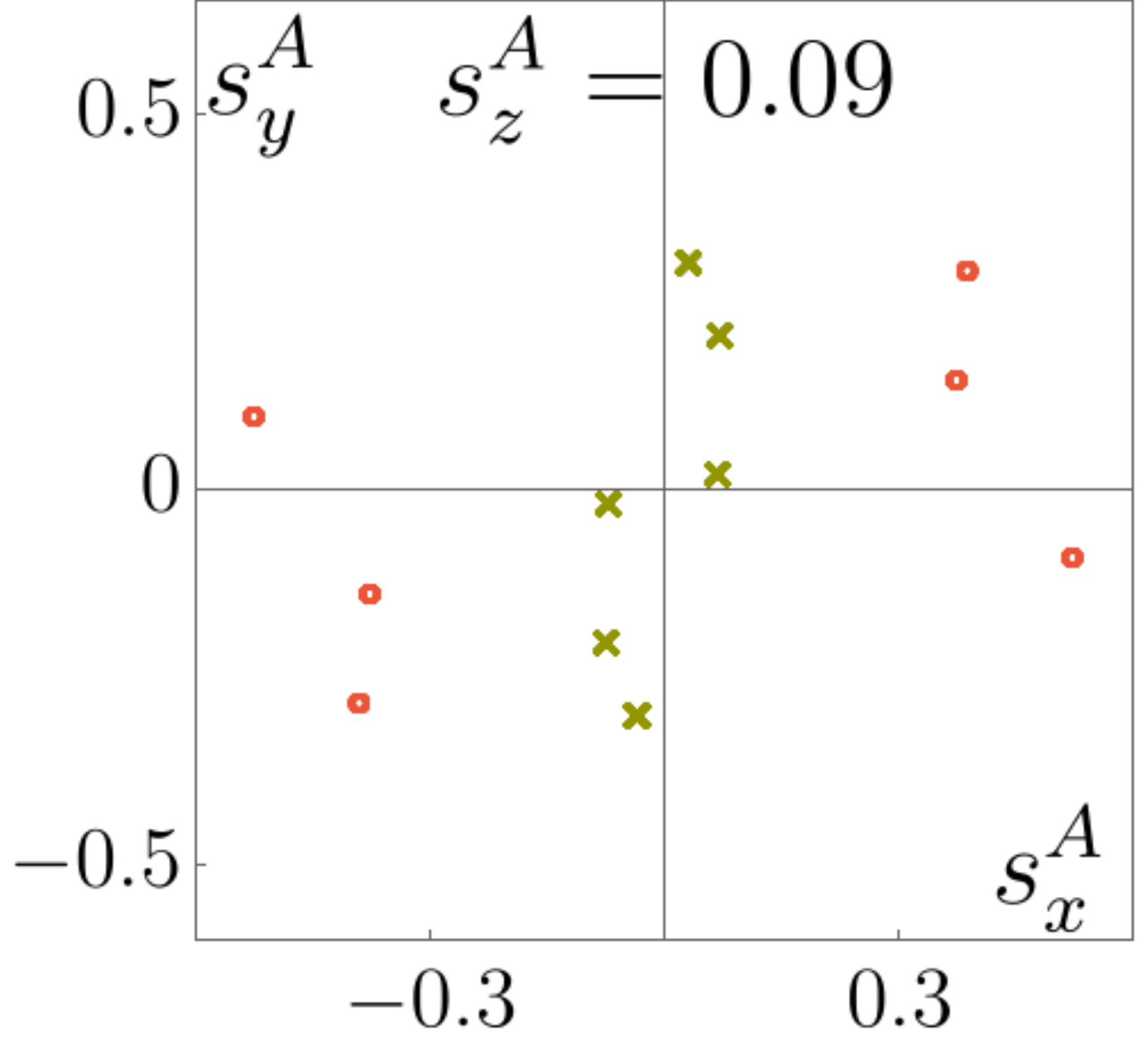}}}
\qquad\qquad
\subfloat[\qquad\textbf{(b)}]{\label{Coexist_Full_QP_1}\includegraphics[scale=0.3]{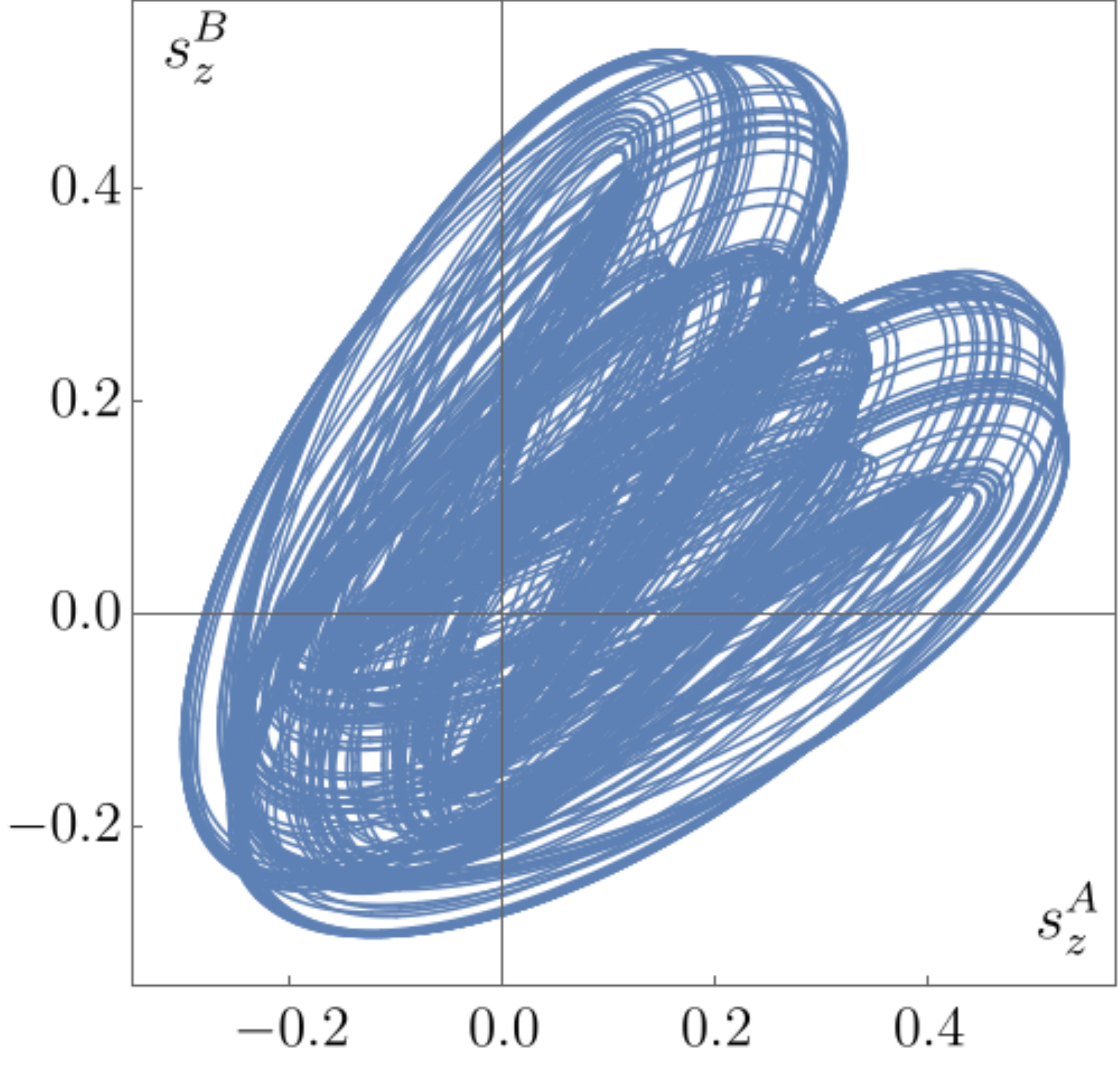}}\llap{\raisebox{1.5cm}{\includegraphics[height=2.4cm]{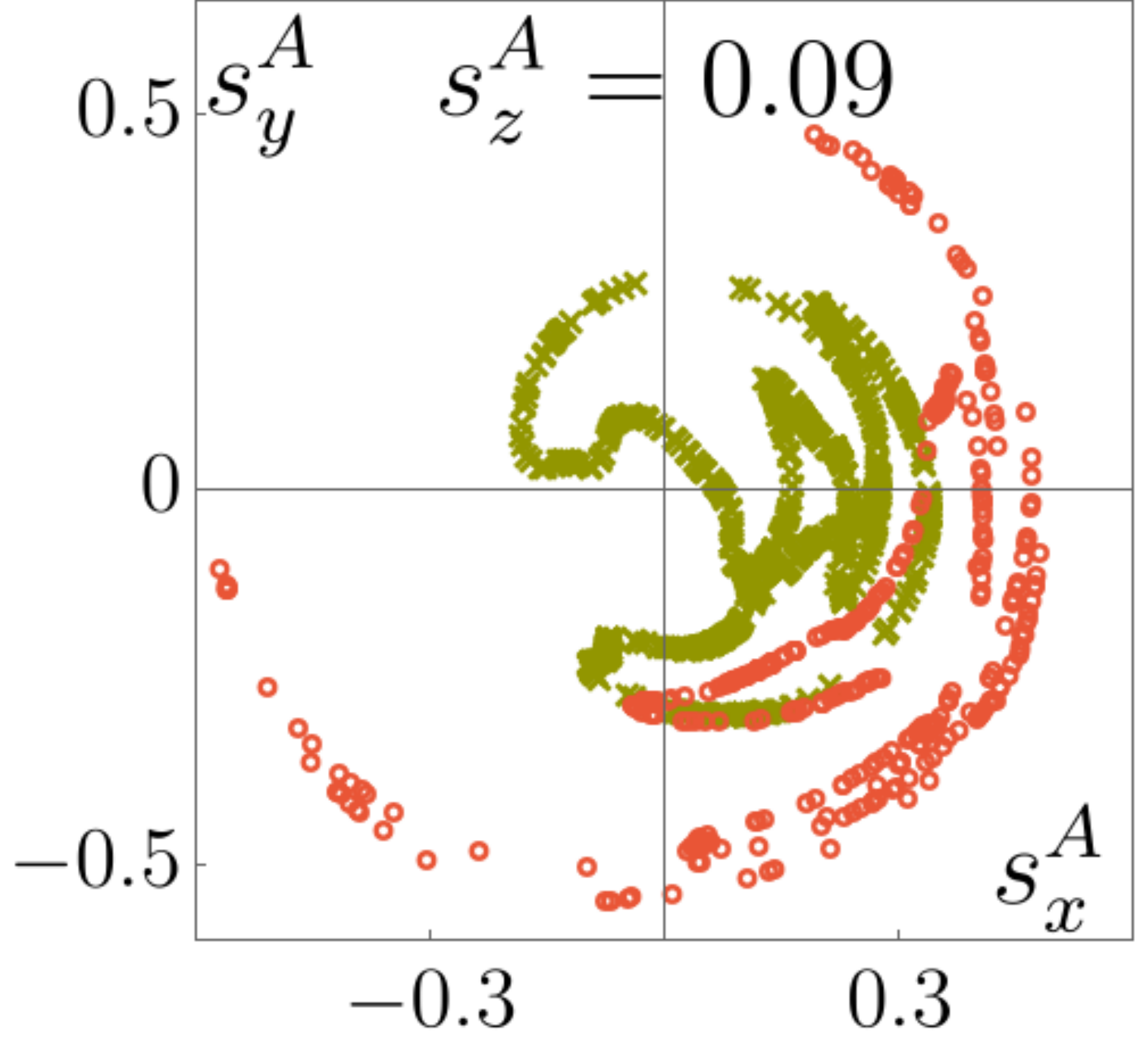}}}
\caption{Coexistence of a $\Z2$-symmetric limit cycle [in \textbf{(a)}] with a quasiperiodic attractor [in \textbf{(b)}] to the left of the $\Z2$ symmetry breaking line (dashed line in \fref{Phase_Diagram}) at $\delta = 0.12, W = 0.055$. The Poincar\'{e} sections in the inset confirm that the $s_{z}^{A}-s_{z}^{B}$ projections in \textbf{(a)} and \textbf{(b)} indeed belong to a periodic (not synchronized chaotic) and a quasiperiodic (not chaotic) attractor, respectively. The initial condition for \textbf{(a)} is $\textbf{K}_{0} = (0.471036, -0.423628, -0.566317,  0.471036, 0.43, -0.566317)$, whereas for \textbf{(b)} it is $\textbf{K}_{0} = (0.378594, -0.0256723, 0.439373, -0.224928, 0.85278, -0.530865)$. (For interpretation of the references to color in this figure legend, the reader is referred to the web version of this article.)}\label{Coexist_SLC_QP}
\end{figure*}

However, not all limit cycles lose their stability thus. Some in fact lose it via subcritical Neimark-Sacker bifurcation, which brings about an unstable torus \textit{before} the bifurcation, that acts as a separatrix for the basin of attraction for the limit cycle -- any initial condition inside the torus leads to the periodic attractor, whereas the ones outside are repelled away. This causes coexistence between the extant limit cycle and other stable attractors that exist for same parameter values. In \fsref{Coexist_(R)SBLC_QP} and \ref{Coexist_SLC_QP} we show two such examples.

Any subcritical (catastrophic) bifurcation gives rise to coexistence. We expect (and observe) more instances of such coexistence near the boundaries between different amplitude-modulated superradiances in the subregion of Phase III circumscribed by the symmetry-breaking line (black dashed line in \fref{Phase_Diagram}) on the right and the subcritical Hopf bifurcation line (II-III boundary) on the left.

\subsection{Manifestation of Chaos via Quasiperiodicity}\label{QP->Chaos}

In \fref{Phase_Diagram}, subregions with chaotic superradiance are always adjacent to the quasiperiodic ones. This indicates appearance of chaos via quasiperiodicity. In fact, the transformation of a chaotic trajectory into a quasiperiodic one, as we keep $W$ fixed and decrease $\delta$, is also understood in terms of the quasiperiodic route to chaos by increasing $\delta$.   

Quasiperiodicity, i.e., motion on a torus is quite fragile, where small perturbations can make the trajectory spill out of the torus bringing about chaos. It is known that twice differentiable infinitesimal perturbations lead to uniformly hyperbolic chaotic attractors (also known as strange axiom A attractors) from a $3$-tori. On the other hand, for an $m$-tori with $m>3$ only infinitely differentiable perturbations give rise to similar chaotic attractors. \cite{QP_Chaos_1, QP_Chaos_2}

In our system, however, we could only observe two frequency quasiperiodicity bringing about chaos. It is hard either to extract more than two incommensurate frequencies from the quasiperiodic spectrum (e.g., \fref{Spectrum_QP}), or to visualize the full 6D attractor to ascertain its genus. 

Here we must cite the examples of systems (two strongly coupled nonlinear oscillator) where chaos manifests after two frequency quasiperiodicity \cite{Hilborn}. Moreover, the theorem quoted above \cite{QP_Chaos_1, QP_Chaos_2} applies to uniformly hyperbolic chaotic attractors only. For a more general chaotic attractor similar theorem is lacking.

\begin{figure*}[tbp!]
\centering
\subfloat[\qquad\textbf{(a)}]{\includegraphics[scale=0.3]{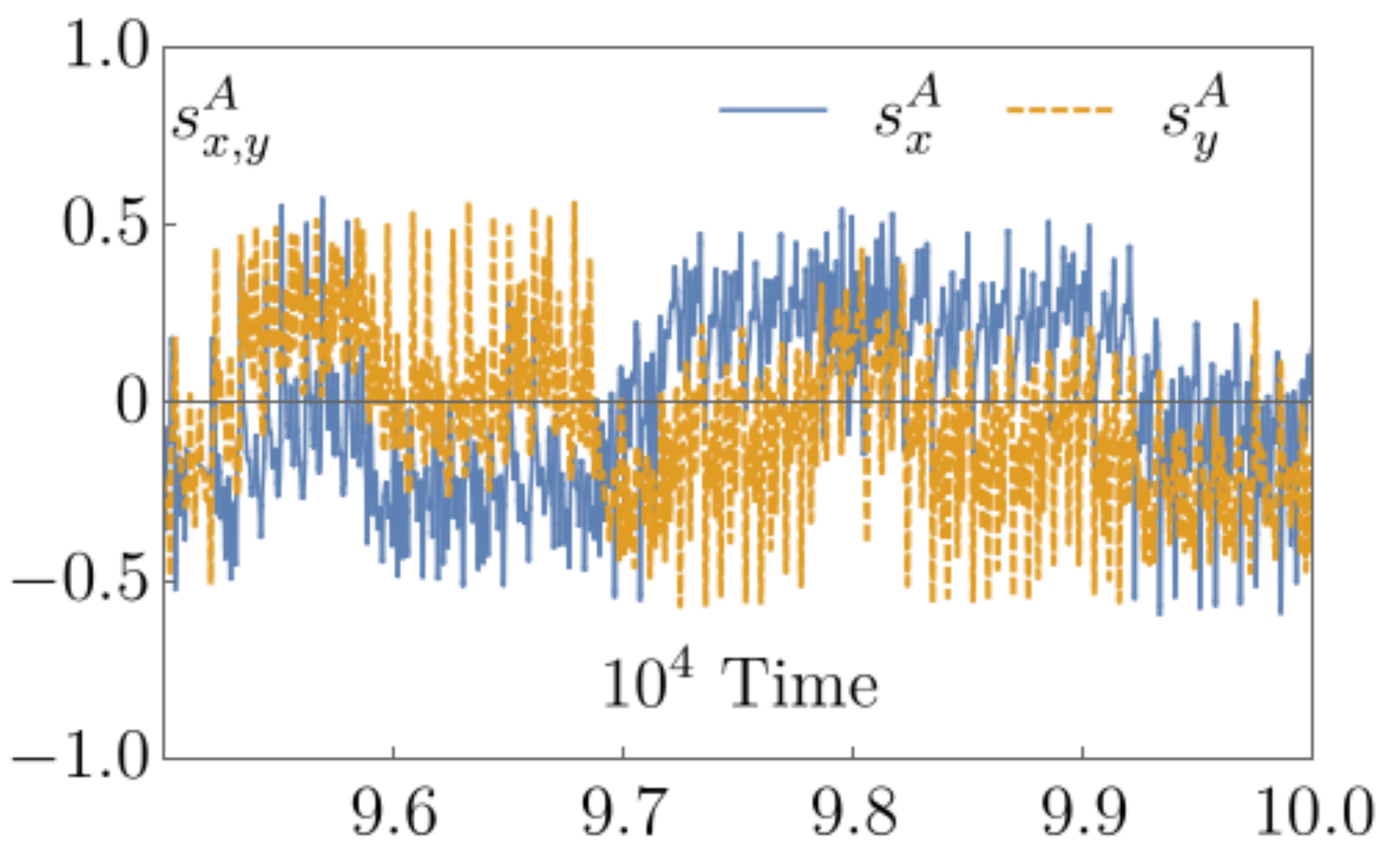}}\qquad\qquad
\subfloat[\qquad\textbf{(b)}]{\includegraphics[scale=0.3]{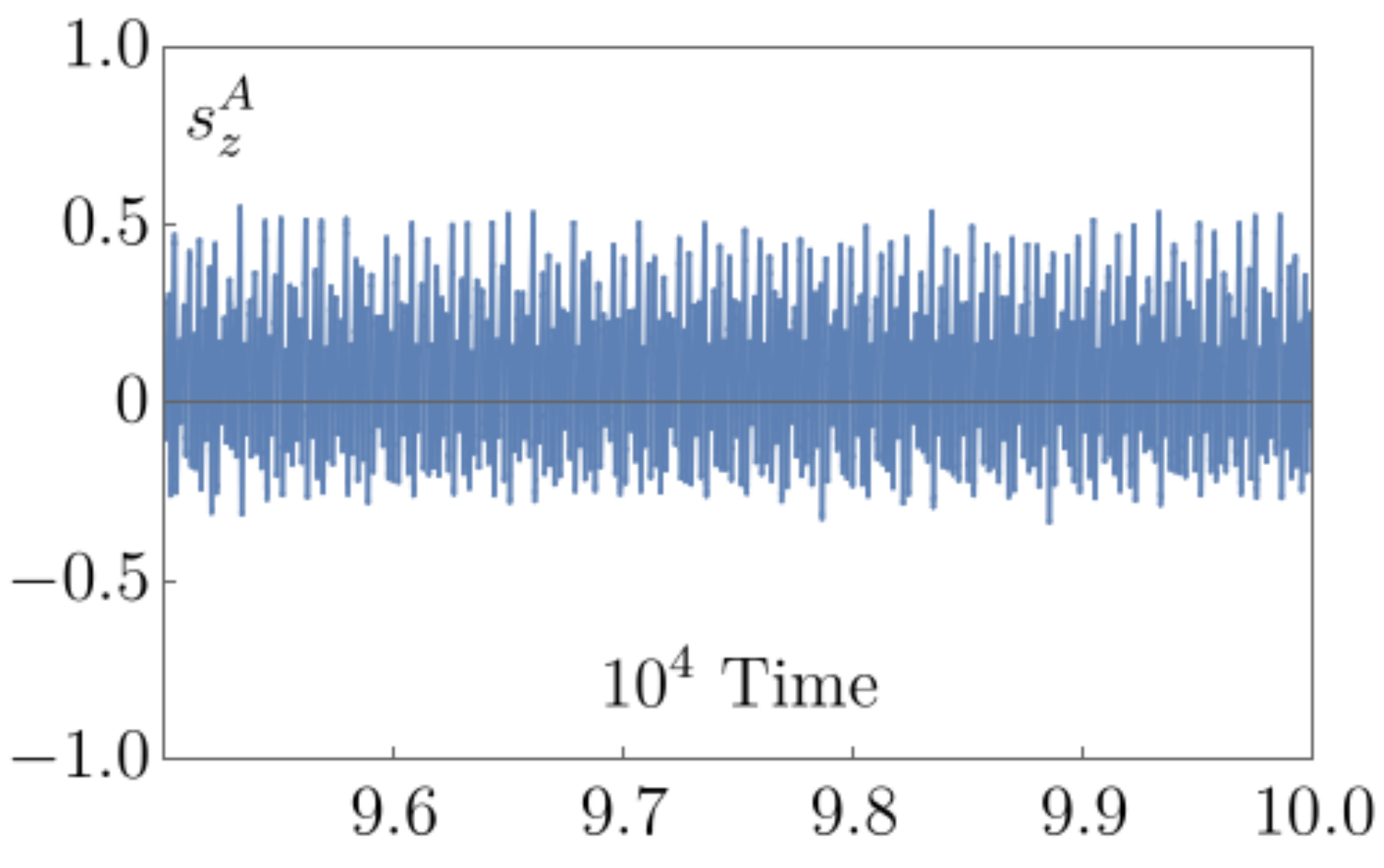}}\\
\subfloat[\qquad\textbf{(c)}]{\includegraphics[scale=0.3]{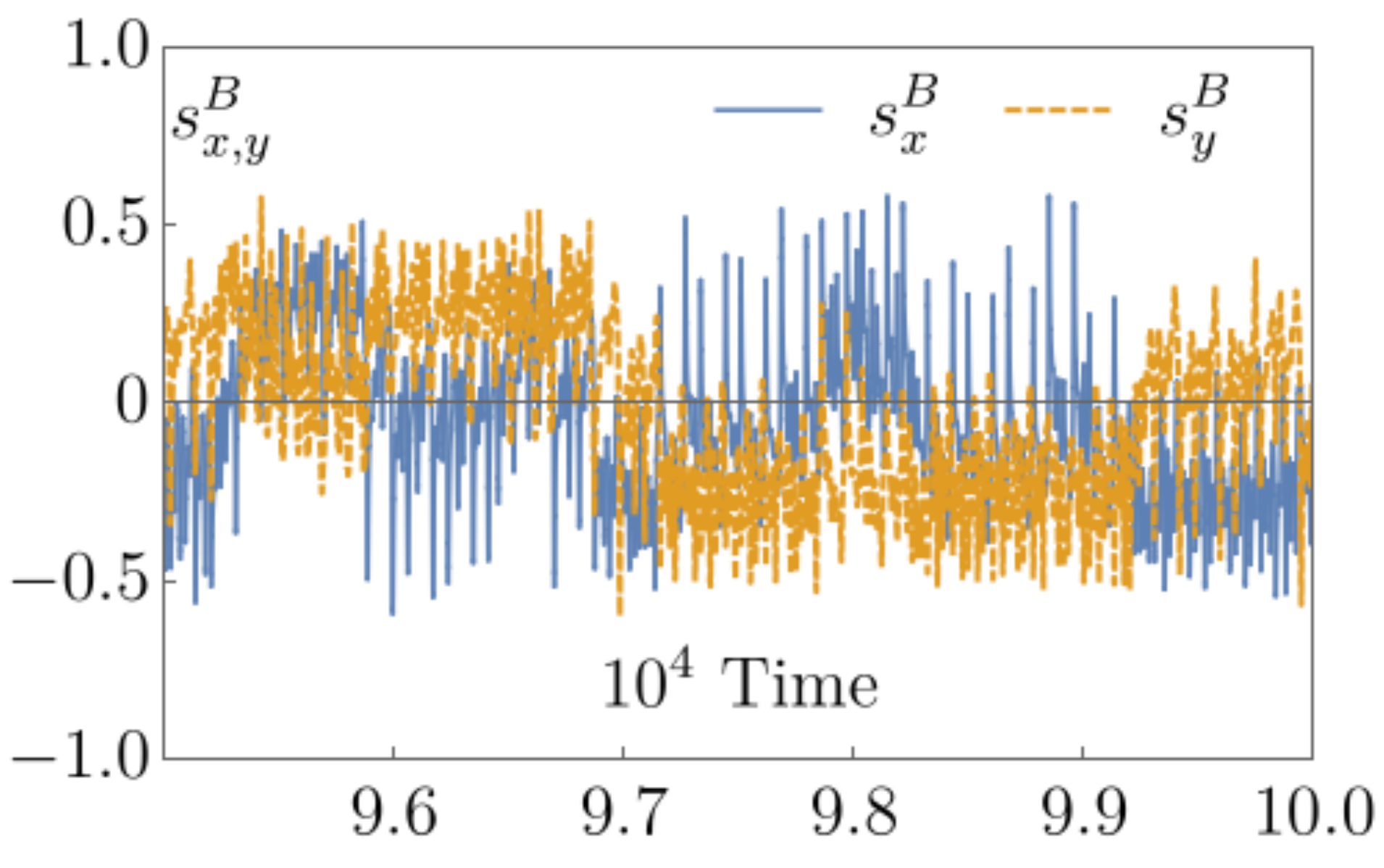}}\qquad\qquad
\subfloat[\qquad\textbf{(d)}]{\includegraphics[scale=0.3]{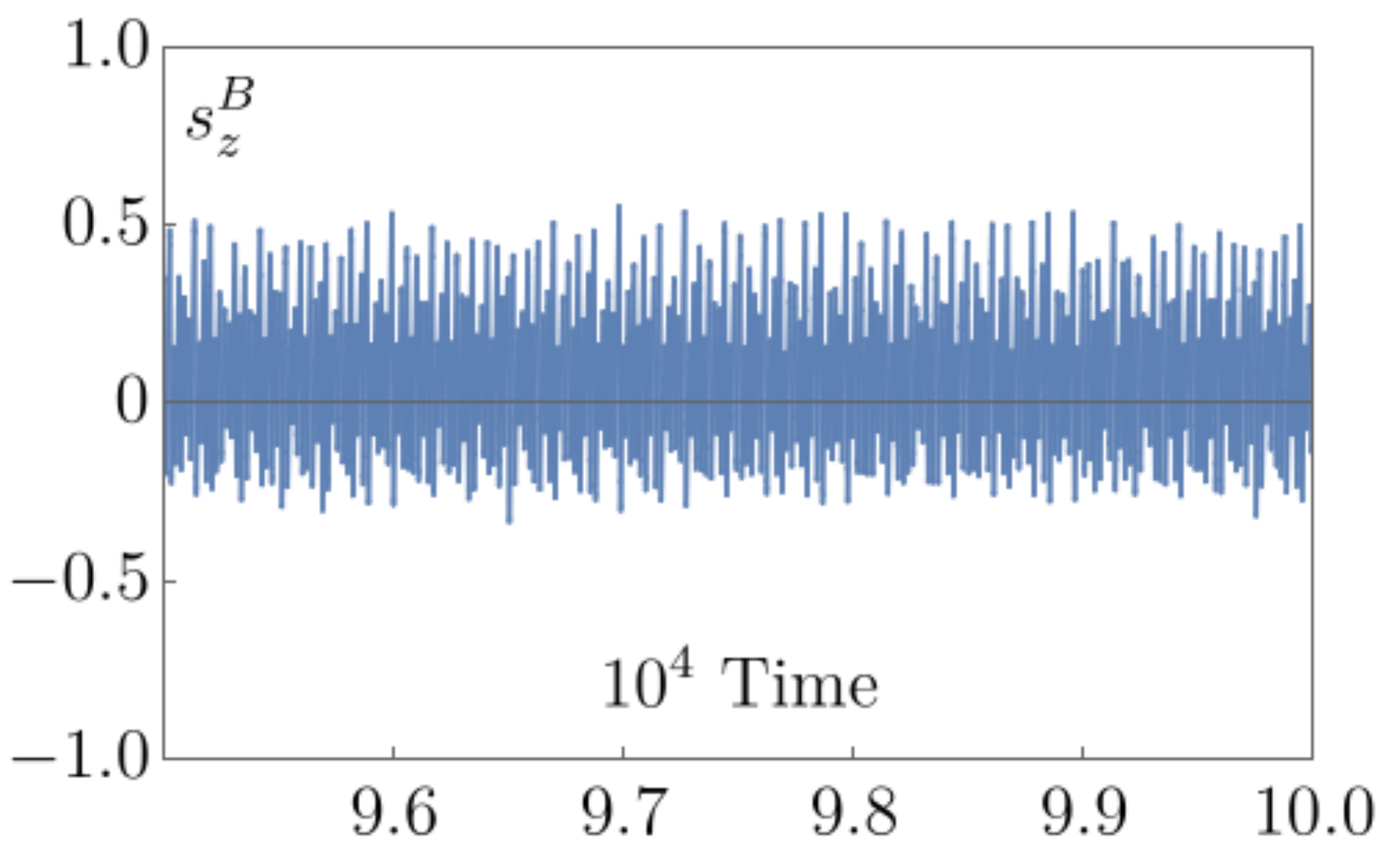}}\\
\subfloat[\qquad\textbf{(e)}]{\includegraphics[scale=0.3]{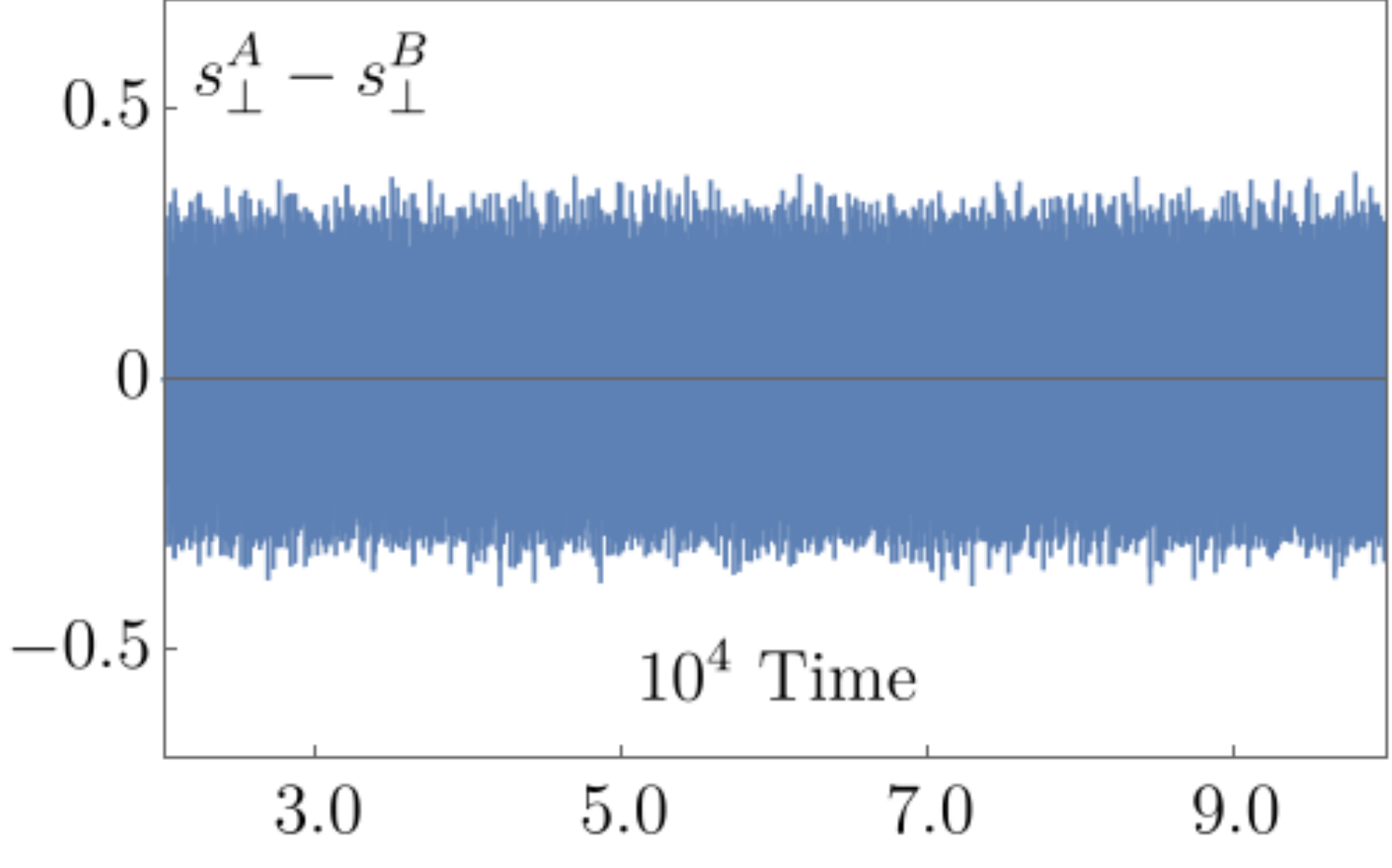}}\llap{\raisebox{2.4cm}{\includegraphics[height=2.0cm]{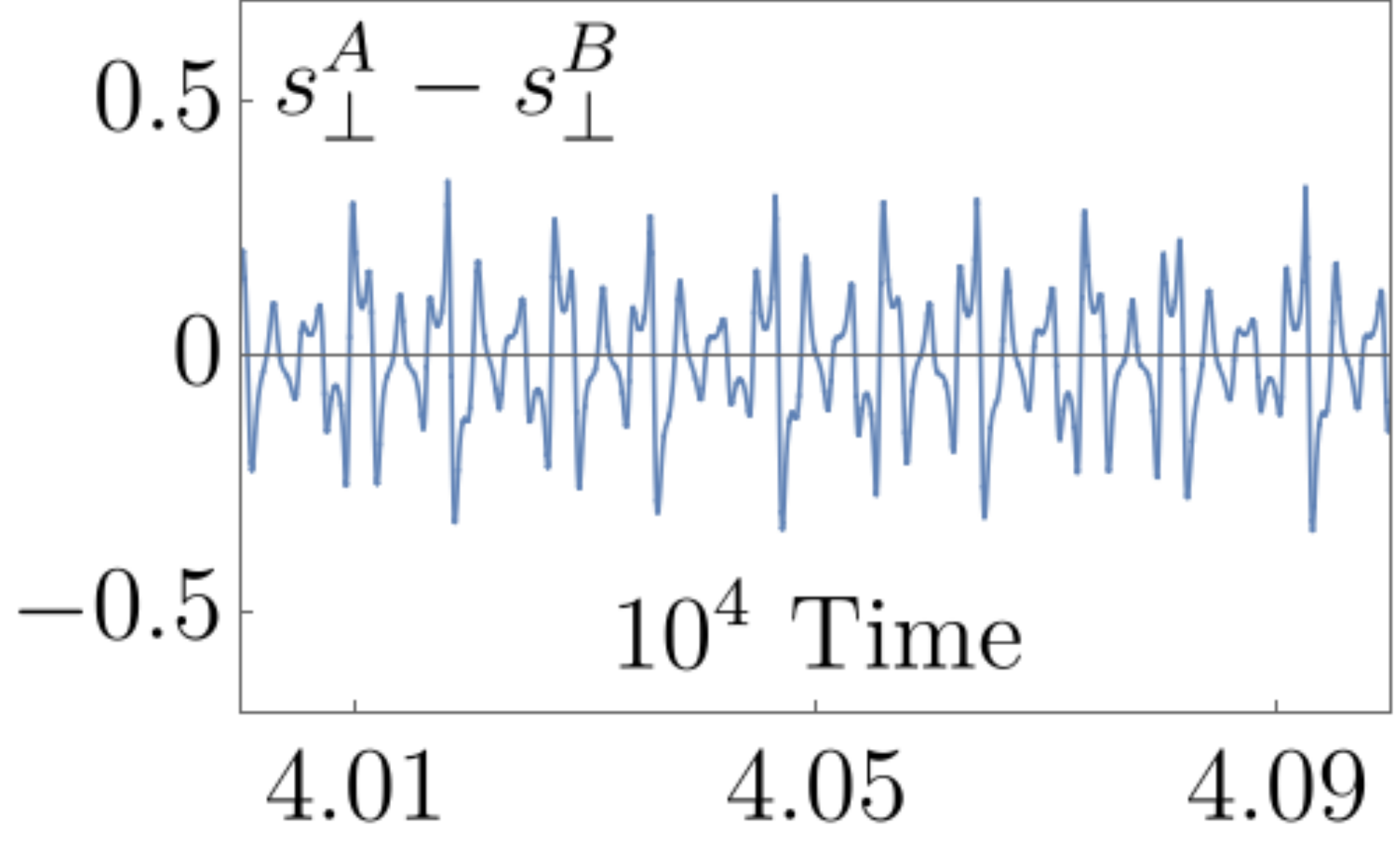}}}\qquad\qquad
\subfloat[\qquad\textbf{(f)}]{\includegraphics[scale=0.3]{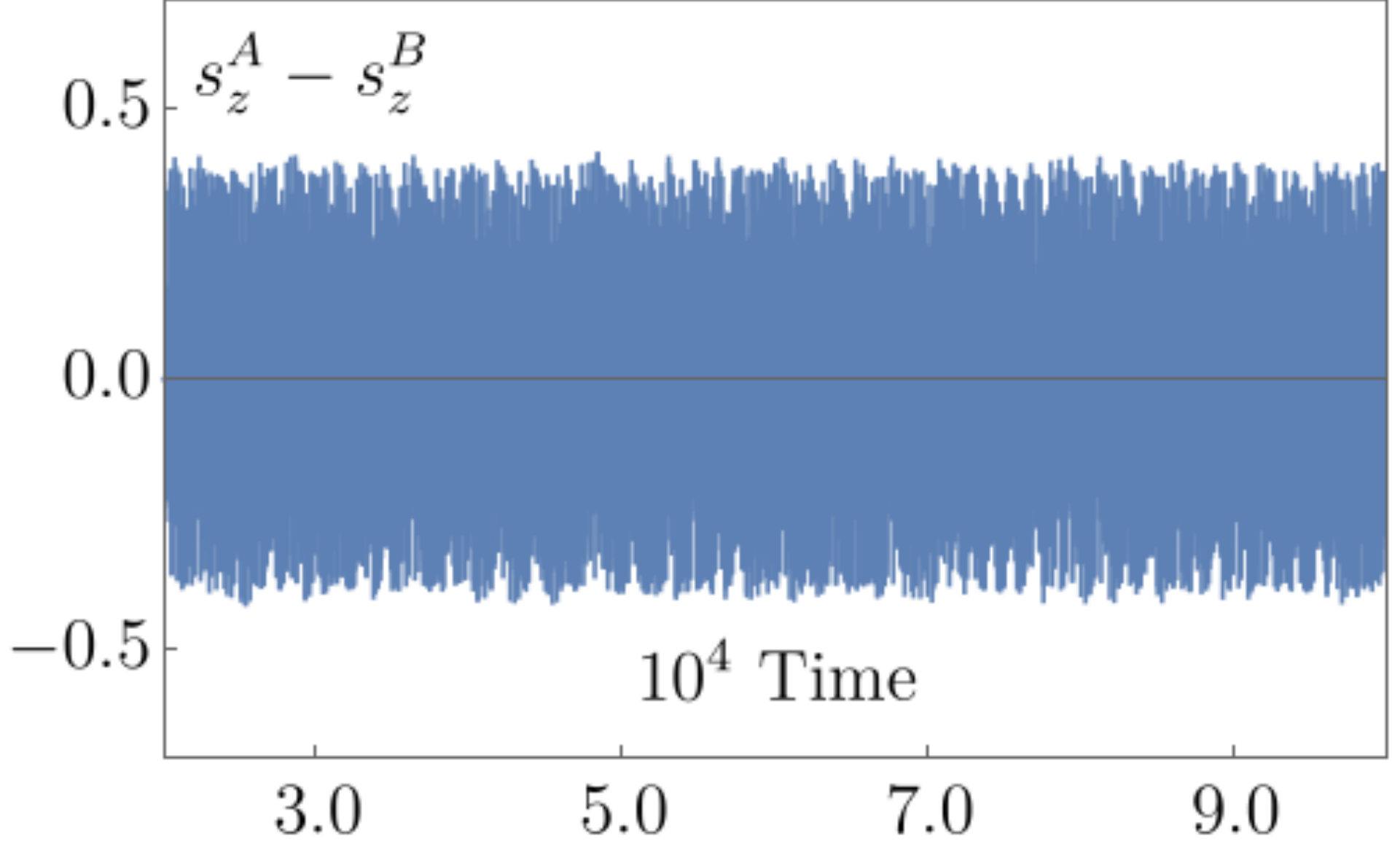}}\llap{\raisebox{2.4cm}{\includegraphics[height=2.0cm]{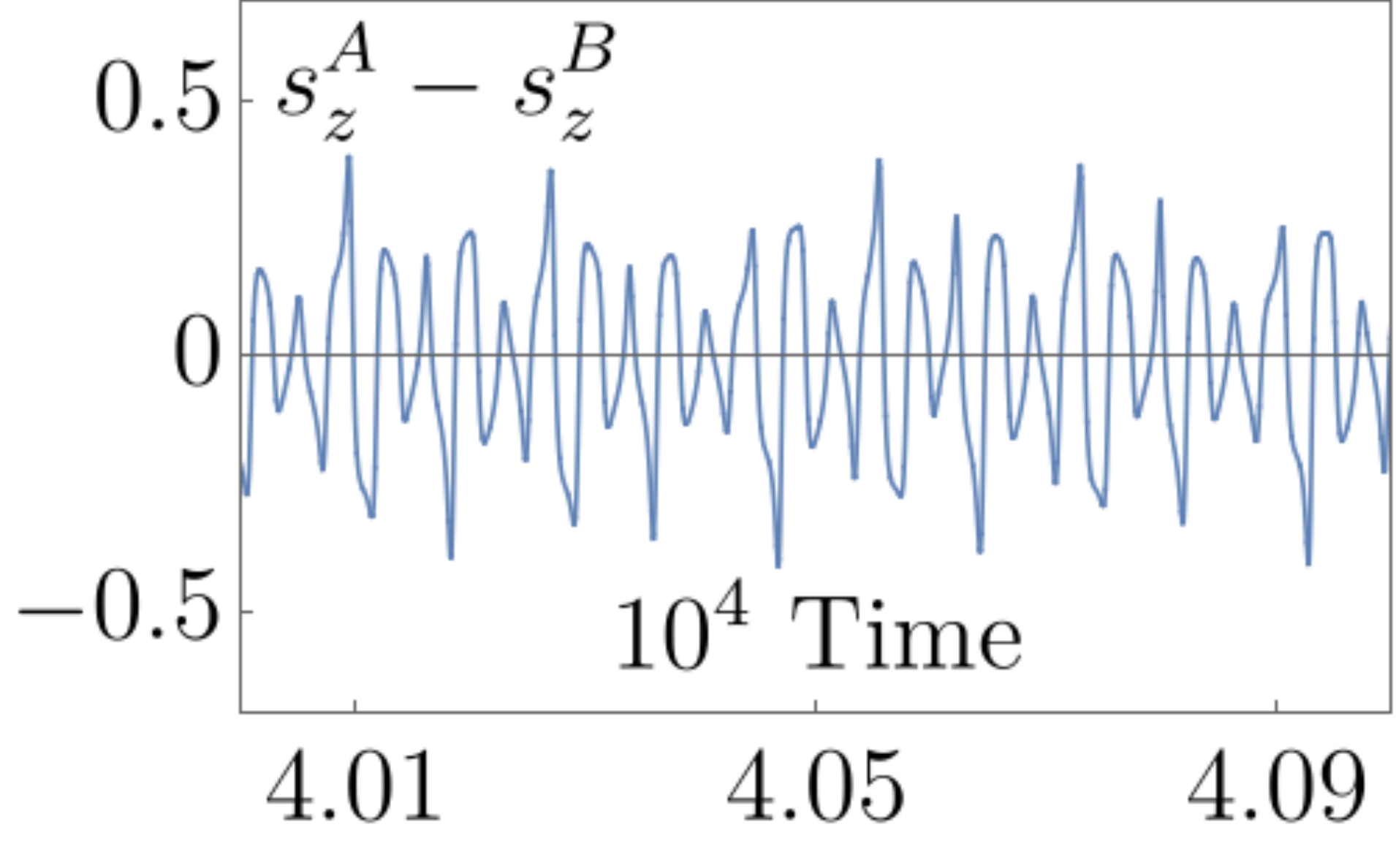}}}
\caption{Evolution of the classical spins  for a chaotic attractor at $\delta = 0.100, W = 0.055$. Conventions are the same as in \fref{QP_Pictures}. Note that just by observing these pictures and comparing these to \fref{QP_Pictures}, it is hard to distinguish between chaotic and quasiperiodic trajectories. However, the corresponding Poincar\'{e} section (\fref{Poincare_C}) and the spectrum (\fref{Spectrum_C_Full}) for this chaotic attractor are distinct from the Poincar\'{e} section (\fref{Poincare_QP}) and the spectrum (\fref{Spectrum_QP}) for the quasiperiodic one in \fref{QP_Pictures}. We also note in \fref{Lyapunov} that the maximum Lyapunov exponent $\lambda$ for this attractor is positive, unlike the one for the quasiperiodic attractor, for which $\lambda = 0$.}
\label{C_Pictures}
\end{figure*}

\begin{figure*}[tbp!]
\centering
\subfloat[\qquad\textbf{(a)}]{\label{C_SA_vs_SB}\includegraphics[scale=0.3]{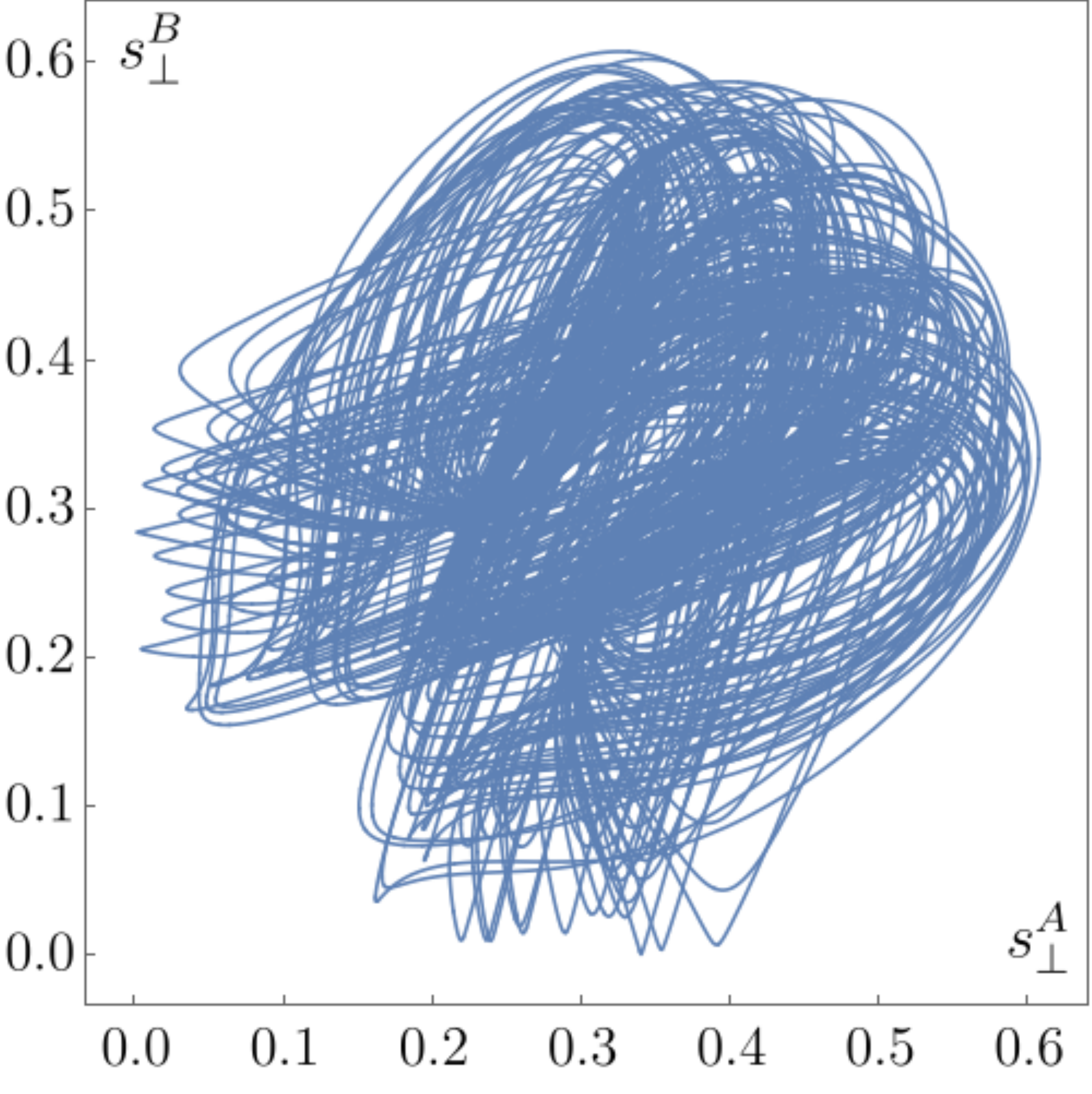}}\qquad\qquad
\subfloat[\qquad\textbf{(b)}]{\label{C_SZA_vs_SZB}\includegraphics[scale=0.3]{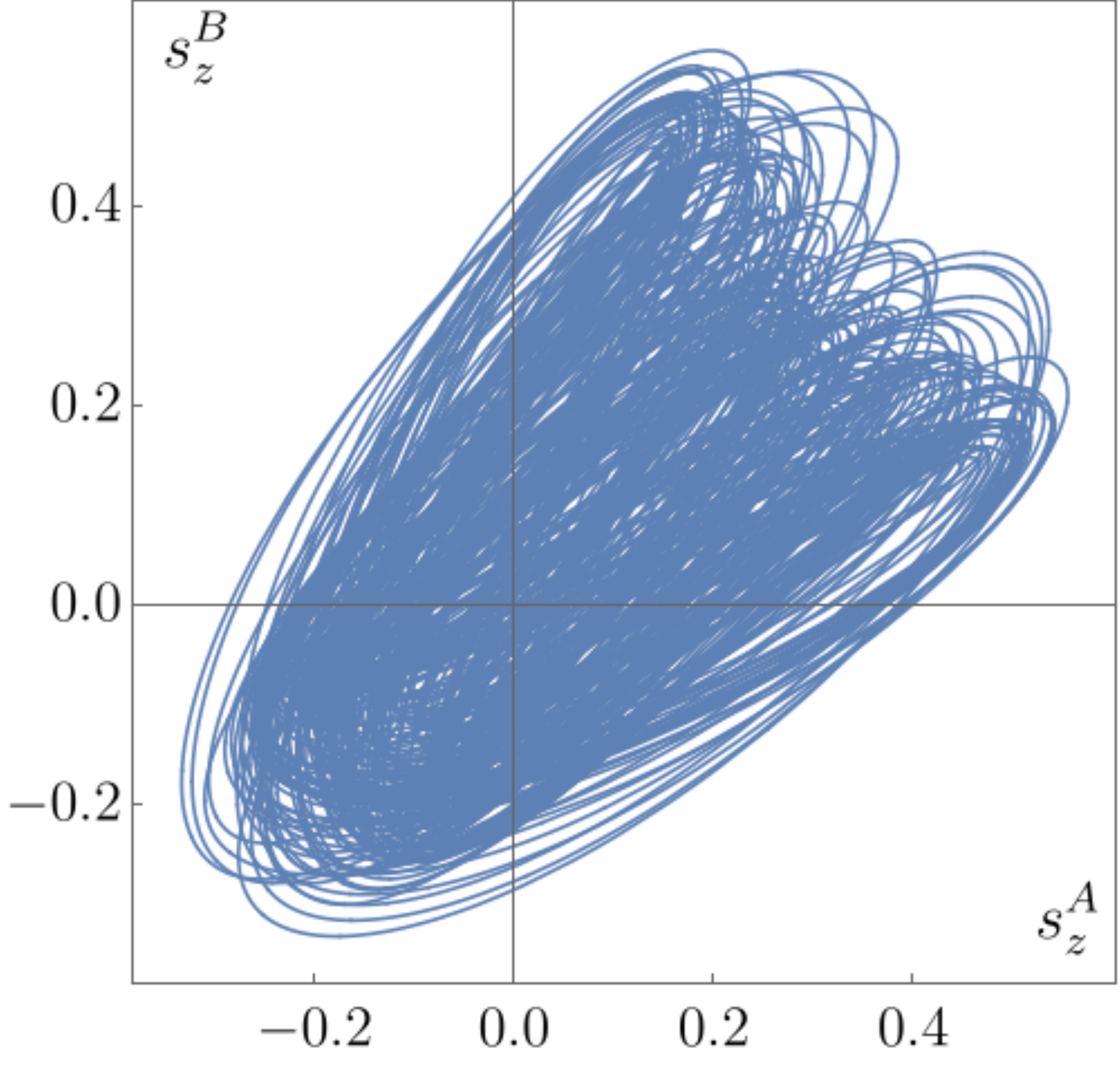}}
\caption{Different projections of the 6D chaotic attractor at $\delta = 0.100$ and $W = 0.055$. In \textbf{(a)} and \textbf{(b)}, we show $s_{\perp}^{A}$ vs. $s_{\perp}^{B}$ and $s_{z}^{A}$ vs. $s_{z}^{B}$, respectively. Note that these projections fill up finite regions more densely compared to the ones in \fref{QP_X_Sect}.}
\label{C_X_Sect}
\end{figure*}

\begin{figure*}[tbp!]
\centering
\subfloat[\textbf{(a)}]{\label{Traj_Sect_C}\includegraphics[scale=0.33]{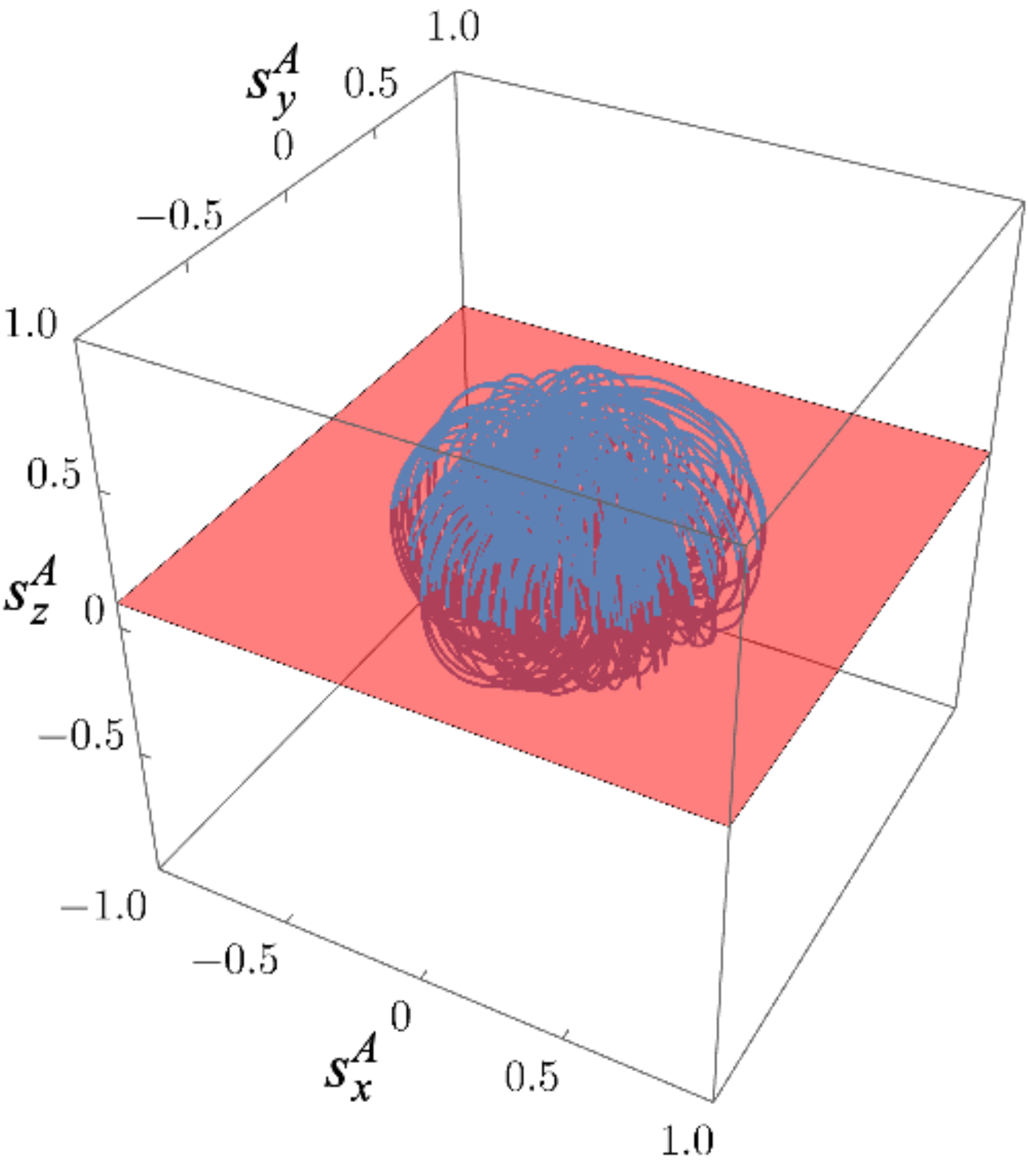}}\qquad\qquad
\subfloat[\qquad\textbf{(b)}]{\label{Poincare_C1}\includegraphics[scale=0.31]{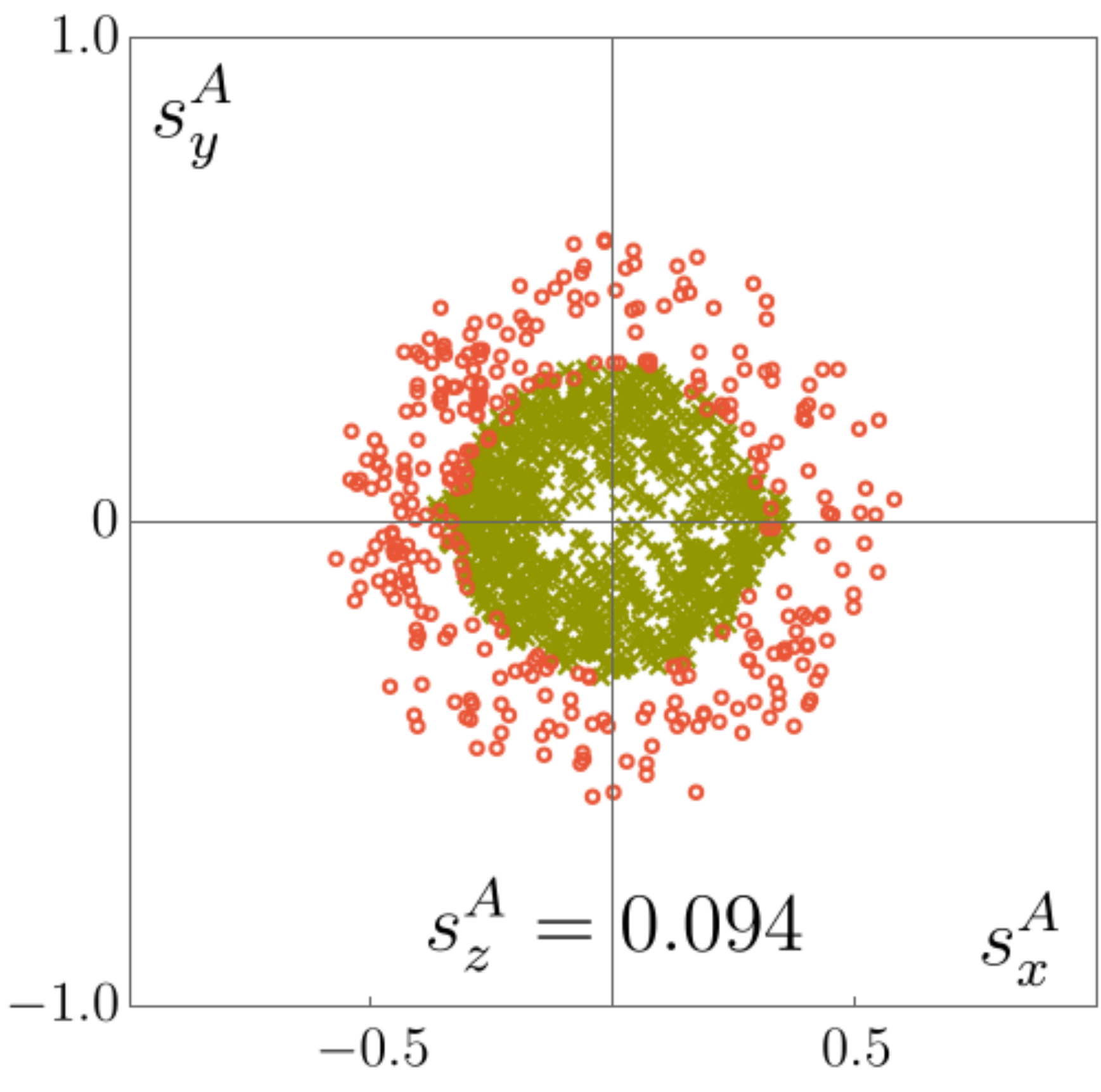}}z
\caption{Poincar\'{e} section of the A spin, when the system is in the chaotic superradiant phase. The parameters $(\delta, W)$ are the same as in \fsref{C_Pictures} and \ref{C_X_Sect}. We show the orbit crossing the transverse plane (translucent red in \fref{Traj_Sect_C}) from above and below in green crosses and red circle, respectively. The scattered nature of \fref{Poincare_C1} corroborates the chaotic nature of the attractor. (For interpretation of the references to color in this figure legend, the reader is referred to the web version of this article.)}
\label{Poincare_C}
\end{figure*}

\begin{figure*}[tbp!]
\centering
\subfloat[\qquad\textbf{(a)}]{\label{Lyapunov_Exponent}\includegraphics[scale=0.35]{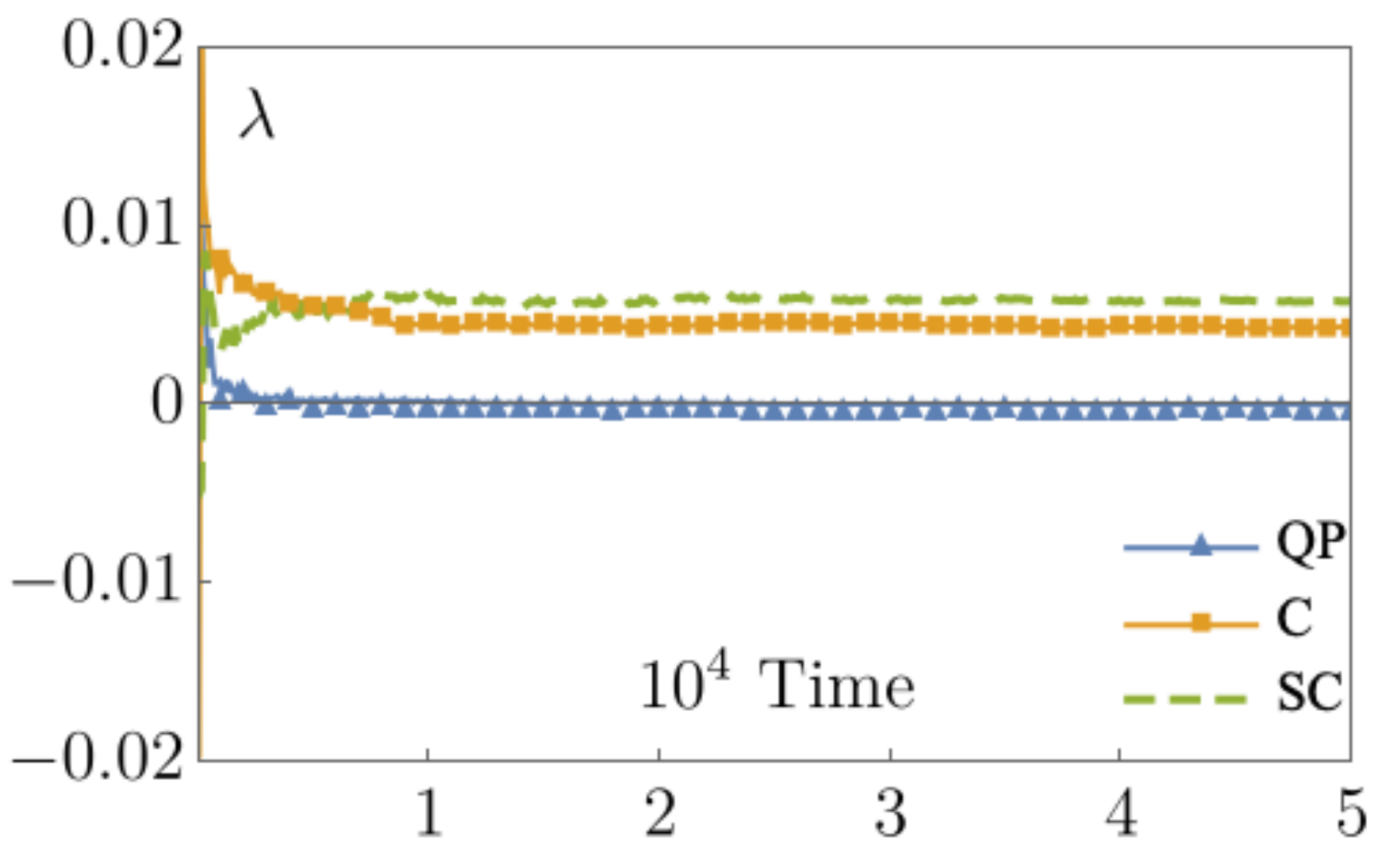}}\qquad\qquad
\subfloat[\qquad\textbf{(b)}]{\label{Log_Lyapunov_Exponent}\includegraphics[scale=0.33]{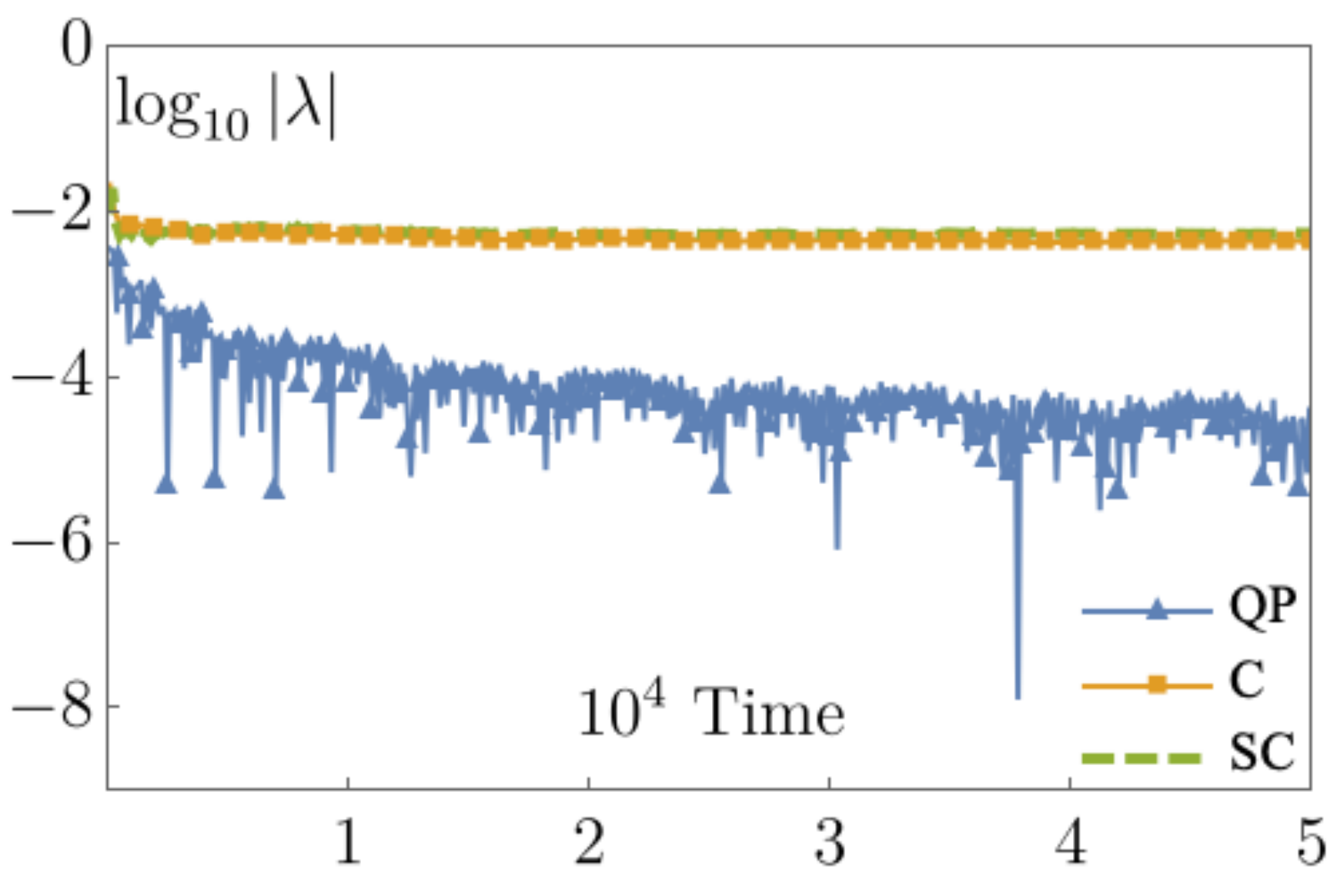}}
\caption{Maximum Lyapunov exponent $\lambda(t)$ (left) and the logarithm of its absolute value (right) for quasiperiodic (blue triangles), chaotic (yellow squares), and synchronized chaotic (green dashed line) attractors at a fixed repump rate $W = 0.055$ and decreasing detuning $\delta$. For the three different behaviors the values of $\delta$ are $0.115$ (cf. \fsref{QP_Pictures}, \ref{QP_X_Sect}, \ref{Poincare_QP} (\blue{c}, \blue{d}) and \ref{Spectrum_QP}), $0.100$ (cf. \fsref{C_Pictures}, \ref{C_X_Sect}, \ref{Poincare_C} and \ref{Spectrum_C_Full}) and $0.080$ (cf. \fsref{SC_Pictures}, \ref{SC_X_Sect}, \ref{Poincare_SC} and \ref{Spectrum_SC_Full}), respectively. In parentheses, we mention the figures that depict the corresponding spin component dynamics, different 2D projections, Poincar\'{e} sections and spectra. In \textbf{(a)} and \textbf{(b)}, we observe that $\lambda$ has saturated to a postive value for chaos and synchronized chaos, whereas for quasiperiodicity it is approaching zero. Note that for quasiperiodicity, unlike the other two attractors, $\log_{10}{|\lambda(t)|}$ has not saturated at the end of simulation time and is still decreasing.} \label{Lyapunov}
\end{figure*}

\begin{figure*}[tbp!]
\centering
\subfloat[\qquad\textbf{(a)}]{\label{Spectrum_C}\includegraphics[scale=0.34]{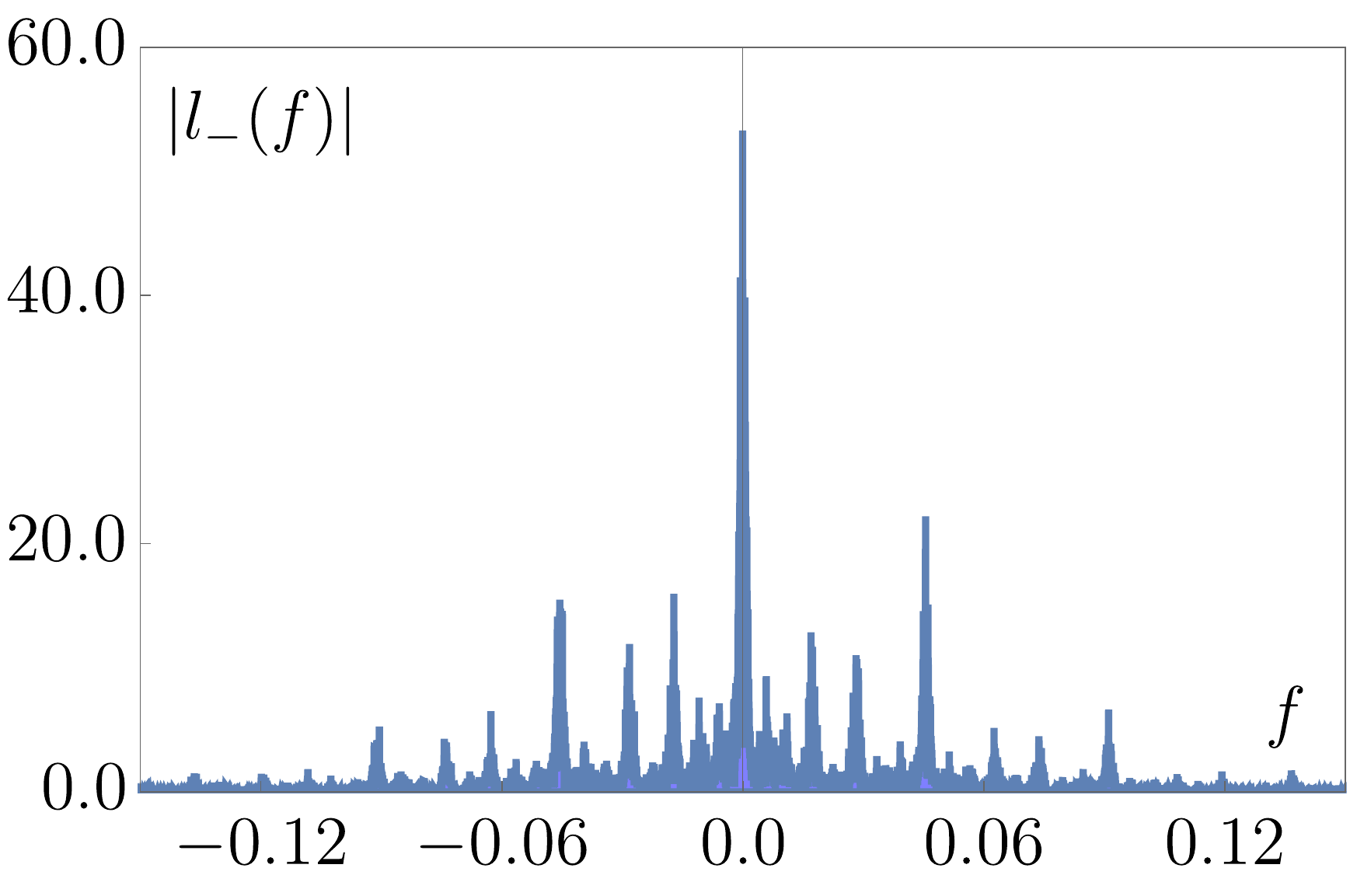}}\qquad\qquad
\subfloat[\qquad\textbf{(b)}]{\label{Spectrum_C_Mag}\includegraphics[scale=0.34]{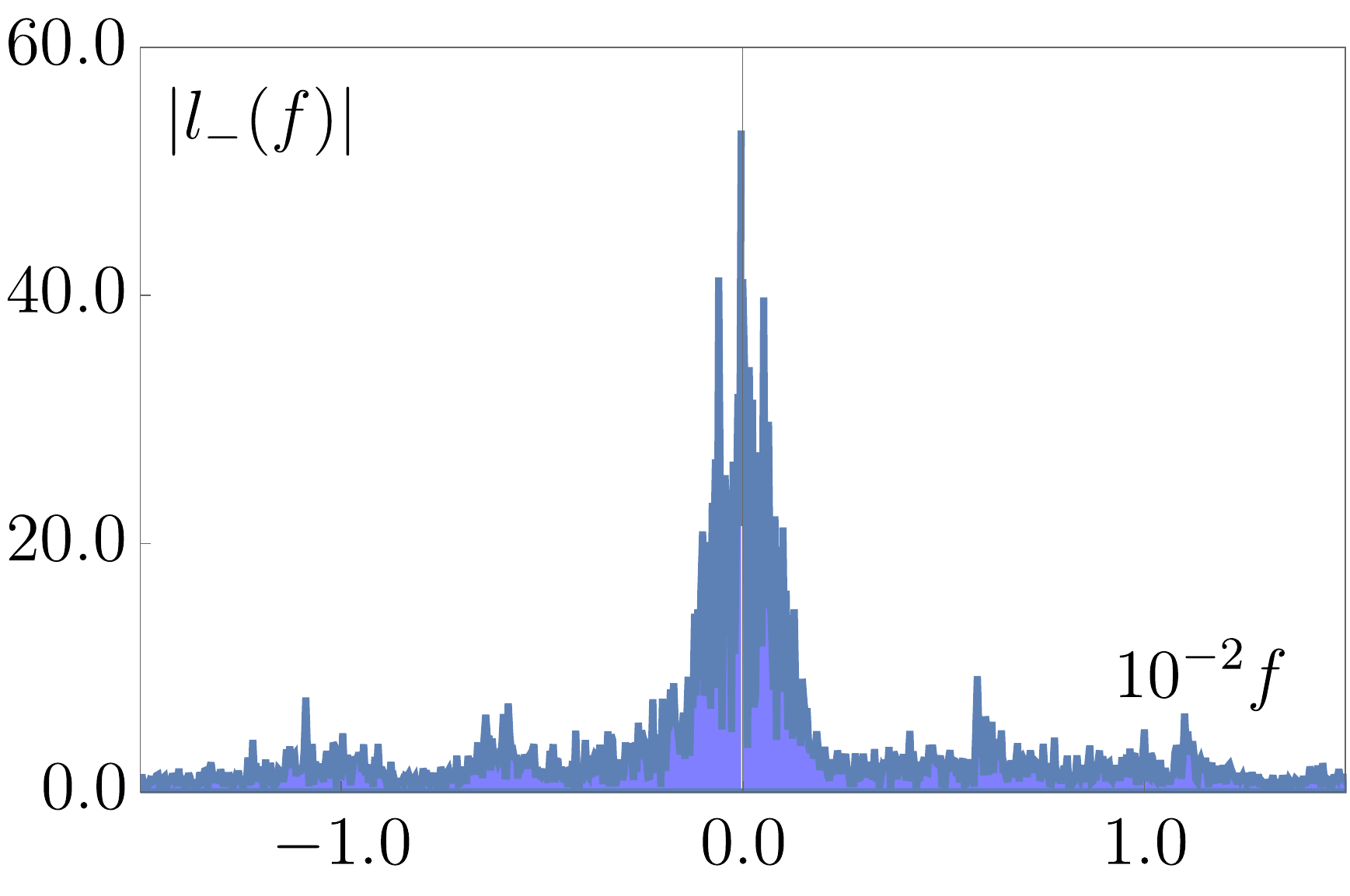}}
\caption{Power spectrum for the chaotic attractor (spin dynamics in \fref{C_Pictures}) at $\delta = 0.100, W = 0.055.$ It is a continuum without any reflection symmetry about the $f = 0$ axis. The spectrum has several peaks. The most prominent of those is located at the origin. In \textbf{(b)}, we magnify the region near $f = 0$ by a factor of $10^{2}$ to accentuate the peak there.} \label{Spectrum_C_Full}
\end{figure*}

\begin{figure*}[tbp!]
\centering
\subfloat[\qquad\textbf{(a)}]{\includegraphics[scale=0.3]{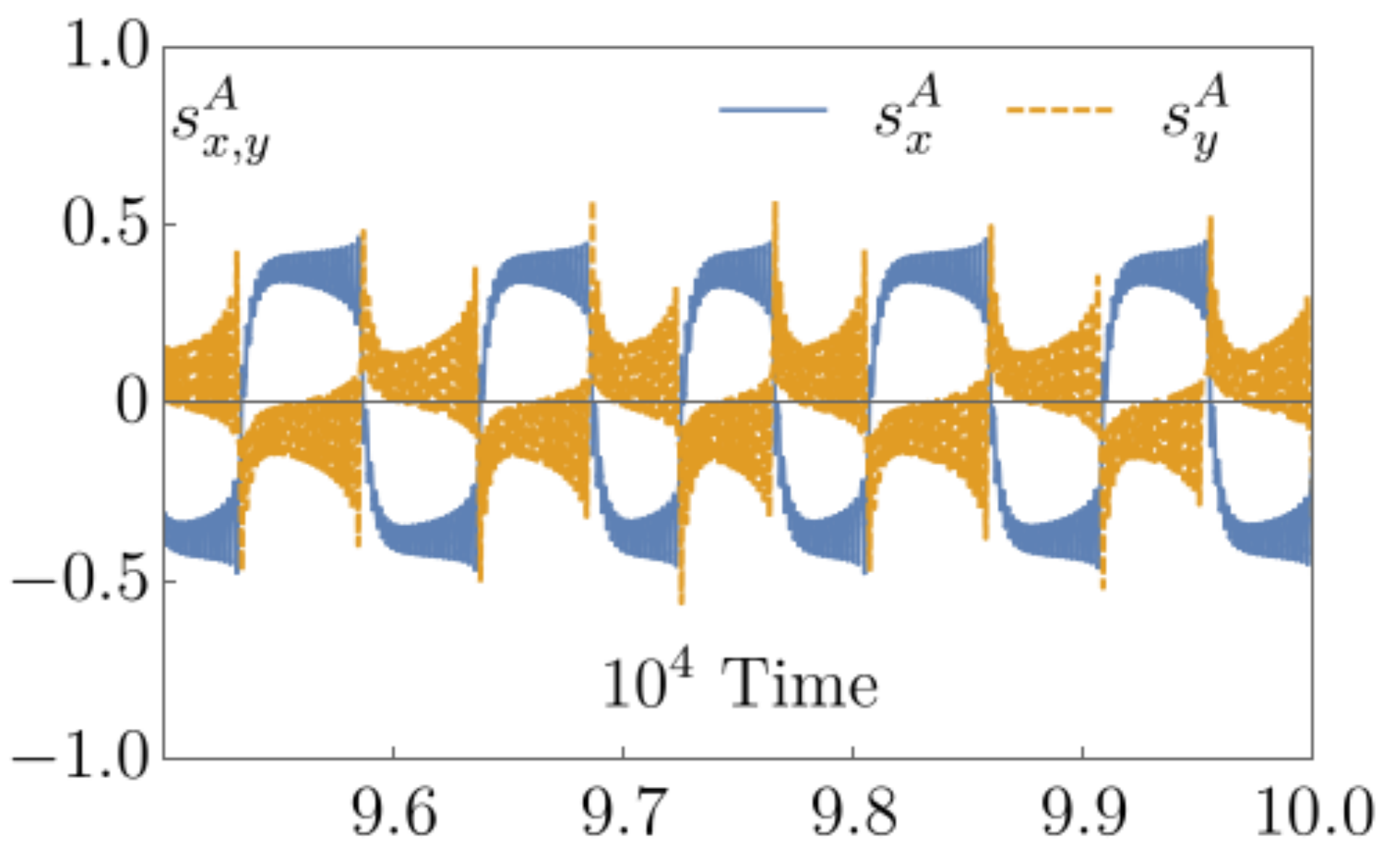}}\qquad\qquad
\subfloat[\qquad\textbf{(b)}]{\includegraphics[scale=0.3]{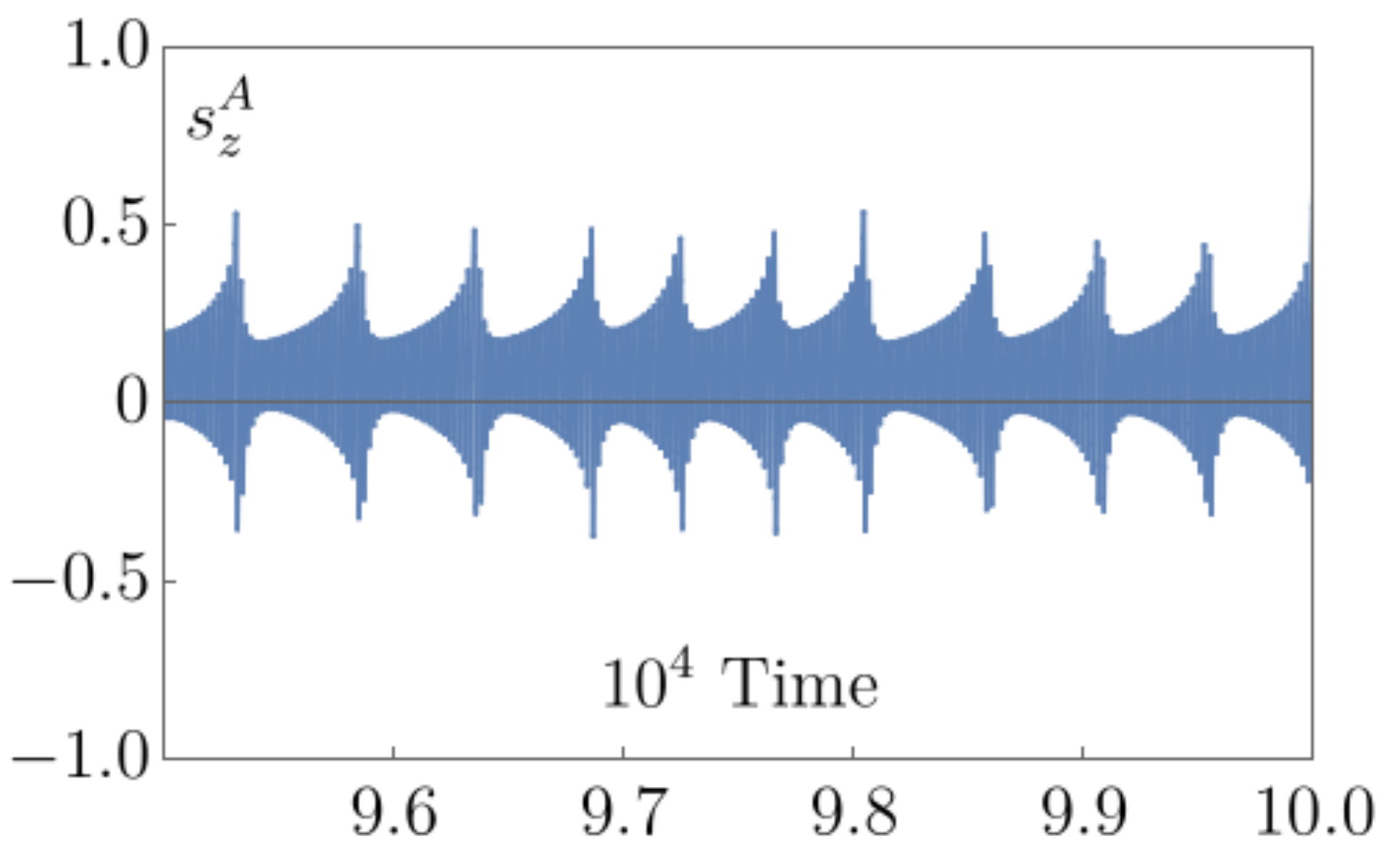}}\\
\subfloat[\qquad\textbf{(c)}]{\includegraphics[scale=0.3]{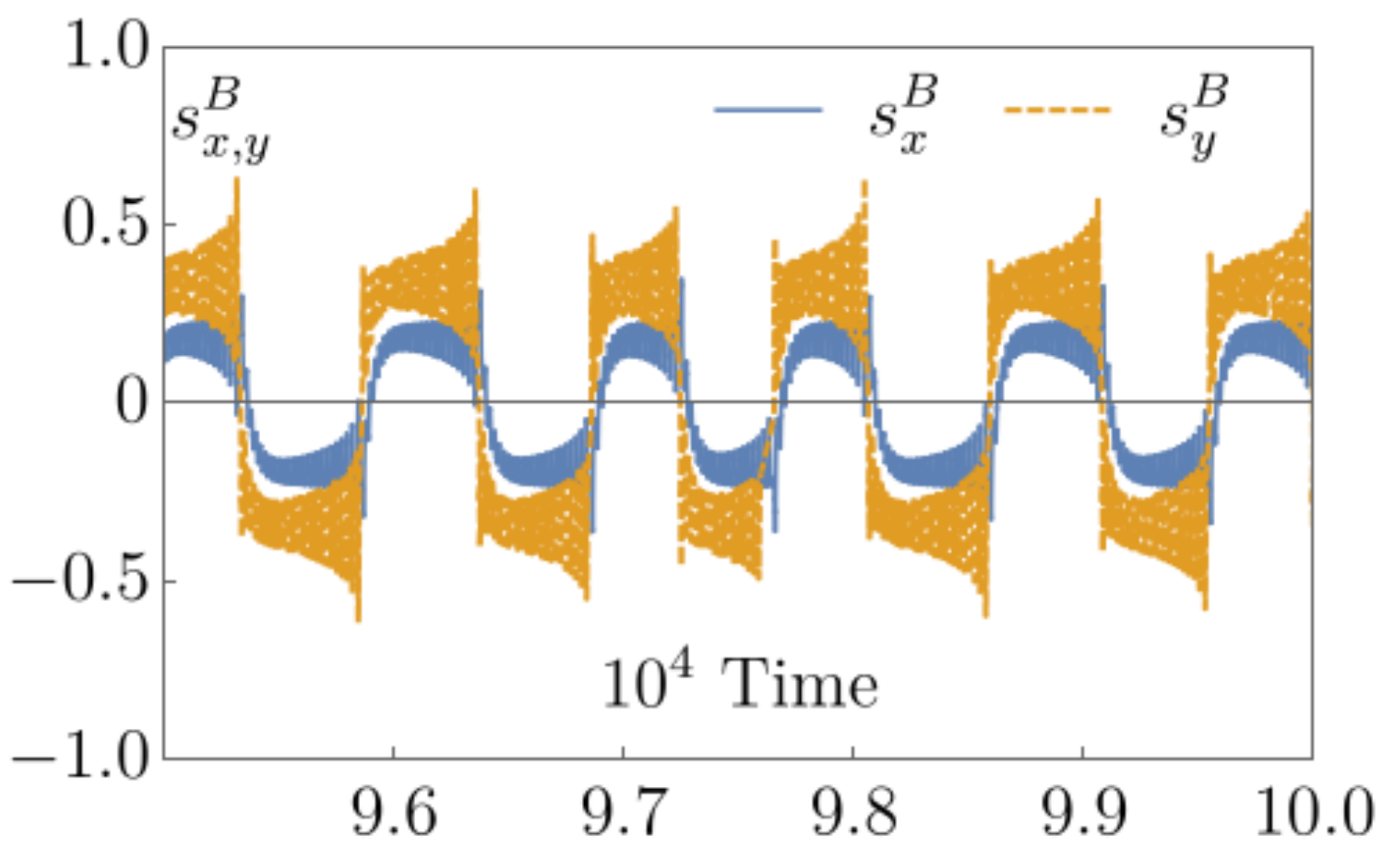}}\qquad\qquad
\subfloat[\qquad\textbf{(d)}]{\includegraphics[scale=0.3]{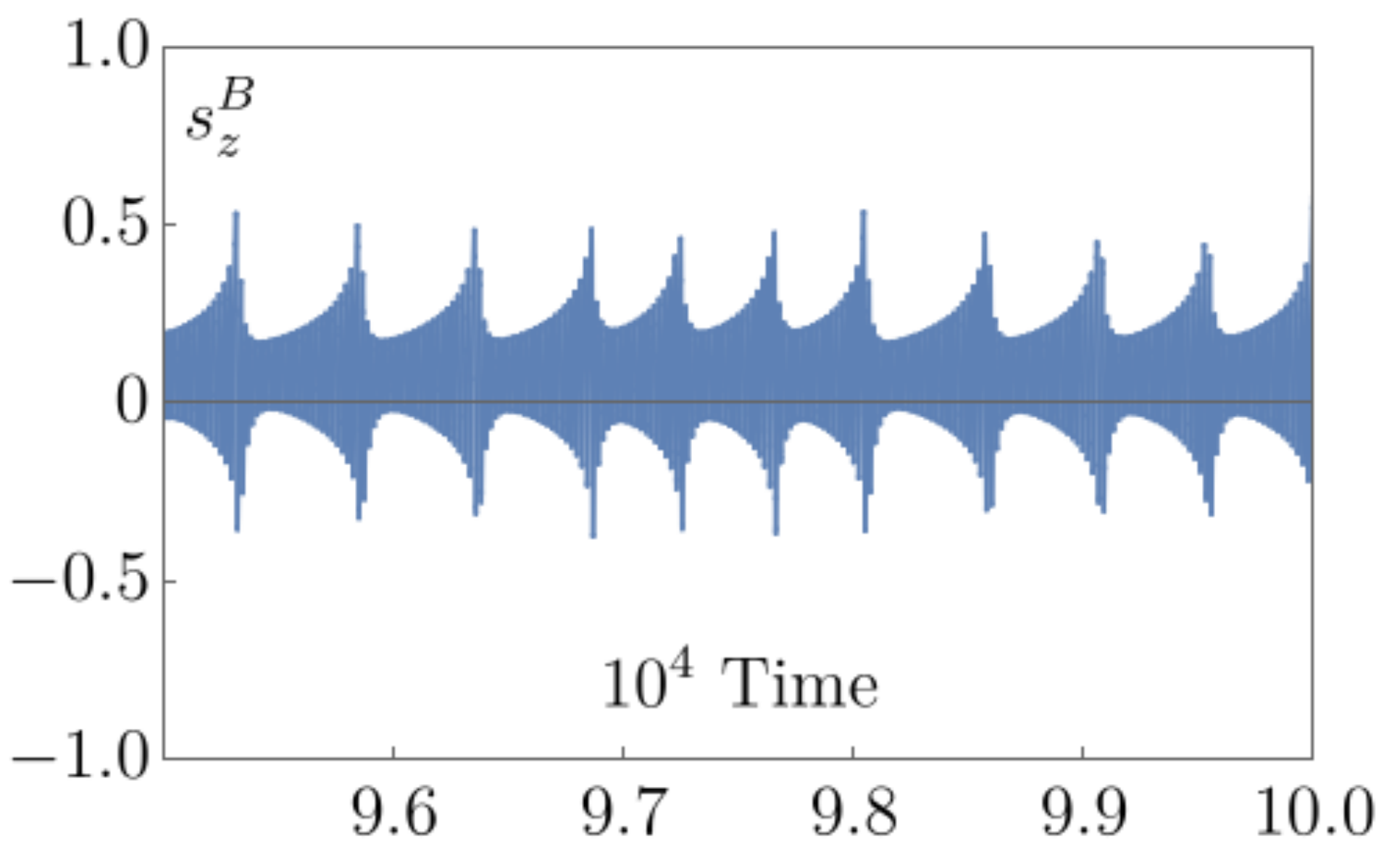}}\\
\subfloat[\qquad\textbf{(e)}]{\includegraphics[scale=0.3]{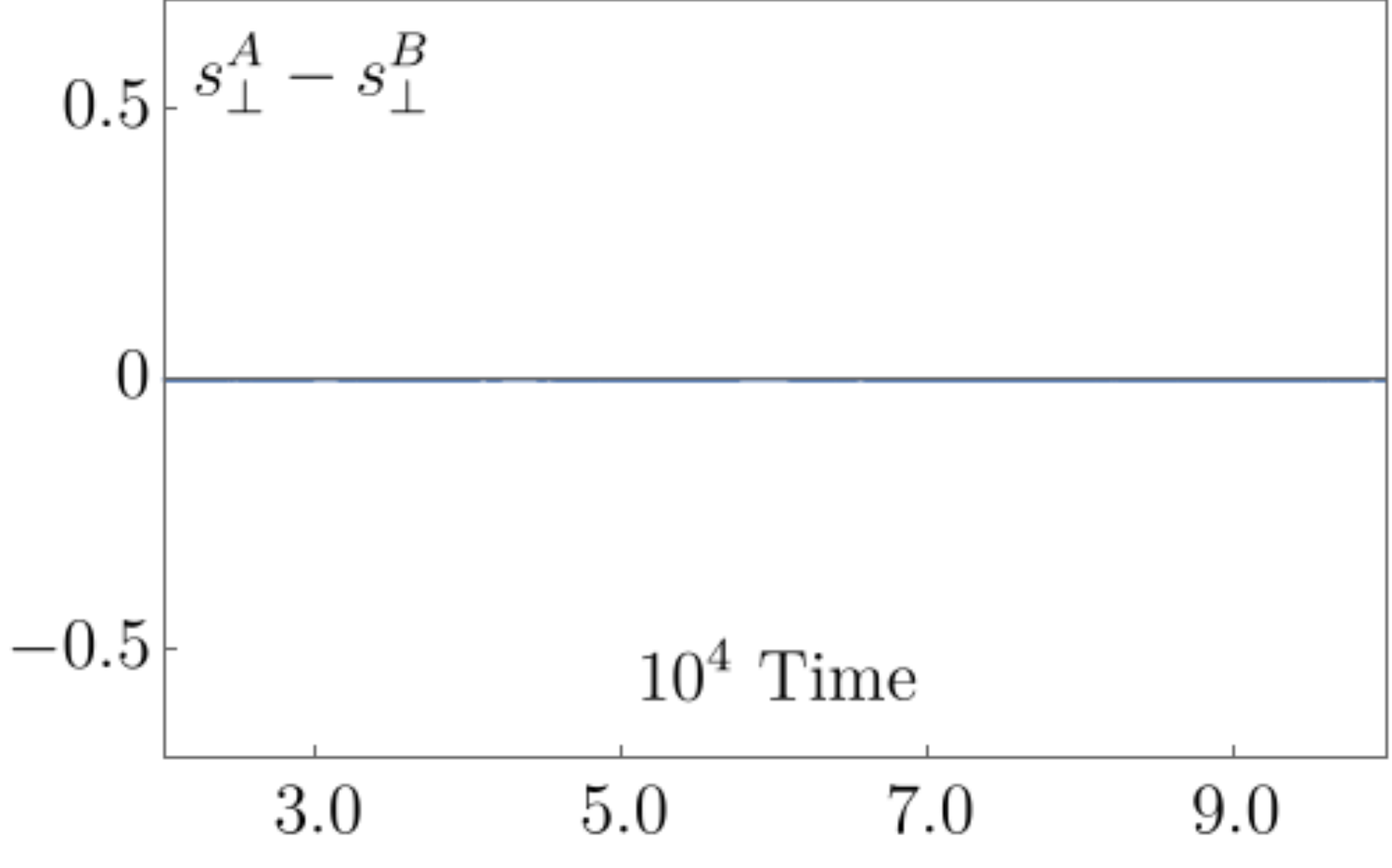}}\qquad\qquad
\subfloat[\qquad\textbf{(f)}]{\includegraphics[scale=0.3]{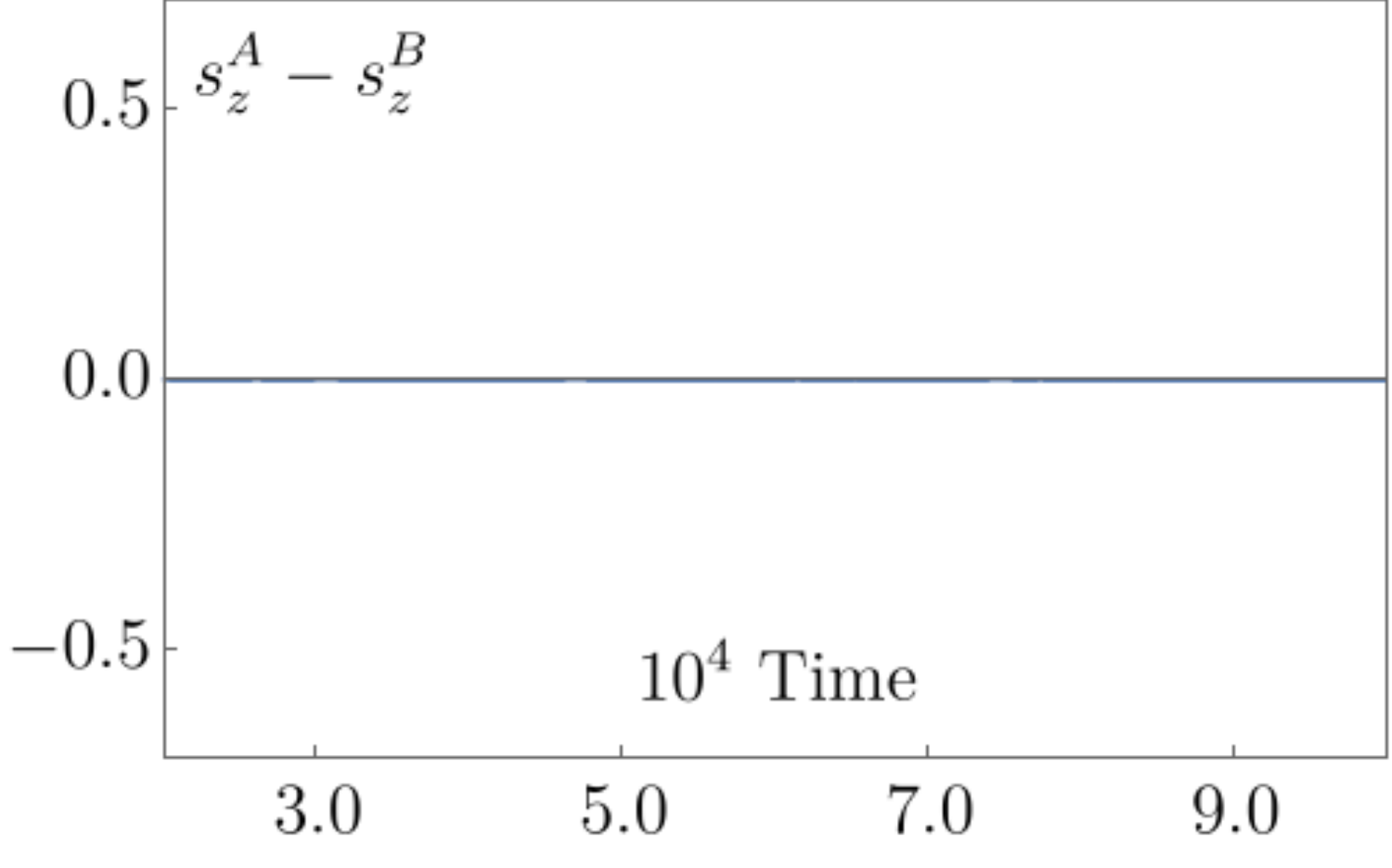}}
\caption{Dynamics of $\bm{s}^{A}$ and $\bm{s}^{B}$ in the synchronized chaotic phase at $\delta = 0.080, W = 0.055$ with the same conventions as in \fsref{QP_Pictures} and \ref{C_Pictures}. The time dependence of the spin components are markedly different from that in other phases. The chaotic nature of the spin dynamics, however, is apparent from \textbf{(a)}, \textbf{(b)}, \textbf{(c)} and \textbf{(d)}. At large times, the differences between $s_{\perp}^{A}$ and $s_{\perp}^{B}$, and $s_{z}^{A}$ and $s_{z}^{B}$ being equal to zero is due to the $\Z2$ symmetry of the chaotic attractor. The reflection symmetry about $f = 0$ axis in the spectrum in \fref{Spectrum_SC_Full} is due to the above symmetry.}
\label{SC_Pictures}
\end{figure*}

Comparing the spin dynamics (\fref{C_Pictures}) and cross sections (\fref{C_X_Sect}) for the chaotic attractors to the quasiperiodic ones (\fsref{QP_Pictures} and \ref{QP_X_Sect}, respectively) is not sufficient for distinguishing the two behaviors. We therefore analyze the Poincar\'{e} sections (chaos: \fref{Poincare_C}, quasiperiodicity: \fref{Poincare_QP} (\blue{c}, \blue{d})), spectra [chaos: \fref{Spectrum_C}, quasiperiodicity: \fref{Spectrum_QP}] and the maximum Lyapunov exponents (\fref{Lyapunov}).        

The Poincar\'{e} section for $\bm{s}^{A}$ for the quasiperiodic attractors are continuous loops, whereas in the ones for chaotic trajectories we observe smearing of points. Note, both the shape of the $\bm{s}^{A}$ trajectory and the Poincar\'{e} section of the chaotic attractor in \fref{Poincare_C} strongly suggest a quasiperiodic (torus shaped) precursor. The main difference between a chaotic spectrum and a quasiperiodic one is that the first one is a continuum unlike the other. Such discernible distinction between chaotic and quasiperiodic spectra is, however, rare \cite{QP_Chaos_Spectra}. 

\begin{figure*}[tbp!]

\begin{center}
\includegraphics[scale=0.45]{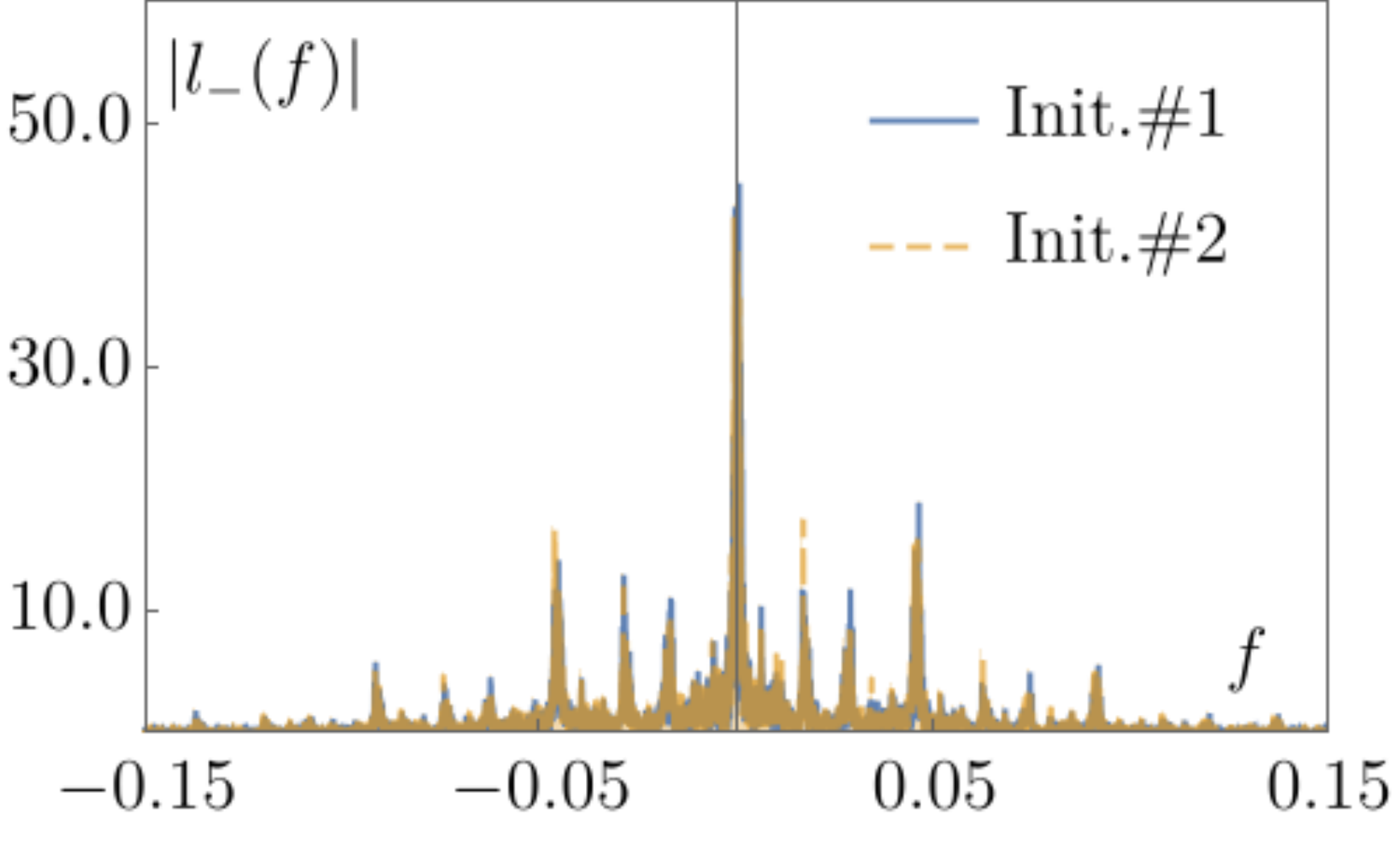}
\caption{Two power spectra in the chaotic superradiant phase for same $\delta$ and $W$ as in \fref{Spectrum_C_Full}, but for two different initial conditions. The peaks belonging to different spectra are still located at the same frequencies.}\label{Spectrum_C_Diff_Ini_Cond}
\end{center}

\end{figure*}

The definition of maximum Lyapunov exponent is      
\beg
\lambda(t) = \lim_{d(0)\to 0} \frac{1}{t}\ln\left[\frac{d(t)}{d(0)}\right].
\label{Lyapunov_Exp_Defn}
\en%
Here $d(t)$ is the distance between two 6D spins $\textbf{K}$ that started out close to each other ($d(0)$ distance apart) for same $\delta$ and $W$. The dependence of $\lambda$ on initial condition is weak. We provide a detailed algorithm for calculating $\lambda(t)$ in \ref{Appnd_Lya_Exp}. In particular, irrespective of initial conditions, we have $\lambda$ saturating to similar (order of magnitude) positive values for a chaotic attractors. However, in \fref{Lyapunov} we note that the values of $\lambda$ for chaotic attractors are small ($\approx 10^{-2} \pm 10^{-5}$ after time $t = 5\times 10^{4}$). This indicates that the chaotic attractors appearing in the orange subregion of Phase III is only weakly chaotic. Moreover, the distinct peaks appearing in the chaotic spectra are robust with respect to initial conditions, see \fref{Spectrum_C_Diff_Ini_Cond}. Therefore we surmise that the chaotic attractors in our system never go too far from other nonchaotic ones (e.g. quasiperiodicity). 

\section{Synchronization of Chaos}\label{Chaotic_Synch}

\begin{figure*}[tbp!]

\begin{center}
\includegraphics[scale=0.45]{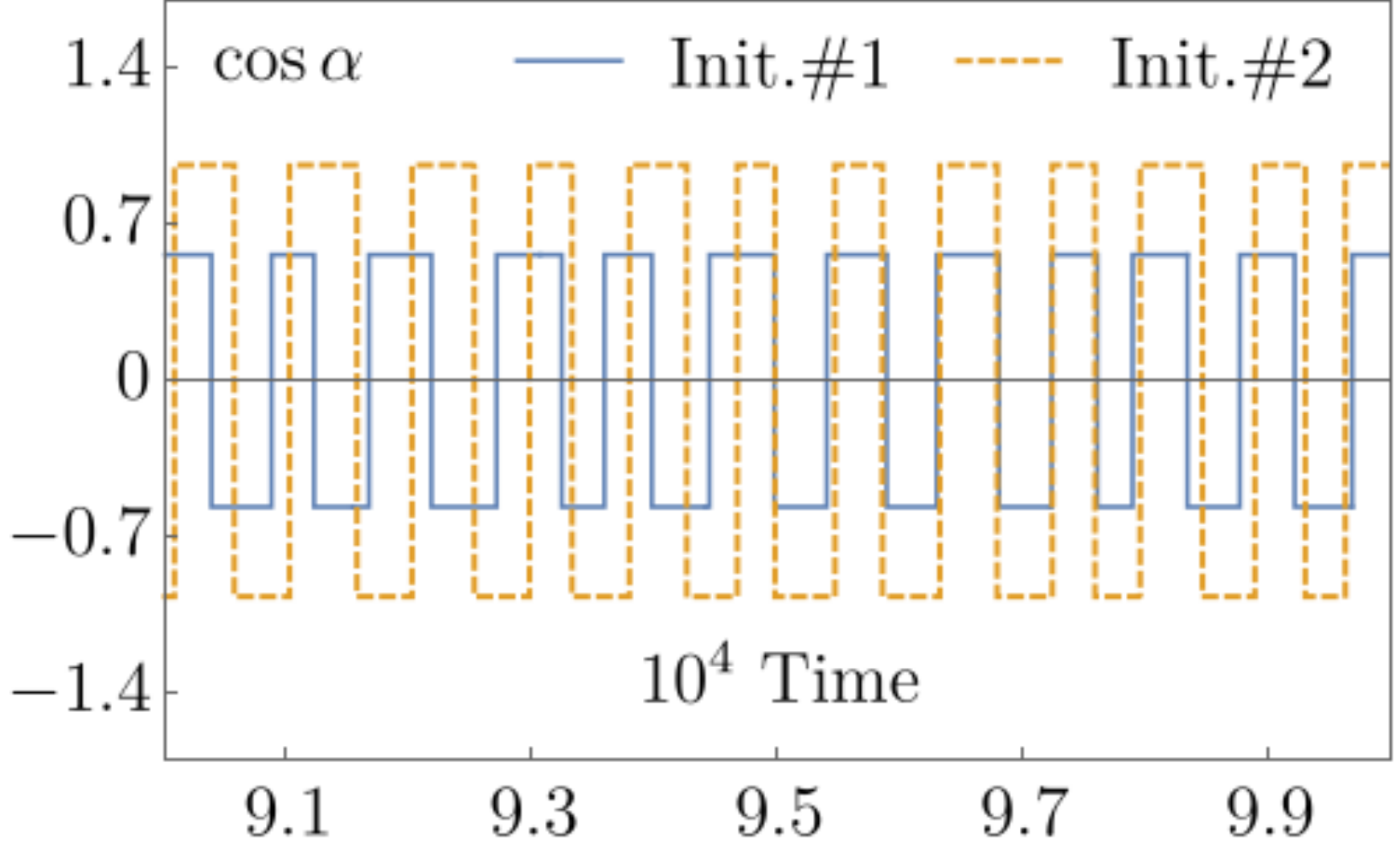}
\caption{Variation of $\cos{\alpha}$ with time for synchronized chaos (same attractor as in \fsref{SC_Pictures} and \ref{SC_X_Sect}), where $\alpha(t)$ is the angle between the 2D vector $\bm{l}_{\perp} = (l_{x}, l_{y})$ and the positive x-axis. Two graphs correspond to two distinct initial conditions. We observe that in both instances the value randomly jumps between $\pm \cos{\alpha_{0}}$, where $\alpha_{0}$ depends on the initial condition, but is independent of time.}\label{Cos_Alph_CS}
\end{center}

\end{figure*}

\begin{figure*}[t]
\centering
\subfloat[\qquad\textbf{(a)}]{\label{SC_SA_vs_SB}\includegraphics[scale=0.3]{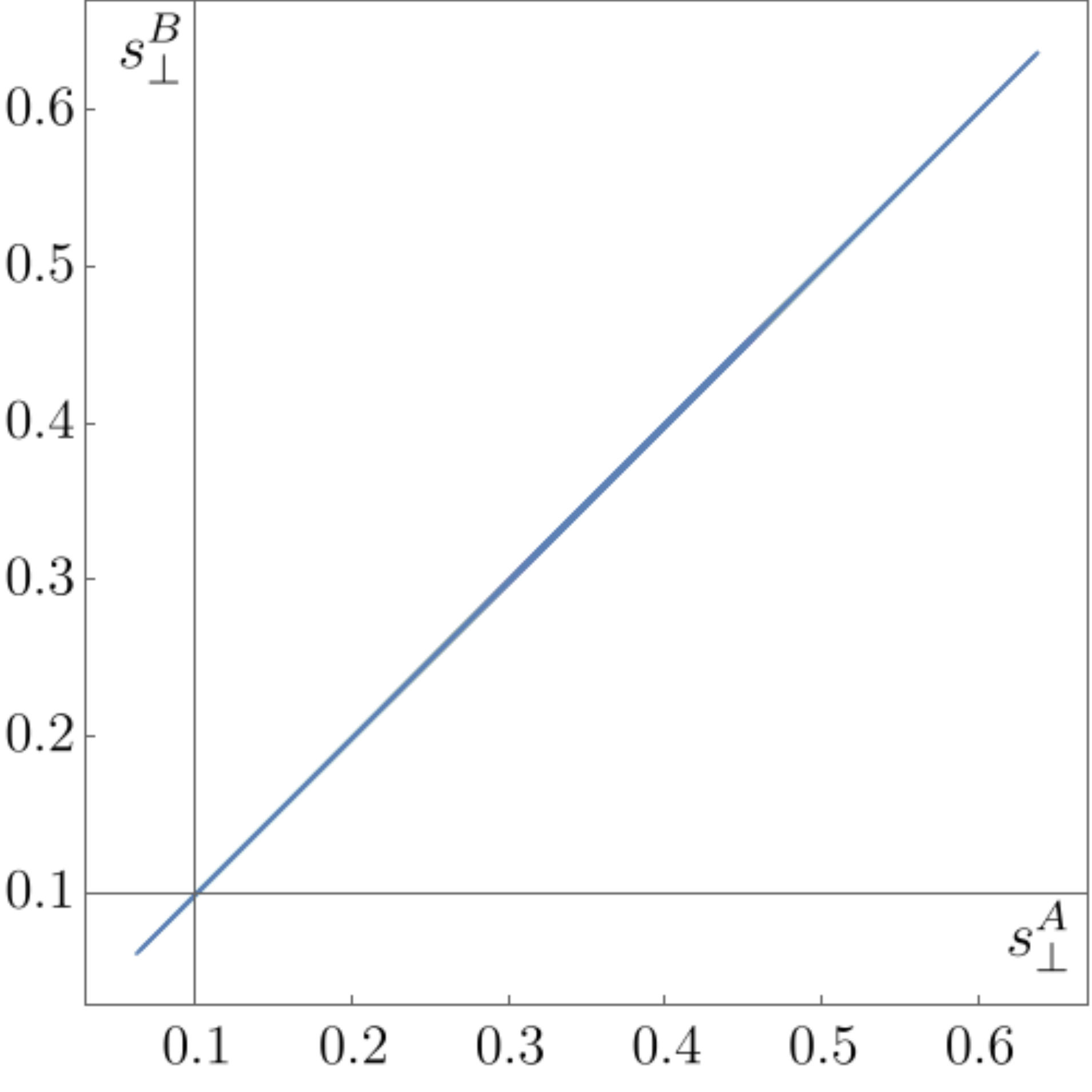}}\qquad\qquad
\subfloat[\qquad\textbf{(b)}]{\label{SC_SZA_vs_SZB}\includegraphics[scale=0.3]{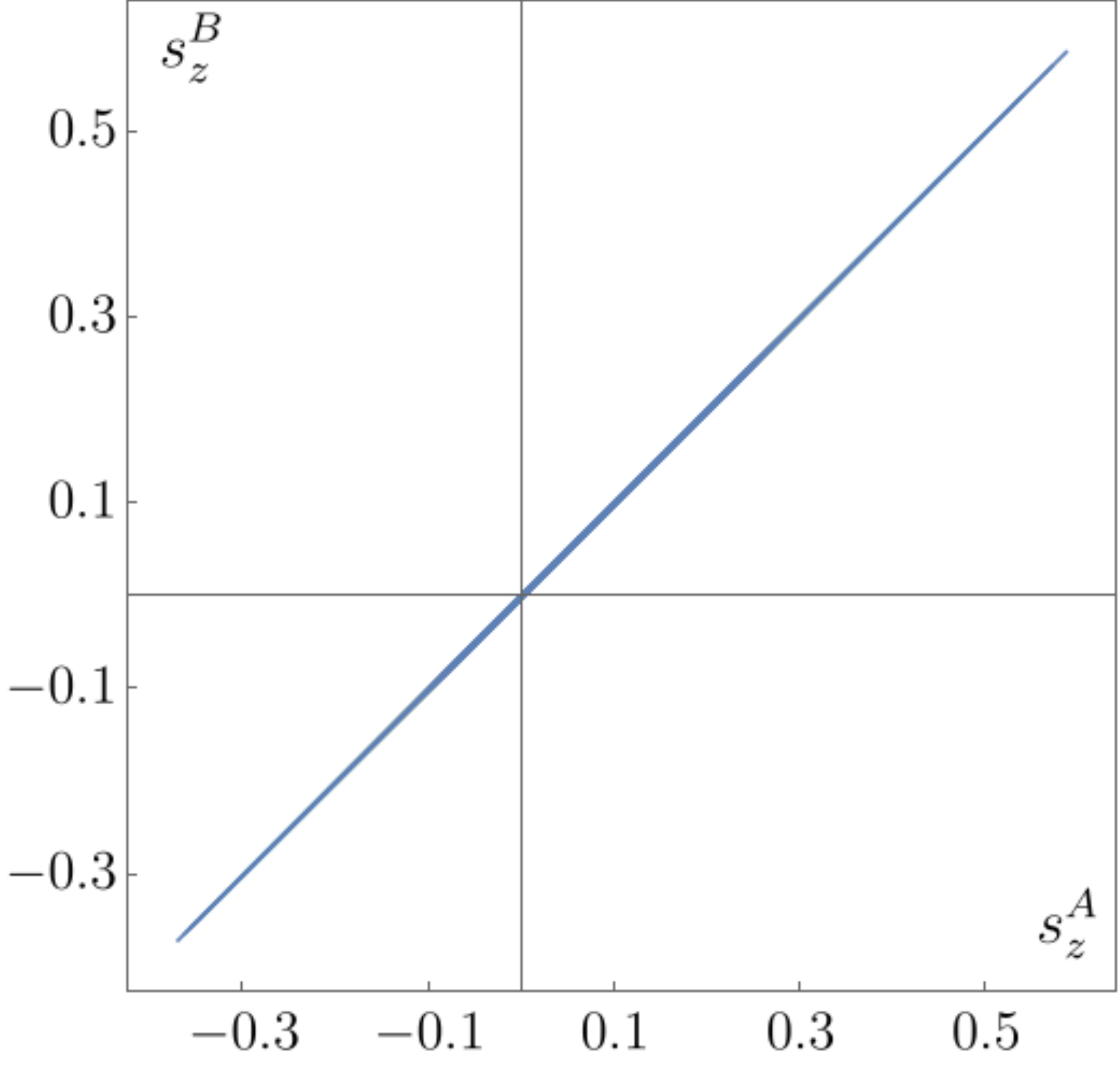}}
\caption{Demonstration of $\Z2$ symmetry of the synchronized chaotic attractor $(\delta = 0.080, W = 0.055)$ in the $s_{\perp}^{A}-s_{\perp}^{B}$ and $s_{z}^{A}-s_{z}^{B}$ projections. Both are straight lines passing through the origin with slopes equal to one.}
\label{SC_X_Sect}
\end{figure*}

\begin{figure*}[tbp!]
\centering
\subfloat[\qquad\textbf{(a)}]{\label{Spectrum_SC}\includegraphics[scale=0.34]{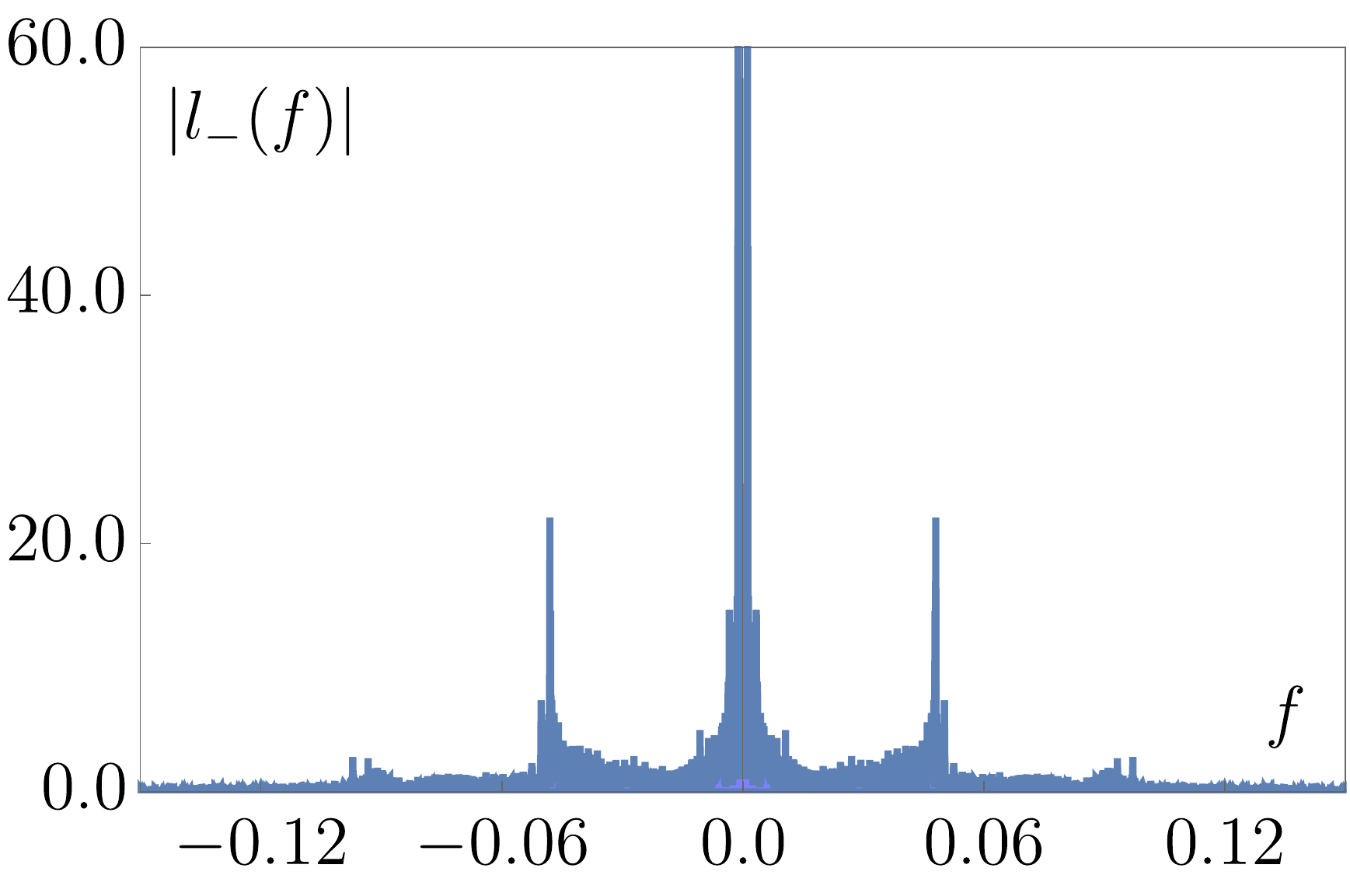}}\qquad\qquad
\subfloat[\qquad\textbf{(b)}]{\label{Spectrum_SC_Mag}\includegraphics[scale=0.34]{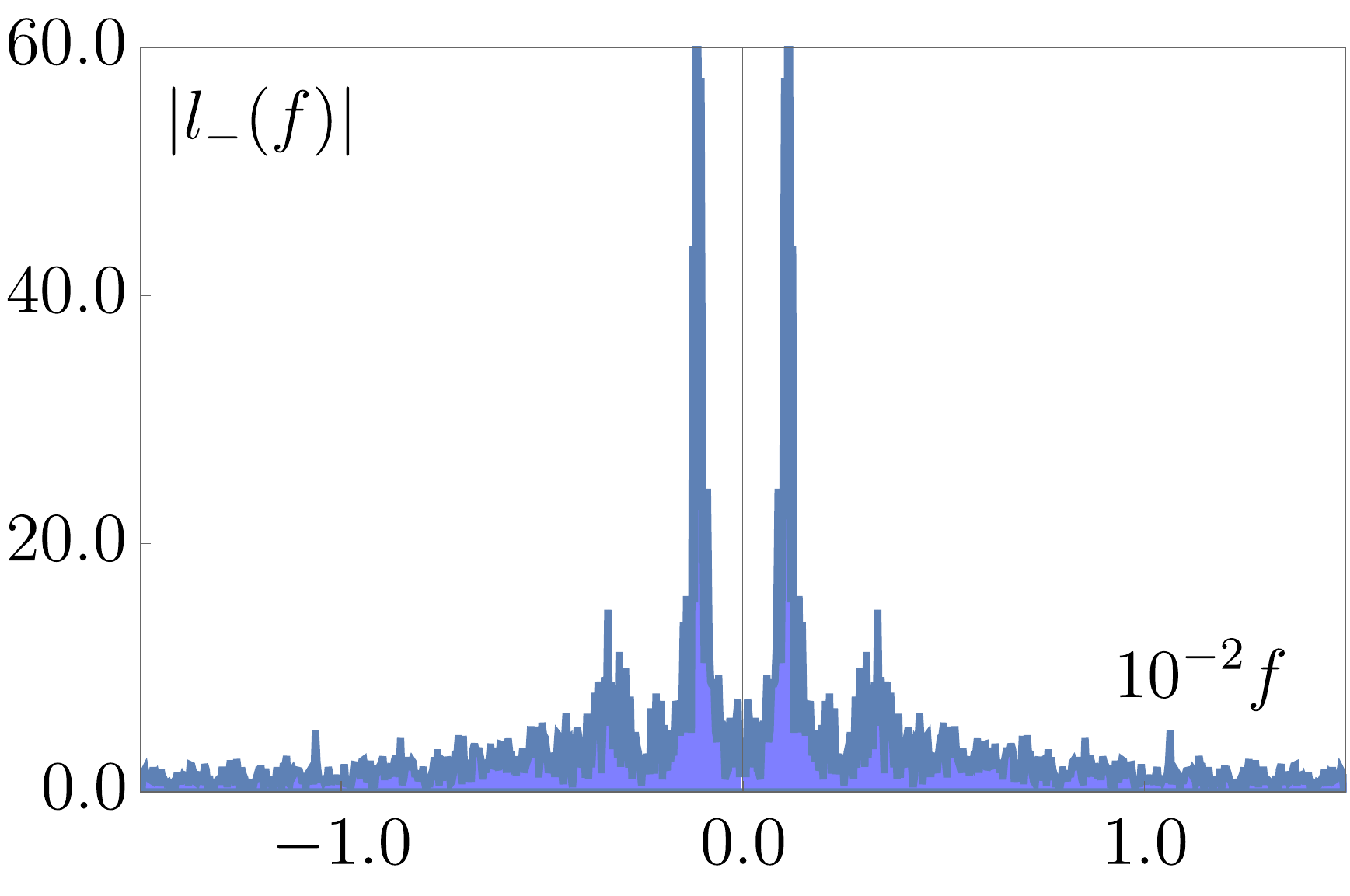}}
\caption{Power spectrum for the synchronized chaotic attractor at $\delta = 0.080, W = 0.055.$ Like the power spectrom of the chaotic attractor in \fref{Spectrum_C_Full}, it is a continuum with several peaks. However, unlike the spectrum there, it does have a reflection symmetry about the $f = 0$ axis. The spectrum has several peaks. Note, unlike \fref{Spectrum_C_Full}, here there is no peak at the origin, see \textbf{(b)} which is a hundred times magnification around $f = 0$ of the original spectrum.} \label{Spectrum_SC_Full}
\end{figure*}

As we move close to the II-III boundary from inside Phase III, a spontaneous restoration of the $\Z2$-symmetry leads to synchronization of chaos. We show the evolution of cosine of the angle between $\bm{l}_{\perp} = (l_{x},l_{y})$ and the x-axis, spin dynamics, different projections,  Poincare section, and spectrum for such an attractor in \fsref{Cos_Alph_CS}, \ref{SC_Pictures}, \ref{SC_X_Sect}, \ref{Poincare_SC} and \ref{Spectrum_SC_Full}, respectively. In particular, \fsref{Cos_Alph_CS} and \ref{SC_X_Sect} demonstrate the $\Z2$ symmetry of the attractor most explicitly. Moreover, recall that depending on the initial conditions it is possible to rotate the spin components for a synchronized chaotic attractor by a constant angle ($=\alpha_{0},$ as shown in \fref{Cos_Alph_CS}), so that the different components obey \eref{Z2_Expl}. Thus $e^{-\imath \alpha_{0}}l_{-}(t)$ is a real function. This produces the reflection symmetry in $|l_{-}(f)|^{2}$ about $f = 0$ axis (see \fref{Spectrum_SC_Full}), where $l_{-}(f)$ is the Fourier transformation of $l_{-}(t)$.

\subsection{Synchronization Manifold}\label{Synch_Manifold}

\begin{figure*}[tbp!]
\centering
\subfloat[\qquad\textbf{(a)}]{\label{Cond_Lyapunov_Exponent}\includegraphics[scale=0.35]{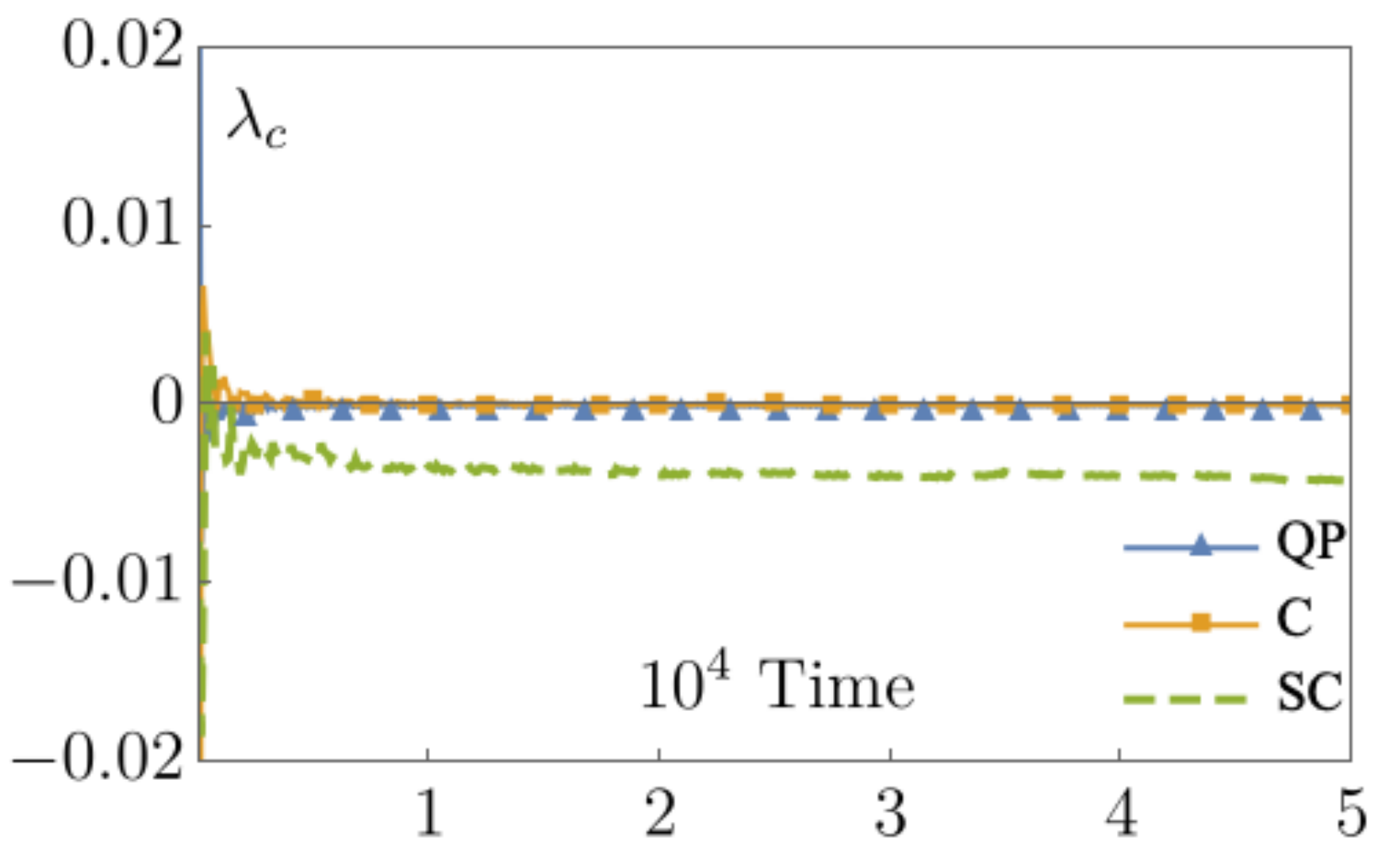}}\qquad\qquad
\subfloat[\qquad\textbf{(b)}]{\label{Log_Cond_Lyapunov_Exponent}\includegraphics[scale=0.33]{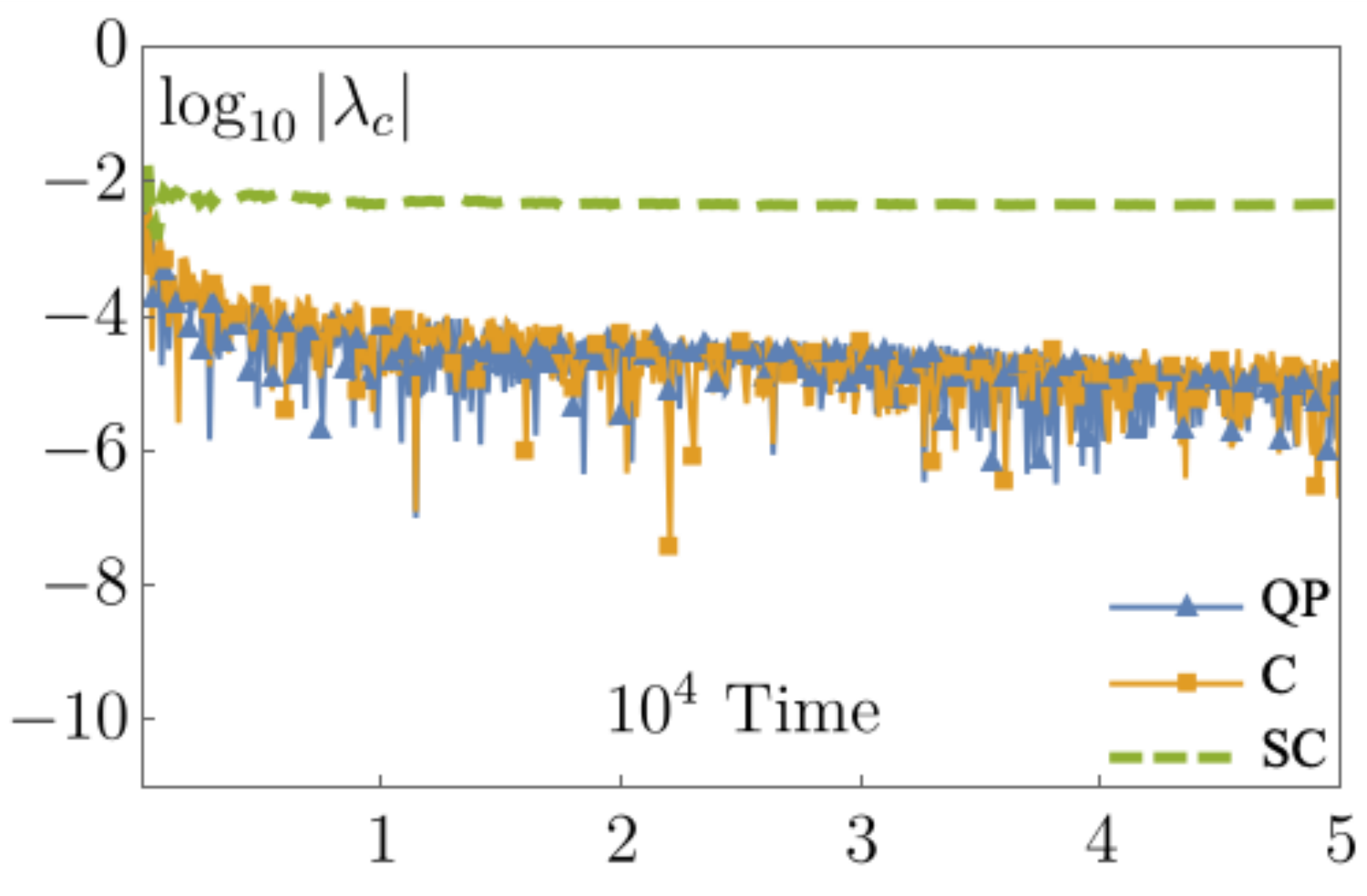}}
\caption{Conditional Lyapunov exponent $\lambda_{c}(t)$ (left) and the logarithm of its absolute value (right) for quasiperiodic (blue triangles), chaotic (yellow squares), and synchronized chaotic (green dashed line) trajectories. The parameters $(\delta, W)$ are the same as in \fref{Lyapunov}. For the above plots we calculate the divergence of two infinitesimally close initial conditions in the transverse manifols \re{Trans_Coord}. In \textbf{(a)}, $\lambda_{c}$ has saturated to a negative value for the synchronized chaotic attractor. This is corroborated in \textbf{(b)} as well. In contrast, for quasiperiodic and chaotic attractors $\lambda_{c}$ asymptotes to zero in \textbf{(a)}. Note, $\lambda_{c} = 0$ implies $\log_{10}{|\lambda_{c}|} = -\infty$. The expected decreasing trends of $\log_{10}{|\lambda_{c}}|$ (limited by machine precision) for both these attractors are indeed apparent in \textbf{(b)}.} \label{Cond_Lyapunov}
\end{figure*}

\begin{figure}[tbp!]

\begin{center}
\includegraphics[scale=0.45, trim=0 0cm -1cm -1cm, clip]{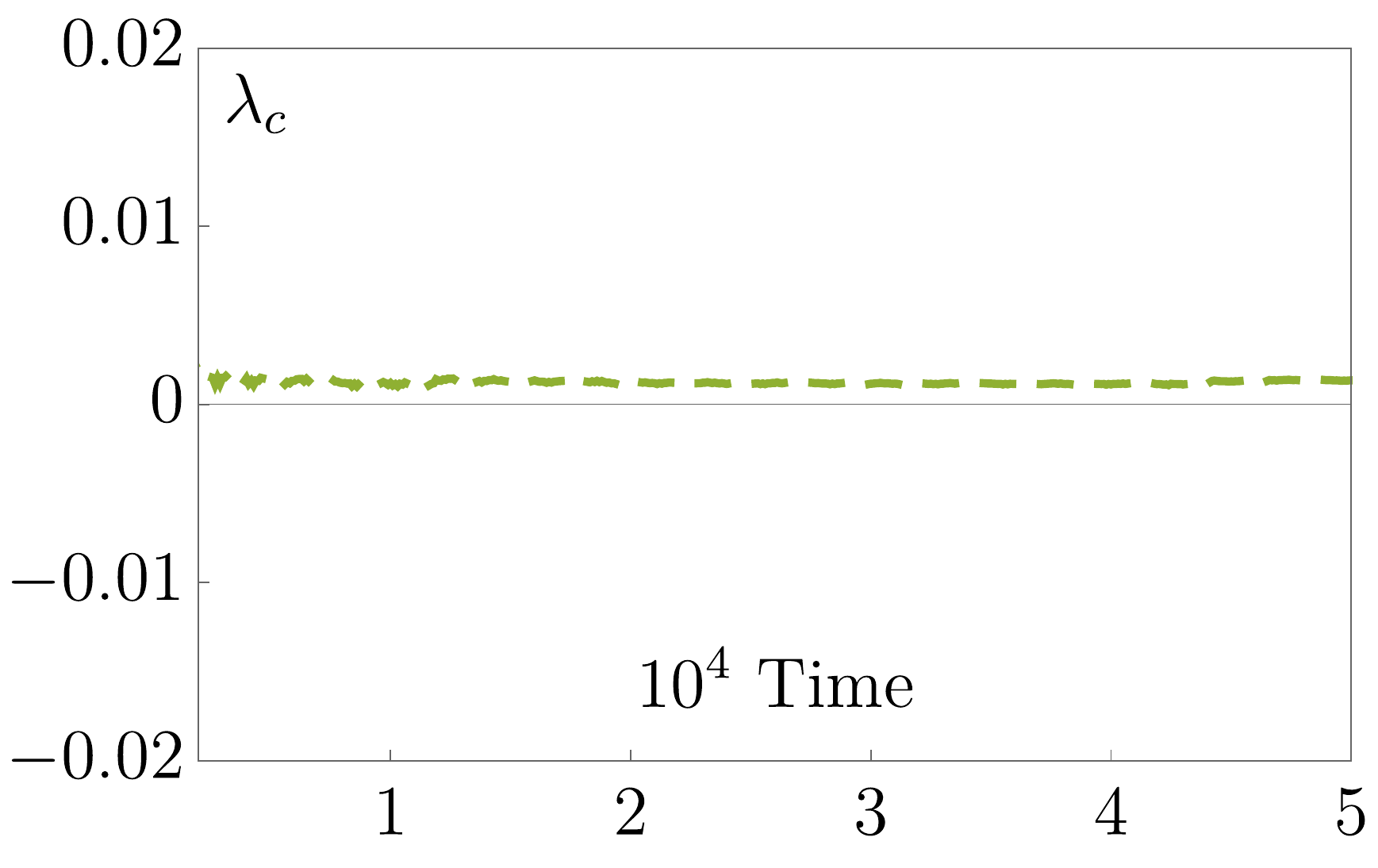}
\caption{Conditional Lyapunov exponent $\lambda_{c}$ for the synchronized chaotic attractor right after it comes to exist at $\delta = 0.106, W = 0.055$. A positive $\lambda_{c}$ demonstrates the unstable nature of the attractor. Contrast this with \fref{Cond_Lyapunov}, where the synchronized chaotic attractor at $\delta = 0.080, W = 0.055$ is stable and hence leads to $\lambda_{c}<0$.}\label{Cond_Lyapunov_0106}
\end{center}

\end{figure}

In \sref{Review} we discussed that a $\Z2$-symmetric attractor remains confined in the synchronization manifold \re{Gen_Z2_Symm_MF}. The attraction toward this submanifold determines the stability of the synchronized chaotic attractor \cite{Chaotic_Synch_X}. We quantify this attraction by calculating the conditional Lyapunov exponent (see \fsref{Cond_Lyapunov_0106} and \ref{Cond_Lyapunov}). 

We start by defining the coordinates that cover the ``transverse submanifold", which is complimentary to the synchronization manifold, as follows:
\beg 
\begin{split} 
\mathsf n_{1} \equiv  \big(s_{x}^{A}\big)^{2} + \big(s_{y}^{A}\big)^{2} -  \big(s_{x}^{B}\big)^{2} - \big(s_{y}^{B}\big)^{2}\!\!,\\
\mathsf n_{2} \equiv s_{z}^{A} - s_{z}^{B}.
\end{split} 
\label{Trans_Coord}%
\en%
These coordinates obey the following equations:   
\begs
\bea
\dot{\mathsf{n}}_{1} &=& \frac{1}{2}\big(l_{z} - 2W\big)\mathsf n_{1} + \frac{1}{2}\big(l_{x}^{2} + l_{y}^{2}\big)\mathsf n_{2}, \\[0pt] 
\dot{\mathsf{n}}_{2} &=& -\frac{\mathsf  n_{1}}{2}-W\mathsf n_{2},
\eea 
\label{Transverse_Coordinates_Eqn}%
\ens%
which we derived using \eref{Mean-Field_1}. We obtain $\bm{l}$ also from \eref{Mean-Field_1}. Recall, the conditional Lyapunov exponent is nothing but the maximum among the Lyapunov exponents corresponding to the transverse directions. The linearity of \eref{Transverse_Coordinates_Eqn} guarantees that $\Delta \mathsf{n}_{1,2}$ obeys the same equation. In \eref{Lyapunov_Exp_Defn}, defining $d = \sqrt{\Delta\mathsf n_{1}^{2} + \Delta\mathsf n_{2}^{2}}$, we then compute the conditional Lyapunov exponent (a detailed algorithm is provided in \ref{Appnd_Lya_Exp}). 

At its inception, the synchronized chaotic attractor is unstable in the full phase space, and so it leads to $\lambda_{c} \geqslant 0$, see \fref{Cond_Lyapunov_0106}. It is, however, stable in the synchronization submanifold, see \fref{Phase_Diagram_Z2}. Close to the II-III boundary (red subregion) this attractor becomes sufficiently attractive such that any initial condition, even the ones without the $\Z2$-symmetry, give rise to this attractor. As a result, deep in the red subregion $\lambda_{c} < 0$.

\subsubsection{The $\Z2$-symmetric Limit Cycle} \label{Z2_LC_Synch_MF}  

\begin{figure*}[tbp!]

\centering
\subfloat[\large (a)]{\label{Lambda_c_SLC_044}\includegraphics[scale=0.33]{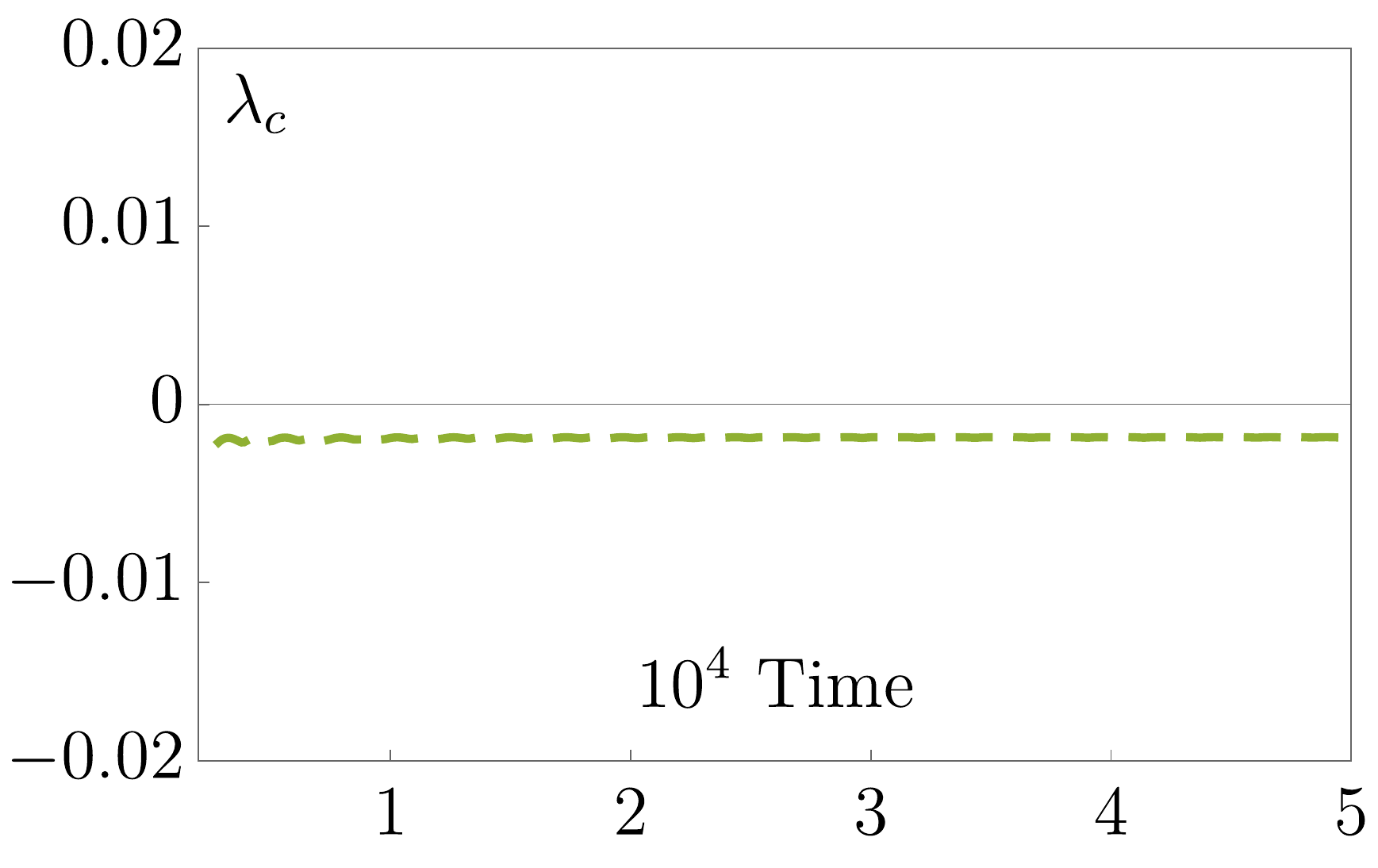}
\begin{picture}(0,0)
\put(-80,52){\includegraphics[height=1.7cm]{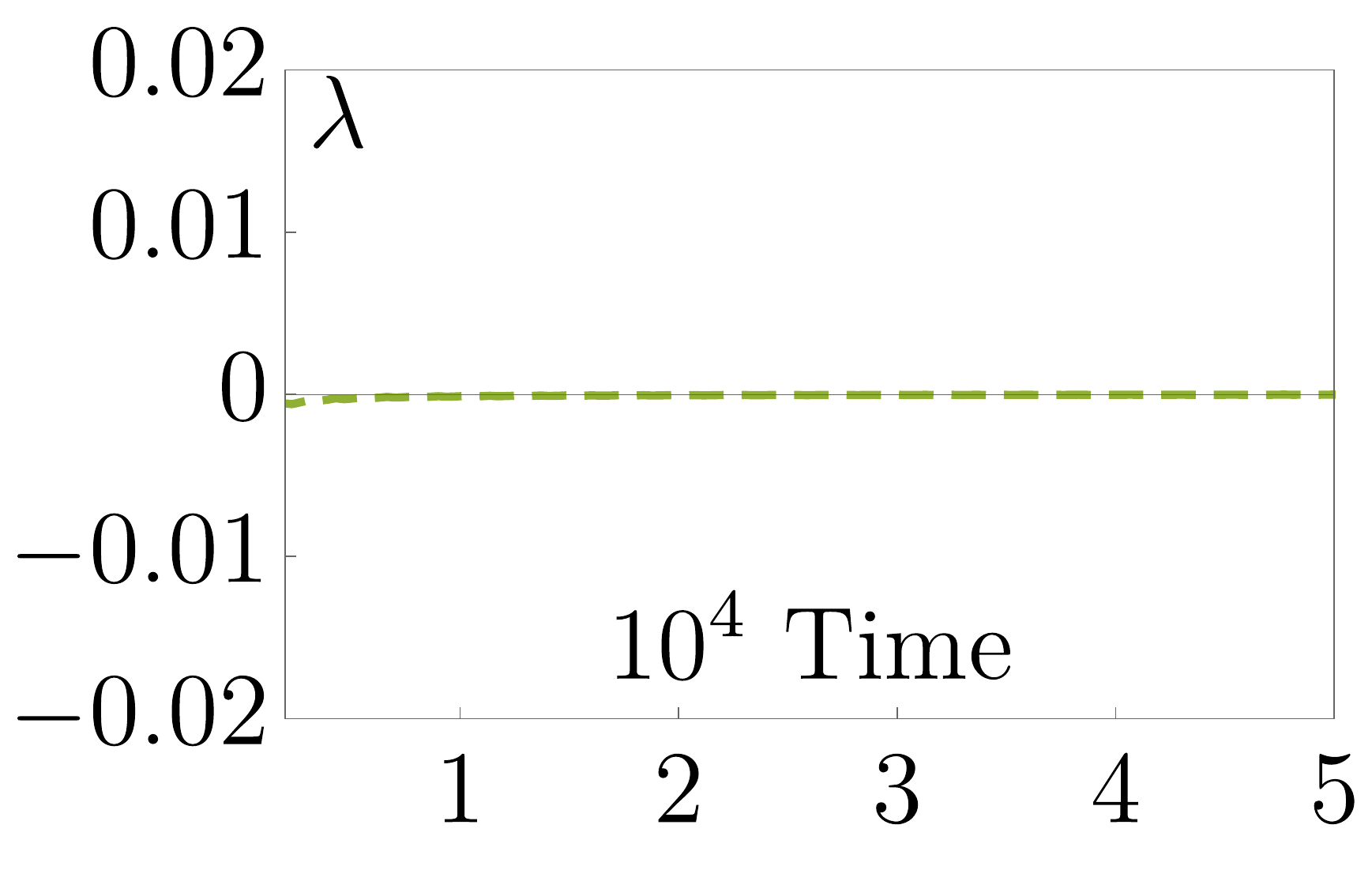}}
\end{picture}}\qquad\qquad
\subfloat[\large (b)]{\label{Lambda_c_SLC_042}\includegraphics[scale=0.33]{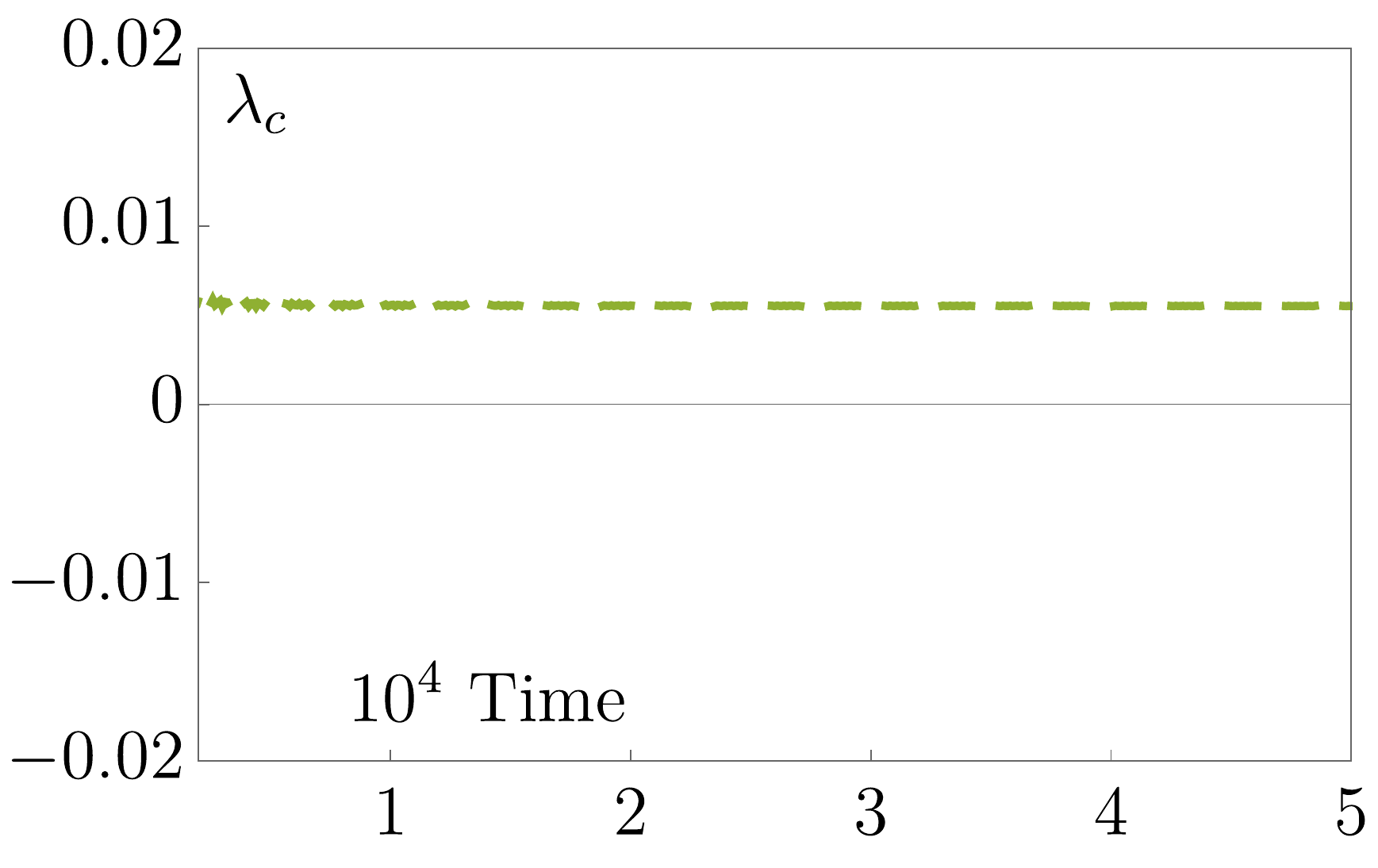}
\begin{picture}(0,0)
\put(-80,10){\includegraphics[height=1.7cm]{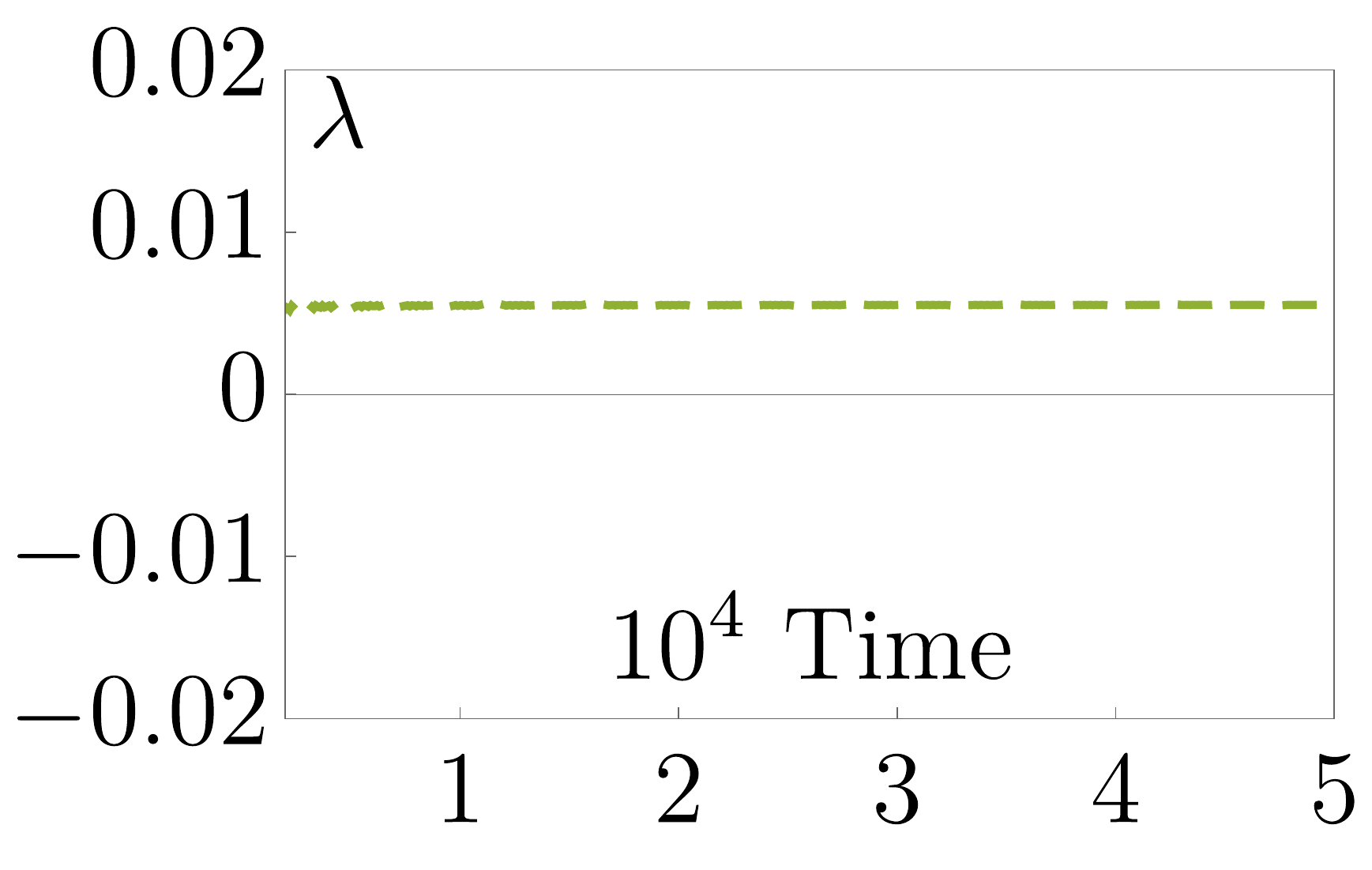}}
\end{picture}}
\caption{Conditional Lyapunov exponents $\lambda_c(t)$ computed using \eref{Transverse_Coordinates_Eqn} for a stable $(\lambda_{c}<0)$, and an unstable $(\lambda_{c}>0)$ $\Z2$-symmetric limit cycle at \textbf{(a)} $\delta = 0.44, W = 0.056$, and \textbf{(b)} $\delta = 0.42, W = 0.056$, respectively. As expected -- since Lyapunov exponents are closely related to the Floquet exponents -- the maximum Lyapunov exponent $\lambda$ is zero for the stable limit cycle, whereas $\lambda = \lambda_{c}>0$ for the unstable one (see insets). The latter scenario is unlike the synchronized chaotic attractor, where $\lambda_{c}<0$ but $\lambda>0$, see \fref{Cond_Lyapunov}.} \label{Lya_SLC}

\end{figure*}

As an aside, recall that the $\Z2$ symmetric limit cycles also belong to the synchronization manifold \re{Gen_Z2_Symm_MF}. Therefore in the green regions of Phase III, similar computation yields negative $\lambda_{c}$ for the $\Z2$-symmetric periodic attractors. In the yellow region, however, the $\Z2$-symmetric limit cycle is unstable. In particular, the Floquet exponents corresponding to the asymmetric perturbations \re{Floquet_Asymm_Var} is positive. Since the Lyapunov exponent is closely related to the Floquet exponent, we have $\lambda = \lambda_{c}>0$ in this subregion. This shows that one can determine the $\Z2$-symmetry breaking line (the dashed line of \fref{Phase_Diagram}) by noting the sign change of $\lambda_{c}$ as well (see \fref{Lya_SLC}). 

\begin{figure}[tbp!]

\begin{center}
\includegraphics[scale=0.45, trim=0 0cm -1cm -0.5cm, clip]{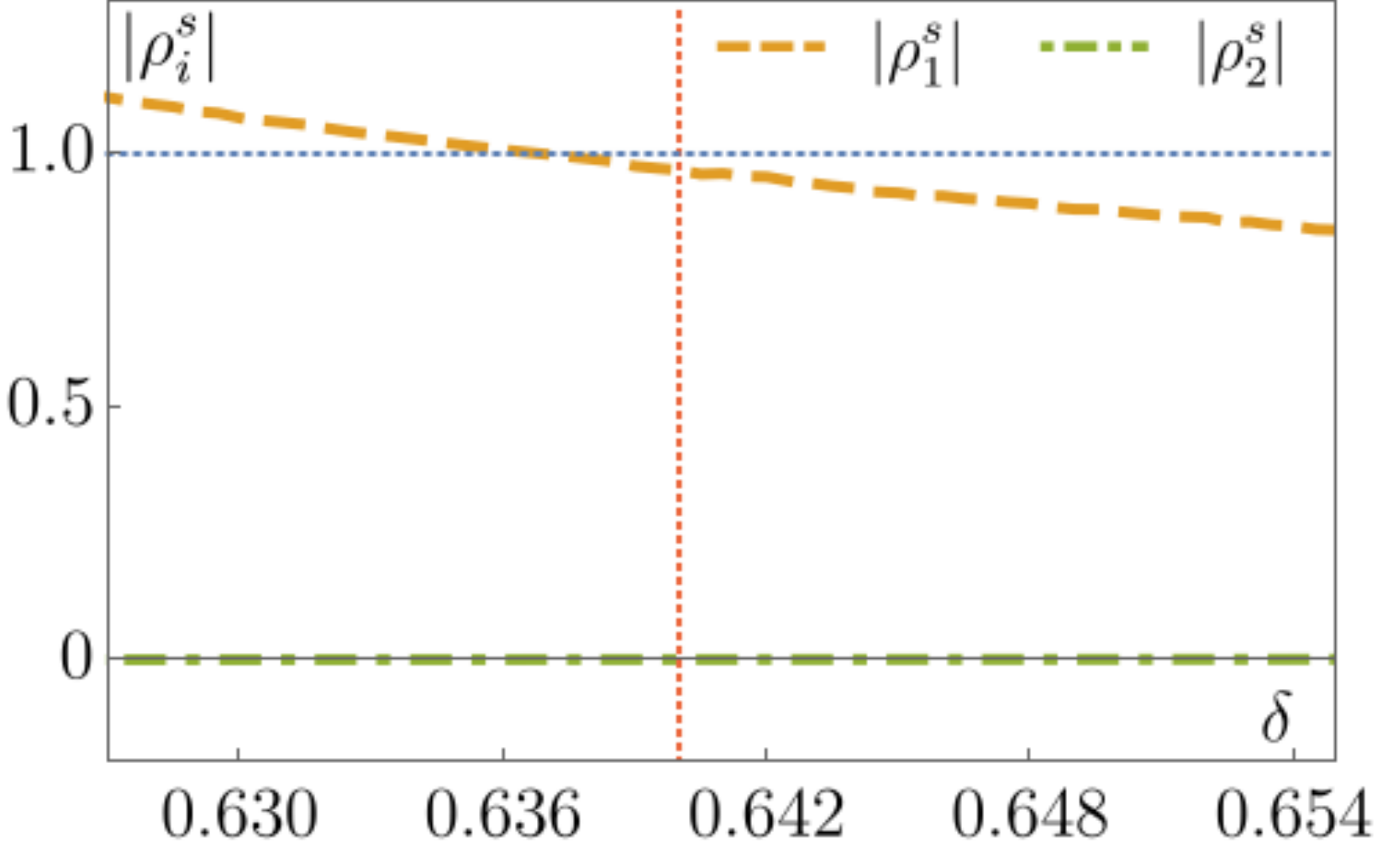}
\caption{The absolute values of the Floquet multipliers $|\rho_{1,2}^{s}|$ (shown in orange dashed and green dot-dashed lines, respectively) as a function of $\delta$ using \eref{Transverse_Coordinates_Eqn} for $W = 0.40$. From \Rref{Patra_1}, we know that the $\Z2$-symmetric limit cycle loses stability at $\delta \approx 0.64$ (depicted as a vertical red dotted line) for the $W$ considered here. According to this analysis, however, the stability loss occurs at $\delta\approx 0.637$, where $|\rho_{1}^{s}|$ crosses the $y = 1$ line. The latter is shown as a horizontal blue dotted line. Also, note that in this analysis one does not obtain any multiplier, whose magnitude remains one across the bifurcation.}\label{Tranverse_Floquet_n1_n2_W=0_40}
\end{center}

\end{figure}

Yet another way to obtain the $\Z2$-symmetry breaking line would be to perform Floquet analysis with the linearized equations for $\Delta \mathsf{n}_{1,2}$, which are the same as \eref{Transverse_Coordinates_Eqn} with $\mathsf{n}_{1,2}$ replaced by $\Delta \mathsf{n}_{1,2}$. We similarly construct the $2 \by 2$ monodromy matrix $\mathbb{B}_{\mathsf{n}}$, whose eigenvalues are the Floquet multipliers: $\rho^{s}_{1,2}$. Since both $\mathsf{n}_{1}$ and $\mathsf{n}_{2}$ are invariant under rotations $\R(\theta)$ about the z-axis, $(\Delta \mathsf{n}_{1},\Delta \mathsf{n}_{2})$ does not have a particular solution similar to \eref{Special_m_TF}. As a result, none of the multipliers remain identically one across criticality. The multipliers here are in fact some functions of $\rho_{2,3}$ of \eref{Asymm_Floquet_Soln}. Note, similar to $\rho_{2,3}$, $\rho^{s}_{1,2}$ are real near criticality. Finally, for a fixed $W$, we obtain the $\delta$ on the dashed line ($\Z2$-symmetry breaking line) for which $\rho_{1}^{s}(\delta)>1$ (see \fref{Tranverse_Floquet_n1_n2_W=0_40}).

\subsection{Mechanism for Synchronization of Chaos}\label{Mech_SC}

In this section, we study the emergence of synchronized chaos from $\Z2$-symmetric limit cycle via tangent bifurcation intermittency. Both of these attractors obey \eref{Z2_Expl} in a suitably rotated (about z-axis) coordinate system. We, therefore, start by obtaining the nonequilibrium phase diagram \ref{Phase_Diagram_Z2} for \eref{Symm_One_Spin_Eqn}. The Phase III of this diagram only possesses two amplitude-modulated superradiances. These correspond to the $\Z2$-symmetric limit cycles and the synchronized chaotic attractors of the original phase diagram \ref{Phase_Diagram}.

\begin{figure*}[tbp!]
\centering
\subfloat[\textbf{(a)}]{\label{Traj_Sect_SC}\includegraphics[scale=0.35]{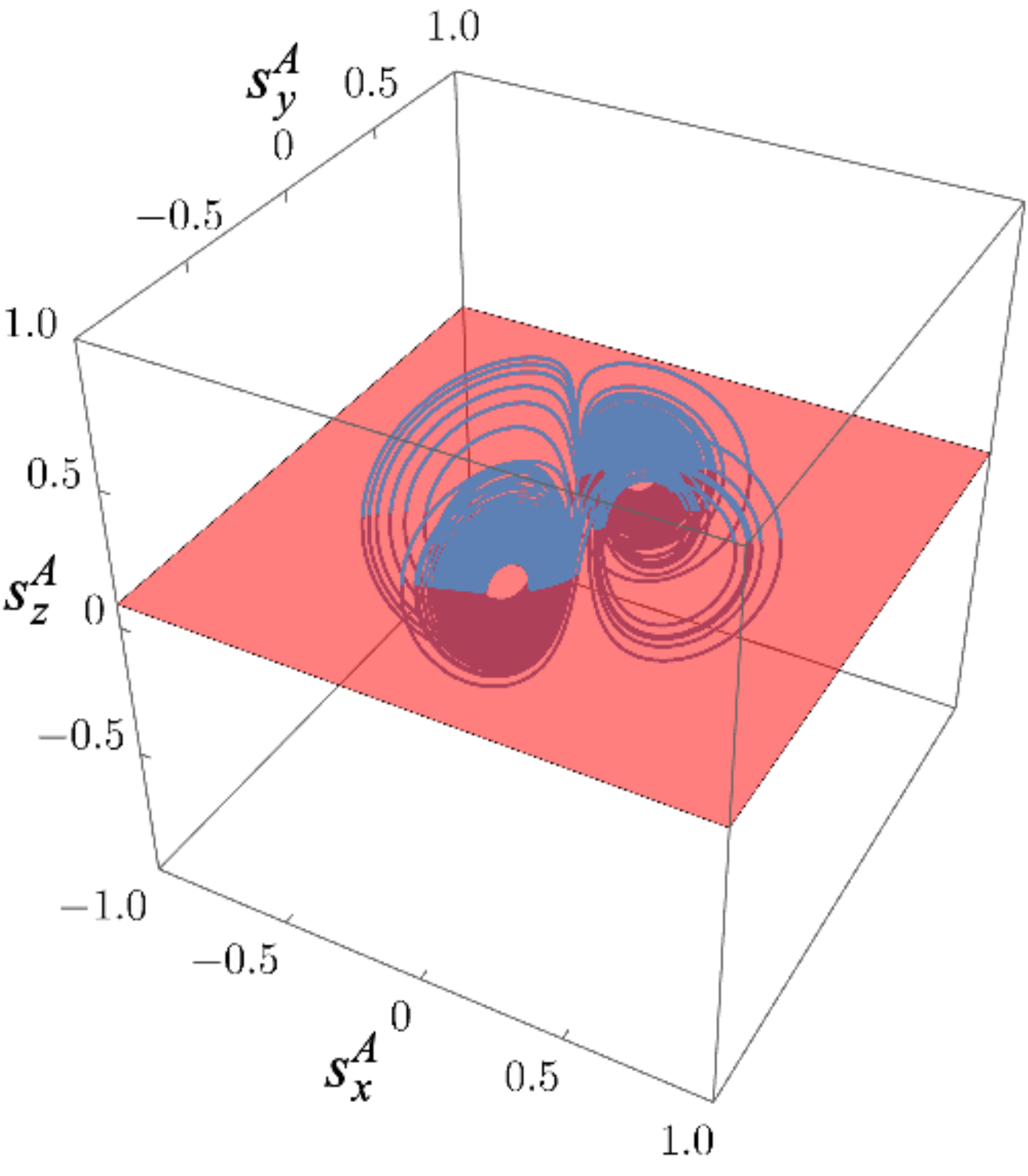}}\qquad\qquad
\subfloat[\qquad\textbf{(b)}]{\label{Poincare_SC}\includegraphics[scale=0.33]{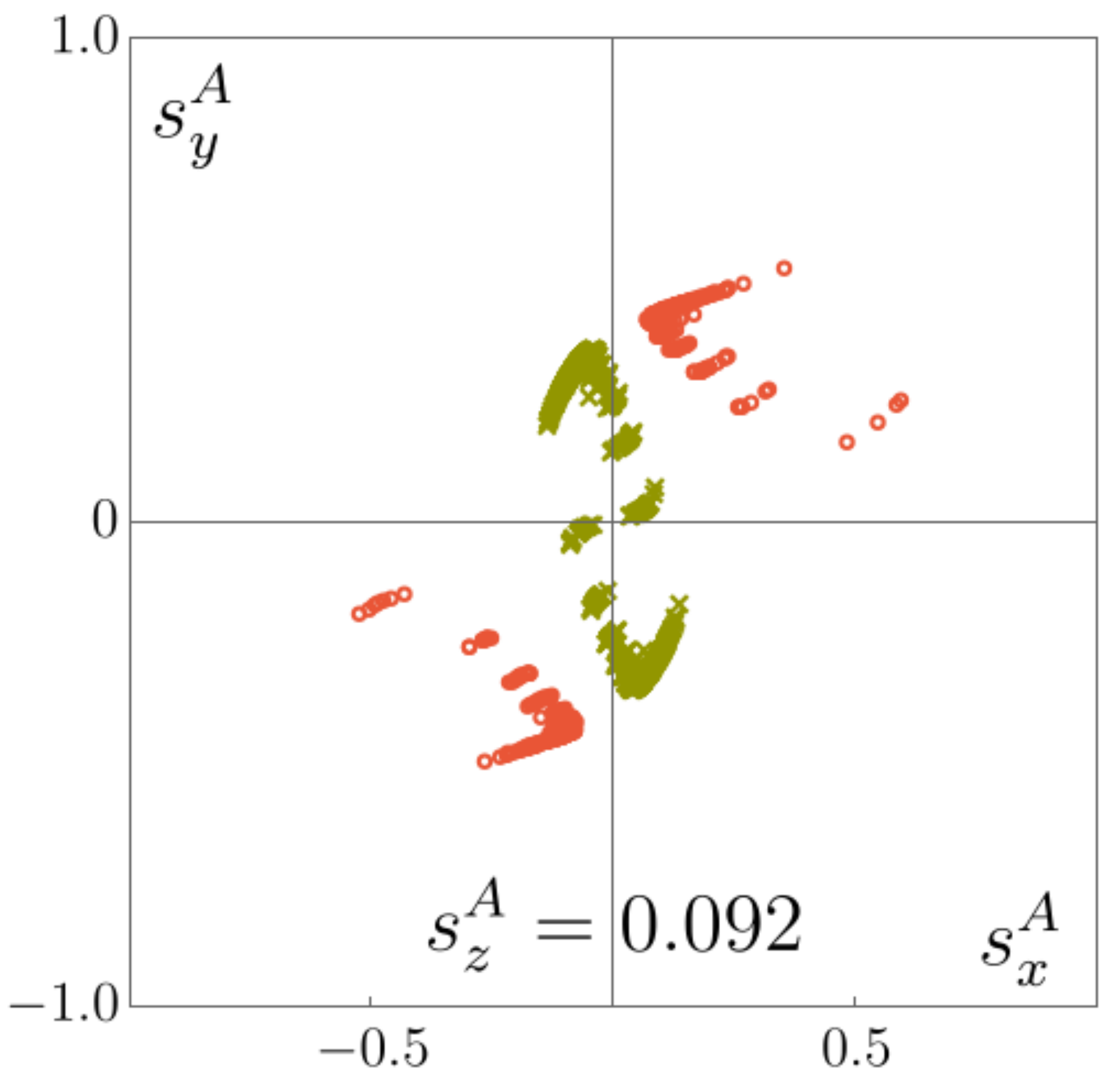}}
\caption{Poincar\'{e} section of $\bm{s}^{A}$ at $\delta = 0.080, W = 0.055$. The attractor is synchronized chaotic. Conventions are the same as in the earlier figures showing Poincar\'{e} sections. The dimension of the Poincar\'{e} section in \textbf{(b)} is less than the one in \fref{Poincare_C1}. The constricted nature of the Poincar\'{e} section is due to the $\Z2$ symmetry of the attractor. (For interpretation of the references to color in this figure legend, the reader is referred to the web version of this article.)}
\label{Poincare_SC}
\end{figure*}

\begin{figure}[tbp!]

\begin{center}
\includegraphics[scale=0.45]{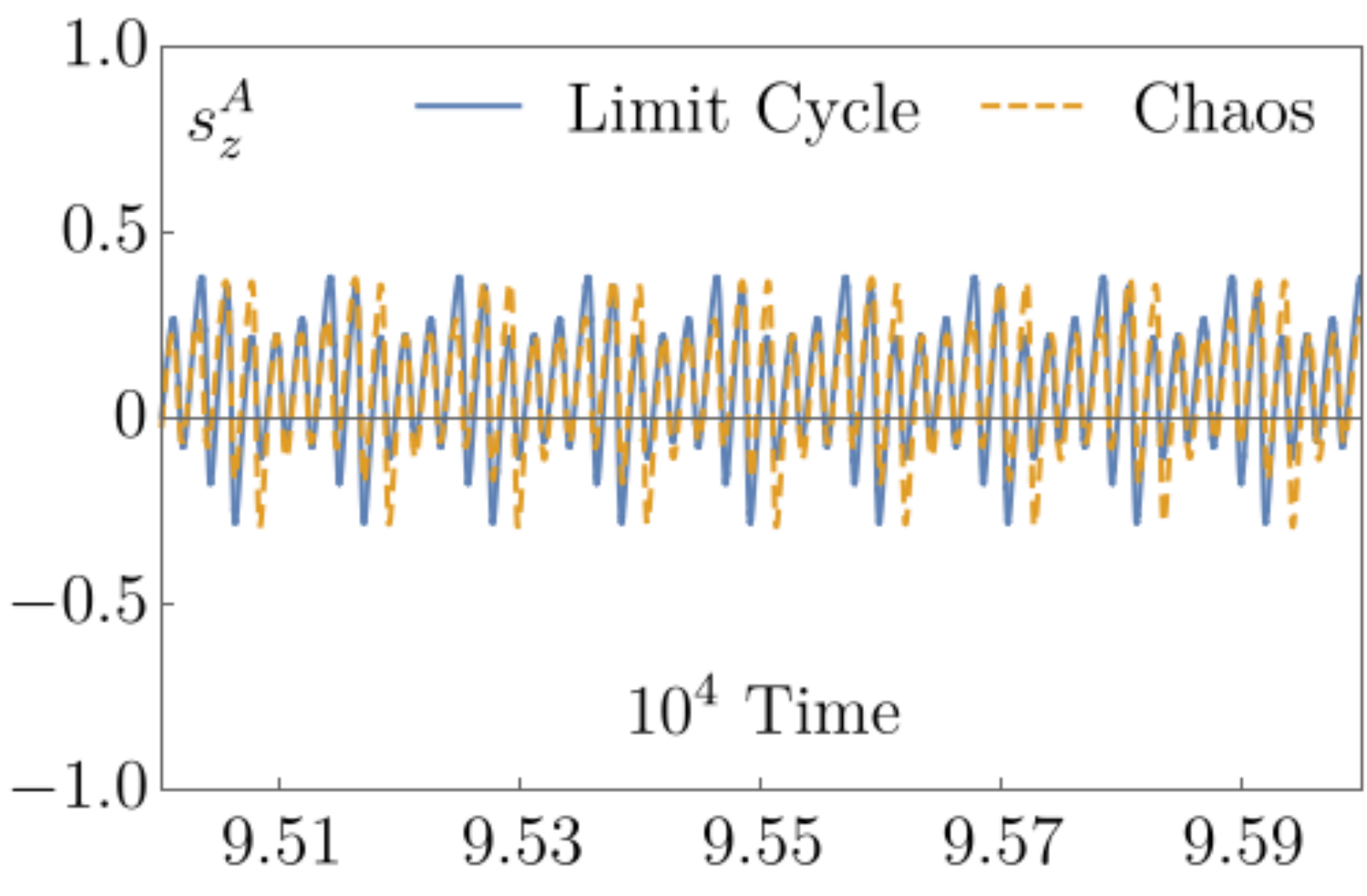}
\caption{Comparison of the $s_{z}^{A}$ vs. time plots for the $\Z2$-symmetric periodic attractor at $\delta = 0.107, W = 0.055$ (solid blue) and the synchronized chaotic one (dashed orange) at $\delta = 0.106, W = 0.055$. The Floquet analysis with the help of \eref{Symm_One_Spin_Eqn} reveals that in the $\Z2$-symmetric submanifold the synchronized chaotic attractor originates from the limit cycle via tangent bifurcation intermittency between the above two values of $\delta$. The closeness of the two trajectories indeed corroborates occurrence of the above bifurcation. However, numerical analysis reveals that  both these parameter-sets with a non-$\Z2$-symmetric initial condition results in quasiperiodicity.}
\label{SC_SLC_ZA}
\end{center}

\end{figure}

\begin{figure}[tbp!]

\begin{center}
\includegraphics[scale=0.45]{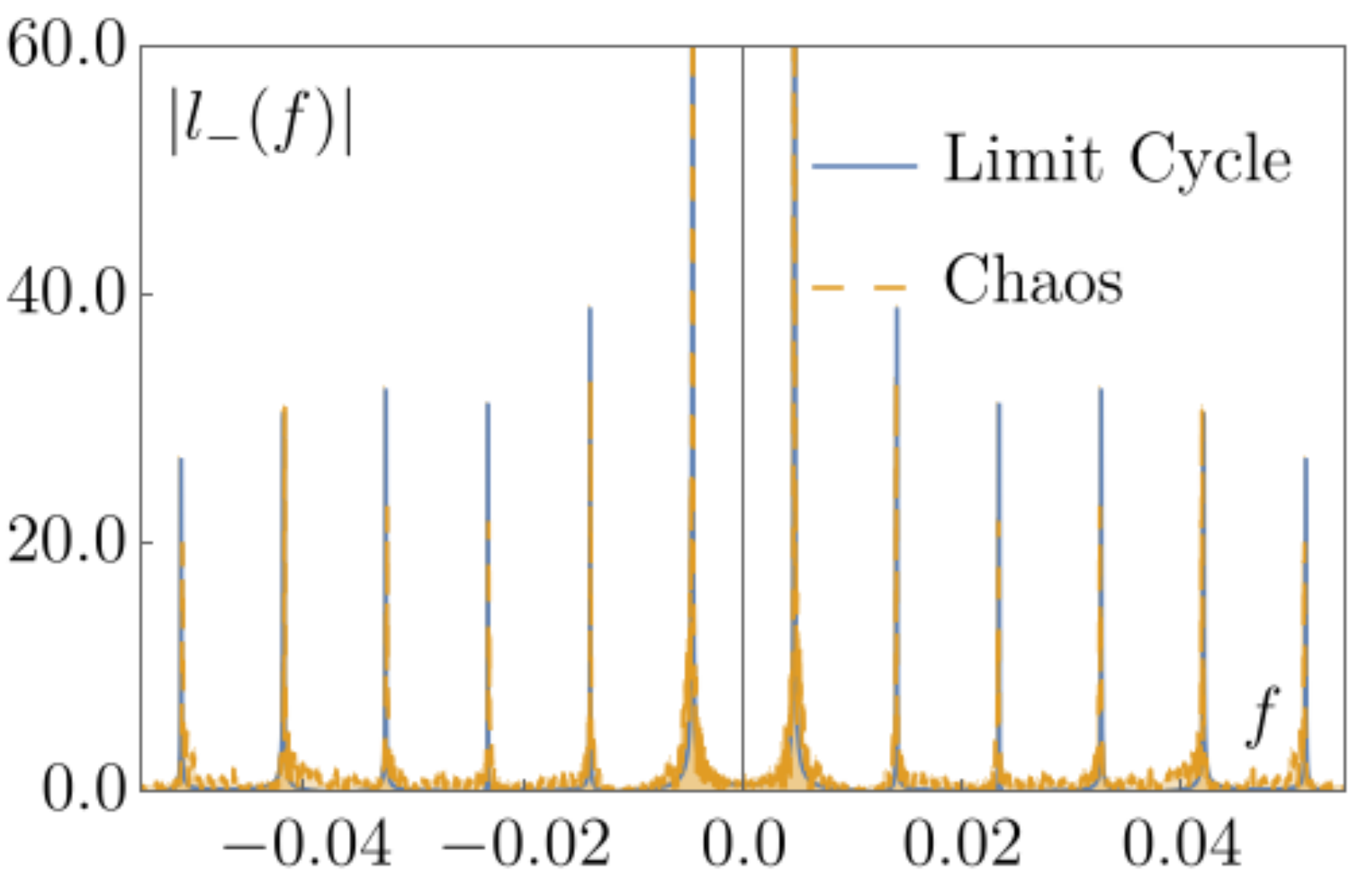}
\caption{Power spectra of the $\Z2$-symmetric limit cycle and the synchronized chaotic attractor that it gives rise to via tangent bifurcation intermittency. The values of $(\delta, W)$ are the same as in \fref{SC_SLC_ZA}. Note the closeness of the two spectra. The main peaks of both the attractors are given by the fundamental tone $f_{0} \approx 4.7\times 10^{-3}$ and its higher odd harmonics. The synchronized chaotic spectrum, unlike the one for the limit cycle, is a continuum where frequencies between the main peaks acquire small weights.} 
\label{Spectrum_SC_SLC}
\end{center}

\end{figure}

\subsubsection{Floquet Analysis: Stability against Perturbations in the $\Z2$-Symmetric Submanifold}\label{Sect_Symm_Floquet}

\begin{figure*}[tbp!]
\centering
\subfloat[\qquad\textbf{(a)}]{\label{Lyapunov_Exponent_Z2}\includegraphics[scale=0.35]{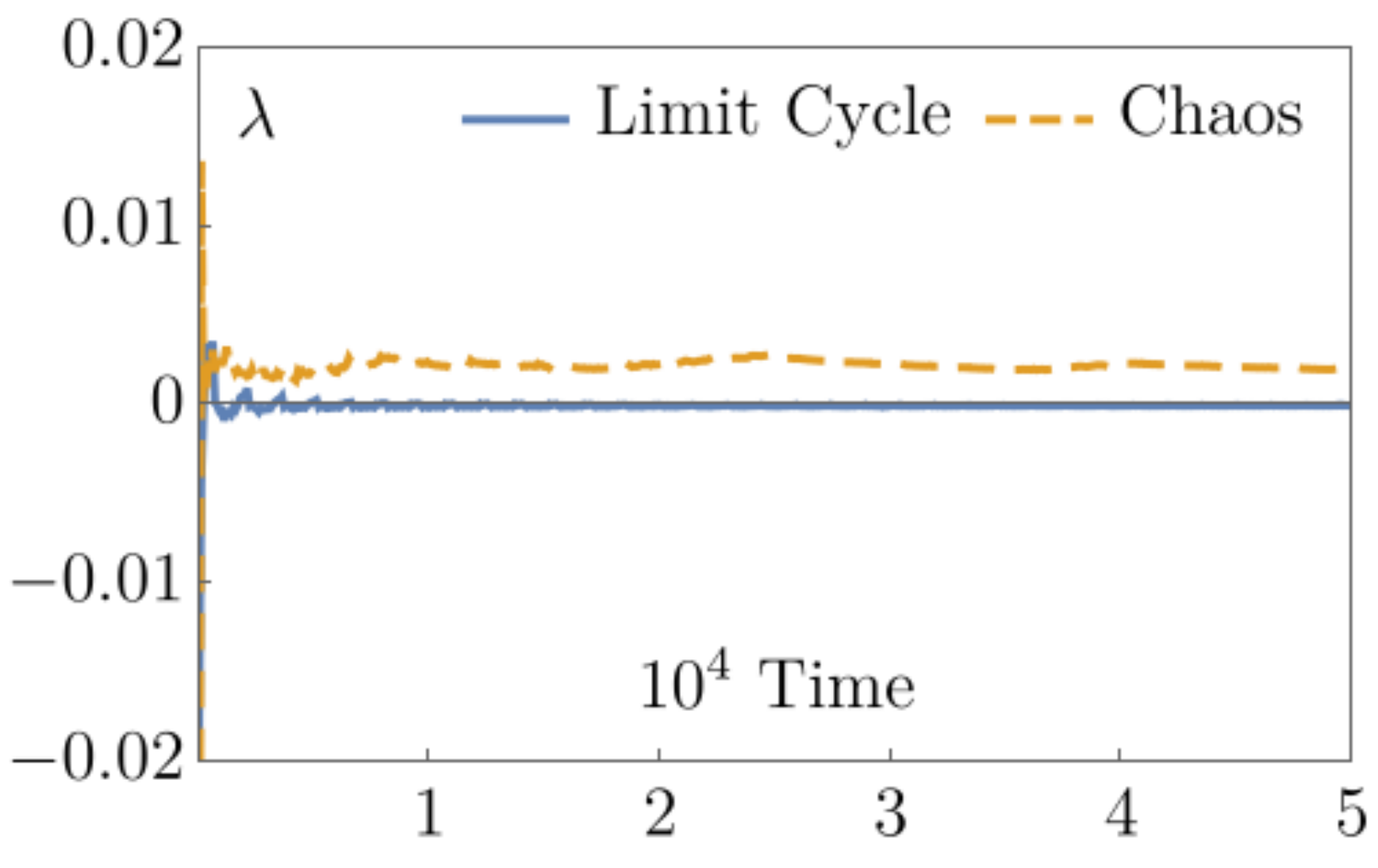}}\qquad\qquad
\subfloat[\qquad\textbf{(b)}]{\label{Log_Lyapunov_Exponent_Z2}\includegraphics[scale=0.33]{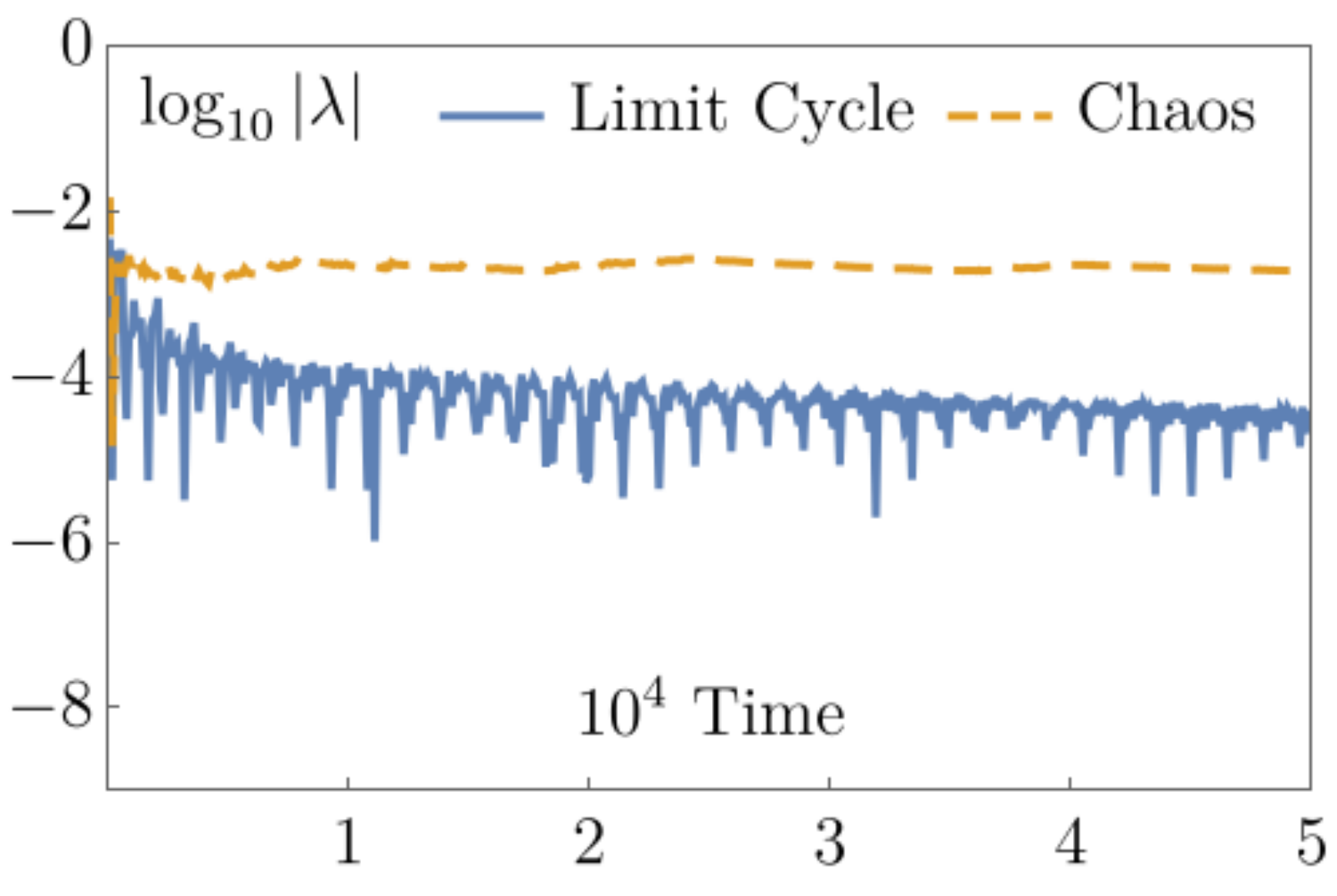}}
\caption{Comparison of the maximum Lyapunov exponent $\lambda(t)$ (left) and the logarithm of its absolute value (right) for the $\Z2$-symmetric periodic and the synchronized chaotic attractors for the same values of the detuning $\delta$ and repump rate $W$ as in \fref{SC_SLC_ZA}. We calculate the divergence of two infinitesimally close initial coditions in \eref{Symm_One_Spin_Eqn} for these figures. The behaviors of $\lambda(t)$ and $\log_{10}{|\lambda(t)|}$ for the two attractors are similar to the periodic attractor in the inset of \fref{Lambda_c_SLC_044}, and chaotic and synchronized chaotic ones in \fref{Lyapunov}, respectively.} \label{Lyapunov_Exponent_Z2_106_107_055}
\end{figure*}

The dot-dashed line in \fref{Reduced_Phase_Diagram} denotes the advent of chaos in the $\Z2$-symmetric submanifold \re{Gen_Z2_Symm_MF}. We obtain this by noting when $\lambda$ computed with \eref{Symm_One_Spin_Eqn} for the $\Z2$-symmetric attractors become positive, see \fref{Lyapunov_Exponent_Z2_106_107_055}. 

One can also determine this line using the Floquet analysis. Consider the following linearized equations for different components of $\Delta\bm{s}$: 
\begs
\bea
\frac{\mathrm{d}\Delta s_{x}}{\mathrm{d}t} &=& \big(s_{z} - \frac{W}{2}\big)\Delta s_{x} - \frac{\delta}{2}\Delta s_{y} + s_{x}\Delta s_{z}, \\ [7pt]
\frac{\mathrm{d}\Delta s_{y}}{\mathrm{d}t} &=& \frac{\delta}{2}\Delta s_{x} - \frac{W}{2}\Delta s_{y}, \\[7pt] 
\frac{\mathrm{d}\Delta s_{y}}{\mathrm{d}t} &=& -2s_{x}\Delta s_{x} - W\Delta s_{z}. 
\eea 
\label{Floquet_Eqn_Z2}
\ens%
The time-dependent coefficients, which are functions of different components of $\bm{s}(t)$, are obtained from \eref{Symm_One_Spin_Eqn}.

Similar to the three instances of Floquet analyses described before, for $i = 1,2,3$ one obtains the Floquet multipliers $r_{i}$, which are the eigenvalues of the $3 \by 3$ monodromy matrix $\mathbb{B}_{s}$ corresponding to \eref{Floquet_Eqn_Z2}. The Floquet multipliers $r_{i}$ are also related to the Floquet exponents $\chi_{i}$ and the period $T$ of $\bm{s}(t)$ as $r_{i} = e^{\chi_{i}T}$, for all $i$. According to the Floquet theorem we obtain a general solution for $\Delta \bm{s}$ as follows:
\beg 
\Delta \bm{s}(t) = C_{s,1}\dot{\bm s}(t)+ C_{s,2}e^{\chi_{2}t}\bm{p}_{2}(t) + C_{s,3}e^{\chi_{3}t}\bm{p}_{3}(t). 
\label{Floquet_Gen_Perturb}
\en%
Similar to \eref{m01_FF}, here $\Delta \bm{s} = \dot{\bm s}$ is a particular solution of \eref{Floquet_Eqn_Z2} with period $T$ that corresponds to multiplier $r_{1} = 1$. Also $C_{s,i}$ for $i = 1,2,3$ are independent constants, and $\bm{p}_{2,3}(t + T) = \bm{p}_{2,3}(t)$. 

All three multipliers are real close to criticality. As shown in \fref{Symmetric_Floquet_W=0_055}, $|r_{2}|$ becoming greater than one signals the $\Z2$-symmetric limit cycle losing stability to usher in chaos. Note, the eigendirection corresponding to $|r_{3}|<1$ always remains stable.

\subsubsection{Tangent Bifurcation Intermittency in the $\Z2$ Symmetric Sub-manifold} \label{Tan_Bifurc_Int}

\begin{figure*}[tbp!]
\centering
\subfloat[\textbf{(a)}]{\label{Traj_Sect_SLC_Comp_SC_del=0107}\includegraphics[scale=0.35]{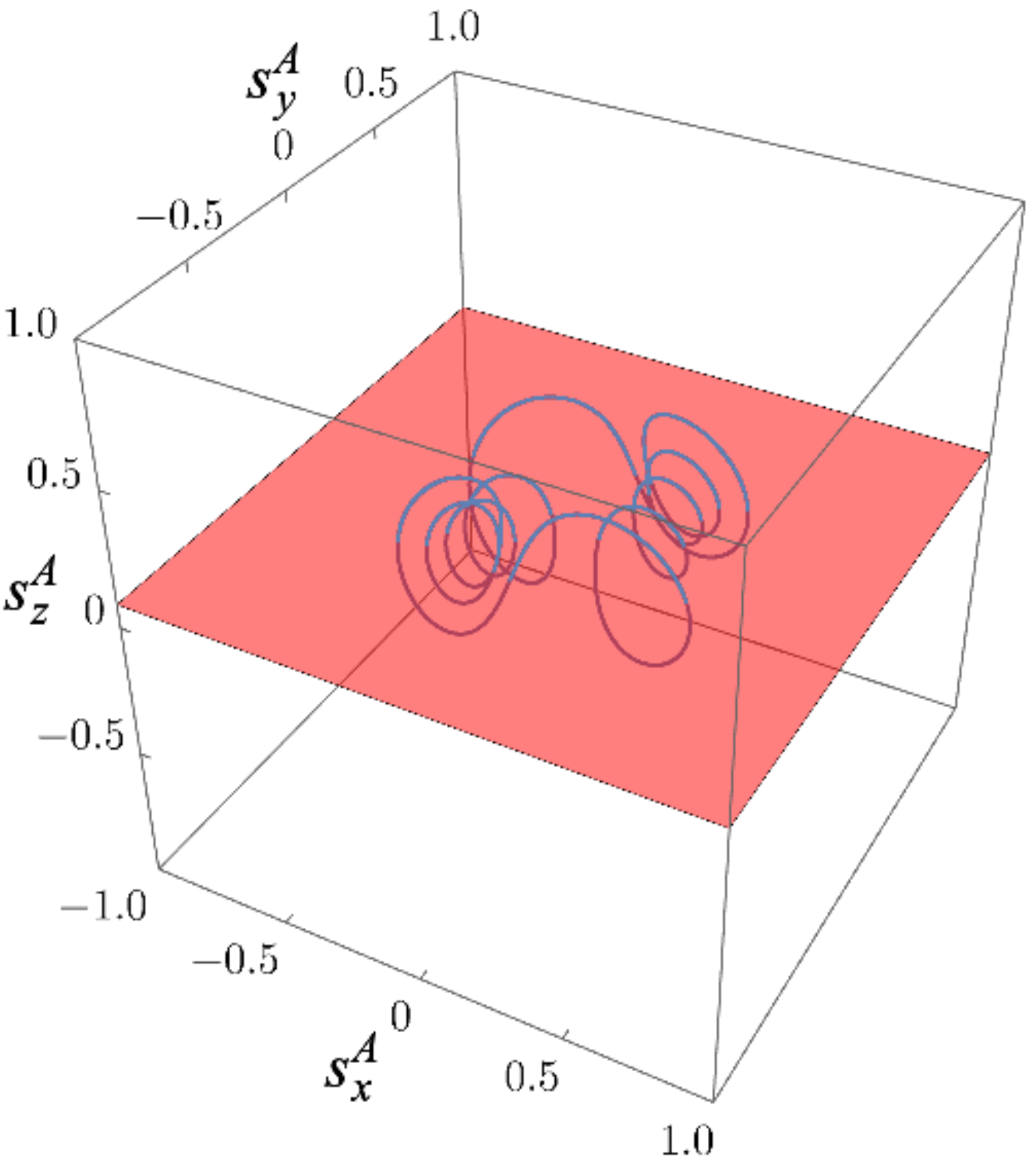}}\qquad\qquad
\subfloat[\qquad\textbf{(b)}]{\label{Poincare_SLC_Comp_SC_del=0107}\includegraphics[scale=0.33]{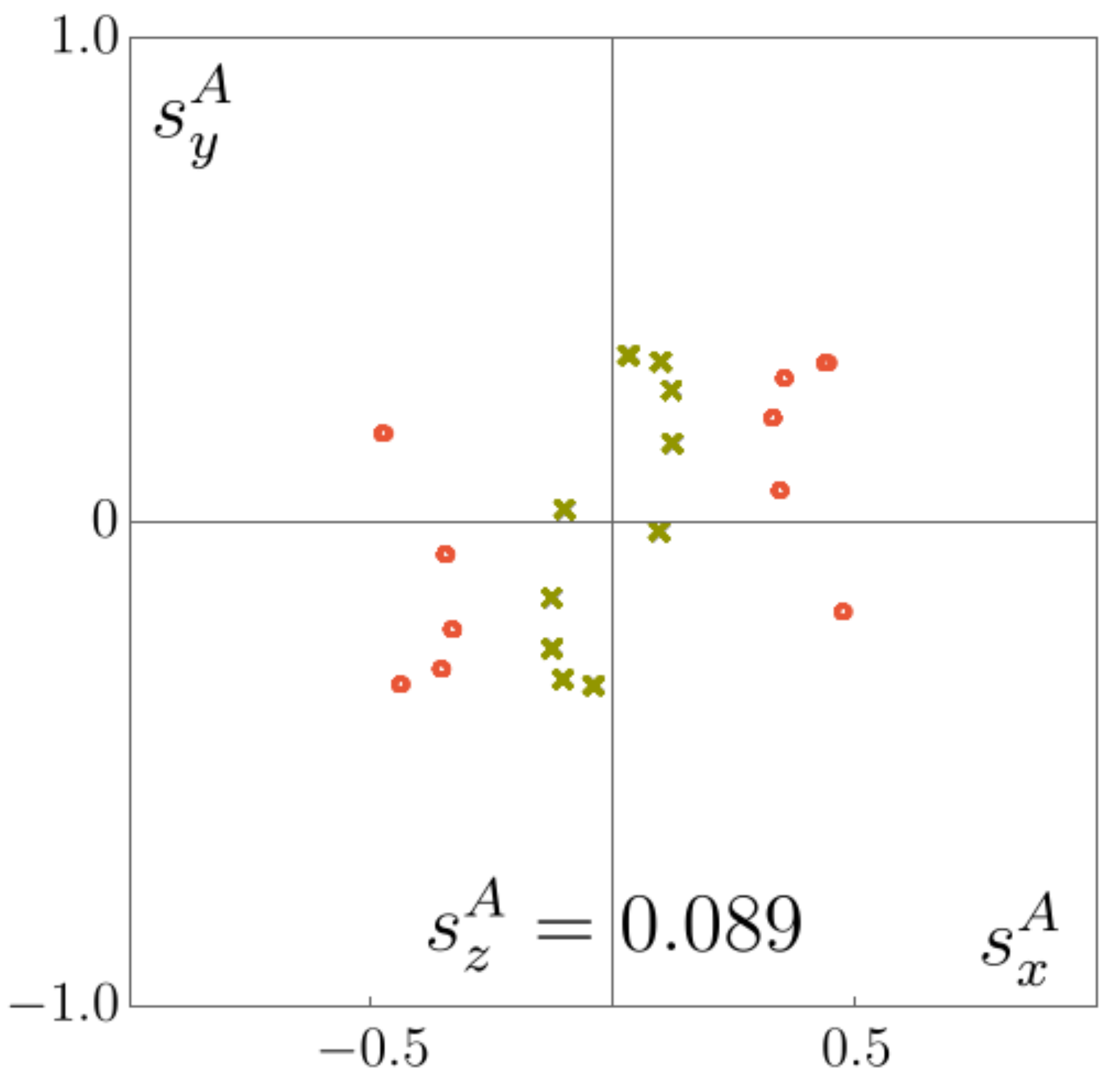}}\\
\subfloat[\textbf{(c)}]{\label{Traj_Sect_SLC_Comp_SC_del=0106}\includegraphics[scale=0.35]{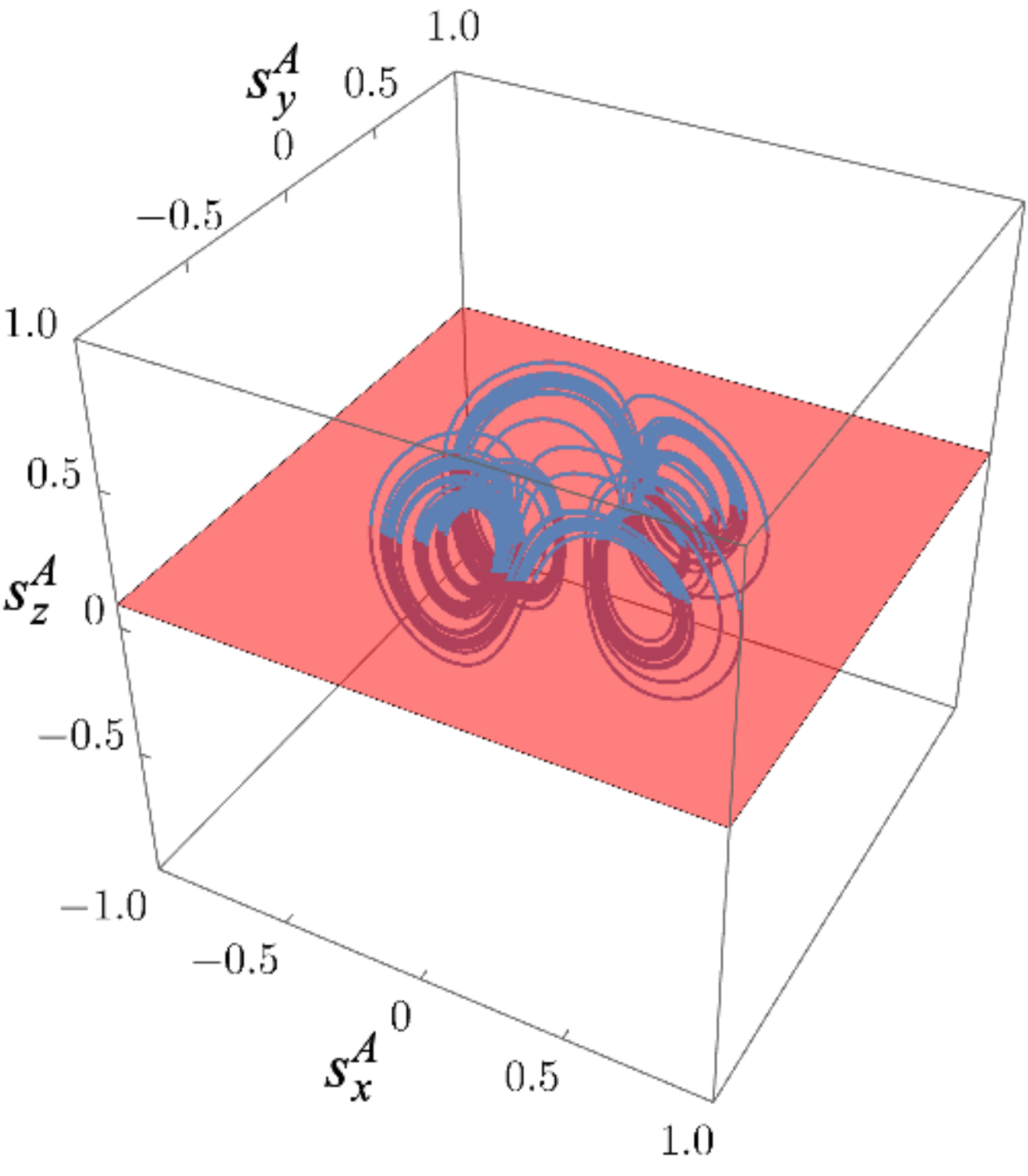}}\qquad\qquad
\subfloat[\qquad\textbf{(d)}]{\label{Poincare_SLC_Comp_SC_del=0106}\includegraphics[scale=0.33]{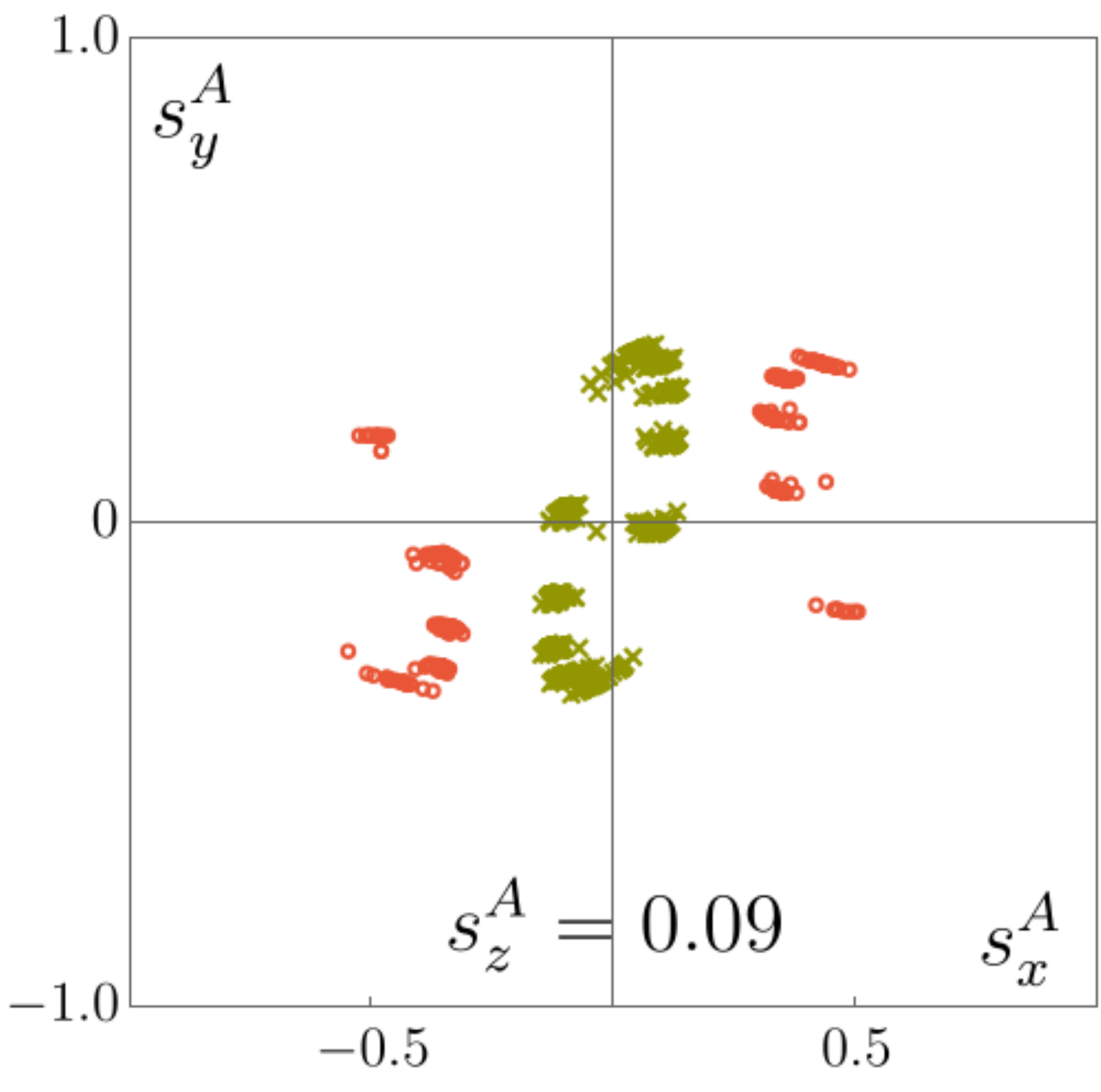}}
\caption{Comparison of Poincar\'{e} sections for the $\Z2$-symmetric periodic attractor (first row) and the synchronized chaotic one (second row). The parameters are the same as in \fsref{SC_SLC_ZA}, \ref{Spectrum_SC_SLC} and \ref{Lyapunov_Exponent_Z2_106_107_055}. Both the trajectories (first column) and the Poincar\'{e} sections (second column) illustrate how the synchronized chaos evolves out of the $\Z2$-symmetric limit cycle. (For interpretation of the references to color in this figure legend, the reader is referred to the web version of this article.)}
\label{Poincare_SC_SLC}
\end{figure*}

Comparing the spin dynamics (\fref{SC_SLC_ZA}), spectra (\fref{Spectrum_SC_SLC}) and the Poincar\'{e} sections (\fref{Poincare_SC_SLC}) pertaining to a synchronized chaotic attractor and a nearby (in the phase diagram) $\Z2$ symmetric limit cycle reveals that the chaotic attractor spends relatively large amount of time near the periodic attractor. In this section, we argue that this fact along with the behavior of the Floquet multipliers $r_{i}$ (of the $\Z2$-symmetric limit cycle) across criticality proves that the limit cycle brings about synchronized chaos via tangent bifurcation intermittency.     

To the right of the dot-dashed line in \fref{Phase_Diagram_Z2}, the $\Z2$-symmetric limit cycle has a well-defined period. Consider a Poincar\'{e} plane that contains the origin and is perpendicular to the  $s_{x} - s_{y}$ plane. Starting from a perturbed initial condition, which is not on the limit cycle, we record the successive (after each period) crossing points $\bm{s}_{n}$ of the perturbed trajectory and the Poincar\'{e} plane. Had we started on the limit cycle, every time the  trajectory would have crossed the plane at the same point $\bm{s}^{*}$. We introduce the Poincare\'{e} map in terms of the following successive displacement vectors: 
\beg 
\Delta \bm{s}_{n} \equiv \bm{s}_{n}-\bm{s}^{*}.
\label{Disp_Poin_Map}
\en
The map is then locally approximated near the fixed point as:
\beg 
\Delta \bm{s}_{n+1} = \mathbb{B}_{s}\cdot \Delta \bm{s}_{n},
\label{TBI_Floquet}
\en%
where $\mathbb{B}_{s}$ is the monodromy matrix in the Floquet analysis from \sref{Sect_Symm_Floquet}. 

We write the above map in the eigenbasis as follows:
\beg 
\bm{x}_{n+1} = \textrm{diag}(r_{1}, r_{2}, r_{3})\cdot \bm{x}_{n},
\label{TBI_Floquet_diag}
\en%
where $\bm{x}$ is a shorthand for $(x, y, z)$, and $r_{i}$ are the Floquet multipliers as defined before. Note, $r_{1} = 1$ corresponds to the eigendirection along the limit cycle. Since this direction is orthogonal to the Poincar\'{e} plane, it is irrelevant to the Poincar\'{e} map function. This indeed makes the above map \re{TBI_Floquet} effectively 2D. 

In fact, we neglect the eigendirections ($x$ and $z$) for which the absolute values of the Floquet multipliers are less than or equal to one (i.e., $r_{1}$ and $r_{3}$) across criticality, since these directions do not take part in destabilizing the limit cycles. Consider the iterative map:
\beg 
y_{n+1} = f(y_{n}).
\label{TBI_Poincare_Map}
\en%
The linear approximation to the above map near the fixed point $\bm{x}^{*}$ is written as:
\beg 
\left(y_{n+1}-y^{*}\right) = r_{2}\left(y_{n}-y^{*}\right), \quad r_{2} = f^{\prime}\left(y^{*}\right)
\label{TBI_Poincare_Map_Lin}
\en%

One can obtain similar 1D maps for other eigenmanifolds -- the ones that correspond to eigendirections $x$ and $z$ near the fixed point.

\begin{figure*}[tbp!]

\begin{center}
\includegraphics[scale=0.35, trim=0 0 0 0.5cm, clip]{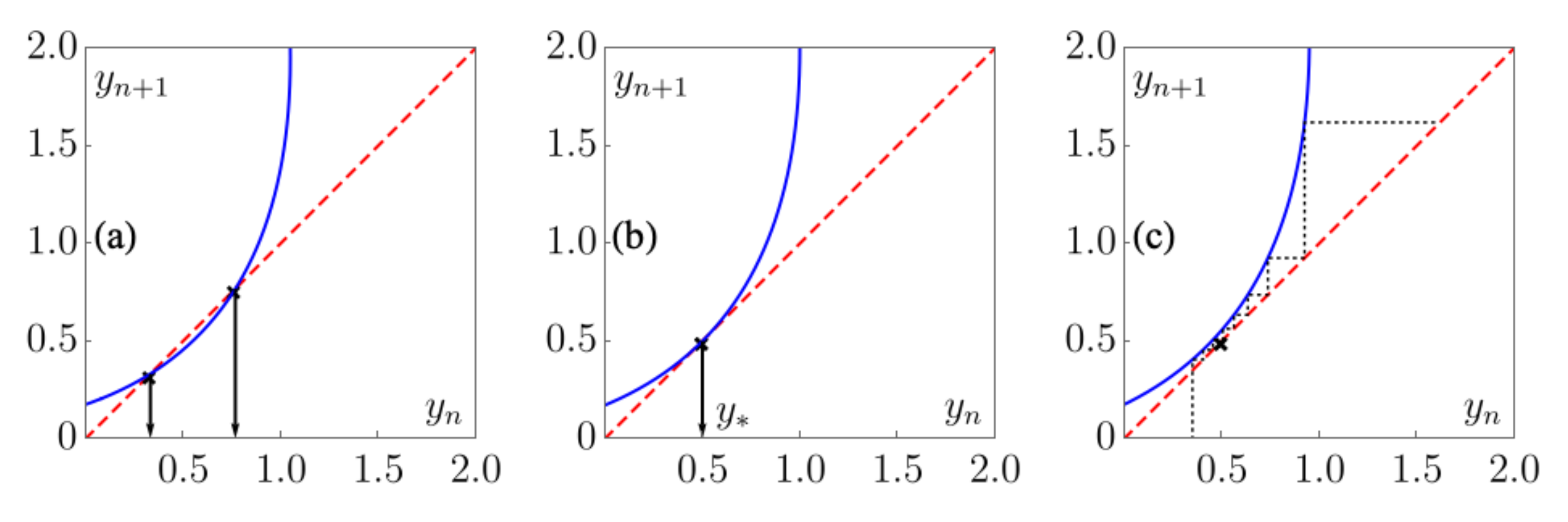}
\caption{Cartoon of tangent bifurcation intermittency. Consider the Poincar\'{e} map function: $f(y_{n}) = y_{n+1}$ that is implicitly defined as $\big(f(y) - y + 1\big)^{2}/2 = \big(y-a\big)^{2} + \big(f(y) - 1\big)^{2}$ (solid blue) and the diagonal line: $y_{n+1} = y_{n}$ line (dashed red). From left to right we gradually decrease $a$ ($a = 0.1, 0.0$ and $-0.1$ in \textbf{(a)}, \textbf{(b)} and \textbf{(c)}, respectively.) In \textbf{(b)}, for $y^{*} = 0.5$ the map obeys \eref{TBI_Poincare_Map_Prop}, and the diagonal line becomes tangential to the graph of the map there. Notice the small gap between the line and the graph near $(y^{*}, y^{*})$ in \textbf{(c)}. With the thin black dashed line we depict successive iterations. This shows that indeed they spend considerable amount of time near the point $y = y^{*}$ even though this is no longer a fixed point.}\label{Tan_Bifurc_Map}
\end{center}

\end{figure*}

At the time of the bifurcation we have
\beg 
\frac{\mathrm{d} f(y^{*})}{\mathrm{d}y} = 1, \qquad f(y^{*}) = y^{*}.
\label{TBI_Poincare_Map_Prop}
\en%
To wit, at $y = y^{*}$ the diagonal line $y_{n+1} = y_{n}$ is tangential to the map function in \eref{TBI_Poincare_Map}, see \fref{Tan_Bifurc_Map}\blue{b}. Hence the name -- tangent bifurcation intermittency. 

Past the bifurcation, the fixed point in the map disappears leading to opening up of a small gap  near $y^{*}$, as shown in \fref{Tan_Bifurc_Map}\blue{c}. Although, this forces the limit cycle to abruptly lose its stability and give rise to chaos, several successive iterates of the Poincar\'{e} map of \eref{TBI_Poincare_Map} remain close to $y^{*}$. This explains the proximity of the $\Z2$-symmetric periodic and the synchronized chaotic attractors close to the bifurcation line (dot-dashed line in \fref{Reduced_Phase_Diagram}). It is known that although after some time the iterate is pushed away from $y^{*}$, eventually it is reinjected near the same fixed point. The rate of reinjection depends on the global featutures of the map $f(y)$ \cite{Hilborn}.

\subsubsection{On-Off Intermittency in $\R^{6}$} \label{On_Off_Int}

\begin{figure*}[tbp!]
\centering
\subfloat[\qquad\textbf{(a)}]{\label{Phase_Diagram_Z2}\includegraphics[scale=0.34]{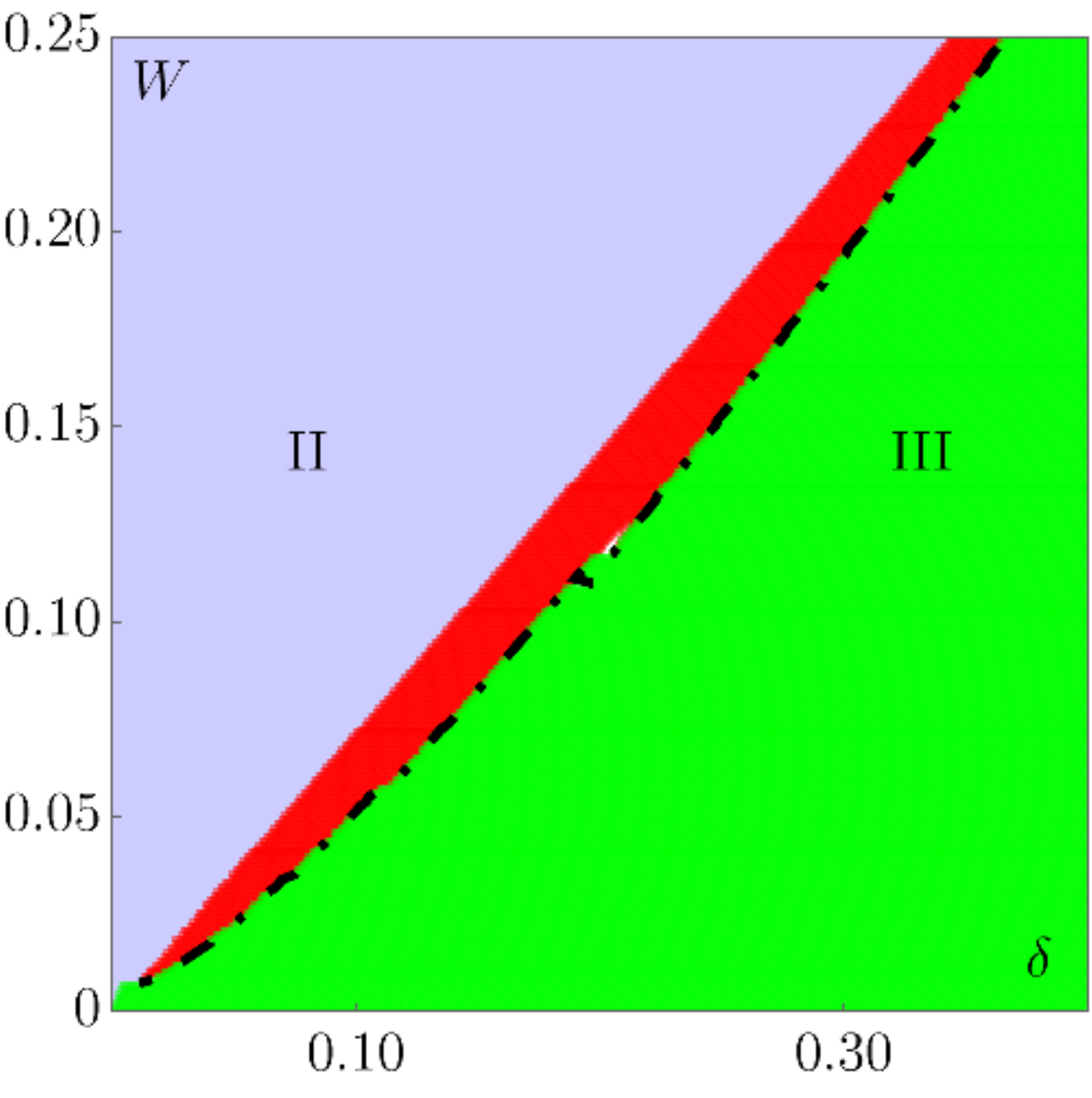}}\qquad\qquad
\subfloat[\qquad\textbf{(b)}]{\label{Start_SC}\includegraphics[scale=0.34]{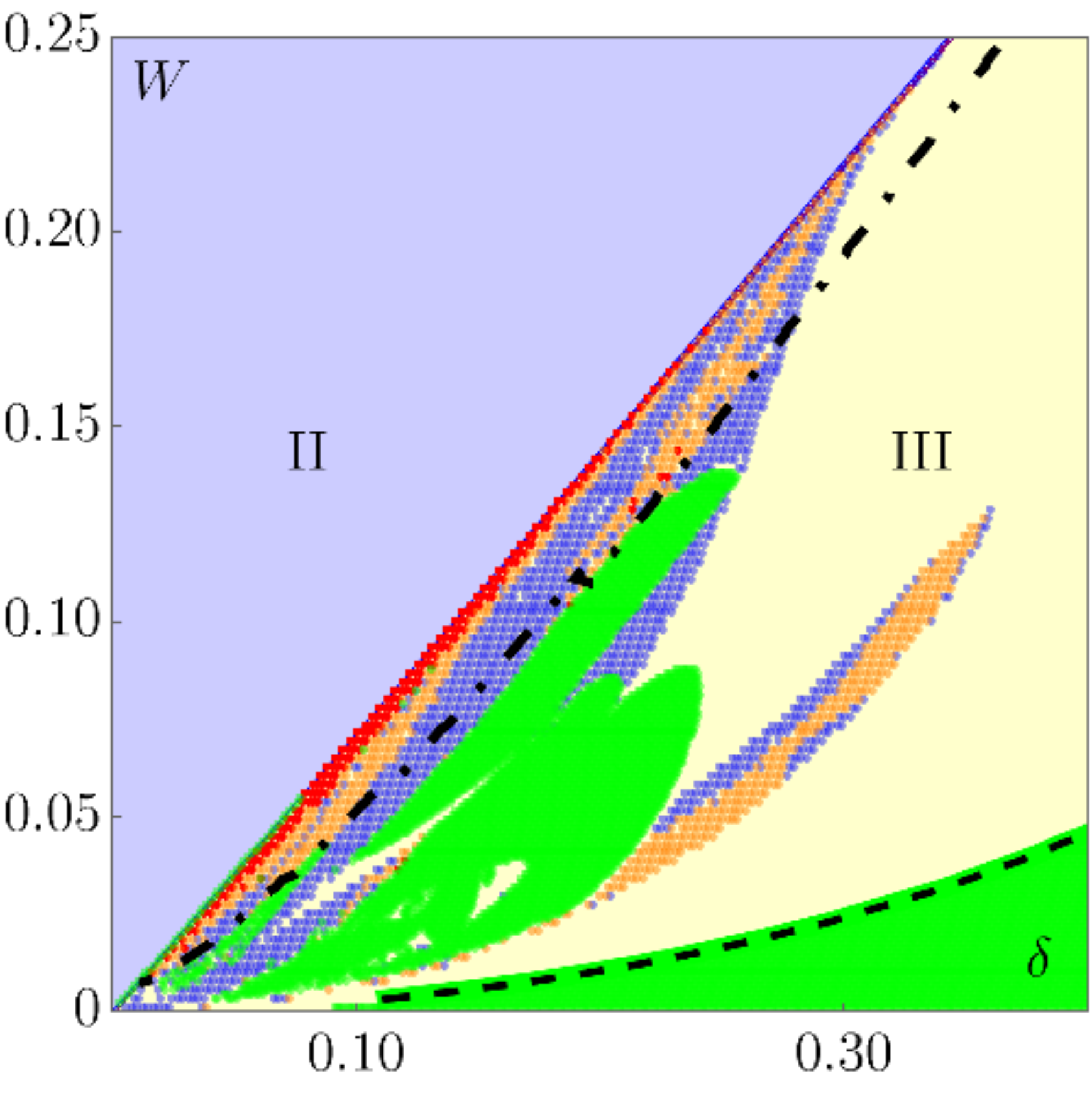}}
\caption{Comparison of the nonequilibrium phase diagram for the $\Z2$-symmetric attractors that are asymptotic solutions of \eref{Symm_One_Spin_Eqn} (left) with the one containing all the attractors living on the unrestrained 6D manifold (right). The latter are asymptotic solutions of \eref{Mean-Field_1}. The coloring scheme is the same as in \fref{Phase_Diagram}. Both \textbf{(a)} and \textbf{(b)} above are comprised of three similar phases. In particular, Phase II is identical in both instances. In \textbf{(a)}, Phase III has only two amplitude-modulated sub-phases compared to five in \textbf{(b)}. The dot-dashed lines in both the diagrams indicate the commencement of the $\Z2$-symmetric chaotic attractor. It is not stable in the full phase space immediately. Only close to the II-III boundary it becomes sufficiently attracting, so that any initial condition leads to this attractor. Note, in \textbf{(a)}, one has different kinds of $\Z2$-symmetric periodic attractors in the green island to the left of the dashed ($\Z2$-symmetry breaking) line. Sometimes they coexist in the $\Z2$-symmetric submanifold \re{Z2_Expl}. Only one of them is stable in the full phase space \cite{Patra_1}. (For interpretation of the references to color in this figure legend, the reader is referred to the web version of this article.)}
\label{Reduced_Phase_Diagram}
\end{figure*}

\begin{figure*}[tbp!]

\centering
\subfloat[\qquad\textbf{(a)}]{\label{Symmetric_Floquet}\includegraphics[scale=0.232]{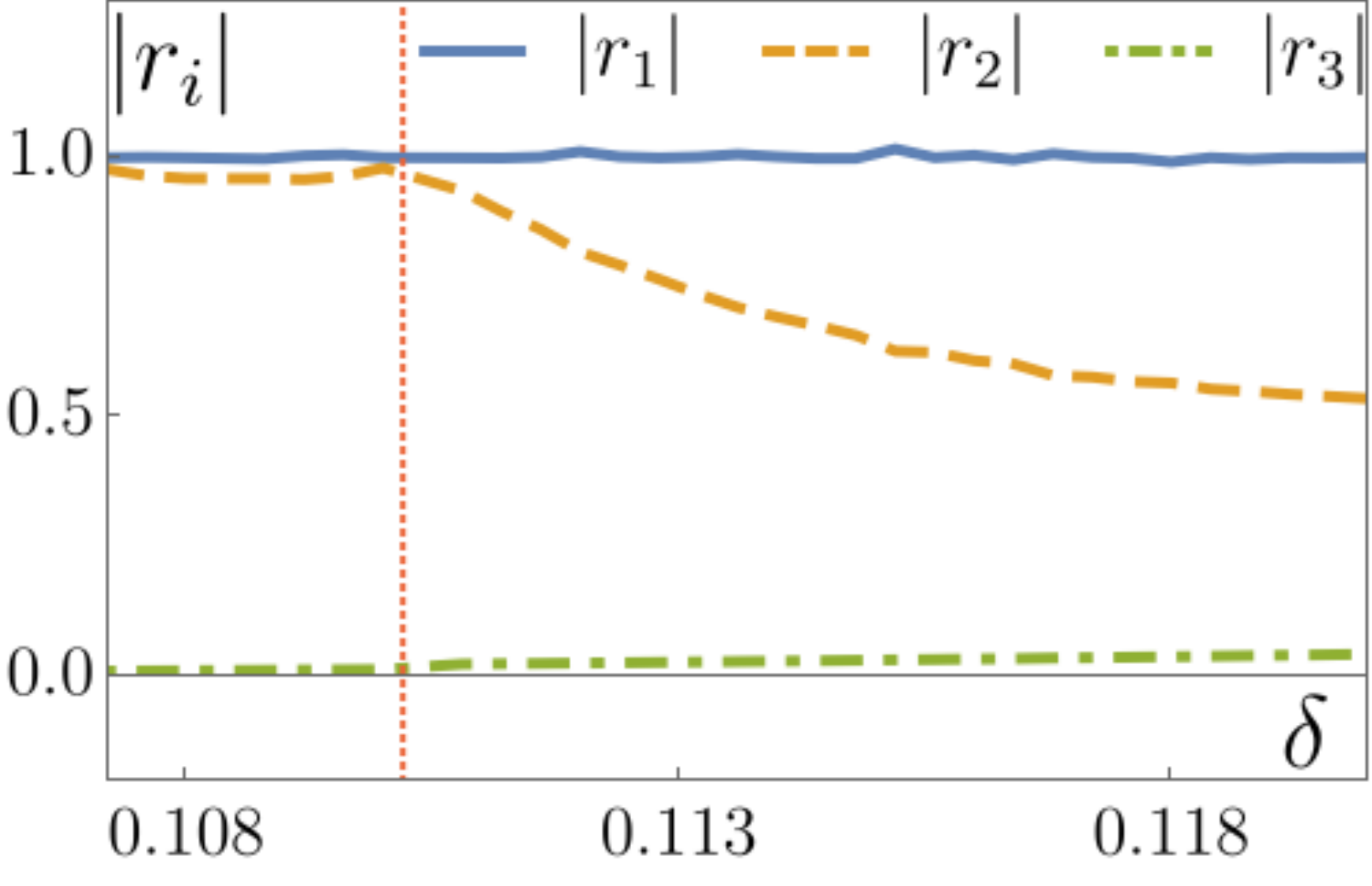}}\quad
\subfloat[\qquad\textbf{(b)}]{\label{Floquet_118}\includegraphics[scale=0.225]{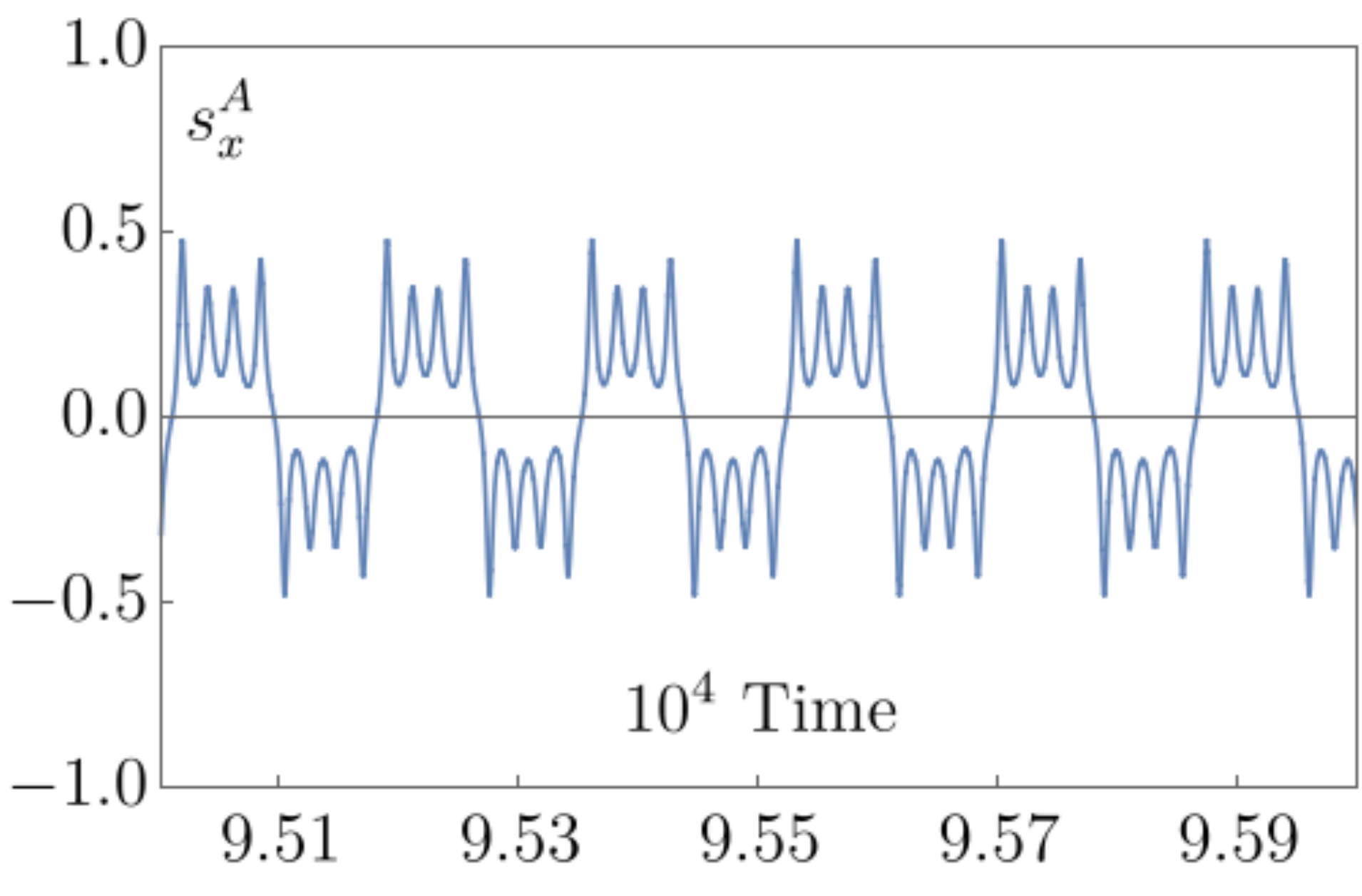}}\quad
\subfloat[\qquad\textbf{(c)}]{\label{Floquet_11}\includegraphics[scale=0.225]{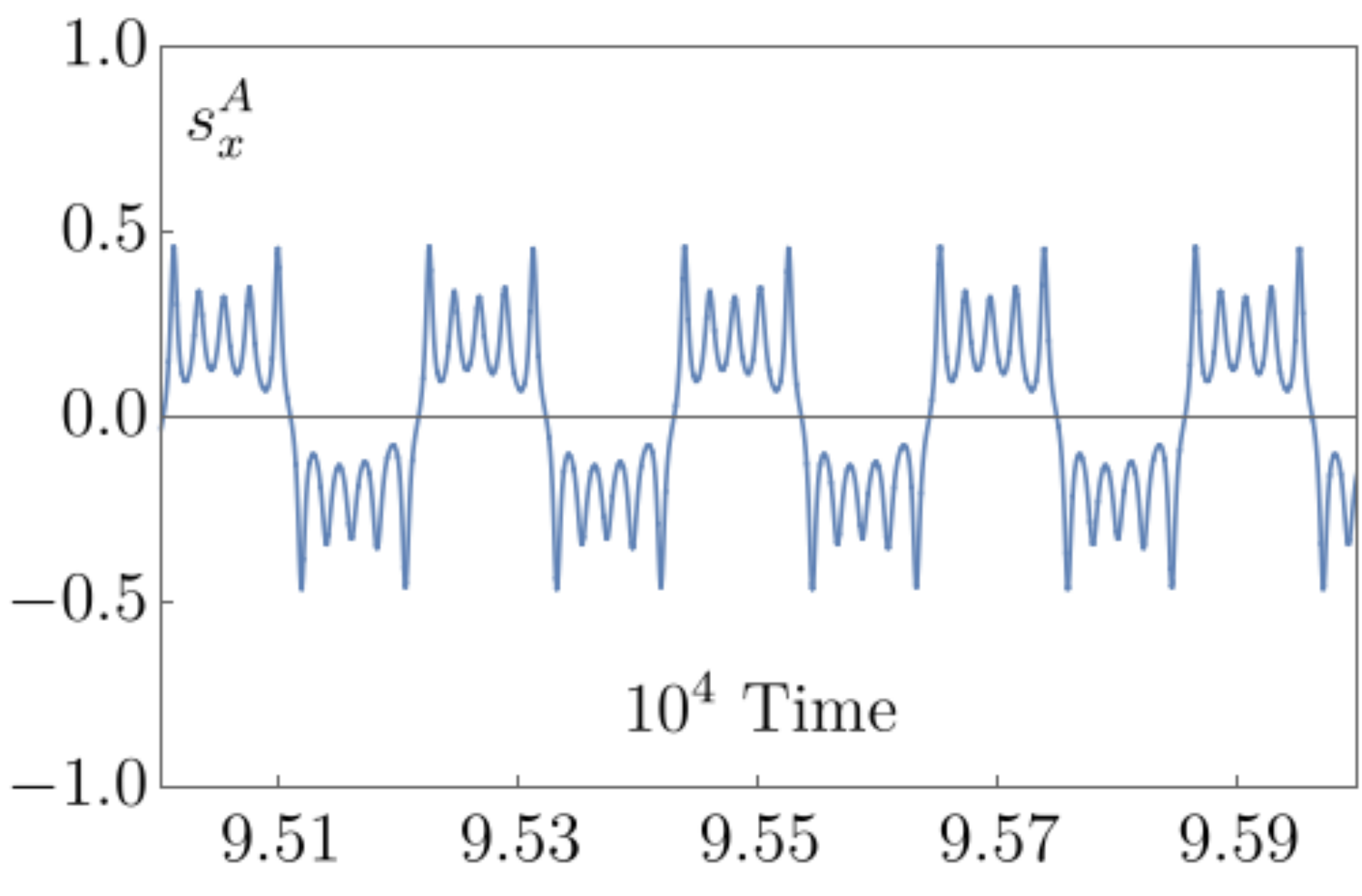}}
\caption{Different periodic attractors in the $\Z2$-symmetric submanifold to the right of the dot-dashed line of \fref{Reduced_Phase_Diagram}. We show the absolute values of three Floquet multipliers as functions of $\delta$ for $W = 0.055$. Same initial condition $\bm{s}_{0} = (0.4, -0.469, 0.7)$ leads to different periodic attractors to the right (in \textbf{(b)} $\delta = 0.118$) and left (in \textbf{(c)} $\delta = 0.11$) of $\delta \approx 0.1104$ (red dotted line). Thus we linearize about different attractors in \eref{Floquet_Eqn_Z2} while in different ranges of $\delta$. Our analysis shows the loss of stability of \textbf{(b)} and \textbf{(c)} at $\delta \approx 0.1104$ and $\delta \approx 0.107$, respectively. For $\delta\lessapprox 0.107$ the periodic attractor ceases to exist and so continuing the Floquet analysis is impossible past this point.}
\label{Symmetric_Floquet_W=0_055}

\end{figure*}

\begin{figure*}[tbp!]
\centering
\subfloat[\qquad\textbf{(a)}]{\includegraphics[scale=0.3]{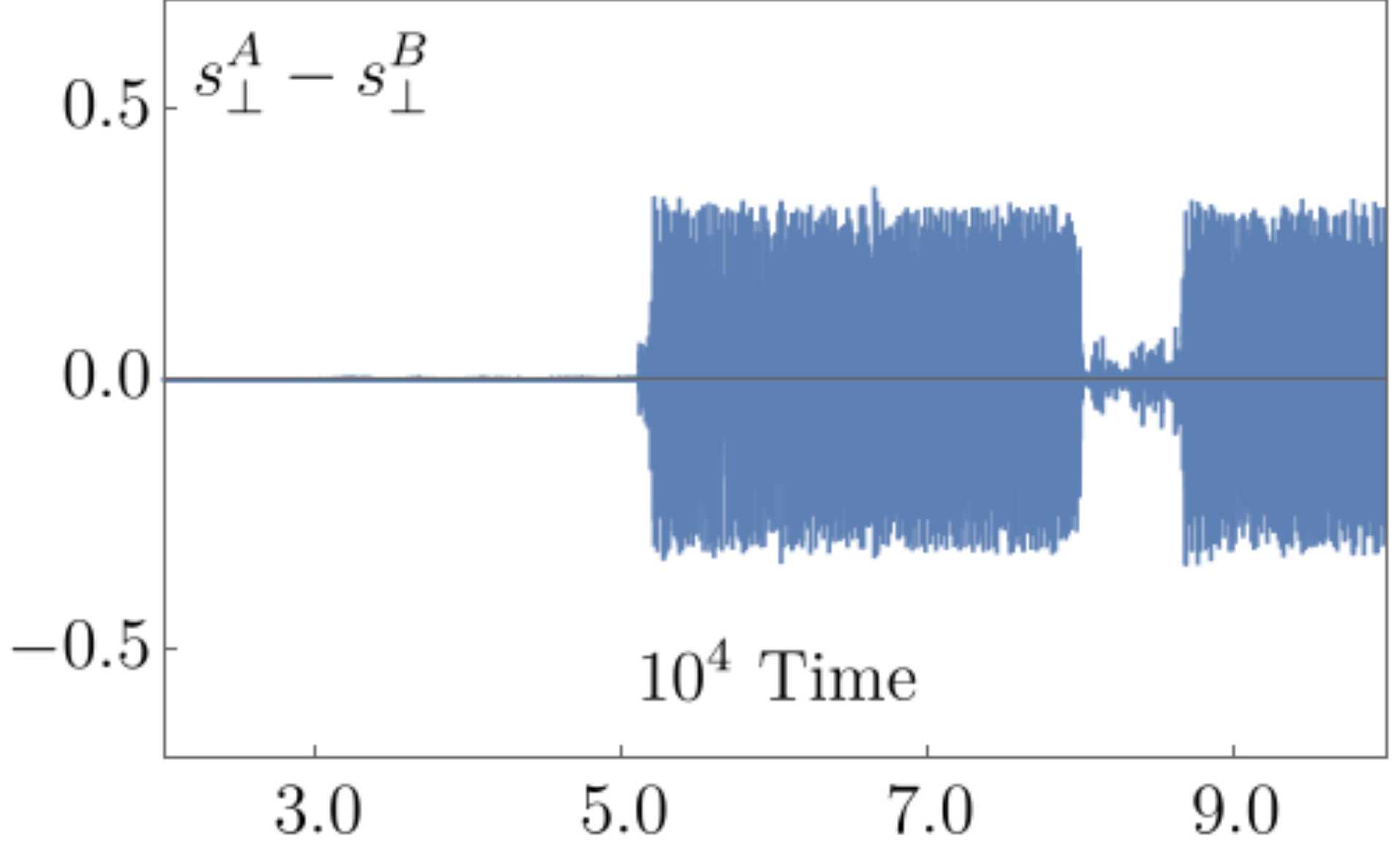}}\qquad\qquad
\subfloat[\qquad\textbf{(b)}]{\includegraphics[scale=0.3]{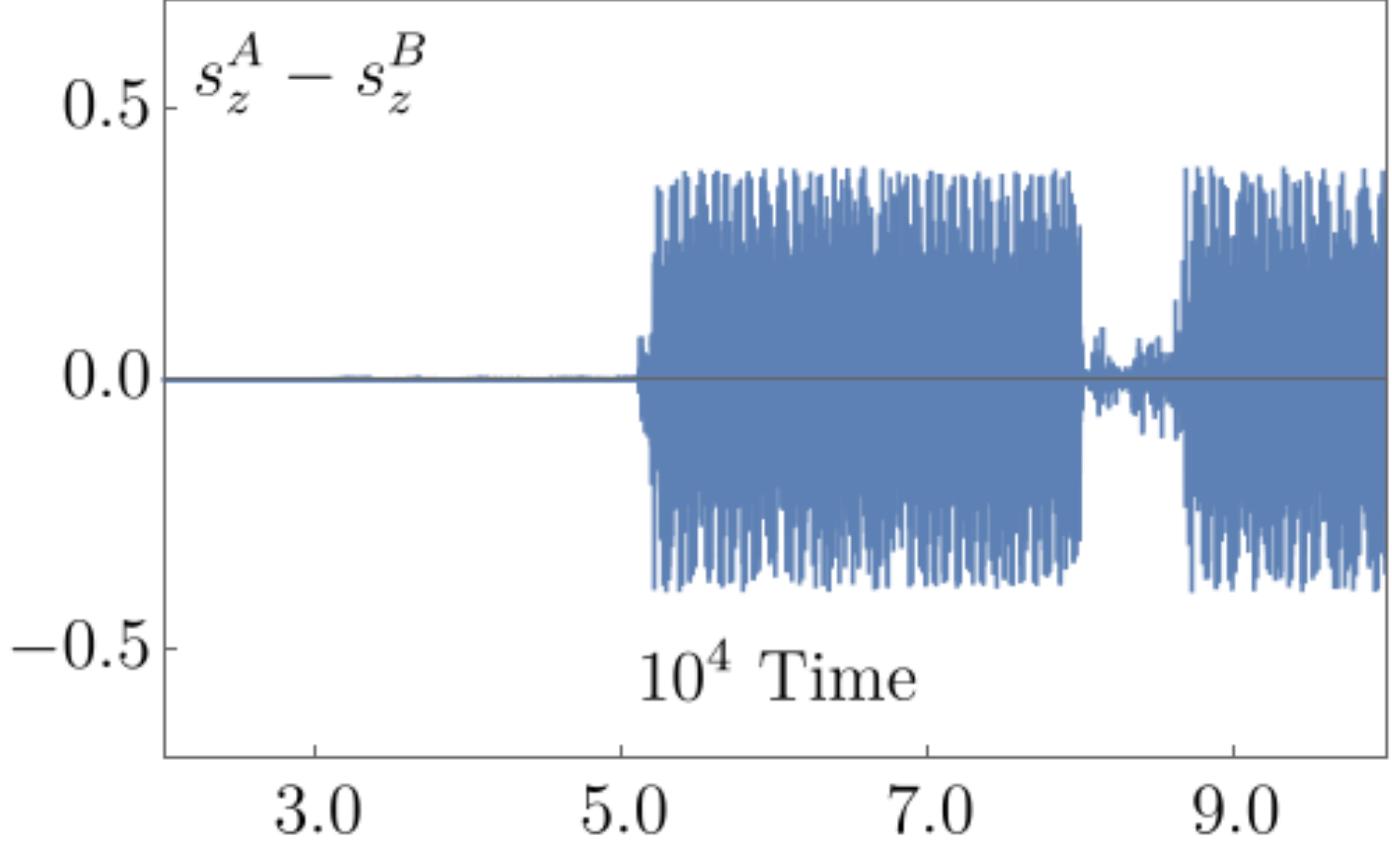}}\\
\subfloat[\qquad\textbf{(c)}]{\includegraphics[scale=0.3]{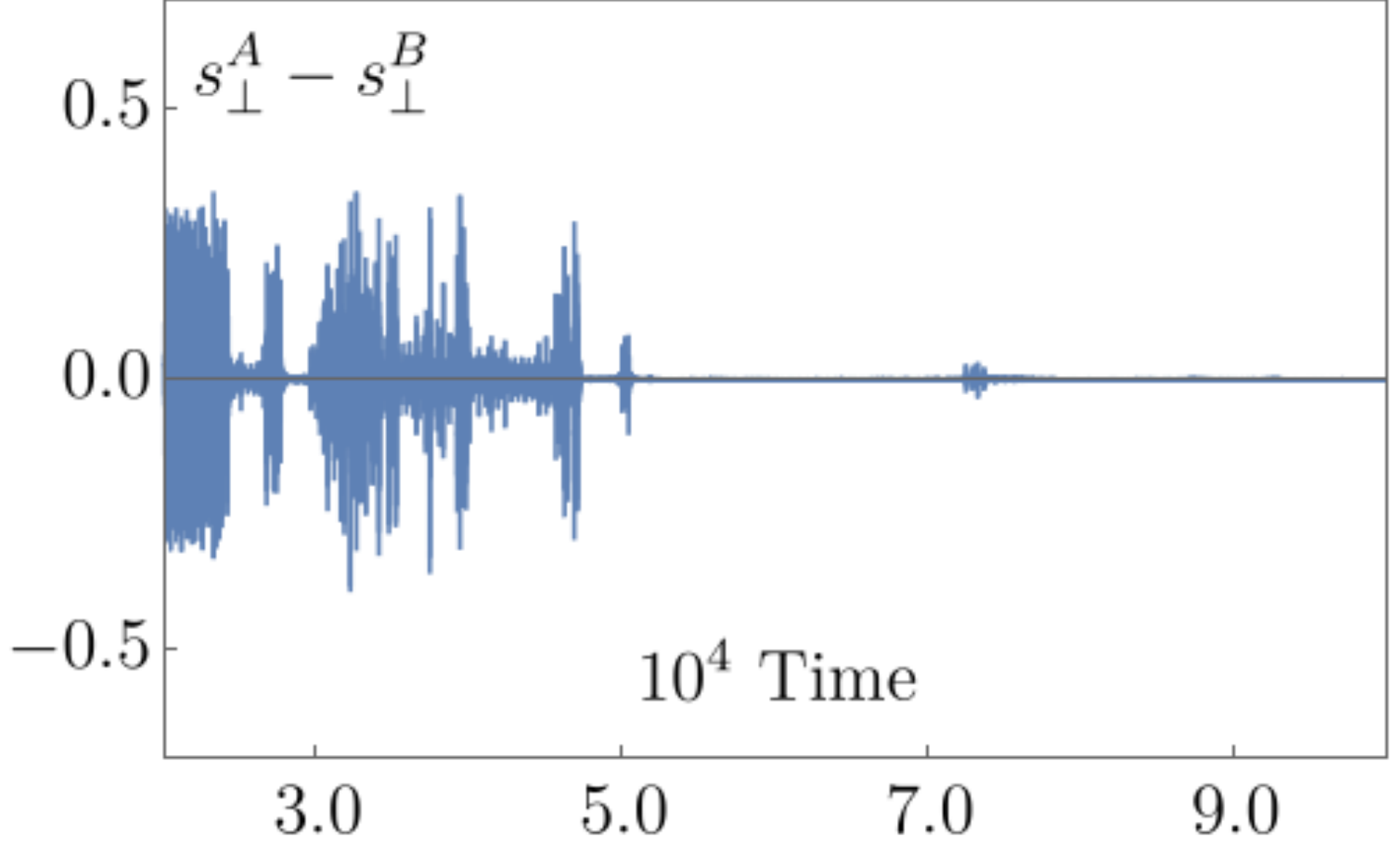}}\qquad\qquad
\subfloat[\qquad\textbf{(d)}]{\includegraphics[scale=0.3]{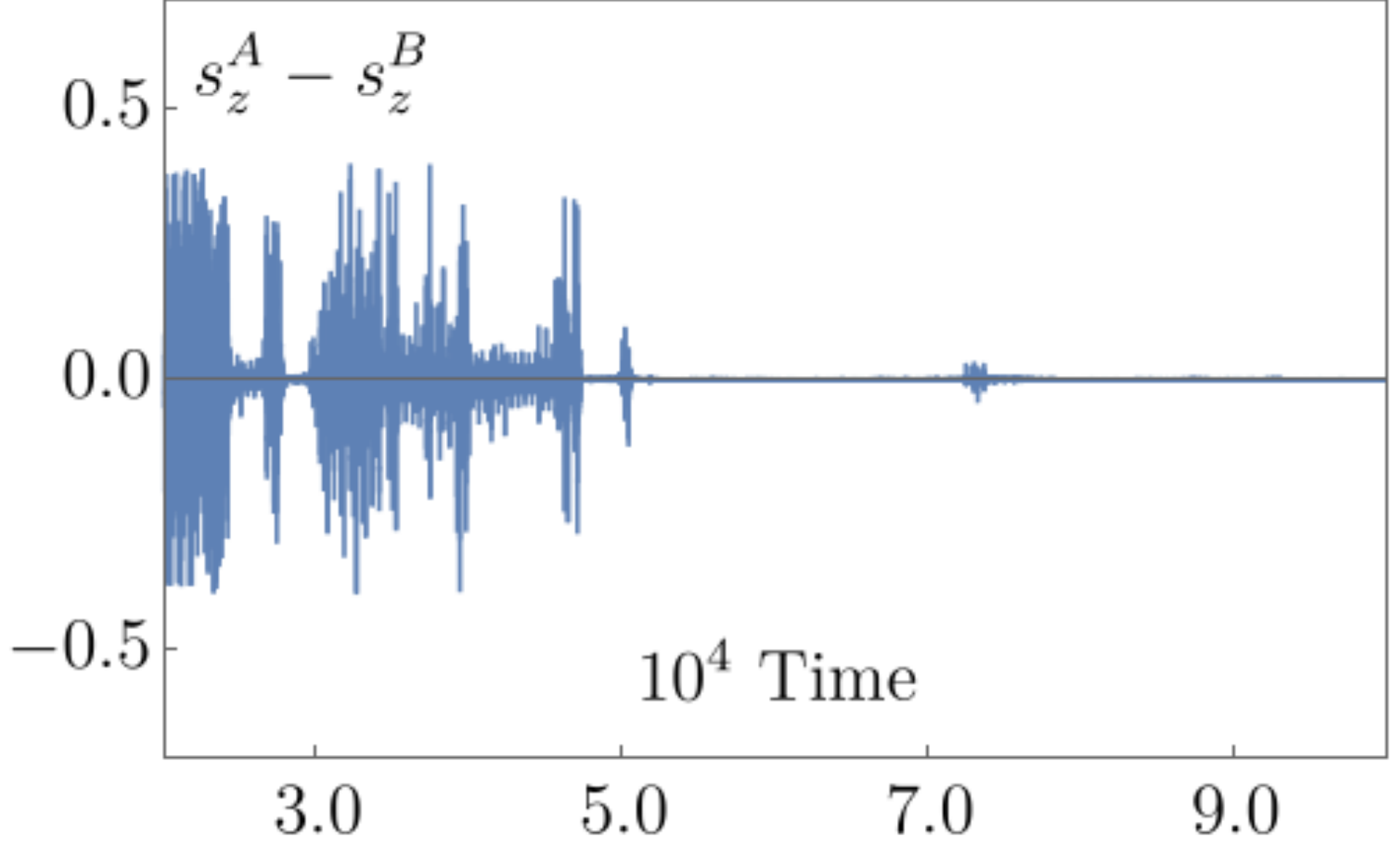}}\\
\subfloat[\qquad\textbf{(e)}]{\includegraphics[scale=0.3]{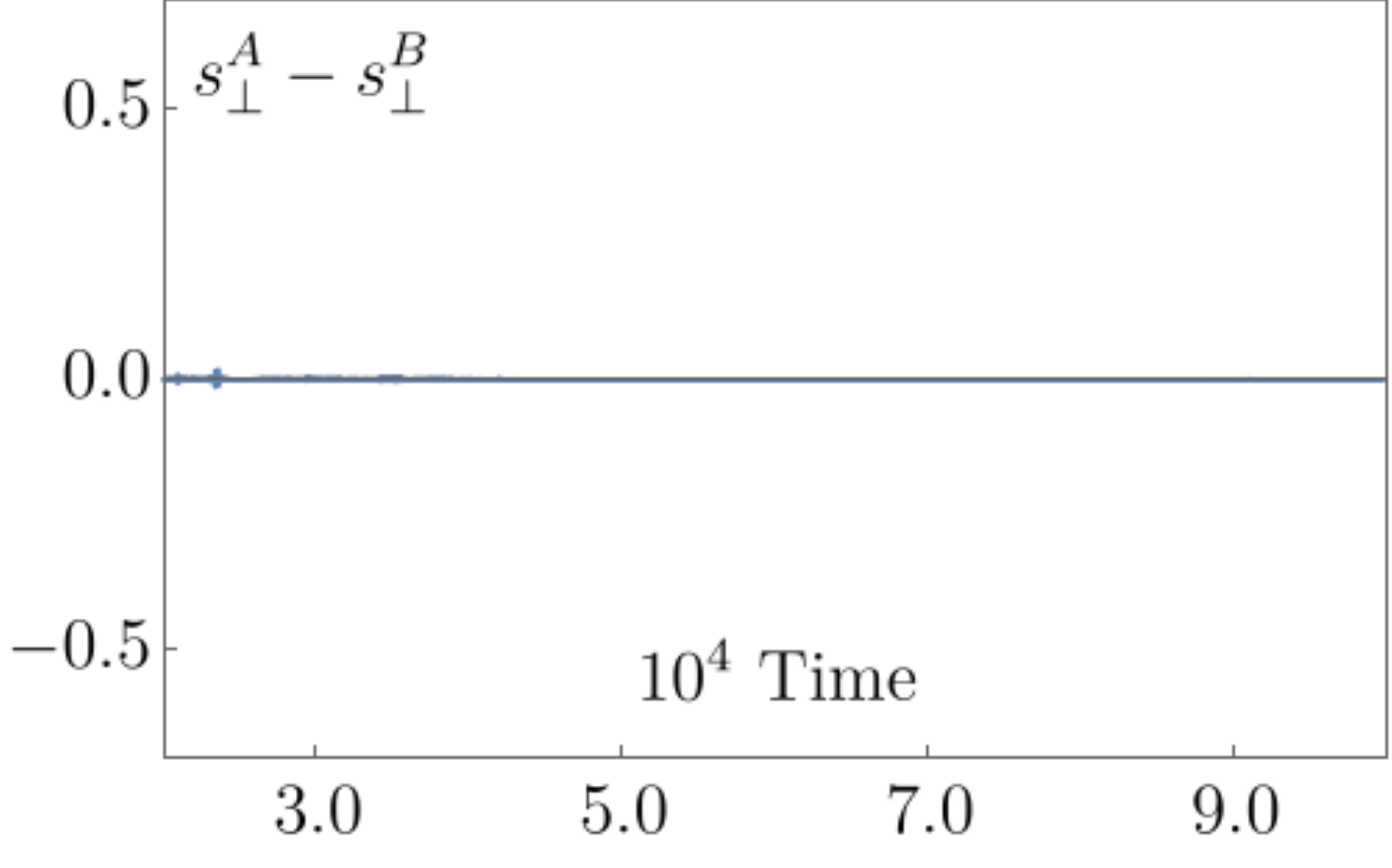}}\qquad\qquad
\subfloat[\qquad\textbf{(f)}]{\includegraphics[scale=0.3]{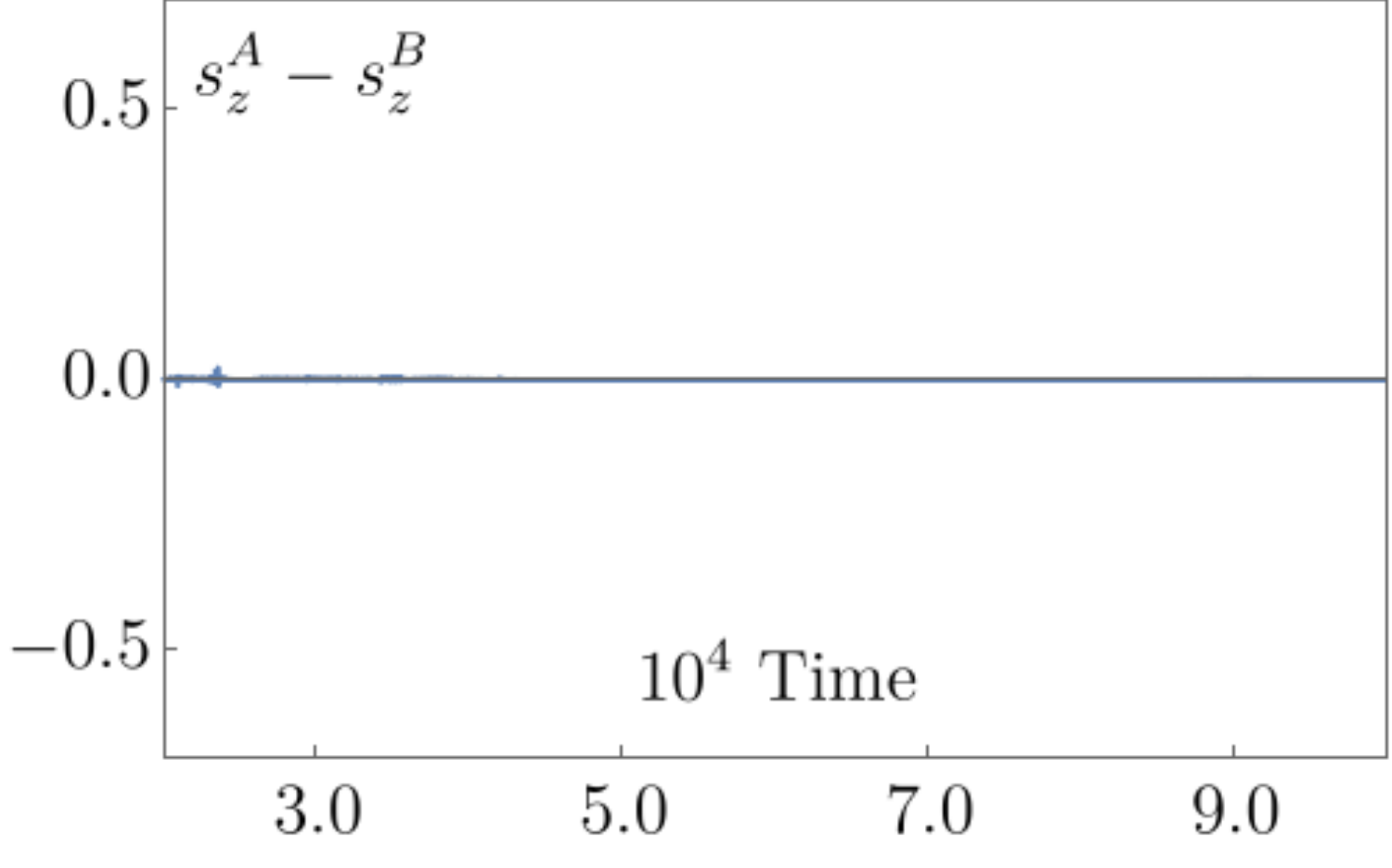}}
\caption{On-off intermittency at a constant repump rate $W = 0.05$ as one decreases the detuning $\delta$ ($ = 0.08021,\,\, 0.08010$ and $0.08000$ in the first, second and third rows, respectively) within Phase III near II-III boundary. We plot $s_{\perp}^{A}-s_{\perp}^{B}$ vs. time and $s_{z}^{A}-s_{z}^{B}$ vs. time in the first and second columns, respectively. For all the figures  initial condition is $\textbf{K}_{0} = ~(-0.729897, 0.538791, -0.298485, -0.117912, -0.0486406, -0.668281)$. The frequency and duration of chaotic outbursts decrease as we decrease the detuning, and eventually become zero at the synchronized chaotic phase (red subregion of Phase III in \fref{Phase_Diagram}).}
\label{Intermitt}
\end{figure*}

As mentioned before, there exist characteristic directions that push the chaotic attractors away from the synchronization submanifold near the dot-dashed line in \fref{Phase_Diagram_Z2}. This prevents the spins to get fully synchronized. As we move closer to the red subregion by decreasing $\delta$ while keeping $W$ fixed, the frequency and amplitude of chaotic outbursts decrease, see \fref{Intermitt}. Eventually the attraction to the $\Z2$-symmetric submanifold wins and we end up with full synchronization \cite{Chaotic_Synch_X}. This explains the on-off intermittency in the transverse variables.

\section{Discussion}

In this paper, we studied the  emergence of chaos via quasiperiodicity in the  dynamics of two atomic ensembles collectively Rabi coupled to a bad cavity mode. Employing  Floquet analysis, we showed that quasiperiodic attractors are born out of the periodic ones via Neimark-Sacker bifurcation. In our system, we could only discern two frequency quasiperiodicity. Eventually the chaotic  dynamics of the atomic ensembles synchronize due to the mutual coupling.  

Within the mean-field description,   accurate when the numbers of atoms in the ensembles are large, the ensembles are represented by two classical spins $\bm{s}^{A}$ and $\bm{s}^{B}$. Mean-field equations of motion posses a $\Z2$-symmetry with respect to the interchange of the two ensembles $A \leftrightarrow B$, while flipping the sign of $s_y$.
 We distinguish different attractors by simultaneously analyzing the properties of spin dynamics,  projections of attractors onto various planes, Poincar\'{e} sections, Lyapunov exponents, and radiated power spectra. The power spectra  are measurable with Michelson interferometry and can be utilized to distinguish the different phases experimentally. 
 
 Finally, we focused on the dynamics in the $\Z2$-symmetric submanifold, where there are only two types of amplitude-modulated superradiance -- limit cycles ($\Z2$-symmetric limit cycles) and chaos (synchronized chaos). In parentheses we indicate attractors in the full phase-space that correspond to these attractors in the $\Z2$-symmetric subspace. The synchronized chaotic attractor originates from the $\Z2$-symmetric limit cycle.  Near instability all three Floquet multipliers of the $\Z2$-symmetric limit cycle are real. One of them is monotonically increasing becoming greater than one after criticality. Moreover, with the help of the spin dynamics, Poincar\'{e} sections, and power spectra we demonstrated that right after coming into existence the synchronized chaotic attractors spend most of the time close to the $\Z2$-symmetric limit cycles. This in turn proves that  synchronized chaos indeed arises from the $\Z2$-symmetric limit cycle via tangent bifurcation intermittency.

It is interesting to investigate the potential  of applications of our system in the synchronized chaotic phase in secure communication, i.e.
 to come up with a viable ``steganography" protocol (instead of hiding the meaning of the transmitted message, hide the existence of the message itself). In particular, it is not apparent how to send a message over a long distance. It  is also of interest to see how synchronized chaos is affected by quantum fluctuations or coupling to multiple cavity modes and to analyze a system with more than two atomic ensembles coupled through a bad cavity mode.     
  
\section*{Acknowledgment}

We thank Y. G. Rubo and D. Ruelle for helpful discussions. This work was supported by the National Science Foundation Grant DMR-1609829. A.P. was also supported by the Rutgers University Louis Bevier Dissertation Completion Fellowship.

\appendix 

\section{Algorithm for Obtaining the Full Non-equilibrium Phase Diagram}\label{Algo_Phase_Diagram}

In this section we briefly explain the algorithm for obtaining \fref{Phase_Diagram}. 

\begin{enumerate}

\item Determine the supercritical and the subcritical Hopf bifurcation lines.
 
\item Follow the $\Z2$ symmetric limit cycle and determine when it becomes unstable. This identifies the symmetry breaking line. One can do so either by performing a Floquet analysis as described in \sref{Review}, or simply by observing if the asymptotic absolute value of one of the transverse coordinates is greater than zero. Here we implemented the later procedure. The criterion we used was $|\mathsf{n}_{2}(t = 2\times 10^{4})|>10^{-2}$.
 
\item Start with a fine grid in $(\delta, W)$ space in the subregion (of Phase III) delimited by the symmetry breaking line and the sub-critical Hopf bifurcation line. In our algorithm we consider $\Delta \delta = \Delta W = 0.0011$.

\item By calculating the maximum Lyapunov exponent we filter out $(\delta, W)$ points that lead to chaos. We consider a trajectory to be chaotic if the maximum Lyapunov exponent saturates to a positive value that is more than $10^{-3}$ after time $t = 5 \times 10^{4}$.

\item From the chaotic $(\delta, W)$ points, separate out the ones where the two spins are synchronized. One does this either by calculating the conditional Lyapunov exponent and obtaining a ``sufficiently" negative value after ``sufficiently" long time, or by making sure that coordinates transverse to the synchronization manifold remain zero after the transients have died down. We implemented the latter option. In particular, we choose the chaotic points where the maximum $|\mathsf{n}_{2}(t)|$ for $1.5<t\times 10^{-4}<2$ is less than $10^{-3}$. 

\item After subtracting the chaotic and synchronized chaotic points from the subregion described in step 3, obtain the $(\delta, W)$ points where a $\Z2$ symmetric limit cycle reappears. We obtain this by again finding out (same procedure as above) where inside this subregion the asymptotic transverse coordinates are zero.

\item We subtract the synchronized periodic points further from the leftover points, and as a result end up with $(\delta, W)$ points that are either non-$\Z2$-symmetric periodic or are quasiperiodic. To distinguish between these two types of attractors we use of the Poincar\'{e} section as explained in \sref{Q-P}. 

Also, note that the lines of the 2D periodic projections (e.g., in $s_{z}^{A} - s_{z}^{B}$) acquire finite thicknesses when the quasiperiodic attractors are born via the Neimark-Sacker bifurcation. Consider a transverse straight line (e.g. $s_{z}^{A} = \textrm{const.}$) that divides the projections into two halves, and note the intersections of the cross sections with the above line. This is a simplified 1D projected Poincar\'{e} section. For the periodic attractors this section has a finite number of discrete points, whereas for quasiperiodic attractors this is a union of a few disjointed line segments and has infinitely many points. Using this latter property, it is relatively easy to distinguish the periodic and quasiperiodic attractors. Note, although we can distinguish between quasiperiodicity and non-$\Z2$-symmetric limit cycles using the 1D Poincar\'{e} section, it is impossible to differentiate chaos from quasiperiodicity, or synchronized chaos from $\Z2$-symmetric limit cycle using this approach.
   
\end{enumerate}

\section{Calculation of Maximum Lyapunov Exponent and Conditional Lyapunov Exponent}\label{Appnd_Lya_Exp}

In this section we list all the necessary steps for the calculation of the maximum Lyapunov exponent $\lambda$ for the attractors of \eref{Mean-Field_1}.

\begin{enumerate}

\item Start with a random initial condition and evolve the aforementioned mean-field equations for sufficiently long time $t_{0}$, where $t_{0}$ is large enough so that the transient dynamics have died down and we are left with the attractor.

\item Consider the stretch of the trajectory with the initial condition $\textbf{K}_{0} = \textbf{K}(t_{0})$ (i.e., start time is $t_{0}$) and the end time for the simulation $(t_{0} + t_{1})$. We consider this to be a ``fiduciary trajectory". The maximum Lyapunov exponent measures how a point in the configuration space close to this trajectory diverges on average.
  
\item Redefine $t_{0}$ to be the origin and divide the time interval $(t_{0}, t_{0} + t_{1})$ in $N$ equal steps. Thus, $t_{0}\rightarrow 0$ and $(t_{0} + t_{1})\rightarrow N\Delta t$, where $\Delta t$ is the length of the time steps.
 
\item At each time step we simulate the mean-field equations \re{Mean-Field_1} and the linearized (about the fiduciary trajectory) equations \re{Full_Floquet_Eqn} simultaneously.
 
\item In the first step we start with the initial condition $(\textbf{K}_{0}, \Delta \textbf{K}_{0})$ for the combined system of equations (six nonlinear and six linearized equations), where we defined  $\textbf{K}_{0}$ in step 2 above, and $ \Delta \textbf{K}_{0}$ are any small six random numbers. Also define $d_{0} = \sqrt{|\Delta \textbf{K}_{0}|^{2}} \ll 1$. By implementing step 4 we obtain $d_{1} = \sqrt{|\Delta \textbf{K}(\Delta t)|^{2}}$, and calculate $\ln{\big[\frac{d_{1}}{d_{0}}\big]}$. 

\item At each successive steps we need to choose initial conditions with sufficiently small norms for \eref{Full_Floquet_Eqn}. Therefore in the next step, we choose $\textbf{K}(t_{0}+\Delta t)$ and $\Delta \textbf{K}(\Delta t)\frac{|\Delta \textbf{K}_{0}|}{|\Delta \textbf{K}(\Delta t)|}$ as the initial conditions for \eref{Mean-Field_1} and \eref{Full_Floquet_Eqn}, respectively. Note, the norm of the initial condition for the linearized equation is still $d_{0}$. Moreover, the spins are still on the fiduciary attractor. As we did in step 5, we calculate the norm of the linearized variables at the end of the second time step as  $d_{2} = \sqrt{|\Delta \textbf{K}(2\Delta t)|^{2}}$, and obtain $\ln{\big[\frac{d_{2}}{d_{0}}\big]}$.

Here one needs to normalize the initial condition for the linearized equations \re{Full_Floquet_Eqn} properly. In particular, it would be wrong to take $\frac{\Delta \textbf{K}(\Delta t)}{|\Delta \textbf{K}(\Delta t)|}$ to be the initial condition at the beginning of the second time step, as suggested in \cite{Uchida}, since, for $|\Delta \textbf{K}(\Delta t)|\ll 1$ one ends up having an initial condition that is far away from the fiduciary trajectory. Other than this alteration, the algorithm discussed here is the same as in \cite{Uchida}.

\item In fact, at the $i$-th time step any initial condition $\Delta \textbf{K}_{i}$ for \eref{Full_Floquet_Eqn} (see 6 above) will suffice, as long as $\sqrt{|\Delta \textbf{K}_{i}|^{2}}\ll 1$. For a chaotic attractor, in particular, one neglects the time evolution in other eigendirections compared to the one where the points are repelled the most from the attractor, i.e., the direction corresponding to $\lambda$. As a result, one ends up with the same $\ln{\big[\frac{d_{i}}{d_{0}}\big]}$.         

\item Continuing in this fashion for $N$ steps we calculate the maximum Lyapunov exponent as follows:
\beg 
\lambda(N\Delta t) = \frac{1}{N\Delta t}\sum_{i = 1}^{N}\ln{\bigg[\frac{d_{i}}{d_{0}}\bigg]}.
\label{Cond_Lya_Exp_Uchida}
\en
   
\end{enumerate}

Note, \eref{Lyapunov_Exp_Defn} is an equivalent way to estimate $\lambda$. Similar to the algorithm above, consider $N$ time steps of length $\Delta t (\ll 1)$ each. At each time step we start with two nearby initial conditions for \eref{Mean-Field_1} so that the distance between the two is $d(i\Delta t)\ll 1$. Given that $d(i\Delta t)$ increases exponentially (with slightly different rates $\lambda_{i}$) at each time step, we have
\begin{multline}
d(N\Delta t) \approx e^{\lambda_{N}\Delta t}d((N-1)\Delta t) \approx e^{(\lambda_{N} + \lambda_{N-1})\Delta t}d((N-2)\Delta t)\approx \\ \approx\cdots \approx e^{(\lambda_{N} + \lambda_{N-1} + \cdots \lambda_{1})\Delta t}d(0).    
\end{multline}%
Finally, 
\beg 
\frac{1}{N\Delta t}\ln{\bigg[\frac{d(N\Delta t)}{d(0)}\bigg]} \approx \frac{1}{N}\sum_{i = 1}^{N}\lambda_{i}.
\label{Wiki_Defn_Lya_Exp}
\en
Writing $N\Delta t$ as $t$, we arrive at \eref{Lyapunov_Exp_Defn}.

As explained in \sref{Synch_Manifold}, the conditional Lyapunov exponent $\lambda_{c}$ is a measure of points diverging from an attractor in the transverse directions, $\mathsf n_{1}$ and $\mathsf n_{2}$. Recall that the linearized equations for small transverse deviations ($\Delta\mathsf n_{1}$ and $\Delta\mathsf n_{2}$) from the fiduciary trajectory is given by the same equation \re{Transverse_Coordinates_Eqn} with $\mathsf n_{1,2}$ replaced by $\Delta\mathsf n_{1,2}$. This is so, because \eref{Transverse_Coordinates_Eqn} is linear to begin with. To calculate $\lambda_{c}$, we use the mean-field equations \re{Mean-Field_1} and \eref{Transverse_Coordinates_Eqn} where $n_{1,2}\ra \Delta\mathsf n_{1,2}$. We then essentially follow the same algorithm as the one used for the calculation of $\lambda$, but at each time step we simulate \eref{Mean-Field_1} along with \eref{Transverse_Coordinates_Eqn}. Moreover we define $d = \sqrt{\Delta\mathsf n_{1}^{2} + \Delta\mathsf n_{2}^{2}}$. With the above two modifications we obtain $\lambda_{c}$ from \eref{Cond_Lya_Exp_Uchida}. Moreover, \eref{Wiki_Defn_Lya_Exp} with the stipulated modifications also obtains $\lambda_{c}$. 
    


\begin{thebibliography}{999}
\bibitem{Sup_PF} Here the TSS (one fixed point) gives rise to the NTSS, which is a collection of fixed points lying on a circle parameterized by $\Phi$. If we restrict our choice of initial conditions such that the y-component of the total spin $l_{y0} = 0$, after the bifurcation we are indeed left with two NTSS points that lie diametrically opposite to each other on the above circle, as one expects in a supercritical pitchfork bifurcation. They correspond to $\Phi = \Phi_{0}$ and $\pi + \Phi_{0}$. In the rest frame ($\omega_{A}\neq -\omega_{B}$), the NTSS is a limit cycle with $\Phi = (\omega_{A}+\omega_{B})t/2$, and the TSS $\ra$ NTSS bifurcation is a supercritical Hopf one.
\end{thebibliography}
\end{document}